\documentclass[prd,aps,superscriptaddress,nofootinbib,amsmath,amssymb,9pt]{revtex4}
\usepackage{graphicx}
\usepackage{braket}
\usepackage{nicefrac}
\usepackage{float}%
\usepackage{amsmath}
\usepackage{hyperref}
\usepackage{multirow}
 \usepackage{bbold}
 \usepackage{slashed}
 \usepackage{graphics}
\usepackage{graphicx}
\usepackage{dcolumn}
\usepackage{bm}
\usepackage{multirow}
\usepackage{adjustbox}
\usepackage{textcomp}

\usepackage[usenames, dvipsnames]{color}

\def\slashchar#1{\setbox0=\hbox{$#1$}
   \dimen0=\wd0 \setbox1=\hbox{/} \dimen1=\wd1
   \ifdim\dimen0>\dimen1 \rlap{\hbox to \dimen0{\hfil/\hfil}} #1
   \else  \rlap{\hbox to \dimen1{\hfil$#1$\hfil}} / \fi}

\graphicspath{{./eps/}}

\begin{document}

\title{Weak charged current induced electron and positron scattering off proton at JLab and MAMI energies}

\author{A. \surname{Fatima}}
\affiliation{Department of Physics, Aligarh Muslim University, Aligarh-202 002, India}
\author{M. Sajjad \surname{Athar}}
\email{sajathar@gmail.com}
\affiliation{Department of Physics, Aligarh Muslim University, Aligarh-202 002, India}
\author{S. K. \surname{Singh}}
\affiliation{Department of Physics, Aligarh Muslim University, Aligarh-202 002, India}

\begin{abstract}
The development of next-generation, high-luminosity, and high-precision charged lepton beam facilities at the Thomas Jefferson National Accelerator Facility~(JLab) and the Mainz Microtron~(MAMI) has opened, in recent years, a new frontier in the exploration of weak interaction processes induced by electrons and positrons in the neutral current~(NC) sector, which can also be used to study weak interaction processes induced by charged currents~(CC). In particular, these processes in the intermediate energy regime, spanning from a few hundred MeV to a few GeV, play a crucial role in understanding electroweak dynamics, nucleon structure, and hadronic response functions. 

In this review, we present a comprehensive analysis of some of the electron induced weak CC processes on the free proton due to quasielastic scattering in the strangeness-conserving~($\Delta S=0$) and strangeness-changing~($|\Delta S|=1$) channels. We also study the inelastic production of spin $\frac{1}{2}$ and $\frac{3}{2}$ resonances such as $P_{33}(1232)$, $P_{11}(1440)$, $S_{11}(1535)$, as well as the inelastic production of $\eta$ and $K$ mesons, and the associated particle production of strange particles in electron and positron induced weak CC processes on the free proton. The results for both the differential and total scattering cross sections for these processes are presented and discussed. Furthermore, in the quasielastic scattering sector, the polarization observables of the produced baryons as well as the possibility of using a polarized proton target, where spin asymmetries may be observed are investigated, highlighting their sensitivity to the underlying weak interaction dynamics. Any possible manifestation of the second class currents is explored by computing both the cross sections and the polarization observables with and without the assumption of time reversal invariance, thereby providing a stringent test of  G-  and T- invariance.

This review provides essential theoretical benchmarks for interpreting precision measurements of various observables in the electron and positron interaction with the proton target in the few-GeV energy domain of weak interaction physics in the charged current sector, directly relevant to the present and future experiments at JLab and MAMI. Beyond its phenomenological relevance, the explored kinematic region offers a unique and independent opportunity to constrain the axial vector sector of the weak interaction. In particular,
 it provides a discussion of alternative ways  to determine the axial dipole mass in the quasielastic scattering region, a fundamental parameter that has remained under persistent debate for nearly two decades, as well as the axial vector form factors in the case of $P_{33}(1232)$ resonance excitation in a manner that is free from the uncertainties present in their determination from the study of (anti)neutrino induced weak processes.
\end{abstract}
\maketitle

\section{Introduction}
The discovery of neutrino oscillations established that at least two neutrino mass eigenstates possess non-zero masses, providing one of the clearest indications of physics beyond the Standard Model. The experimenters are making great efforts to determine very precisely the various neutrino oscillation parameters, namely, the three mixing angles $\theta_{12}$, $\theta_{13}$, and $\theta_{23}$; the mass squared differences $\Delta m_{12}^2$, $\Delta m_{13}^2$, and $\Delta m_{23}^2$;  and the CP violating phase $\delta_{CP}$. Moreover, the question of the neutrino mass hierarchy, whether neutrino masses follow the normal hierarchy or the inverted hierarchy, remains unresolved~\cite{ParticleDataGroup:2024cfk, SajjadAthar:2021prg, Denton:2022een}.

To determine these fundamental parameters, current and next-generation neutrino scattering experiments are geared toward high-precision measurements of neutrino-nucleus cross sections. This task is inherently challenging. Most neutrino scattering experiments operate in the few-GeV energy range, employ medium- to heavy-mass nuclear targets, and rely on accelerator-produced neutrino beams with a broad, continuous energy spectra. As a result, the extraction of neutrino-nucleus cross sections from these experiments is subject to significant systematic uncertainties, whereas the present precision goals demand control of systematics at the level of only a few percent~\cite{Ruso:2022qes, Alvarez-Ruso:2022ctb, Dolan:2026nlr}.
In theoretical simulations of the neutrino-nucleus interaction cross section, the neutrino-nucleon scattering cross section serves as a fundamental input. Various nuclear medium effects and final-state interaction effects are then incorporated to account for the modifications in the nuclear cross sections arising due to the complexity of the nuclear environment.  Recently, there has been a strong and growing emphasis on the study of electron-proton, electron-deuteron, and the electron-nucleus scattering experiments across a range of nuclear targets to understand the weak interaction properties and probe the nuclear medium effects, in the few GeV energy region~\cite{Ankowski:2022thw, Mosel:2016cwa, Benhar:2014qaa, Amaro:2019zos}. These studies are particularly crucial, as they offer a powerful means to test and constrain nuclear models, while also providing benchmarks for the validation of event generators employed in predicting the neutrino interaction rates.

Moreover, any imprecision in the knowledge of the basic weak interaction vertex at the nucleon level propagates directly into the nuclear cross sections.
Beyond this practical necessity, there is a compelling intrinsic motivation to understand the weak interaction hadronic vertex in the charged current neutrino and antineutrino scattering off free nucleons. This vertex encodes the underlying structure of the weak interaction and is described in terms of the weak nucleon form factors in the vector and axial vector sectors, for describing the matrix elements for the vector--axial vector~($V-A$) current, making their precise determination essential for both modelling the nuclear cross sections and broader understanding of the weak interaction.
The weak nucleon vector form factors~($f_1(Q^2)$ and $f_2(Q^2)$) are parameterized in terms of the nucleon electromagnetic form factors using isospin symmetry and the conserved vector current~(CVC) hypothesis.
The axial vector contribution is parameterized by the axial vector~($g_1(Q^2)$) and the induced pseudoscalar~($g_3(Q^2)$) form factors of the nucleon. Generally, the axial vector form factor is assumed to have a  dipole form, i.e. $g_1(Q^2)= \frac{g_1(0)}{(1 + Q^2/M_A^2)^2}$, where $M_A$ is a free parameter, known as the axial dipole mass, which is determined from the (anti)neutrino-nucleus scattering experiments. 
While $g_1(Q^2)$, at $Q^2=0$ i.e. $g_{1}(0)$ is the Gamow-Teller strength of the $\beta$ decays, obtained from the high precision data on the $\beta$ decays of nucleons and nuclei. The pseudoscalar form factor $g_{3}(Q^2)$ is determined in terms of $g_{1}(Q^2)$ using the partially conserved axial current~(PCAC) hypothesis and the pion pole dominance of divergence of axial vector current. A precise determination of $M_A$ suffers from  uncertainties arising due to the use of nuclear targets and accelerator produced neutrino beams, which do not have a well defined energy but have a continuous energy spectrum. These uncertainties can be significant as recent values of $M_{A}$ reported from the neutrino-nucleus experiments in the intermediate energy region are 25\% larger than the world average value of $M_{A}=1.026 \pm 0.021$~GeV~\cite{Bernard:2001rs}. Moreover, different neutrino experiments, in recent years, using nuclear targets have not been fully consistent in determining the value of $M_A$, especially when comparing the values determined from low energy data sets with those obtained at higher energies~\cite{SajjadAthar:2022pjt, Benhar:2009wi, NuSTEC:2017hzk}. Clearly, there is a pressing need to determine $M_{A}$ with better precision.

With the development of high luminosity and high precision electron beam facilities at the Thomas Jefferson National Accelerator Laboratory~(JLab) in the USA and the Mainz Microtron~(MAMI) in Germany, there has been growing interest in studying  weak interaction processes induced by the electron scattering off nucleons in the intermediate energy range, from a few hundred MeV to a few GeV~\cite{Ankowski:2022thw, Amaro:2019zos}.  These laboratories have already studied the weak interaction processes in the neutral current~(NC) sector and are particularly well suited for exploring the charged current~(CC) induced reactions in a kinematic region that complements traditional neutrino experiments. Moreover, these laboratories are also planning to develop positron beams~\cite{Afanasev:2019xmr}. Thus, the ongoing and proposed electron and positron beam programs at JLab~\cite{JeffersonLabSoLID:2022iod, Arrington:2021alx, Accardi:2023chb}, especially those focused on making precision electroweak measurements, provide an exceptionally powerful platform to carry out such studies with unprecedented accuracy. The combination of well-controlled kinematics, high beam intensity, and minimal nuclear effects achieved by using hydrogen and deuterium targets enables a level of precision that is difficult to achieve in neutrino sector.

A particularly striking illustration of this capability is provided by the proposed experimental configuration at JLab for the measurement of the charged current weak reaction induced by electrons i.e. $e^- + p \rightarrow \nu_e + n$~\cite{Averett}, which envisages the use of a 2.2~GeV electron beam with an intensity of $100 \mu\text{A}$ (corresponding to $\sim 6.3 \times 10^{14}$ electrons per second) incident on a 10~cm long liquid hydrogen target containing approximately $4.2 \times 10^{23}$ protons, over an exposure time of 500 hours. Such an arrangement would yield events of the order of $10^6$ despite the small weak interaction cross section~($\sim 10^{-40}~ \text{cm}^2$) in this energy region. Furthermore, there are also proposals~\cite{JLab:axial} to employ a 1.1~GeV electron beam with even higher luminosity, specifically aimed at achieving superior control over both systematic and statistical uncertainties. Notably, these uncertainties are expected to be smaller than those associated with the 2.2~GeV electron beam, thereby making such configurations particularly advantageous for making precision measurements. The above developments clearly underscore the extraordinary statistical reach of these facilities, making them ideally suited for precision studies of  the weak interaction processes. A similar experiment is also under consideration at JLab, which would use polarized positron beam on  deuterium target to
extract the axial vector form factor~\cite{Androic}. Such measurements using the electron and positron beams available in the future at JLab will be particularly valuable in determining the axial vector form factor $g_1(Q^2)$. By studying the quasielastic weak interaction induced processes in the same $Q^2$ region with these beams, one may be able to address some of the pressing issues in the neutrino oscillation sector. 
 
 The experimental developments at JLab and MAMI have already opened the door to a wide range of precision experiments involving the scattering of unpolarized and polarized electrons from nucleon and nuclear targets. These setups have already felicitated researchers to study the existence of some  phenomena in the weak interaction processes, predicted by the standard model~(SM) of the electroweak interaction such as the existence of NC interaction in the charged lepton sector. Another example is that of  the measurement of parity violating electron asymmetries in the scattering of polarized electron from nucleon and nuclear targets that arise when the contribution from weak neutral currents through the $Z$ exchange interferes with the parity conserving contribution from the electromagnetic scattering due to the photon exchange. Indeed, the pioneering experiments in 1978 by the E-122 collaboration~\cite{Prescott:1978tm, Prescott:1979dh} at the Stanford Linear Accelerator Center~(SLAC) on the measurement of the parity violating electron asymmetry~(PVES) off a deuteron target in the energy region of the deep inelastic  scattering played a leading role in validating the SM of electroweak interaction.

Since those early milestones, several experimental works have focused on the measurement of PVES across a wide spectrum of electroweak processes using  polarized electron beams in the kinematic region of elastic, inelastic, and deep inelastic scattering, from nucleon and nuclear targets~\cite{Beck:2001yx, Armstrong:2012bi, Maas:2003xp, Wang:2014guo, Beck:2001dz, Adderley:2022uql, Wang:2013kkc, QWeak:2019kdt, Cahn:1977uu}. These measurements have been carried out over relevant energy ranges at several leading accelerator facilities, including the Mainz Microtron~\cite{A4:2004gdl, Maas:2003xp} in Germany, the MIT-Bates Linear Accelerator Center~\cite{SAMPLE:1999mku, Beise:2004py, BatesFPP:1997rpw}, SLAC~\cite{SLACE158:2003onx} and JLab~\cite{HAPPEX:1998epc, Qweak:2018tjf, G0:2011rpu} in the United States. These experiments have
successfully demonstrated the utility of using the polarized electron beams to study weak interaction processes. In order to reduce the systematic 
errors and improve the precision of PVES measurements, future experiments in the neutral current sector have already been planned using electron beams of  
even higher luminosities of the order $10^{39}\;\text{cm}^{-2}\;\text{sec}^{-1}$, in the low energy region of a few hundred MeV at the Mainz Energy-recovering Superconducting Accelerator~(MESA)~\cite{Cadeddu:2024baq, Schlimme:2024eky}.  At higher energies, the Solenoidal Large Intensity Device~(SoLID) at JLab~\cite{JeffersonLabSoLID:2022iod, Arrington:2021alx, Accardi:2023chb} is planned to explore PVES in the few GeV regime using electron beams of high luminosity and acceptance. Theoretically, the PVES processes have been analyzed using the effective Lagrangian for the NC interaction predicted by the SM~\cite{Souder:2015mlu, Beck:2001yx, Armstrong:2012bi}. 

Moreover, the development of a polarized positron beam at JLab is expected to open new areas for studying both the electromagnetic and weak interaction processes~\cite{Arrington:2021alx}. In particular, comparing scattering processes induced by electrons and positrons allows for a clean separation of different electroweak contributions, including vector and axial vector couplings, through changes in the interference structure of the amplitudes. This capability is especially significant in the context of processes such as  $e^- - e^+$   annihilation, where axial vector currents play a central role. The interference between photon mediated~(purely vector) and $Z$-boson mediated~(vector and axial vector) amplitudes gives rise to observables such as forward-backward asymmetries, which are directly sensitive to the axial vector couplings of the fermions. The combined program of high luminosity polarized electron and positron scattering running presently at facilities such as MAMI and JLab, marks a major advancement in our ability to further explore processes in the weak interaction induced neutral current sector.

Another class of rare weak interaction processes in the NC sector with $\Delta S=1$ currents, the flavor-changing neutral current~(FCNC) processes have been observed through the decays of strange $K$-mesons and hyperons, such as $K \rightarrow \mu^+ \mu^-$, $K \rightarrow \pi \nu \bar\nu$ and $\Sigma^+ \rightarrow p e^+ e^-$, etc. These processes are highly suppressed in the SM at the tree level but can occur in higher orders under the combined effect of electromagnetic and weak interactions. The basic process which involves $s \rightarrow d$ quark transition can also induce electron-proton scattering leading to the strange particle production through the reaction $e^- p \rightarrow e^- \Sigma^+$. The feasibility of studying such reactions at MAMI electron accelerator with high luminosity electron beams was theoretically explored by the Mainz and MIT groups~\cite{Drechsel:1997jm, Jin:1996vu} almost 30 years ago, but the scattering cross sections were found to be very small as compared to the $e^- p \rightarrow e^- p$ scattering, making its measurements quite unlikely. However, in light of the recent observation of $\Sigma^+ \rightarrow p \mu^+ \mu^-$ decays at  LHCb~\cite{LHCb:2025evf, LHCb:2018hne}, earlier evidence for this decay from HyperCP~\cite{HyperCP:2005mvo}, and the hope of observing $\Sigma^+ \rightarrow p e^+ e^-$ decays in the near future at BESIII~\cite{Li:2016tlt} and LHCb~\cite{LHCb:2025evf, LHCb:2018hne}, it will be very interesting to study theoretically the cross sections for the weak strange particle production induced by the FCNC through the reaction $ e^- p \rightarrow e^- \Sigma^+$ using the monoenergetic electron beams of very high luminosities available at JLab and MAMI laboratories. The early theoretical calculations of Drechsel and Giannini~\cite{Drechsel:1997jm}, and Jin and Jaffe~\cite{Jin:1996vu} need to be updated by using the transition form factors for the $s \rightarrow d$ transition employed in the recent theoretical studies of $\Sigma^+ \rightarrow p l^+ l^-~(l=e,\mu)$ decays by making reasonable modifications for their high $q^2$-behaviour~\cite{Goudzovski:2022vbt, Erben:2022igb, He:2026chn, Roy:2024hqg, PANDA:2020zwv, He:2018uey, He:2018yzu}. However, these reactions in the NC sector of $\Delta S=1$ processes leading to FCNC as well as the extensive work on NC reactions in the  $\Delta S=0$ sector leading to the PVES effects, in the electron scattering mentioned earlier will not be discussed further in the present review.

With the increase in center-of-mass energy~($W$) and the four-momentum transfer squared~($Q^2$), particularly in the region of high $Q^2~(>4$~GeV$^2$) and $W$~($> 3$ GeV), the dominant contribution to the scattering cross sections arises from the deep inelastic scattering~(DIS) regime. In this region, the electron-proton scattering is described in terms of the incoherent scattering of charged leptons off the partonic constituents of the nucleon. Here, perturbative QCD~(pQCD) provides a systematic and well established theoretical framework, wherein the cross sections are expressed in terms of structure functions built from parton distribution functions (PDFs). This description has achieved remarkable predictive success, firmly establishing DIS as the underlying mechanism of high-energy lepton-nucleon scattering. In contrast, the region of moderate $Q^2$ and $W$ constitutes a nontrivial transition region between the resonance production and DIS. This kinematic window is particularly rich and challenging, as nonperturbative QCD effects become indispensable. A comprehensive treatment of this transition region remains an open problem~\cite{SajjadAthar:2020nvy}. Accordingly, it lies beyond the scope of the present review, which is focused on the quasielastic sector, excitation of some low-lying resonances, and inelastic channels leading to $K$ and $\eta$ meson production.

The present review focuses on the status of theoretical work in light of experimental proposals~\cite{Averett, Androic} for charged current induced weak scattering processes of electrons and positrons from a proton target, relevant for future experiments. The unavailability of data is mainly due to the 
small cross sections of pure weak interaction processes compared to those of  electromagnetic processes induced by the electrons. However, by making a  suitable choice
of those charged current induced weak processes in which the particles produced in the final state can be different and distinguished from the particles
produced in the electromagnetic processes, it may be possible to study the weak interaction processes induced by the electrons in the charged current sector. This distinction is not 
feasible in the case of weak electron-proton scattering processes induced by pure weak neutral current, where the particles produced in the final state are same as those produced in  
electromagnetic interaction. However, the weak NC production is highly suppressed by a factor $\Big(\frac{G_F M^2}{\alpha}\Big)^2\approx 10^{-6}$ around $Q^2\sim1$~GeV$^2$, where $G_F$ is the Fermi coupling constant, $M$ is the nucleon mass, and $\alpha$ is the fine structure constant. 

Recent advances made in the development of the unpolarized and polarized electron beams with very high luminosity in the low- as well as the high- energy regimes alongside the 
development of polarized targets, specifically at MAMI and JLab, open the possibility of measuring very small cross sections of the order of $10^{-39}-10^{-40}$~cm$^{2}$ and facilitate the study of the weak
charged current processes induced by the electrons and positrons.  This potential has motivated numerous theoretical calculations to be made for the various quasielastic and inelastic
processes induced by the electrons in CC weak sector~\cite{Fatima:2018gjy, Akbar:2017qsf, Alvarez-Ruso:1998ais, Mintz:2001jc, Mintz:2002cj, Mintz:2004eu, Fatima:2024hlu, Hwang:1987sd, Hwang:1988fp}. Moreover, some of the calculations have  been extended to explore the CC weak interaction processes 
induced by positrons, in view of the advances made at JLab in developing both polarized and unpolarized positron beams~\cite{Alvarez-Ruso:1998ais, Fatima:2024hlu}.

It is worth noting that the theoretical studies of the electron-induced weak interaction processes in the charged current sector are not entirely new. Early investigation of these processes date back several decades but were largely confined to very low-energy processes, such as weak electron capture by the free nucleons and nucleons bound in nuclei, $e^- +p \to \nu_e + n$. This process is of profound importance in astrophysical contexts, particularly in understanding the stellar energy production. An early hypothesis proposed by Eddington in 1926~\cite{Eddington} attributed stellar energy generation to the complete annihilation of matter into radiation. Although this idea was later refined, it paved the way for recognizing that energy release does occur partially through weak CC interaction processes such as weak electron capture from nucleons and nuclei in stellar environments~\cite{Raffelt:1996wa}.

Subsequent theoretical work, notably by Bethe and Bacher~\cite{Bethe:1936zz} within the framework of Fermi’s theory of beta decay, laid the foundation for quantitative studies of these processes. Later contributions by Fowler, Raffelt and others~\cite{Burbidge:1957vc, Raffelt:1996wa, Wagoner:1966pv}, further expanded the scope to include a wide range of the weak interaction processes involving electrons. Many of these processes produce neutrinos that escape stellar interiors, carrying away significant amounts of energy, particularly in the pre-supernova phase. Most of these mechanisms arise from the interaction of electrons
with nucleons and nuclei induced by weak charged currents and are studied in the very low energy region of a few eV relevant to astrophysical applications.

In contrast, the exploration of such processes at accelerator energies remains largely uncharted territory. This dichotomy underscores both the challenge and the immense scientific potential of studying electron induced charged current weak interaction in laboratory settings, which could provide further insights into our understanding of weak interaction dynamics, nucleon structure, and neutrino physics.

Experimentally, the study of weak charged current processes in the high energy region of a few hundreds of MeV and a few GeV started with the availability of the (anti)neutrino beams at particle accelerators
at CERN, ANL and BNL using light to heavy nuclear targets. Around the same time, the idea of performing such experiments with electron beams was discussed by Schwartz, Schiff and
many others~\cite{LlewellynSmith:1971uhs, Winter:1992iyh}. Electron beams have the advantage over the (anti)neutrino beams of possessing high luminosity and monochromatic energy with a well
defined direction; however, no further progress was made in pursuing this idea experimentally at that time. The theoretical calculations based on this idea relevant for high energies were performed earlier by Fearing et al.~\cite{Fearing:1969nr} for an electron scattering process in the 
charged current sector induced by $\Delta S=1$ currents, i.e., $e^-+p \to \Lambda + \nu_e$. Since then, many calculations of the quasielastic and 
inelastic reactions induced by electrons and positrons via weak charged currents in the $\Delta S=0$ and $\Delta S=1$ sectors have been done and the feasibility
of doing such experiments at MAMI and JLab has been discussed in the literature~\cite{Hwang:1987sd, Hwang:1988fp, Alvarez-Ruso:1998ais, Fatima:2026mac, Akbar:2017qsf, Fatima:2018gjy, Mintz:2001jc, Mintz:2002cj, Mintz:2004eu}. 

Recently, { Klest~\cite{Klest:2025bfl}, and independently Yang and Kumar~\cite{Yang:2026vuf}, have demonstrated that at very high energies corresponding to the center of energy $\sqrt{s}=140$~GeV, relevant to the proposed Electron Ion Collider~(EIC) at BNL~\cite{Deshpande:2005wd}, and Electron Ion Collider in China~(EIcC) experiments, the electron scattering on a longitudinally polarized proton target induced by the weak charged currents offers a viable avenue to constrain the nucleon axial-vector form factor. They emphasize that target spin asymmetries in the low-$Q^2$ region, where the axial coupling overwhelmingly dominates, constitute a sensitive probe.

Beyond their role in studying nucleon structure, the weak interaction processes induced by  electrons and positrons also provide an excellent testing ground for various symmetry properties such as the T-invariance and the G-invariance. These symmetries are used to determine the weak nucleon-nucleon~($N-N$), nucleon-hyperon~($N-Y$), and nucleon-resonance~($N-R$) transition form factors. They also enable direct validation of the CVC hypothesis by comparing weak and electromagnetic form factors, as well as the PCAC hypothesis, and SU(3) symmetry in the case of the production of strange particles. 

In view of these developments, we present in this paper a review of the current status of the theoretical work and discuss the scope for future work in studying weak interaction processes induced by electrons and positrons in the charged current sector. Specifically, we present results for the total and differential scattering cross sections for the various quasielastic and inelastic processes listed in Tables-\ref{tab1:processes} and \ref{tab2:processes}, assuming electron and positron beams incident on unpolarized and polarized proton targets.

Numerical results are presented for quasielastic processes leading to a neutrino and an unpolarized or polarized neutron, $\Lambda$, or $\Sigma$ hyperon in the final state. A comprehensive set of spin observables, including the longitudinal and perpendicular spin asymmetries of a polarized proton target, as well as the longitudinal, perpendicular, and transverse components of the final baryon polarization in the quasielastic processes listed in Table-\ref{tab1:processes}, has also been investigated. These observables provide independent and enhanced sensitivity to the underlying weak nucleon form factors. While most of our analyses assume time-reversal ($T$) invariance, we also explore possible $T$-violating effects, which correspond to physics beyond the standard model. These effects can lead to a non-zero transverse polarization component of the final baryon, perpendicular to the reaction plane in case, if T-invariance is violated in weak interactions.

\begin{table}\centering
 \begin{tabular*}{120mm}{@{\extracolsep{\fill}} c  c c  }\hline \hline
 Process & Observable & Section\\ \hline
$e^- + p \longrightarrow \nu_e + n$ & &  \ref{QE:n} \\
 $e^- + p \longrightarrow \nu_e + \Lambda$ & Cross section &  \ref{QE:lambda} \\
 $e^- + p \longrightarrow \nu_e + \Sigma^0$ & &  \ref{QE:sigma} \\ \hline
 $e^- + \vec{p} \longrightarrow \nu_e + n $ & &  \ref{ALAP:n} \\
 $e^- + \vec{p} \longrightarrow \nu_e + \Lambda$ & Spin asymmetry of the target proton &  \ref{AlAp:lambda} \\
 $e^- + \vec{p} \longrightarrow \nu_e + \Sigma^0$ & &  \ref{AlAp:sigma} \\ \hline
 $e^- + p \longrightarrow \nu_e + \vec{n} $ & &  \ref{PlPp:n} \\
 $e^- + p \longrightarrow \nu_e + \vec{\Lambda}$ & Polarization of the final baryon &  \ref{polarization:lambda} \\
 $e^- + p \longrightarrow \nu_e + \vec{\Sigma}^0$ & &  \ref{polarization:sigma} \\ \hline \hline
 \end{tabular*}
\caption{Processes and the observables discussed in different sections for the $\Delta S=0$ and $\Delta S=1$ quasielastic processes. The cross sections are calculated for the unpolarized electron beam and unpolarized proton target, while the spin asymmetries are calculated for the polarized target proton, and the polarization observables are calculated for the final polarized baryon.}
\label{tab1:processes}
\end{table}

\begin{table}\centering
 \begin{tabular*}{120mm}{@{\extracolsep{\fill}} c  c c  }\hline \hline
 Process & Observable & Section\\ \hline
  $e^- + p \longrightarrow \bar{\nu}_e + \Delta^{0} (1232)$ & Cross section &  \ref{sec:Delta}\\
 $e^+ + p \longrightarrow \bar{\nu}_e + \Delta^{++} (1232)$ & &  \ref{sec:Delta} \\ \hline
  $e^- + p \longrightarrow \nu_e + P_{11}(1440)$ & Cross section &  \ref{sec:Nstar} \\
 $e^- + p \longrightarrow \nu_e + S_{11}(1535)$ & &  \ref{sec:Nstar} \\ \hline
 $e^- + p \longrightarrow \nu_e + p + \eta$ & Cross section &  \ref{weak:eta} \\ \hline
 $e^- + p \longrightarrow \nu_e + \Lambda + K^0$ & Cross section &  \ref{weak:associated} \\ \hline
  $e^- + p \longrightarrow \nu_e + n + \bar{K}^0$ & & \ref{sec2:kaon} \\
  $e^- + p \longrightarrow \nu_e + p + K^-$ & Cross section &  \ref{sec2:kaon} \\
  $e^+ + p \longrightarrow \bar{\nu}_e + p + K^+$ & &  \ref{sec2:kaon} \\ \hline \hline
 \end{tabular*}
\caption{Processes and the observables discussed  in different sections for the production of spin $\frac{1}{2}$ and $\frac{3}{2}$ resonant states like $P_{33}(1232)$, $P_{11}(1440)$, and $S_{11}(1535)$ resonances, and some inelastic processes such as $\eta$ production, and $K$ production in the $\Delta S=0$ and $\Delta S=1$ sectors.}
\label{tab2:processes}
\end{table}

Furthermore, we also present results for the total and differential scattering cross sections for inelastic electron and positron scattering off a proton target leading to the resonant states such as $P_{33}(1232)$~(a $\Delta$ resonant state), $P_{11}(1440)$ and $S_{11}(1535)$~(positive and negative parity spin $\frac{1}{2}$ nucleon resonances), listed in Table-\ref{tab2:processes}. These resonant states then subsequently decay into a nucleon/hyperon and a single meson or multiple mesons in the final state. The results are also presented for electron induced $\eta$ production, associated particle production of strange particles off a proton target, and  electron and positron induced single kaon production off a proton target as outlined in Table-\ref{tab2:processes}. 

In Section~\ref{sec2:QE}, we describe the quasielastic weak charged current scattering induced by electrons on the unpolarized and polarized proton target for the $\Delta S=0$ and $|\Delta S|=1$ processes and obtain the results for the cross section, spin asymmetries of the initial proton, and polarization observables of the final baryon~($n,~\Lambda$, and $\Sigma^0$). 
 In Section~\ref{sec2:Delta}, we describe the weak charged current processes
induced by electrons and positrons leading to the resonance excitations 
in the $\Delta S=0$ sector and discuss the feasibility of measuring the cross 
sections for these processes. In Section~\ref{sec2:eta}, we describe the inelastic production of $\eta$ and $K$ mesons as well as the associated particle production of strange particles in weak charged current processes
induced by electrons and positrons. Finally, Section~\ref{sec:summary} provides a summary of the work.

\section{Quasielastic scattering: Cross section, spin asymmetries, and polarization observables}\label{sec2:QE}
\subsection{Cross section}
 
The general expression of the differential scattering cross section corresponding to the processes 
\begin{eqnarray}\label{nuc-rec}
 e^- (k) + p (p) &\longrightarrow& \nu_e (k^\prime) + n (p^{\prime}),
 \end{eqnarray}
 induced by the $\Delta S=0$ weak charged current, and
 \begin{eqnarray} \label{hyp-rec}
 e^- (k) + p (p) &\longrightarrow& \nu_e (k^\prime) + Y (p^{\prime}), ~~~~~ Y = \Lambda,\Sigma^0,
 \end{eqnarray}
induced by the $\Delta S=1$ weak charged current, 
depicted
in Fig.~\ref{fyn_hyp}, in the rest frame of the initial proton, is written as:
 \begin{eqnarray}
 \label{crosv.eq}
 d\sigma&=&\frac{1}{(2\pi)^2}\frac{1}{4ME_e }\delta^4(k+p-k^\prime-p^\prime) \frac{d^3k^\prime}{2E_{k^\prime}}  
 \frac{d^3p^\prime}{2E_{p^\prime}} \overline{\sum} \sum |{\cal{M}}|^2,
 \end{eqnarray}
 where the quantities in the brackets (Eqs.~(\ref{nuc-rec}) and (\ref{hyp-rec})) represent the four momenta of the corresponding 
 particles, $E_e$ is the electron energy and $M$ is the proton mass. ${\cal M}$ is the matrix element for the quasielastic processes in Eqs.~(\ref{nuc-rec}) and (\ref{hyp-rec}) and is defined explicitly in Section~\ref{sec:FF:QE}.

 \begin{figure}
 \begin{center}
    \includegraphics[height=3cm,width=6cm]{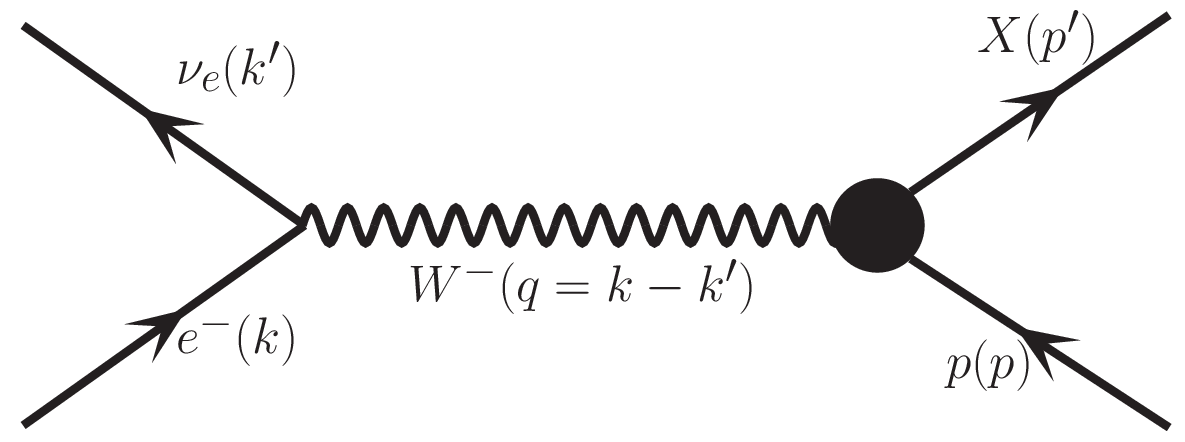}
  \caption{Feynman diagram  for the process $ e^-(k) + p(p) \rightarrow \nu_e(k^\prime) + X(p^{\prime})$; $X = n, \Lambda, \Sigma^0$. 
  The quantities in the bracket represent 
  four momenta of the corresponding particles.}\label{fyn_hyp}
   \end{center}
 \end{figure}
 
 The transition matrix element square $|{\cal M}|^2$, averaged over the initial spin states and summed over the final spin states, is expressed as:
\begin{equation}\label{matrix}
  \overline{\sum} \sum |{\cal{M}}|^2 = \frac{G_F^2 a^2}{2} \cal{L}^{\alpha \beta} \cal{J}_{\alpha \beta},
\end{equation}
where $G_F$ is the Fermi coupling constant, $a = \cos \theta_c$ for $\Delta S = 0$ processes, $a = \sin \theta_c$ for 
 $\Delta S = 1$ processes, and $\theta_c~(=13.1^\circ)$ is the Cabibbo mixing angle.
 
The leptonic and the hadronic tensors are given by
\begin{eqnarray}\label{L}
\cal{L}^{\alpha \beta} &=& \frac{1}{2}\mathrm{Tr}\left[\gamma^{\alpha}(1-\gamma_{5})\Lambda(k)
\gamma^{\beta}(1-\gamma_{5})\Lambda(k^\prime)~\right], \\ 
\label{J}
\cal{J}_{\alpha \beta} &=& \frac{1}{2} \mathrm{Tr}\left[\Lambda(\not{p^\prime}) J_{\alpha}
  \Lambda(\not{p}) \tilde{J}_{\beta} \right], 
\end{eqnarray} 
where the factor $\frac{1}{2}$ in the above expressions arises due to the spin averaging over the initial state particles~(if both the beam and target are unpolarized),  $\tilde{J}_{\beta} =\gamma^0 
J^{\dagger}_{\beta} \gamma^0$ with $J_{\beta}$ being the weak hadronic current. 

In Eqs.~(\ref{L}) and (\ref{J}), $\Lambda(P)=(P\!\!\!/+M_P)$ is the projection operator of an unpolarized particle with momentum $P$ and mass $M_{P}$. To take into account the spin polarization of a particle, the projection operator $\Lambda(P)$ in  Eqs.~(\ref{L}) and (\ref{J}) is replaced by the spin projection operator $\Lambda(P,s_P)$, defined as
\begin{equation}
 \Lambda (P,s_P) = (P\!\!\!/+M_P) \left(\frac{1+\gamma_5 \slashed{s}_P}{2} \right),
\end{equation}
where $s_P$ is the spin 4-vector of the particle with momentum $P$. 
Moreover, for the polarized initial particle scattering, one has to drop the factor of $\frac{1}{2}$ in Eqs.~(\ref{L}) and (\ref{J}) and use $\Lambda (P,s_P)$ instead of $\Lambda(P)$.

The $Q^2$ distribution for the cross section is written as
\begin{equation}\label{dsig}
 \frac{d\sigma}{dQ^2}=\frac{G_F^2 ~a^2}{16 \pi M^2 {E_e}^2} N(E_e,Q^2),
\end{equation}
where the expression of $N(E_e,Q^2) = \cal{L}_{\alpha \beta} \cal{J}^{\alpha \beta}$ is given in Appendix-I.

\subsubsection{Matrix element and form factors}\label{sec:FF:QE}
The transition matrix element for the quasielastic processes given in Eqs.~(\ref{nuc-rec}) and (\ref{hyp-rec}), is written as
 \begin{eqnarray}\label{matrix}
  {\cal{M}} = \frac{G_F}{\sqrt{2}} a~ l^\mu {{J}}_\mu.
 \end{eqnarray}
 The leptonic current $l^\mu$ is given by 
 
\begin{equation}
 l^\mu = \bar{u} (k^\prime) \gamma^\mu (1-\gamma_5) u (k),
\end{equation}
and the hadronic current ${J}_\mu$ is expressed in terms of the vector~($V_{\mu}$) and the axial vector~($A_{\mu}$) currents as~\cite{LlewellynSmith:1971uhs}:
\begin{equation}
 {{J}}_\mu =  \bar{u} (p^\prime) (V_\mu - A_\mu) u (p)
\end{equation}
with
\begin{eqnarray}\label{vx}
 V_\mu &=& \gamma_\mu f_1^{pX}(Q^2)+i\sigma_{\mu\nu} \frac{q^\nu}{M+M^\prime} f_2^{pX}(Q^2)
  + \frac{2 ~q_\mu}{M+M^\prime} f_3^{pX}(Q^2),\\
  \label{vy}
  A_\mu &=&  \gamma_\mu \gamma_5 g_1^{pX}(Q^2) + i \sigma_{\mu\nu}\gamma_5 \frac{q^\nu}{M+M^\prime} g_2^{pX}(Q^2)  
   + \frac{2 ~q_\mu} {M+M^\prime} g_3^{pX}(Q^2) \gamma_5 ,
\end{eqnarray}
where X stands for a nucleon $N (=n)$ or a hyperon $Y(=\Lambda, \Sigma^0)$, $M$ and $M^\prime$ are the masses of the initial and 
final baryons with $M^\prime = M$ for the nucleon and $M^\prime = M_\Lambda$ or $M_\Sigma$ for the hyperons. 
$q_\mu (= k_\mu - k_\mu^\prime = p_\mu^\prime -p_\mu)$ is the four momentum transfer with $Q^2 = - q^2, Q^2 \ge 0$. $f_{1}^{pX}(Q^2)$, 
$f_{2}^{pX}(Q^2)$ and $f_{3}^{pX}(Q^2)$ are the vector, weak magnetic and scalar form factors, and $g_{1}^{pX}(Q^2)$, 
$g_{2}^{pX}(Q^2)$ and $g_{3}^{pX}(Q^2)$ are the axial, weak electric and pseudoscalar form factors, respectively.

 The six form factors $f_i^{pX} (Q^2)$ and $g_i^{pX} (Q^2) ~ (i=1-3)$ are determined using the following assumptions about the 
 symmetry properties of the vector and the axial vector currents in the weak interaction~\cite{Pais:1971er,LlewellynSmith:1971uhs, 
 Marshak, Cabibbo:2003cu, Block:1964gj, Cabibbo:1965zza}:
 
 \begin{itemize}
  \item [(i)] The requirement of T-invariance implies that all the vector $f_i^{pX} (Q^2) (i=1-3)$ and the axial vector 
  $g_i^{pX} (Q^2) (i=1-3)$ form factors are real.
  
   \item[(ii)] The hypothesis of CVC and SU(3) symmetry implies that $f_3^{pX} (Q^2) = 0$.
  
  \item[(iii)] In the case of $\Delta S=0$ currents, where $X=n$, the assumption of the weak hadronic currents belonging to the isotriplet implying charge symmetry requires that 
  $f_1^{pX} (Q^2)$, $f_2^{pX} (Q^2)$, $g_1^{pX} (Q^2)$ and $g_3^{pX} (Q^2)$ be real, while $f_3^{pX} (Q^2)$ and $g_2^{pX} (Q^2)$ be 
  imaginary.
  
  \item[(iv)] The requirement of G-invariance dictates that 
  the form factors $f_3^{pX} (Q^2)=0$ and $g_2^{pX} (Q^2)=0$, as they correspond to second class currents or irregular 
  currents and possess the wrong G-parity~\cite{Weinberg:1958ut}. This classification based on G-parity is strictly true for the isotriplet of $\Delta S = 0$ currents assuming SU(2) symmetry, but can be extended to the octet of
  vector and axial vector currents in the limit of exact SU(3) symmetry, in which case, it is applicable to $\Delta S = 1$ 
  currents as well~\cite{Cabibbo:2003cu, Cabibbo:1965zza}.   While the 
  implications of G-invariance are strictly valid for $\Delta S=0$ currents under SU(2) symmetry, they are only approximately applicable to the octet 
  currents due to SU(3) flavor symmetry breaking.

  \item[(v)] The PCAC hypothesis implies that the induced pseudoscalar form factor $g_3^{pX} (Q^2)$ can be related to the axial vector form factor $g_1^{pX} (Q^2)$ in terms of the pion pole dominance of the divergence of the axial vector current, in the case of $\Delta S = 0$ current with $X=n$ or via a kaon pole dominance in the case of $\Delta S = 1$ current with $X=Y$~\cite{Goldberger:1958vp, 
  Marshak, Nambu:1960xd}. However, we set $ g_3^{pX} (Q^2) = 0$ in our numerical calculations because its contribution is proportional 
  to the lepton mass, making it quite negligible for reactions involving electron and electron type neutrino.  
  
  \item[(vi)] It should be noted that while the existence of a purely imaginary $g_2^{pX} (Q^2)$ form factor implies a violation of T-invariance, 
  a purely real $g_2^{pX} (Q^2)$ is consistent with T-invariance while giving rise to G-violation along with the violation of charge symmetry. A complex value of $g_2^{pX} (Q^2)$, therefore, implies the violation of both G-invariance and 
  charge symmetry, and also leads to the violation of T-invariance.  It should be mentioned that while the experimental limits on the real values of
  $g_2^{pn} (Q^2)$ are quite small in the $\Delta S = 0$ sector~\cite{Ahrens:1988rr, Wilkinson:2000gx, Baker:1981su, Belikov:1983kg}, they are not as stringent in the case of 
  $g_2^{pX} (Q^2)$ corresponding to the $\Delta S=1$ semileptonic hyperon decays~\cite{Cabibbo:2003cu, Holstein}.
  
 \end{itemize}
 
 We now elaborate the implications of assuming SU(3) symmetry in determining the vector and axial vector $N-Y$ transition form factors $f_i^{pX} (Q^2) (i=1,2)$ and 
 $g_i^{pX} (Q^2)(i=1,2)$, respectively, for $X=Y$ in terms of the nucleon form factors:
\begin{enumerate}
\item[a)] The weak vector~($V_\mu$) and axial vector~($A_\mu$) currents corresponding to the $\Delta S=0$ and $\Delta S = 1$ 
hadronic currents, whose matrix elements are defined between the states $|N\rangle$ and $|X\rangle$ are assumed to belong to the octet representation of the SU(3), and are 
defined as
\begin{eqnarray}\label{su3}
V^\mu_i&=&\bar{q}F_i\gamma^\mu q\nonumber\\
A^\mu_i&=&\bar{q}F_i\gamma^\mu\gamma^5 q,
\end{eqnarray}
 where $F_i=\frac{\lambda_i}{2}$($i=1-8$) are the generators of flavor SU(3) and $\lambda_i$s are the well known 
 Gell-Mann matrices.
 The generators of the SU(3) group $F_i$ obey the following commutation and anticommutation algebra
 \begin{eqnarray}\label{algebra}
[F_i,F_j]&=&if_{ijk}F_k\nonumber\\
\{F_i,F_j\}&=&\frac{1}{3}\delta_{ij} + d_{ijk}F_k,~~ i,j,k=1-8,
\end{eqnarray}
 where $f_{ijk}$ and $d_{ijk}$ are the structure constants, and are antisymmetric and symmetric, 
 respectively, under the interchange of any two indices~\cite{Fayyazuddin}.

 The electromagnetic current~($J_{em}^\mu$) and the weak vector ($V_{\pm}^\mu$) and the axial vector 
 ($A_{\pm}^\mu$) charged currents are defined in terms of $V_i^\mu$ and $A_i^\mu$; $i=1-8$, as
 \begin{eqnarray}\label{jmu1}
 J^\mu_{em}&=& V^\mu_3~+~ \frac{1}{\sqrt{3}} V^\mu_8, \nonumber\\
 V^\mu_{\pm}&=&\left[V^\mu_1~\pm~ i V^\mu_2\right]\cos\theta_C~+~\left[V^\mu_4~\pm~ i V^\mu_5\right]\sin\theta_C, 
 \nonumber \\
 A^\mu_{\pm}&=&\left[A^\mu_1~\pm~ i A^\mu_2\right]\cos\theta_C~+~\left[A^\mu_4~\pm~ i A^\mu_5\right]\sin\theta_C.
 \end{eqnarray}

 In the Cabibbo theory, $V_{i}^\mu;~(i=1-8)$  transform as an octet of vector currents under SU(3) symmetry. Similarly, the axial vector currents $A_{i}^\mu;~(i=1-8)$  are also 
 assumed to transform as an octet under SU(3) symmetry. The form factors defined in the matrix element of an octet of the vector and 
 axial vector currents taken between the octets of the initial and the final baryon states as defined in 
 Eqs.~(\ref{vx}) and (\ref{vy}) can, therefore, be expressed in terms of the two couplings of the vector and axial 
 vector currents corresponding to the symmetric and antisymmetric octets according to the decomposition:
\begin{equation}
 8 \times 8 = 1 + 8^S + 8^A + 10 + \overline{10} + 27
\end{equation}
with the corresponding SU(3) Clebsch-Gordan coefficients. In general, the expression for the matrix element of the 
transition between the two states of baryons (say $B_i$ and $B_j$), through the SU(3) octet ($V_j$ or $A_j$) of
currents can be written as~\cite{Fayyazuddin, Athar:2020kqn}:
 \begin{eqnarray}\label{bb}
< B_i | V_j | B_k > &=& if_{ijk}F^V + d_{ijk} D^V, \\
\label{b2}
< B_i | A_j | B_k > &=& if_{ijk}F^A + d_{ijk} D^A.
 \end{eqnarray}
  $F^{V}$ and $D^{V}$ are determined from the experimental data on the electromagnetic form factors, and $F^{A}$ and 
  $D^{A}$ are determined from the experimental data on the semileptonic decays of the nucleons and hyperons. Explicitly,
  the form factors $f_i^{pX} (Q^2)$ and $g_i^{pX} (Q^2)$ defined in Eqs.~(\ref{vx})--(\ref{vy}) can be expressed as 
\begin{eqnarray}\label{fi}
f_i^{pX} (Q^2) &=& a F_i^V (Q^2) + b D_i^V  (Q^2) \qquad i=1,2,3 \\
\label{gi}
g_i^{pX} (Q^2) &=& a F_i^A (Q^2) + b D_i^A  (Q^2) \qquad i =1,2,3
\end{eqnarray}
The Clebsch-Gordan coefficients $a$ and $b$ can be calculated for each transition, if we specify the quantum numbers 
($| I,~I_3,~Y\rangle$, where $I$ is the isospin, $I_{3}$ is the 3rd component of isospin, and $Y$ is the hypercharge) of the initial and the final state baryons and the current operators $V_\mu$, $A_\mu$ and $J_\mu^{em}$ 
in the octet representation. A straightforward calculation of the various Clebsch-Gordan coefficients in the case of 
weak $\Delta S=0$, $\Delta S=1$ hadronic currents and the electromagnetic currents (in the case of vector currents) 
gives the values of $a$ and $b$ which are obtained using Eqs.~(\ref{algebra}), (\ref{jmu1}), (\ref{bb}) and (\ref{b2}) 
and are presented in Table~\ref{tabI}.

\begin{table} \centering
\begin{tabular}{|c|c|c|}\hline
 ~~ Transitions ~~ & ~~ $a$ ~~ & ~~ $b$ ~~ \\ \hline
 ~~ $n\rightarrow p$ ~~ & ~~1~~ & ~~1~~ \\
 ~~ $p\rightarrow \Lambda$ ~~ & ~~ $-\sqrt{\frac{3}{2}}$ ~~ & ~~ $-\frac{1}{\sqrt{6}}$ ~~\\
 ~~ $p\rightarrow \Sigma^0$ ~~ & ~~ $-\frac{1}{\sqrt{2}}$ ~~ & ~~ $\frac{1}{\sqrt{2}}$ ~~\\
 ~~ $n\rightarrow \Sigma^-$ ~~ & ~~ $-1$ ~~ & ~~ 1 ~~ \\ \hline
\end{tabular}
\caption{Values of the coefficients $a$ and $b$ given in Eqs.~(\ref{fi})$-$(\ref{gi}).}
\label{tabI}
\end{table}

\item[b)] The two vector form factors \textit{viz.} $f^{pX}_1(Q^2)$ and $f^{pX}_2(Q^2)$ are determined in terms of the 
electromagnetic form factors of the nucleon, \textit{i.e.} $F_{1}^{N}(Q^{2})$ and $F_{2}^{N}(Q^{2}),~ N=(p,n)$. This is done by 
taking the matrix element of the electromagnetic current operator between the nucleon states and determining $F_{i}^{V}(Q^2)$ and 
$D_{i}^{V}(Q^2)$ in terms of $F^{N}_{i}(Q^2);~(i = 1, 2)$. The functions $F_{i}^{V}(Q^2)$ and $D_{i}^{V}(Q^2)$ are, thus, 
expressed in terms of the nucleon form factors $f_{1}^{p,n}(Q^{2})$ and $f_{2}^{p,n}(Q^{2})$ as 
\begin{eqnarray}\label{eq:fiv_div}
 F_i^V(Q^2) &=& F_i^p(Q^2) + \frac12 F_i^n (Q^2),  \\ 
 D_i^V(Q^2) &=& - \frac32 F_i^n (Q^2). 
\end{eqnarray}
The explicit expressions of the vector form factors $f^{pX}_1(Q^2)$ and $f^{pX}_2(Q^2)$ in terms of $F_{1}^{N}(Q^{2})$ and $F_{2}^{N}(Q^{2})$  are given in Table-\ref{tab:formfac}.

 The electromagnetic form factors $F_{1}^{N}(Q^{2})$ and $F_{2}^{N}(Q^{2});~(N=n,p)$ are expressed in terms of the experimentally determined Sachs' electric $G_E^{p,n} (Q^2)$ 
 and magnetic $G_M^{p,n}(Q^2)$ form factors for which various parametrizations are available in the literature. We have used 
 the parametrization given by Bradford et al.~\cite{Bradford:2006yz} for the numerical calculations in this work.

\item[c)]  The axial vector form factors $g^{pX}_{1}(Q^{2})$ and $g^{pX}_{2}(Q^{2})$ are determined using Eq.~(\ref{gi}).
$g^{pX}_{1,2}(Q^2)$ are written in terms of the two functions $F_{1,2}^A(Q^2)$ and $D_{1,2}^A(Q^2)$. Using Table-\ref{tabI} 
for the coefficients $a$ and $b$, we find 
\begin{eqnarray}
   g_{1,2}^{pn}(Q^2)&=& F_{1,2}^A(Q^2)+D_{1,2}^A(Q^2), \\
  g_{1,2}^{p\Lambda}(Q^2)&=&\sqrt{\frac16}\left[3F^A_{1,2}(Q^2)+D^A_{1,2}(Q^2)\right)] \\
   g_{1,2}^{p\Sigma^0}(Q^2)&=& \sqrt{\frac12} \left[ D^A_{1,2}(Q^2)-F^A_{1,2}(Q^2) \right].
\end{eqnarray}\label{Eq:xdep}
$g_{1,2}^{p\Lambda}(Q^2)$ and $g_{1,2}^{p\Sigma^0}(Q^2)$ are, generally, expressed in terms of $g_{1,2}^{pn}(Q^2)$ and the ratios $x_{1,2}(Q^2)$, 
with $x_{1,2}(Q^2)$ defined as 
\begin{equation}
x_{1,2}(Q^2)=\frac{F^A_{1,2}(Q^2)}{F^A_{1,2}(Q^2)+D^A_{1,2}(Q^2)},
\end{equation}
and the expressions for $g_{1,2}^{pX}(Q^2);~(X=\Lambda, \Sigma)$ are given explicitly  in Table-\ref{tab:formfac}.
\begin{table} 
 \begin{center}
\begin{adjustbox}{max width=\textwidth} 
\begin{tabular}{|c|c|c|c|}  \hline 
&$e^- p \rightarrow \nu_e X (=n)$&$e^- p \rightarrow \nu_e X( =\Lambda)$&$e^- p \rightarrow \nu_e X=(\Sigma^0)$\\ \hline  \hline     
 $f_1^{pX}(Q^2)$& $F_1^p(Q^2) - F_1^n(Q^2)$&$ -\sqrt{\frac{3}{2}}~F_1^p(Q^2)$&$-\frac{1}{\sqrt2}\left[F_1^p(Q^2) + 2 F_1^n(Q^2) 
 \right]$ \\ \hline
$f_2^{pX}(Q^2)$& $F_2^p(Q^2) - F_2^n(Q^2)$&$-\sqrt{\frac{3}{2}}~F_2^p(Q^2)$&$-\frac{1}{\sqrt2}\left[F_2^p(Q^2) + 2 F_2^n(Q^2) 
\right]$\\ \hline
$g_1^{pX}(Q^2)$& $g_1^{pn}(Q^2)$&$-\frac{1}{\sqrt{6}}(1+2x_1) g_1^{pn}(Q^2)$&$\frac{1}{\sqrt2}(1-2x_1)g_1^{pn}(Q^2)$ \\ \hline 
 $g_2^{pX}(Q^2)$& 
 $~g_2^{pn}(Q^2)$&$- \frac{1}{\sqrt{6}}(1+2x_2) g_2^{pn}(Q^2)$&$ \frac{1}{\sqrt2}(1-2x_2)g_2^{pn}(Q^2)$\\ \hline 
  \end{tabular}
  \end{adjustbox}
\end{center}
\caption{Vector and axial vector from factors $ f_i^{pX} (Q^2)$ and $g_i^{pX} (Q^2)~ (i=1,2)$, $f_3^{pX}(Q^2)$ and $g_3^{pX} (Q^2)$ 
are taken to be zero and negligible, respectively, for the $e^-(k) + p(p)\rightarrow \nu_e(k^\prime) + X(p^\prime)$ 
processes, where $X=n, \Lambda^0, \Sigma^0$.}
 \label{tab:formfac}
\end{table}

We further assume that $F^A_{1,2}(Q^2)$ and $D^A_{1,2}(Q^2)$ have the same $Q^2$ dependence, such that $x_{1,2}(Q^2)$ become 
constant given by $x_{1,2}(Q^2)=x_{1,2}=\frac{F^A_{1,2}(0)}{F^A_{1,2}(0)+D_{1,2}^A(0)}$. 
%

\item [d)] For the axial vector form factor $g_{1}^{pn}(Q^2)$, a dipole parametrization has been used:
\begin{eqnarray}\label{g1}
 g_{1}^{pn}(Q^2)= g_1(Q^2) = g_{A}(0)\left(1+\frac{Q^2}{M_{A}^2}\right)^{-2},
\end{eqnarray}
where $M_A$ is the axial dipole mass and $g_A(0)$ is the axial vector charge. For the numerical calculations, we have used the world average
value of $M_A=1.026$ GeV~\cite{Bernard:2001rs} unless stated. $g_A(0)$ and $x_1$ are taken to be 1.2723 and 0.364, respectively, as 
determined from the experimental data on the $\beta-$decay of neutron and the semileptonic decay of hyperons~\cite{Cabibbo:2003cu}. 

Recently, the neutrino community has focussed on another parametrization of the axial vector form factor based on the $z$-expansion~\cite{Hill:2010yb}.
The $z$-expansion provides a conformal mapping that transforms $Q^2$ into a small expansion variable $z$ across the entire kinematic region relevant to quasielastic scattering. The conformal variable $z$ is defined as
\begin{equation}
 z=\frac{\sqrt{t_c + Q^2} - \sqrt{t_c - t_0}}{\sqrt{t_c + Q^2} + \sqrt{t_c - t_0}},
\end{equation}
where the parameter $t_c \le (3m_{\pi})^2$, with $m_{\pi}$ being the mass of the pion, is fixed by the particle production threshold for the axial vector current interaction, and the variable $t_0=-0.5$~GeV$^2$ fixes the value of $Q^2$ for which $z=0$ is satisfied.

Within this framework, the axial vector form factor can be expressed as a power series in the variable $z$:
\begin{equation}
 g_1(z) = \sum_{k=0}^{\infty} a_k z^k.
\end{equation}
In general, the sum over $k$ is taken up to some finite order $k_{max}$, which is determined by the fit range and data accuracy. In this work, we have followed Ref.~\cite{MINERvA:2025ygc, Meyer:2026kdl} for this fitting, where the authors concluded that the best fit is obtained with $k_{max}=6$. 
The values of the parameters $a_{k};~(k=0-6)$ for the MINERvA hydrogen data~\cite{MINERvA:2023avz}, LQCD, previous deuterium data, and combined MINERvA hydrogen and LQCD fits as obtained in Ref.~\cite{MINERvA:2025ygc, Meyer:2026kdl} are tabulated in Table~\ref{tab:ak}.
\begin{table*}
\centering
\begin{tabular*}{150mm}{@{\extracolsep{\fill}} c  c c c c }\hline\hline
&MINERvA H fit& LQCD fit & deuterium fit& Combined H-LQCD fit \\ \hline
$a_0$ & 0.61490770 & 0.71742019 & 0.54264533 & 0.71070233 \\
$a_1$ & $-1.64778080$ & $-1.72089706$ & $-2.08493637$ & $-1.74307738$ \\
$a_2$ & 0.94181417 & 0.30982708 & $1.89831616$ & 0.37944565 \\
$a_3$ & 0.41239729 & 1.62125837 & 2.40319245 & 1.69894456 \\
$a_4$ & 0.36611559 & $-0.27506993$ & $-5.88979056$ & $-0.60326876$ \\
$a_5$ & $-1.18722194$ & $-1.25297945$ & 4.14554900 & $-0.95690585$ \\
$a_6$ & 0.49976799 & 0.60044079 & $-1.01497601$ & 0.51415945 \\
\hline\hline
\end{tabular*}
\caption{Values of the parameters $a_k;~(k=0-6)$ as obtained in Ref.~\cite{MINERvA:2025ygc}.}
\label{tab:ak}                                                  
\end{table*}

\begin{figure} 
\begin{center}
\includegraphics[width=8cm,height=6.5cm]{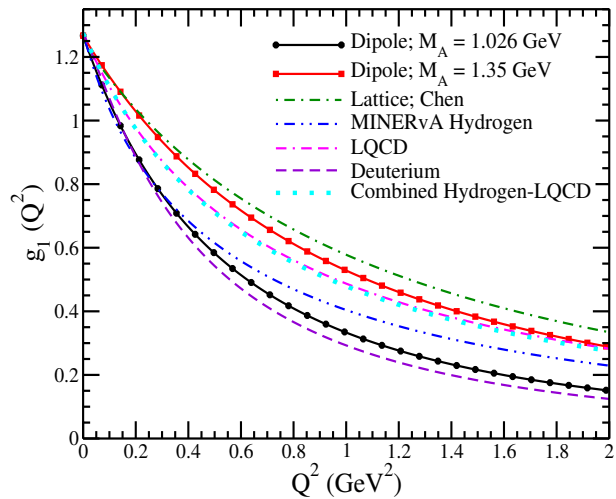}
\caption{$g_{1}(Q^2)$ as a function of $Q^2$ using the dipole parametrization with $M_{A} = 1.026~(1.35)$~GeV shown by solid lines with circle~(square). The results obtained using the lattice gauge parametrization of Chen and Roberts~\cite{Chen:2022odn} are shown by the dash-dotted line. The results obtained using the $z$-expansion~\cite{Meyer:2026kdl, MINERvA:2025ygc} for the MINERvA hydrogen, LQCD, deuterium, and combined hydrogen-LQCD fits are represented by double-dot-dashed line, double dash-dotted line, dashed line and dotted line, respectively.}\label{axial_FF}
\end{center}
\end{figure}

More recently, Chen and Roberts~\cite{Chen:2022odn} used lattice gauge formalism to determine the axial vector form factor and parameterized it as a function of $x$, where $x=Q^2/m_x^2$, with $m_x=1.18$~GeV:
\begin{equation}
 g_1(x) = \frac{1.248 + 0.039x}{1 + 1.417x + 0.318x^2 + 0.071x^3}.
\end{equation}

In Fig.~\ref{axial_FF}, we show the results of axial vector form factor $g_{1}(Q^2)$ as a function of $Q^2$ using (i)~the dipole parametrization with two values of $M_{A}$ viz. $M_{A}=1.026$ and 1.35~GeV, (ii)~the $z$-expansion parametrization~\cite{MINERvA:2025ygc} using (a)~MINERvA hydrogen data, (b)~LQCD, (c)~deuterium, and (d)~combined hydrogen-LQCD fits, and (iii)~the lattice gauge parametrization given by Chen and Roberts~\cite{Chen:2022odn}. 
It may be observed from the figure that the results obtained with the $z$-expansion fit of combined LQCD and MINERvA data are almost 30\% larger than the results obtained with the traditional dipole parametrization with $M_{A}=1.026$~GeV in the entire region of $Q^2$ considered in this work. However, the results obtained using $M_{A}=1.35$~GeV in the traditional dipole parametrization are comparable to the results obtained with the $z$-expansion fit of combined LQCD and MINERvA data.

\item [e)] The weak electric form factor $g_2^{pn} (Q^2)$ is taken to be of dipole form, i.e., 
\begin{eqnarray}\label{g2} 
 g_{2}^{pn}(Q^2)=g_{2}^{pn}(0)\left(1+\frac{Q^2}{M_{2}^2}\right)^{-2}.
 \end{eqnarray} 
There is limited experimental information about $g_2^{pn} (Q^2)$, which has been obtained from analyses of nuclear $\beta$ decays~\cite{Oka:1979bw, Minamisono:2001cd, Minamisono:2011zz, Wilkinson:2000gx}, muon capture~\cite{Commins:1983ns} and (anti)neutrino quasielastic scattering performed at BNL~\cite{Baker:1981su, Ahrens:1988rr} and SKAT~\cite{Belikov:1983kg}. While nuclear $\beta$ decay and muon capture experiments obtained small values for $g_2^{pn} (0)$ namely $g_2^{pn}(0) = (0.504\pm 1.134)$~\cite{Commins:1983ns} and $g_2^{pn}(0) = 0 \pm 0.075$~\cite{Day:2012gb, Oka:1979bw, Minamisono:2001cd, Minamisono:2011zz, Wilkinson:2000gx},  the limits on $g_2^{pn}(0)$ reported by neutrino experiments are much larger. For example, neutrino induced quasielastic scattering using  bubble chamber detector~\cite{Baker:1981su} and antineutrino quasielastic scattering using scintillator detector~\cite{Ahrens:1988rr} recommended, respectively, the real values of $g_2^{pn}(0)$ to be  $-3.69 \pm 1.26$ and $-1.008$. Furthermore, Belikov et al.~\cite{Belikov:1983kg} at SKAT used neutrino and antineutrino quasielastic scattering data to estimate the upper limit for $g_2^{pn}(0)$ to be $-1.63$ at 90\% C.L.
This information is useful in 
estimating the range of the real values of $g_2^{pn} (Q^2)$ i.e. Re[$g_2^{pn}(Q^2)$]. 
Our present knowledge about the imaginary values of  $g_2^{pX} (Q^2)$ i.e. $Im[g_2^{pX} (Q^2)]; (X=n,\Lambda,\Sigma^0)$ is almost non-existent. The analysis of T-violating polarization 
observables in neutrino scattering from the CERN experiment~\cite{Erriquez:1978pg} quotes a value for the T-violating polarization observable that is 
consistent with zero, but no explicit limits on $Im[g_2^{p \Lambda} (Q^2)]$ are reported. Similarly, an older analysis of the T-violating spin 
correlations of electrons in the beta decay of polarized $\Lambda$ implies a large value of $Im[g_2^{p \Lambda} (Q^2)]$, though without providing a 
quantitative estimate~\cite{Oehme:1971rd}. 
Some of the earlier theoretical calculations for the (anti)neutrino-nucleon/nucleus scattering were performed assuming nonzero values for real as well as imaginary values of $g_2^{pn}(0)$. For example; Fujii and Yamaguchi~\cite{Fujii1, Fujii2} considered real and imaginary values of $g_2^{pn}(0)$ taken to be 1.92, while Berman and Veltman~\cite{Berman:1964zza} used  Im[$g_2^{pn}(0)]=$~3.7 and 6.
The older calculations of T-violating effects in weak processes have phenomenologically used values of $Im[g_2^{p \Lambda} (Q^2)]$ across a wide range, specifically $1<Im~g_2^{p\Lambda}(0)<10$~\cite{Fearing:1969nr}. 

In Eq.~(\ref{g2}), $M_{2}$ is the axial dipole mass corresponding to the weak electric form factor $g_2 (Q^2)$.
 Since no experimental information on the value of $M_{2}$ is currently available, therefore, for simplicity, in the numerical calculations, we have set $M_{2}=M_{A}=1.026$~GeV~\cite{Fatima:2018tzs}, unless stated otherwise.

In view of the above discussion, in the numerical calculations, we have taken into account both purely real and purely imaginary values for $g_2^{pX}(0)$. Here a purely real value of $g_2^{pX}(0)$ implies G-violation with T-invariance, while a purely imaginary value of $g_2^{pX}(0)$ implies both G- and T- violation.  Clearly a knowledge of $g_2^{pn} (0)$ is very important in the study of G- and T- invariance in weak interaction physics.
\end{enumerate}

\subsection{Spin asymmetries of the polarized target proton}
Traditionally, the expressions for spin asymmetries of the target proton are obtained using the spin projection operator formalism, as demonstrated recently, for example, by Graczyk et al.~\cite{Graczyk:2017rti, Graczyk:2023lrm, Graczyk:2004uy}, Tomalak et al.~\cite{Tomalak:2023pdi, Borah:2024hvo}, in the case of neutrino scattering processes off nucleons. In this work, we use the covariant density matrix formalism following Bilenky and Christova~\cite{Bilenky:2013fra, Bilenky:2013iua, Bilekny}, which has been successfully employed in our previous calculations~\cite{Fatima:2018gjy, Akbar:2016awk, Akbar:2017qsf, Fatima:2018tzs, Fatima:2018wsy, Fatima:2020pvv, Fatima:2022tlf} for studying the polarization components of the final baryon. Although these are two distinct approaches, they lead to the same results for the target asymmetries of an initial polarized proton.

Using the covariant density matrix formalism, the polarization 4-vector~($\zeta^\tau$) of the initial proton of momentum $p^{\sigma}$ is written as~\cite{Bilekny}:
\begin{eqnarray}\label{polar4:i}
\zeta^{\tau}&=&\left( g^{\tau\sigma}-\frac{p^{\tau}p^{\sigma}}{M^2}\right) \frac{  {\cal L}^{\alpha \beta}  \mathrm{Tr}
\left[\gamma_{\sigma}\gamma_{5}\Lambda(p')J_{\alpha} \Lambda(p)\tilde{J}_{\beta} \right]}
{ {\cal L}^{\alpha \beta} \mathrm{Tr}\left[\Lambda(p')J_{\alpha} \Lambda(p)\tilde{J}_{\beta} \right]},
\end{eqnarray}
where in the rest frame $\zeta^\tau = (0,\vec{\zeta})$ and $\vec{\zeta}$ is expressed in terms of the orthogonal vectors $\hat{e}_{i}~(i=L,P,T)$, i.e.,
 \begin{equation}\label{polarLab:i}
\vec{\zeta}=\zeta_{L} \hat{e}_{L} + \zeta_{P} \hat{e}_{P} + \zeta_{T} \hat{e}_{T},
\end{equation}
where $\hat{e}_{L}$, $\hat{e}_{P}$, and $\hat{e}_T$  are chosen to be the set of orthogonal unit vectors corresponding to the 
longitudinal, perpendicular, and transverse directions with respect to the momentum of the initial electron, shown in Fig.~\ref{TRI}, 
and are written as~\cite{Graczyk:2023lrm}:
\begin{equation}\label{vectors:i}
\hat{ e}_{L}=\frac{\vec{ k}}{|\vec{ k}|}, \qquad \quad 
\hat{ e}_{T}=\frac{\vec{ q}\times \vec{ k}}{|\vec{ q}\times \vec{ k}|}, \qquad \quad 
\hat{ e}_{P}=\frac{\vec{ k}}{|\vec{ k}|} \times \frac{\vec{ q}\times \vec{ k}}{|\vec{ q}\times \vec{ k}|} .
 \end{equation}
 Since the transverse component lies perpendicular to the reaction plane and vanishes when T-invariance is assumed, therefore, the polarization vector is expressed only in terms of $\hat{ e}_{L}$ and $\hat{ e}_{P}$.
 
The longitudinal and perpendicular components of the polarization vector $\vec{\zeta}_{L,P} (Q^2)$ using Eqs.~(\ref{polarLab:i}) and (\ref{vectors:i}) may be written as
\begin{equation}\label{PL:i}
 \zeta_{L,P}(Q^2)=\vec{\zeta} \cdot \hat{e}_{L,P}~.
\end{equation}
In the rest frame of the initial proton, the polarization vector $\vec{\zeta}$ is expressed as
\begin{equation}\label{pol2:i}
 \vec{\zeta} = A_1(E_e,Q^2)~ \vec{k} + B_1(E_e,Q^2)~ \vec{q},
\end{equation}
and is explicitly calculated using Eq.~(\ref{polar4:i}). The expressions for the coefficients $A_1(E_e,Q^2)$, and $B_1(E_e,Q^2)$ are given in the Appendix-I.

The longitudinal~($A_L(Q^2)$) and perpendicular~($A_P(Q^2)$) spin asymmetries of the initial proton are written as:
\begin{equation}\label{PlPp:i}
 A_L (Q^2) = \zeta_L (Q^2), ~~~~~~~ A_P (Q^2) = \zeta_P (Q^2).
\end{equation}
Using Eqs.~(\ref{vectors:i}), (\ref{PL:i}) and 
(\ref{pol2:i}) in Eq.~(\ref{PlPp:i}), the expressions for $A_L (Q^2)$ and $A_P (Q^2)$ are obtained as
\begin{eqnarray}
  A_L (Q^2) &=& \frac{1}{E_e} \frac{A_1(E_e,Q^2) |\vec{k}|^2 + B_1 (E_e,Q^2) \vec{k} \cdot \vec{q}}
  {N(E_e,Q^2)},
  \label{Al} \\
 A_P (Q^2) &=& \frac{B_1(E_e,Q^2) [|\vec{q}| \sin\beta_k]}{N(E_e,Q^2) },
 \label{Ap} 
\end{eqnarray}
where $\beta_k$ is the angle between $\vec{k}$ and $\vec{q}$ and the expression for $N(E_e,Q^2) = {\cal L}_{\alpha\beta}{\cal J}^{\alpha \beta}$  is given in Appendix-I.

\subsection{Polarization observables of the final baryon}\label{polarization}
\begin{figure}
  \begin{center}
    \hspace{-1cm}
    \includegraphics[height=6cm,width=8cm]{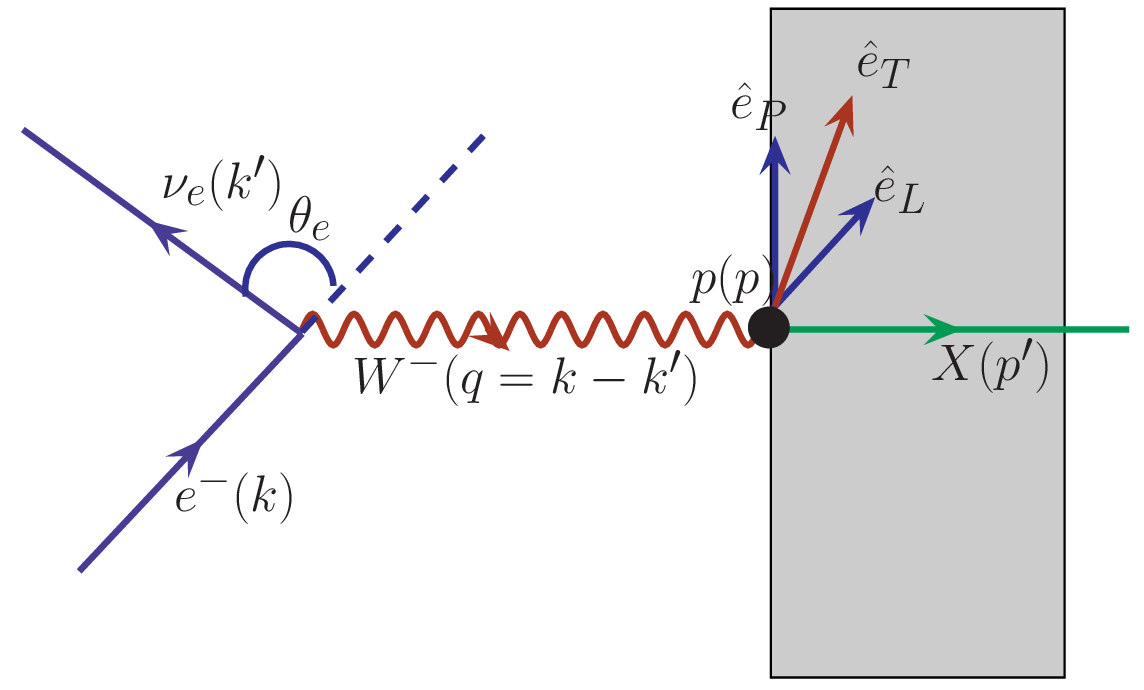}
    \includegraphics[height=6cm,width=8cm]{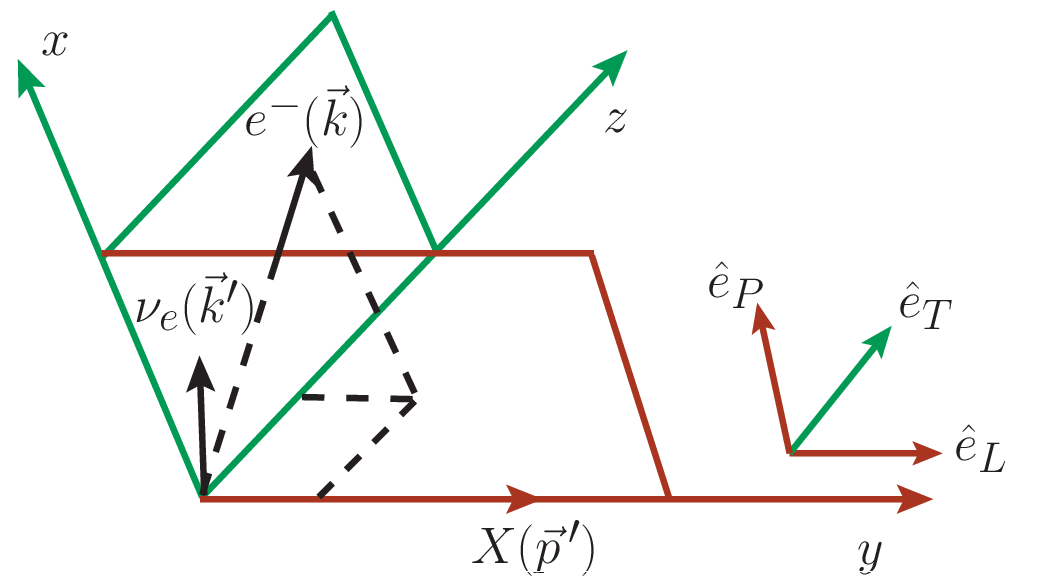}
   \caption{Diagrammatic representation of the process $ e^-(\vec{k}) + p(\vec{p}=0) \rightarrow \nu_e(\vec{k^\prime}) + 
   X(\vec{p}^{~\prime})$, and the longitudinal and perpendicular directions of the polarized proton~(left panel). The longitudinal, perpendicular and transverse directions with respect to 
  the momentum of the final baryon~(right panel).}\label{TRI}
    \end{center}
  \end{figure}
  Using the covariant density matrix formalism, the polarization 4-vector~($\xi^\tau$) of the final baryon of momentum ${p^{\prime}}^\sigma$ produced in the reactions 
  (\ref{nuc-rec}) and (\ref{hyp-rec}) is written as~\cite{Bilekny}: 
\begin{eqnarray}\label{polar4}
\xi^{\tau}&=&\left( g^{\tau\sigma}-\frac{p'^{\tau}p'^{\sigma}}{{M^\prime}^2}\right) \frac{  {\cal L}^{\alpha \beta}  \mathrm{Tr}
\left[\gamma_{\sigma}\gamma_{5}\Lambda(p')J_{\alpha} \Lambda(p)\tilde{J}_{\beta} \right]}
{ {\cal L}^{\alpha \beta} \mathrm{Tr}\left[\Lambda(p')J_{\alpha} \Lambda(p)\tilde{J}_{\beta} \right]}.
\end{eqnarray}

One may write the polarization vector $\vec{\xi}$ in terms of the three orthogonal vectors $\hat{e}_{i}~(i=L,P,T)$, i.e.,
 \begin{equation}\label{polarLab}
\vec{\xi}=\xi_{L} \hat{e}_{L} + \xi_{P} \hat{e}_{P}+\xi_{T} \hat{e}_{T} ,
\end{equation}
where $\hat{e}_{L}$, $\hat{e}_{P}$ and $\hat{e}_{T}$ are chosen to be the set of orthogonal unit vectors corresponding to the 
longitudinal, perpendicular and transverse directions with respect to the momentum of the final baryon, shown in Fig.~\ref{TRI}, 
and are written as
\begin{equation}\label{vectors}
\hat{ e}_{L}=\frac{\vec{ p}^{\, \prime}}{|\vec{ p}^{\, \prime}|},~~~~~
\hat{ e}_{P}=\hat{ e}_{L}\times \hat{ e}_T, ~~~~ 
\hat{e}_T=\frac{\vec{ p}^{\, \prime}\times \vec{ k}}{|\vec{ p}^{\, \prime}\times \vec{ k}|}.
 \end{equation}
The longitudinal, perpendicular and transverse components of the polarization vector $\vec{\xi}_{L,P,T} (Q^2)$ using Eqs. 
(\ref{polarLab}) and (\ref{vectors}) may be written as
\begin{equation}\label{PL}
 \xi_{L,P,T}(Q^2)=\vec{\xi} \cdot \hat{e}_{L,P,T}~.
\end{equation}

In the case of final baryon polarization, we consider two scenarios: (i)~when time reversal invariance is assumed, and (ii)~when time reversal is violated. As discussed earlier, the assumption of T-invariance implies all form factors are real, therefore, in the first scenario, we have taken into account purely real values for $g_2(0)$ and represent it by $g_2^R(0)$. In the second scenario, when time reversal is violated, the numerical calculations are performed by incorporating purely imaginary values for $g_2(0)$, represented as $g_2^{I} (0)$.

\subsubsection{T-invariance:}
In the rest frame of the initial proton, assuming T-invariance, the polarization vector $\vec{\xi}$ is expressed as
\begin{equation}\label{pol2:TI}
 \vec{\xi} = A(E_e,Q^2)~ \vec{k} + B(E_e,Q^2)~ \vec{p}^{\, \prime} ,
\end{equation}
and is explicitly calculated using Eq.~(\ref{polar4}). The expressions for the coefficients $A(E_e,Q^2)$ and $B(E_e,Q^2)$, obtained using the real values of the weak electric form factor, associated with  second class currents, i.e., $g_{2} (Q^2)=g_2^R (Q^2)$, are given in Appendix-II.

The longitudinal~($P_L(Q^2)$) and perpendicular~($P_P(Q^2)$) components of the polarization vector in the rest frame of the final baryon are then obtained by performing a Lorentz boost and  are written
as~\cite{Fatima:2018tzs}:
\begin{eqnarray}\label{PlPp}
 P_L (Q^2) &=& \frac{M}{E^\prime} \xi_L (Q^2), \qquad \quad P_P (Q^2) = \xi_P (Q^2),
\end{eqnarray}
where $E^\prime$ is the energy of the outgoing baryon.

The expressions for $P_L (Q^2)$ and $P_P (Q^2)$ are obtained by substituting Eqs.~(\ref{vectors}), (\ref{PL}), and (\ref{pol2:TI}) in Eq.~(\ref{PlPp}), and are given as:
\begin{eqnarray}
  P_L (Q^2) &=& \frac{M}{E^\prime} \frac{A(E_e,Q^2) \vec{k} \cdot \hat{p}^{\prime} + B (E_e,Q^2)
  |\vec{p}^{\,\prime}|}{N(E_e,Q^2)},
  \label{Pl} \\
 P_P (Q^2) &=& \frac{A(E_e,Q^2) [(\vec{k}.\hat{p}^{\prime})^2 - |\vec{k}|^2]}{N(E_e,Q^2) ~|\hat{p}^{\prime} \times
 \vec{k}|}.\label{Pp} 
\end{eqnarray}

\subsubsection{T-violation:}
To study the effect of T-violation on the polarization observables of the final baryon, 
the polarization vector $\vec{\xi}$, in the rest frame of the initial proton,  is expressed as
\begin{equation}\label{pol2}
 \vec{\xi} = A^\prime(E_e,Q^2)~ \vec{k} + B^\prime(E_e,Q^2)~ \vec{p}^{\, \prime} + C^\prime(E_e,Q^2)~  M (\vec{k} \times \vec{p}^{\,\prime}),
\end{equation}
and is explicitly calculated using Eq.~(\ref{polar4}). The expressions for the coefficients $A^\prime(E_e,Q^2)$, $B^\prime(E_e,Q^2)$
and $C^\prime(E_e,Q^2)$ are given in Appendix-III.

The longitudinal ($P_L(Q^2)$) and perpendicular ($P_P(Q^2)$) components of the polarization
vector in the rest frame of the final baryon are given in Eq.~(\ref{PlPp}), and the transverse~($P_T (Q^2)$) component of polarization is written
as~\cite{Fatima:2018tzs}:
\begin{eqnarray}\label{Ptt}
 P_T (Q^2) = \xi_T (Q^2).
\end{eqnarray}
In the case of T-violation, the explicit expressions for $P_L (Q^2)$ and $P_P (Q^2)$ are identical to those given in Eqs.~(\ref{Pl}) and (\ref{Pp}), except that the coefficients $A(E_e,Q^2)$ and $B(E_e,Q^2)$ are now replaced with   $A^\prime(E_e,Q^2)$ and $B^\prime(E_e,Q^2)$. 
The expression for $P_T (Q^2)$ is obtained as:
\begin{eqnarray} \label{Pt}
  P_T (Q^2) &=& \frac{C^\prime(E_e,Q^2) M |\vec{p}^{\,\prime}|[(\vec{k}.\hat{p}^{\prime})^2 - |\vec{k}|^2]}{N(E_e,Q^2)~
  |\hat{p}^{\prime} \times \vec{k}|}.
\end{eqnarray}

If T-invariance is assumed then all the vector and the axial vector form factors are real, and the expression for $C^\prime(E_e,Q^2)$ 
vanishes. This, in turn, implies that the transverse component of the  polarization $ P_T (Q^2)$,  perpendicular to the production plane, vanishes. 
Conversely, if T-invariance is violated, then while $d\sigma/dQ^2$, $P_L(Q^2)$ and $P_P(Q^2)$ receive only small corrections due to the T-violating 
form factor $g_2^{I}(Q^2)$, the transverse component of the polarization $P_T (Q^2)$ becomes non-zero and could be significant. 
Thus, an experimental observation of $P_T (Q^2)$ can be used to probe the physics of T-noninvariance~\cite{DeRujula:1970ek, 
Fearing:1969nr}.

\subsection{Results and discussion}

We have used Eq.~(\ref{dsig}) to calculate $\frac{d\sigma}{dQ^2}$ for weak charged current induced electron scattering off a proton target in the $\Delta S=0$ and $\Delta S = 1$ sectors~($e^- + p \longrightarrow n + \nu_e$, $e^- + p \longrightarrow \Lambda + \nu_e$, and $e^- + p \longrightarrow \Sigma^0 + \nu_e$). Furthermore, Eqs.~(\ref{Al}) and (\ref{Ap}), respectively, are used to calculate the longitudinal~($A_{L} (Q^2)$) and perpendicular~($A_P(Q^2)$) asymmetries for the polarized proton target, while Eqs.~(\ref{Pl}), (\ref{Pp}), and (\ref{Pt}), respectively, are employed for calculating the longitudinal~($P_L(Q^2)$), perpendicular~($P_P(Q^2)$) and transverse~($P_T (Q^2)$) components of polarization for the final baryon. 

The weak vector form factors $f_{1,2}^{pX} (Q^2)$ are expressed in terms of the electromagnetic nucleon form factors $F_1^{p,n}(Q^2)$ and $F_2^{p,n} (Q^2)$, which, in turn,  are expressed in 
terms of the electric and the magnetic Sachs' form factors, the parametrizations for these have been taken from Bradford 
\textit{et al.}~\cite{Bradford:2006yz}. The axial vector form factor $g_1^{pX}(Q^2)$ and the weak electric form 
factor $g_2^{pX} (Q^2)$ are assumed to have a dipole form, as given in Eqs.~(\ref{g1}) and (\ref{g2}), respectively, with $M_A=1.026$~GeV and $M_2=1.026$~GeV, unless stated otherwise. The numerical calculations are performed assuming T-invariance, which corresponds to a purely real $g_2^{pn}(0)$, i.e. $ g_2^{pn} (0) =  g_2^R(0)$, as well as considering T-violation, which corresponds to a purely imaginary $g_2^{pn} (0)$ i.e. $ g_2^{pn} (0) =   g_2^I(0)$. 

In the case of cross sections, the negative and the positive values of $g_2^R(0)$, and $g_2^I(0)$ yield exactly the same results for $\Delta S=0$ quasielastic scattering process, i.e., $e^- + p \longrightarrow \nu_e +n$. However, for the strangeness changing processes $e^- + p \longrightarrow \Lambda + \nu_e$ and $e^- + p \longrightarrow \Sigma^0 + \nu_e$, the negative and the positive values of $g_2^R(0)$, and $g_2^I(0)$, yield different results for $\frac{d\sigma}{dQ^2}$.

In the case of polarization observables, the negative and the positive values of $g_2^R(0)$ yield different results for the T-invariant quantities such as  $A_L(Q^2)$, $A_P(Q^2)$, $P_L(Q^2)$, and $P_P(Q^2)$. 
However, it may be noted that a negative value of $g_2^I (0)$  does not change the T-invariant quantities such as $P_L(Q^2)$, and $P_P(Q^2)$, but changes the sign 
without changing the magnitude for the T-violating variable $P_T(Q^2)$ as compared to the results obtained using the positive values of $g_2^I (0)$.

To see the dependence of $\sigma(E_e)$, the average spin asymmetries, and the average polarization observables on the electron's energy, we have 
integrated $\frac{d\sigma}{dQ^2}$, $A_L(Q^2)$, $A_P(Q^2)$, $P_L (Q^2), ~P_P (Q^2)$ and $P_T (Q^2)$ over $Q^2$ using the following expressions~\cite{Fatima:2018gjy}: 
\begin{eqnarray}
\sigma(E_e) &=& \int_{Q^2_{min}}^{Q^2_{max}} \frac{d\sigma}{dQ^2} dQ^2, \\
A_{L,P} (E_e) &=& \frac{\int_{Q^2_{min}}^{Q^2_{max}} A_{L,P} (Q^2) \frac{d\sigma}{dQ^2} dQ^2}{\int_{Q^2_{min}}^{Q^2_{max}} 
 \frac{d\sigma}{dQ^2} dQ^2}, \\
 P_{L,P,T} (E_e) &=& \frac{\int_{Q^2_{min}}^{Q^2_{max}} P_{L,P,T} (Q^2) \frac{d\sigma}{dQ^2} dQ^2}{\int_{Q^2_{min}}^{Q^2_{max}} 
 \frac{d\sigma}{dQ^2} dQ^2},
\end{eqnarray}
and show the  numerical results for these observables in Figs.~\ref{sigma:ga:nucleon}--\ref{PlPpPt:q2:g2:neutron} for the $\Delta S=0$ process, i.e., $e^- + p \longrightarrow \nu_e + n$, and in Figs.~\ref{sigma:Lambda:g2I}--\ref{PLPPPT:Q2:g2I:Lambda}~and~Figs.~\ref{sigma:MA:Sigma}--\ref{PLPPPT:Q2:g2I:Sigma}, respectively, for the $\Delta S = 1$ processes $e^- + p \longrightarrow \nu_e + \Lambda$ and $e^- + p \longrightarrow \nu_e + \Sigma^0$. 

\begin{figure}
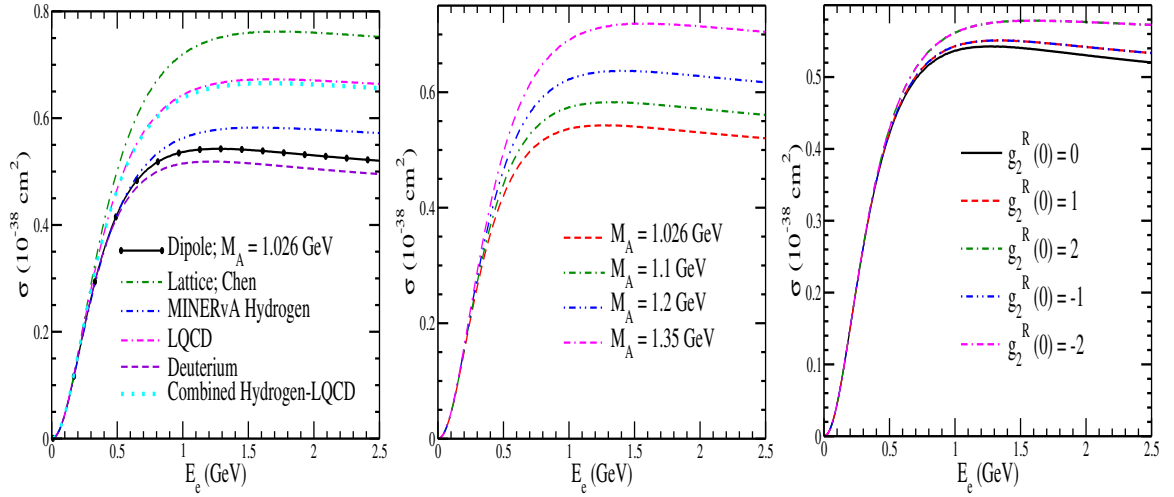
 
\begin{center}
\includegraphics[width=5cm,height=6.5cm]{sigma_ga_variation_unpolarized_electron_proton.eps}
\includegraphics[width=5cm,height=6.5cm]{total_sigma_MA_variation_unpolarized_electron_proton.eps}
\includegraphics[width=5cm,height=6.5cm]{total_sigma_g2_variation_polarized_proton.eps}
\caption{Total cross section $\sigma(E_e)$ as a function of $E_{e}$ for the process $e^- + p \longrightarrow \nu_{e} + n$. Left panel shows the results for different parametrizations of $g_{1}(Q^2)$ viz. the dipole parametrization with $M_{A} = 1.026$~GeV shown by the solid lines with circle, the lattice gauge parametrization of Chen and Roberts~\cite{Chen:2021guo, Chen:2022odn} shown by the dash-dotted line, the $z$-expansion for the MINERvA hydrogen, LQCD, deuterium, and combined hydrogen-LQCD fits are represented by the double-dot-dashed line, double dash-dotted line, dashed line and dotted line, respectively. Middle panel shows the results for $\sigma(E_e)$ using different values of $M_{A}$ viz. $M_{A} =1.026$~GeV~(dashed line), 1.1~GeV~(dash-dotted line), 1.2~GeV~(double-dot-dashed line), and 1.35~GeV~(double-dash-dotted line). Right panel shows the results for $\sigma(E_e)$ obtained using different values of $g_{2}(0)$ assuming T-invariance viz. $g_{2}^{R} (0)=0$~(solid line), +1~(dashed line), +2~(dash-dotted line), $-1$~(double-dot-dashed line), and $-2$~(double-dash-dotted line). }\label{sigma:ga:nucleon}
\end{center}
\end{figure}

\begin{figure} 
 \includegraphics[width=6.5cm,height=6.5cm]{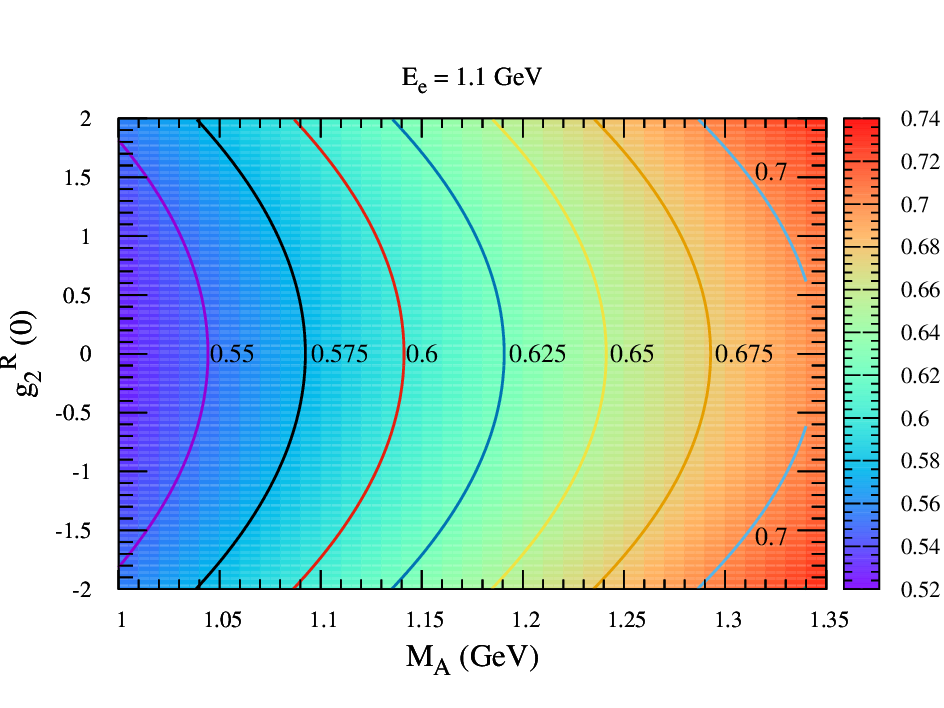} 
 \includegraphics[width=6.5cm,height=6.5cm]{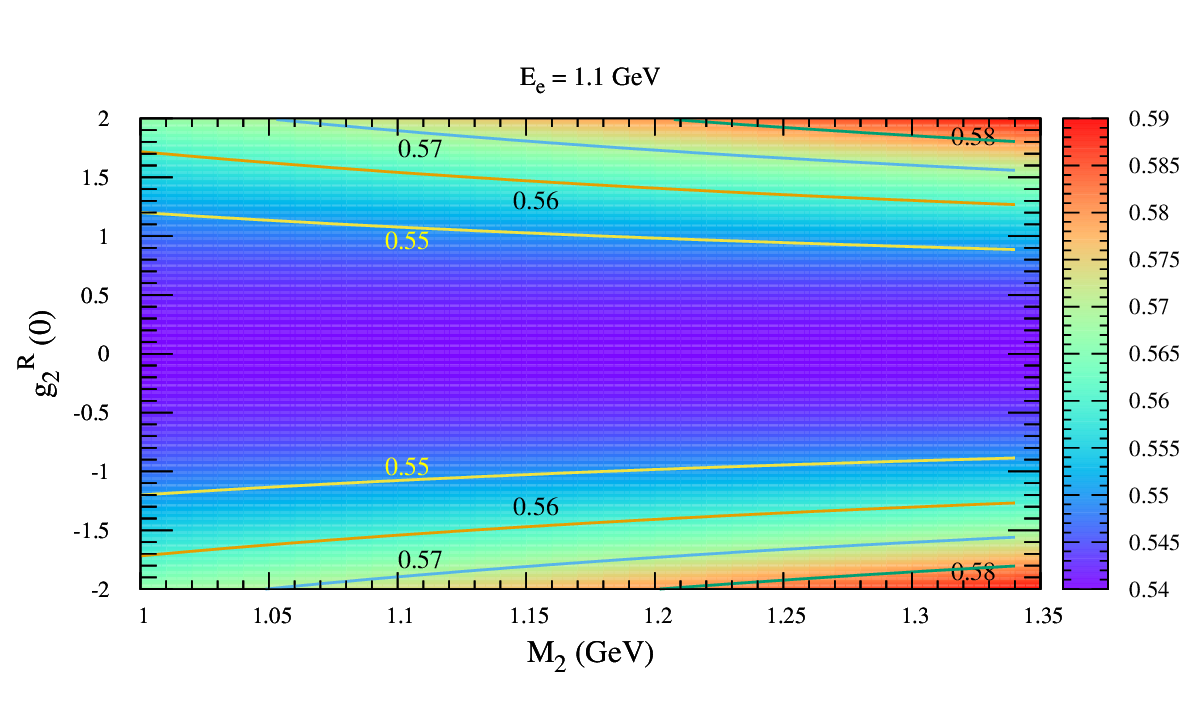}
\caption{$g_2^R(0)-M_A$~(left panel) and $g_2^R(0)-M_2$~(right panel) correlations for $\sigma(E_e)$ for the process $e^- + p \longrightarrow \nu_{e} + n$ at $E_{e}=1.1$~GeV. The solid contour lines are drawn for constant $\sigma(E_e)$~($\times 10^{-38}$~cm$^2$) and their values being quoted in the plot.}\label{sigma:g2:MA:correlation}
\end{figure}

 \begin{figure}
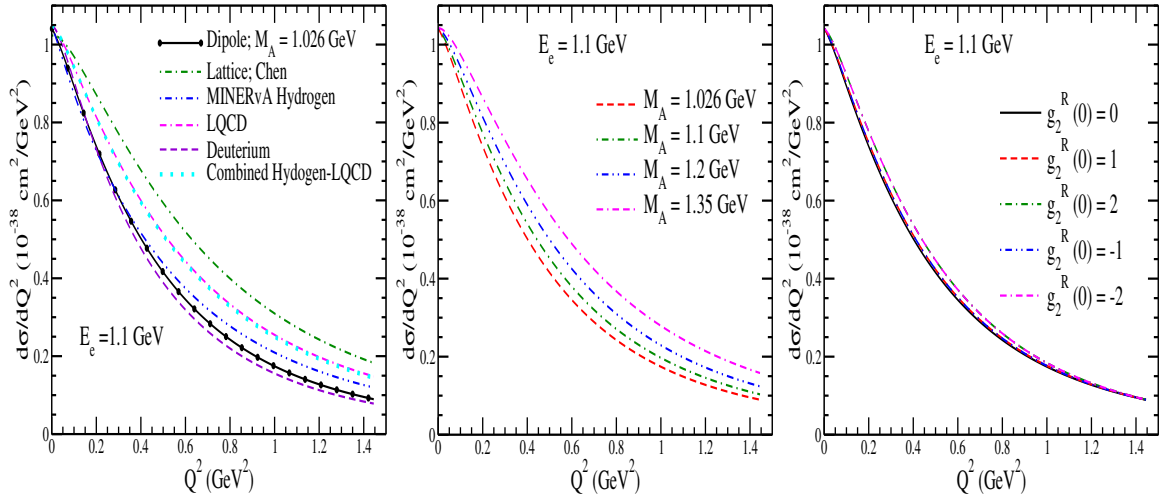
 
\begin{center}
\includegraphics[width=5cm,height=6.5cm]{dsigma_dq2_ga_variation_Ee_11GeV.eps}
\includegraphics[width=5cm,height=6.5cm]{dsigma_dQ2_MA_variation_proton_polarized_Ee_11GeV.eps}
\includegraphics[width=5cm,height=6.5cm]{dsigma_dQ2_g2_variation_proton_polarized_Ee_11GeV.eps}
\caption{$\frac{d\sigma}{dQ^2}$ as a function of $Q^2$ for the process $e^- + p \longrightarrow \nu_{e} + n$ using different parametrizations of $g_1(Q^2)$~(left panel), different values of $M_{A}$~(middle panel), and different values of $g_2^R(0)$~(right panel) at $E_{e}=1.1$~GeV. Lines and points have the same meaning as in Fig.~\ref{sigma:ga:nucleon}.}\label{dsigma:gA:nucleon}
\end{center}
\end{figure}

\begin{figure} 
\begin{center}
\includegraphics[width=7.5cm,height=6.5cm]{Pl_ga_variation.eps}
\includegraphics[width=7.5cm,height=6.5cm]{Pp_ga_variation.eps}
\caption{ $A_{L} (E_e)$~(left panel) and $A_{P} (E_e)$~(right panel) as a function of $E_{e}$ for the process $e^- + \vec{p} \longrightarrow \nu_{e} + n$, when the initial proton is polarized using different parametrizations of $g_{1}(Q^2)$. Lines and points have the same meaning as in Fig.~\ref{sigma:ga:nucleon}.}\label{PlPp:Ee:gA}
\end{center}
\end{figure}

\begin{figure} 
\begin{center}
\includegraphics[width=7.5cm,height=6.5cm]{Pl_Ee_MA_variation_polarized_proton.eps}\includegraphics[width=7.5cm,height=6.5cm]{Pp_Ee_MA_variation_polarized_proton.eps}
\caption{ $A_{L} (E_e)$~(left panel) and $A_{P} (E_e)$~(right panel) as a function of $E_{e}$ for the process $e^- + \vec{p} \longrightarrow \nu_{e} + n$, when the initial proton is polarized using different values of $M_{A}$. Lines and points have the same meaning as in Fig.~\ref{sigma:ga:nucleon}.}\label{PlPp:Ee:MA}
\end{center}
\end{figure}

\begin{figure} 
\begin{center}
\includegraphics[width=7.5cm,height=6.5cm]{Pl_Ee_g2R_variation_polarized_proton.eps}
\includegraphics[width=7.5cm,height=6.5cm]{Pp_Ee_g2R_variation_polarized_proton.eps}
\caption{ $A_{L} (E_e)$~(left panel) and $A_{P} (E_e)$~(right panel) as a function of $E_{e}$ for the process $e^- + \vec{p} \longrightarrow \nu_{e} + n$, when the initial proton is polarized for different values of $g_{2}^R(0)$. Lines and points have the same meaning as in Fig.~\ref{sigma:ga:nucleon}.}\label{Pl:g2}
\end{center}
\end{figure}

\begin{figure}
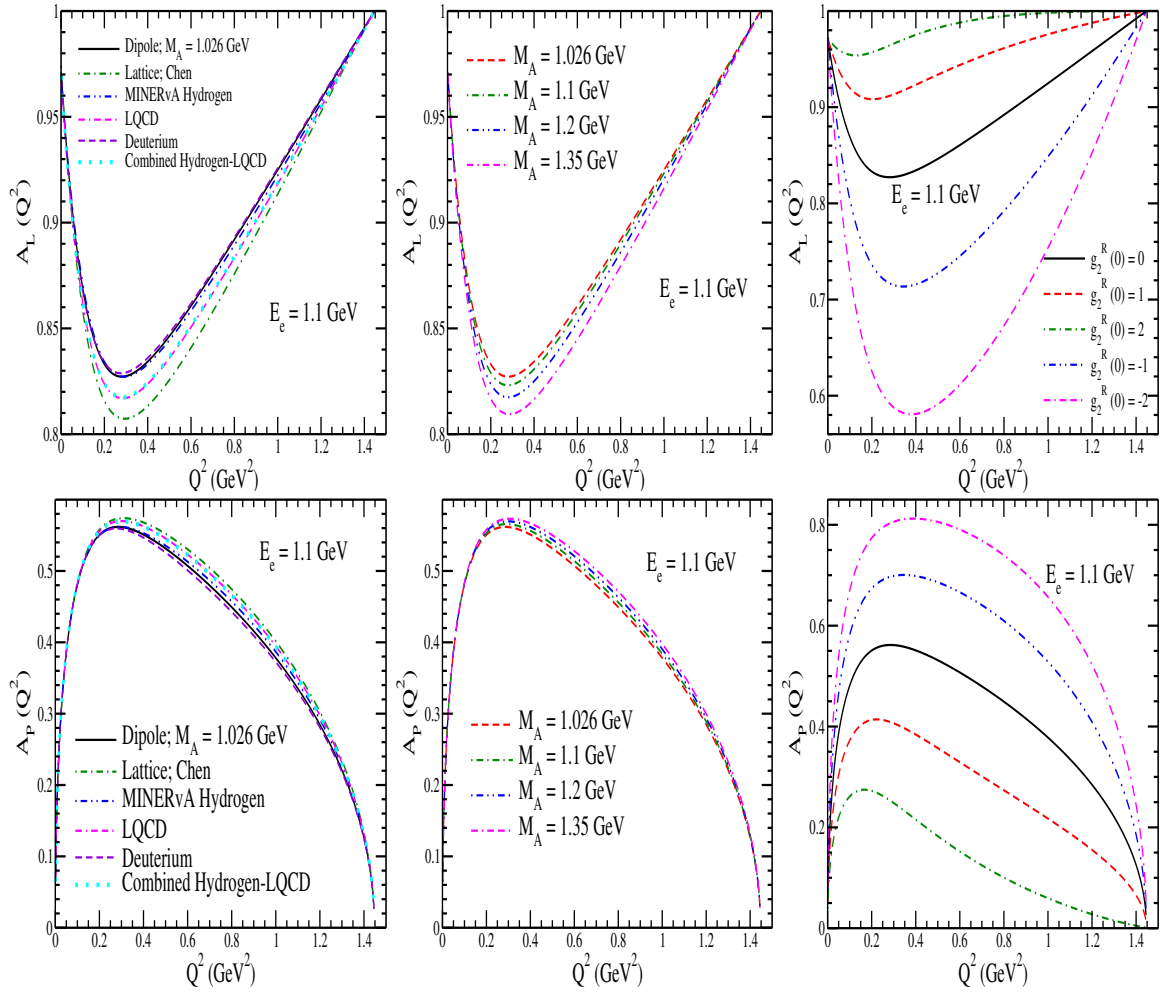
 
\begin{center}
\includegraphics[width=5cm,height=6.5cm]{Pl_q2_ga_variation_Ee_11GeV.eps}
\includegraphics[width=5cm,height=6.5cm]{PL_dQ2_MA_variation_proton_polarized_Ee_11GeV.eps}
\includegraphics[width=5cm,height=6.5cm]{PL_dQ2_g2R_variation_proton_polarized_Ee_11GeV.eps}

\includegraphics[width=5cm,height=6.5cm]{Pp_q2_ga_variation_Ee_11GeV.eps}
\includegraphics[width=5cm,height=6.5cm]{PP_dQ2_MA_variation_proton_polarized_Ee_11GeV.eps}
\includegraphics[width=5cm,height=6.5cm]{PP_dQ2_g2R_variation_proton_polarized_Ee_11GeV.eps}
\caption{$A_{L} (Q^2)$~(top panel) and $A_{P} (Q^2)$~(bottom panel) as a function of $Q^2$ at $E_{e}=1.1$~GeV for the process $e^- + \vec{p} \longrightarrow \nu_{e} + n$, when the initial proton is polarized using different parametrizations of $g_{1}(Q^2)$~(left panel), different values of $M_A$~(middle panel), and different values of $g_2^R(0)$~(right panel). Lines and points have the same meaning as in Fig.~\ref{sigma:ga:nucleon}.}\label{PlPp:q2:ga}
\end{center}
\end{figure}

\begin{figure}
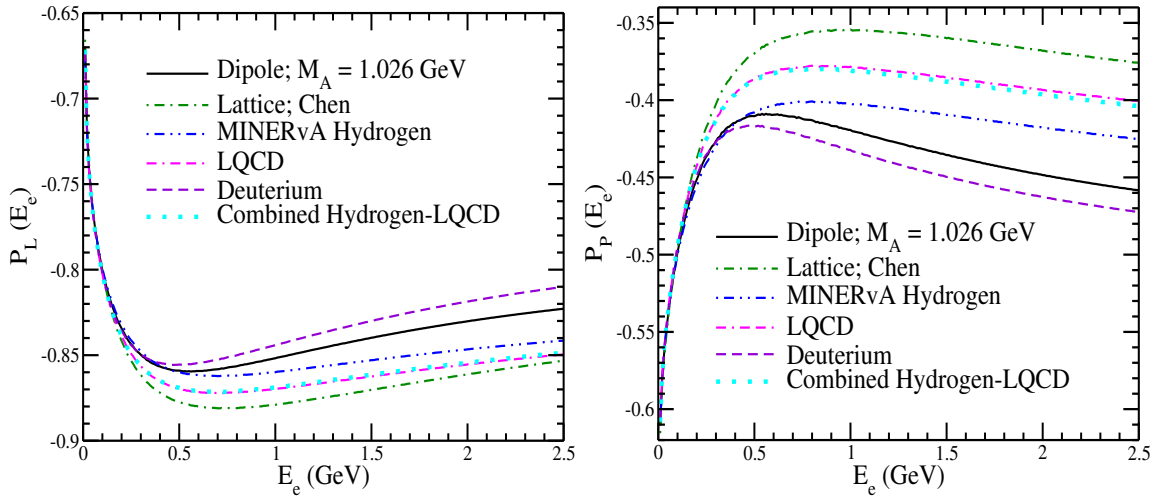
 
  \includegraphics[height=6.5cm,width=7.5cm]{PL_Ee_gA_variation_neutron_polarized.eps}
 \includegraphics[height=6.5cm,width=7.5cm]{PP_Ee_gA_variation_neutron_polarized.eps}
\caption{ $P_L (E_e)$~(left panel) and $P_P (E_e)$~(right panel) as a function of $E_{e}$ for the process $e^- + p \rightarrow \nu_e + \vec{n}$, when the final neutron is polarized, using different parametrizations of the axial vector form factor. Lines and points have the same meaning as in Fig.~\ref{sigma:ga:nucleon}.}\label{PlPp:Ee:ga:neutron}
\end{figure}

\begin{figure}
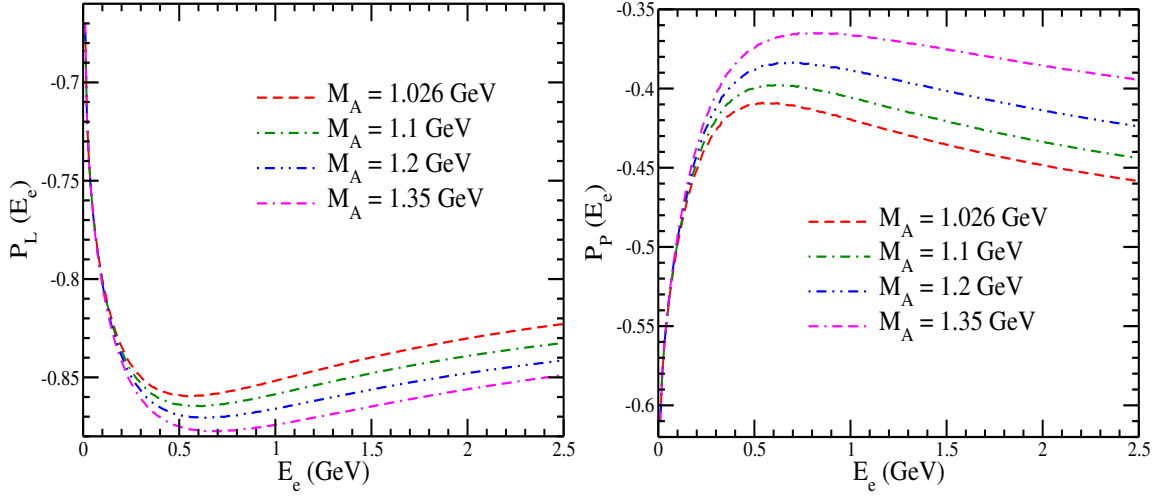
 
  \includegraphics[height=6.5cm,width=7.5cm]{PL_Ee_MA_variation_neutron_polarized.eps}
 \includegraphics[height=6.5cm,width=7.5cm]{PP_Ee_MA_variation_neutron_polarized.eps}
\caption{ $P_L (E_e)$~(left panel) and $P_P (E_e)$~(right panel) for the process $e^- + p \rightarrow \nu_e + \vec{n}$, when the final neutron is polarized, using different values  of $M_{A}$. Lines and points have the same meaning as in Fig.~\ref{sigma:ga:nucleon}. }\label{PlPp:Ee:MA:neutron}
\end{figure}

\begin{figure}
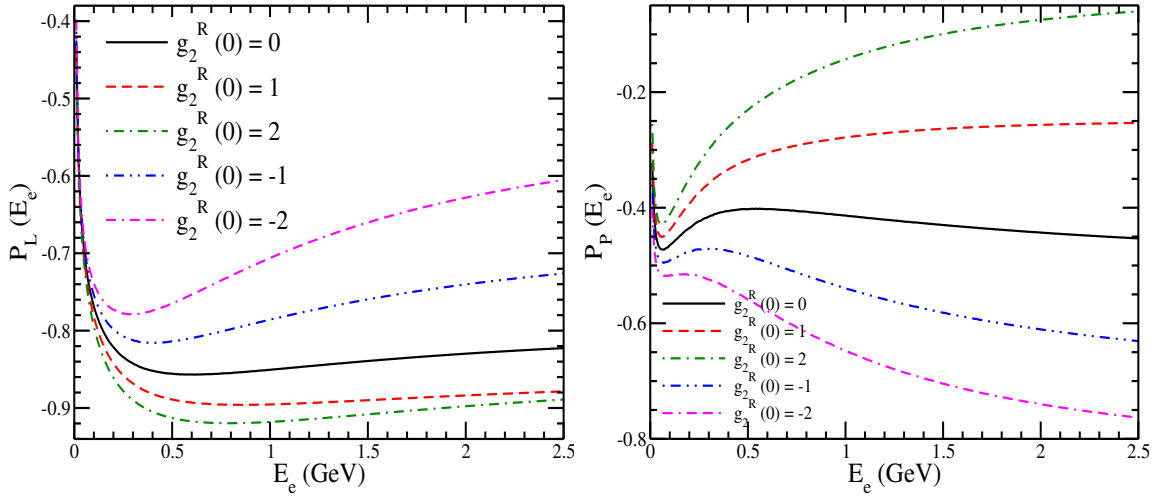
 
  \includegraphics[height=6.5cm,width=7.5cm]{PL_Ee_g2R_variation_neutron_polarized.eps}
 \includegraphics[height=6.5cm,width=7.5cm]{PP_Ee_g2R_variation_neutron_polarized.eps}
\caption{ $P_L (E_e)$~(left panel) and $P_P (E_e)$~(right panel) as a function of $E_{e}$ for the process $e^- + p \rightarrow \nu_e + \vec{n}$, when the final neutron is polarized, using different values  of $g_{2}^{R}(0)$. Lines and points have the same meaning as in Fig.~\ref{sigma:ga:nucleon}.}\label{PlPp:Ee:g2R:neutron}
\end{figure}

\begin{figure}
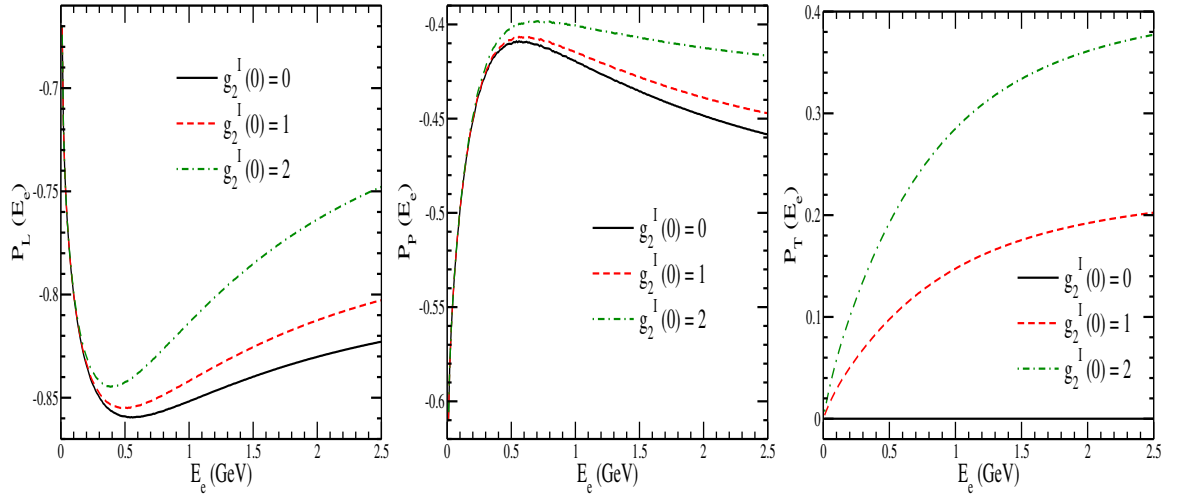
 
  \includegraphics[height=6.5cm,width=5cm]{PL_Ee_g2_variation_neutron_polarized.eps}
 \includegraphics[height=6.5cm,width=5cm]{PP_Ee_g2_variation_neutron_polarized.eps}
 \includegraphics[height=6.5cm,width=5cm]{PT_Ee_g2_variation_neutron_polarized.eps}
\caption{ $P_L (E_e)$~(left panel), $P_P (E_e)$~(middle panel), and $P_T (E_e)$~(right panel) as a function of $E_{e}$ for the process $e^- + p \rightarrow \nu_e + \vec{n}$, when the final neutron is polarized, using different values  of $g_{2}^I(0)$ viz., $g_2^{I}(0) = 0$~(solid line), 1~(dashed line), and 2~(dash-dotted line).}\label{PlPpPt:Ee:g2:neutron}
\end{figure}

 \begin{figure}
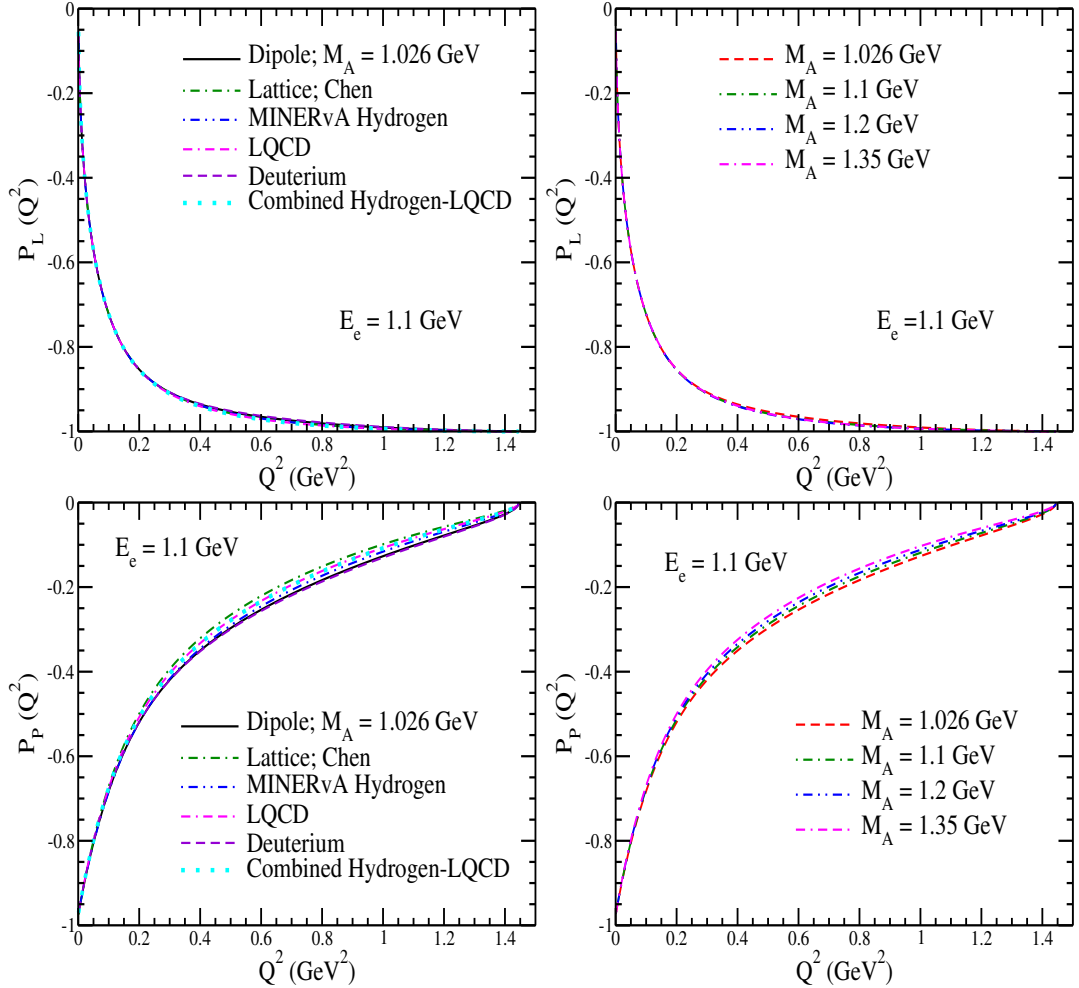
 
\begin{center}
\includegraphics[width=7cm,height=6.5cm]{PL_Q2_gA_variation_neutron_polarized_Ee_11GeV.eps}
\includegraphics[width=7cm,height=6.5cm]{PL_Q2_MA_variation_neutron_polarized_Ee_11GeV.eps}

\includegraphics[width=7cm,height=6.5cm]{PP_Q2_gA_variation_neutron_polarized_Ee_11GeV.eps}
\includegraphics[width=7cm,height=6.5cm]{PP_Q2_MA_variation_neutron_polarized_Ee_11GeV.eps}
\caption{$P_{L}(Q^2)$~(top panel) and $P_{P}(Q^2)$~(bottom panel) as a function of $Q^2$ for the process $e^- + p \longrightarrow \nu_{e} + \vec{n}$, when the final neutron is polarized, using different parametrizations of $g_1(Q^2)$~(left panel) and different values of $M_{A}$~(right panel) at $E_{e}=1.1$~GeV. Lines and points have the same meaning as in Fig.~\ref{sigma:ga:nucleon}.}\label{PlPp:q2:gA:neutron}
\end{center}
\end{figure}

 \begin{figure} 
\begin{center}
\includegraphics[width=5cm,height=6.5cm]{PL_Q2_g2R_variation_neutron_polarized_Ee_855MeV.eps}
 \includegraphics[width=5cm,height=6.5cm]{PL_Q2_g2R_variation_neutron_polarized_Ee_11GeV.eps}
 \includegraphics[width=5cm,height=6.5cm]{PL_Q2_g2R_variation_neutron_polarized_Ee_22GeV.eps}

\includegraphics[width=5cm,height=6.5cm]{PP_Q2_g2R_variation_neutron_polarized_Ee_855MeV.eps}
 \includegraphics[width=5cm,height=6.5cm]{PP_Q2_g2R_variation_neutron_polarized_Ee_11GeV.eps}
\includegraphics[width=5cm,height=6.5cm]{PP_Q2_g2R_variation_neutron_polarized_Ee_22GeV.eps}
\caption{$P_{L}(Q^2)$~(top panel) and $P_{P}(Q^2)$~(bottom panel)  as a function of $Q^2$ for the process $e^- + p \longrightarrow \nu_{e} + \vec{n}$, when the final neutron is polarized, for different values of $g_2^R(0)$ at $E_{e}=855$~MeV~(left panel), 1.1~GeV~(middle panel), and 2.2~GeV~(right panel). Lines and points have the same meaning as in Fig.~\ref{sigma:ga:nucleon}.}\label{PlPp:q2:g2R:neutron}
\end{center}
\end{figure}

 \begin{figure} 
\begin{center}
\includegraphics[width=5cm,height=6.5cm]{PL_Q2_g2_variation_neutron_polarized_Ee_855MeV.eps}
\includegraphics[width=5cm,height=6.5cm]{PL_Q2_g2_variation_neutron_polarized_Ee_11GeV.eps}
\includegraphics[width=5cm,height=6.5cm]{PL_Q2_g2_variation_neutron_polarized_Ee_22GeV.eps}

\includegraphics[width=5cm,height=6.5cm]{PP_Q2_g2_variation_neutron_polarized_Ee_855MeV.eps}
\includegraphics[width=5cm,height=6.5cm]{PP_Q2_g2_variation_neutron_polarized_Ee_11GeV.eps}
\includegraphics[width=5cm,height=6.5cm]{PP_Q2_g2_variation_neutron_polarized_Ee_22GeV.eps}

\includegraphics[width=5cm,height=6.5cm]{PT_Q2_g2_variation_neutron_polarized_Ee_855MeV.eps}
\includegraphics[width=5cm,height=6.5cm]{PT_Q2_g2_variation_neutron_polarized_Ee_11GeV.eps}
\includegraphics[width=5cm,height=6.5cm]{PT_Q2_g2_variation_neutron_polarized_Ee_22GeV.eps}
\caption{$P_{L}(Q^2)$~(top panel), $P_{P}(Q^2)$~(middle panel), and $P_{T} (Q^2)$~(bottom panel)  as a function of $Q^2$ for the process $e^- + p \longrightarrow \nu_{e} + \vec{n}$, when the final neutron is polarized, using different values of $g_2^I(0)$ at $E_{e}=855$~MeV~(left panel), 1.1~GeV~(middle panel), and 2.2~GeV~(right panel). Lines and points have the same meaning as in Fig.~\ref{PlPpPt:Ee:g2:neutron}.}\label{PlPpPt:q2:g2:neutron}
\end{center}
\end{figure}

\begin{figure}
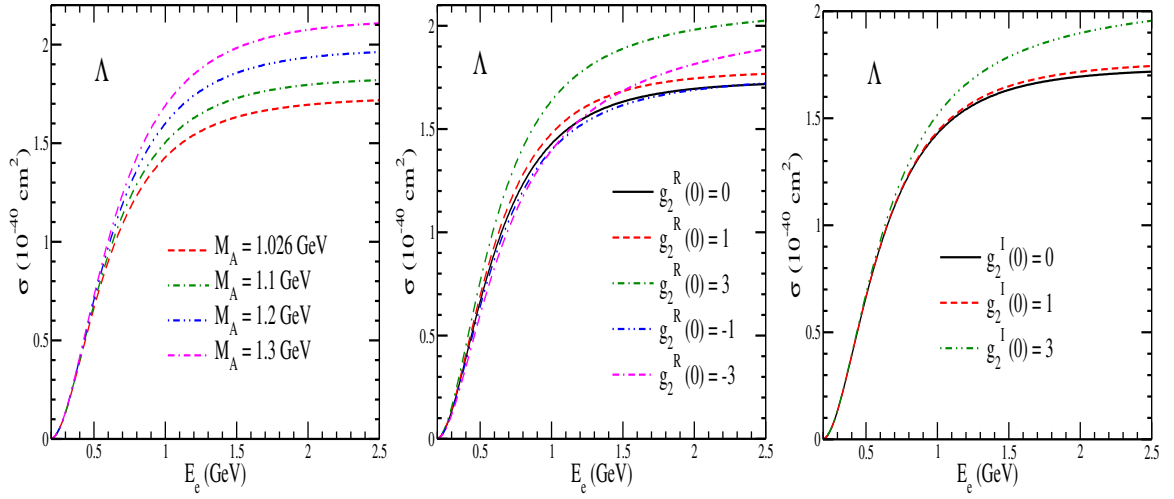
 
\begin{center}
\includegraphics[width=5cm,height=6.5cm]{total_sigma_lambda_MA_variation.eps}
\includegraphics[width=5cm,height=6.5cm]{total_sigma_g2R_variation_Lambda.eps}
\includegraphics[width=5cm,height=6.5cm]{total_sigma_lambda_g2_variation.eps}
\caption{$\sigma(E_e)$ as a function of $E_{e}$ for the process $e^- + p \longrightarrow \nu_{e} + \Lambda$. (Left panel) shows the results for $\sigma(E_e)$ using different values of $M_{A}$ viz. $M_{A} =1.026$~GeV~(dashed line), 1.1~GeV~(dash-dotted line), 1.2~GeV~(double-dot-dashed line), and 1.3~GeV~(double-dash-dotted line). (Middle panel) shows the results for $\sigma(E_e)$ obtained using different values of $g_{2}(0)$ assuming T-invariance viz. $g_{2}^{R} (0)=0$~(solid line), +1~(dashed line), +3~(dash-dotted line), $-1$~(double-dot-dashed line), and $-3$~(double-dash-dotted line). (Right panel) shows the results for $\sigma(E_e)$ obtained using different values of $g_{2}(0)$ assuming T-violation viz. $g_{2}^{I} (0)=0$~(solid line), 1~(dashed line), and 3~(dash-dotted line).}\label{sigma:Lambda:g2I}
\end{center}
\end{figure}

\begin{figure} 
 \includegraphics[width=5cm,height=6.5cm]{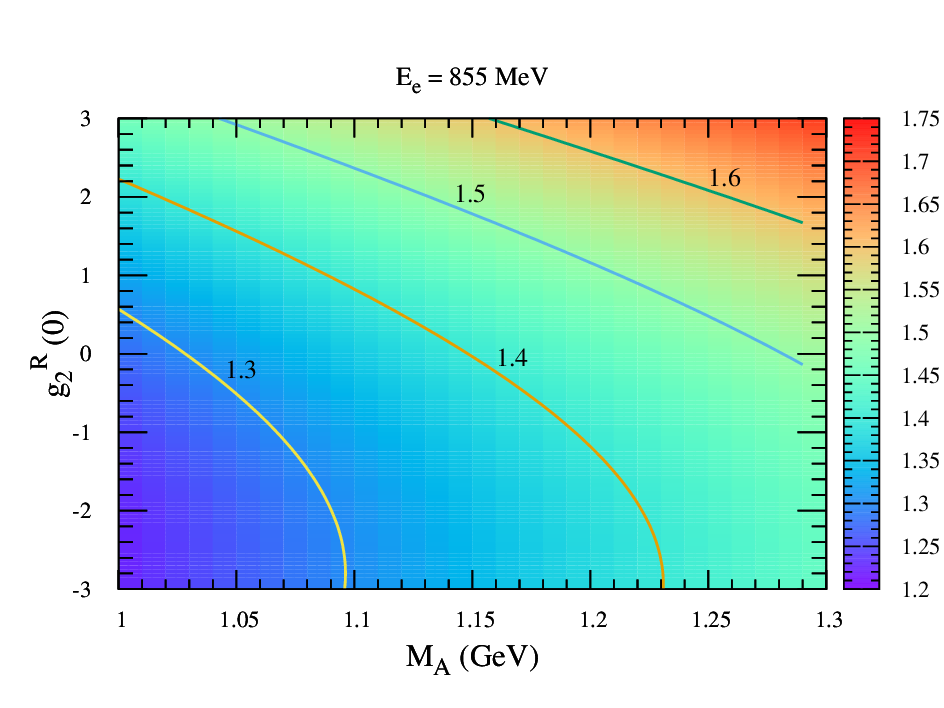} 
 \includegraphics[width=5cm,height=6.5cm]{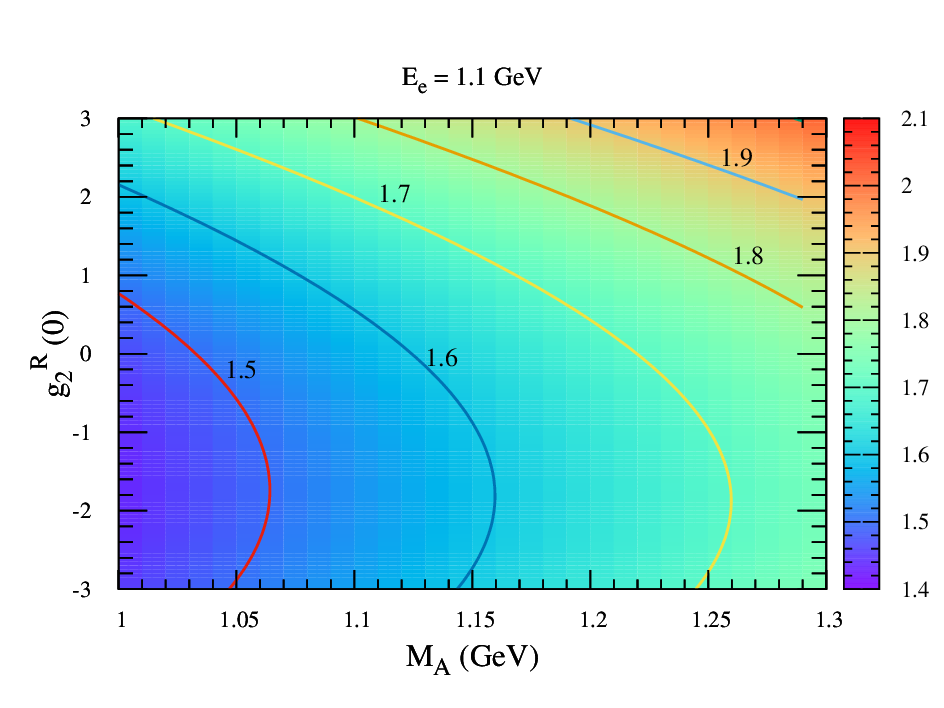}
  \includegraphics[width=5cm,height=6.5cm]{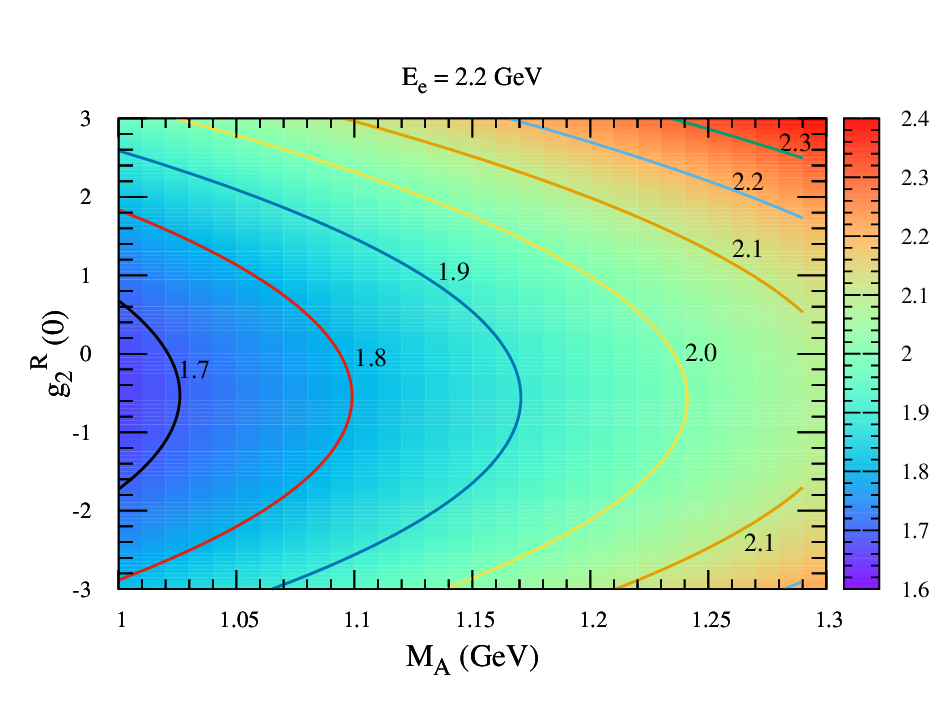}
\caption{$g_2^R(0)-M_A$ correlation for $\sigma(E_e)$ for the process $e^- + p \longrightarrow \nu_{e} + \Lambda$ at $E_{e}=855$~MeV~(left panel), 1.1~GeV~(middle panel), and 2.2~GeV~(right panel). The solid contour lines are drawn for constant $\sigma(E_e)$~($\times 10^{-40}$~cm$^2$) and their corresponding values are indicated in the plot.}\label{sigma:g2:MA:correlation:Lambda}
\end{figure}

\begin{figure}
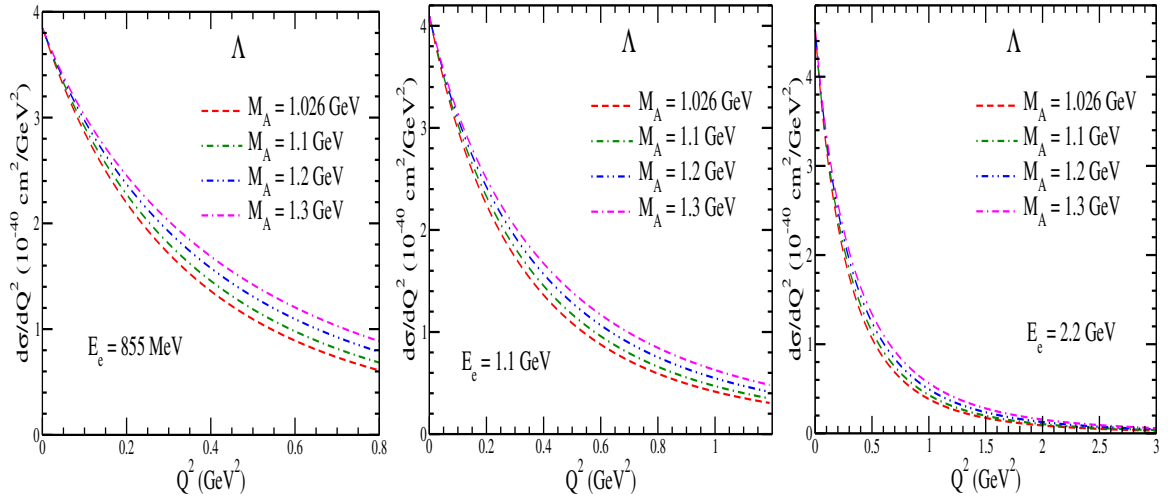
 
\begin{center}
\includegraphics[width=5cm,height=6.5cm]{dsigma_dQ2_Lambda_MA_variation_Ee_855MeV.eps}
\includegraphics[width=5cm,height=6.5cm]{dsigma_dQ2_Lambda_MA_variation_Ee_11GeV.eps}
\includegraphics[width=5cm,height=6.5cm]{dsigma_dQ2_Lambda_MA_variation_Ee_22GeV.eps} 
\caption{$\frac{d\sigma}{dQ^2}$ as a function of $Q^2$ for the process $e^- + p \longrightarrow \nu_{e} + \Lambda$ for different values of $M_A$ at $E_{e}=855$~MeV~(left panel), 1.1~GeV~(middle panel), and 2.2~GeV~(right panel). Lines and points have the same meaning as in Fig.~\ref{sigma:Lambda:g2I}.}\label{dsigma:dQ2:lambda:MA}
\end{center}
\end{figure}

 \begin{figure}
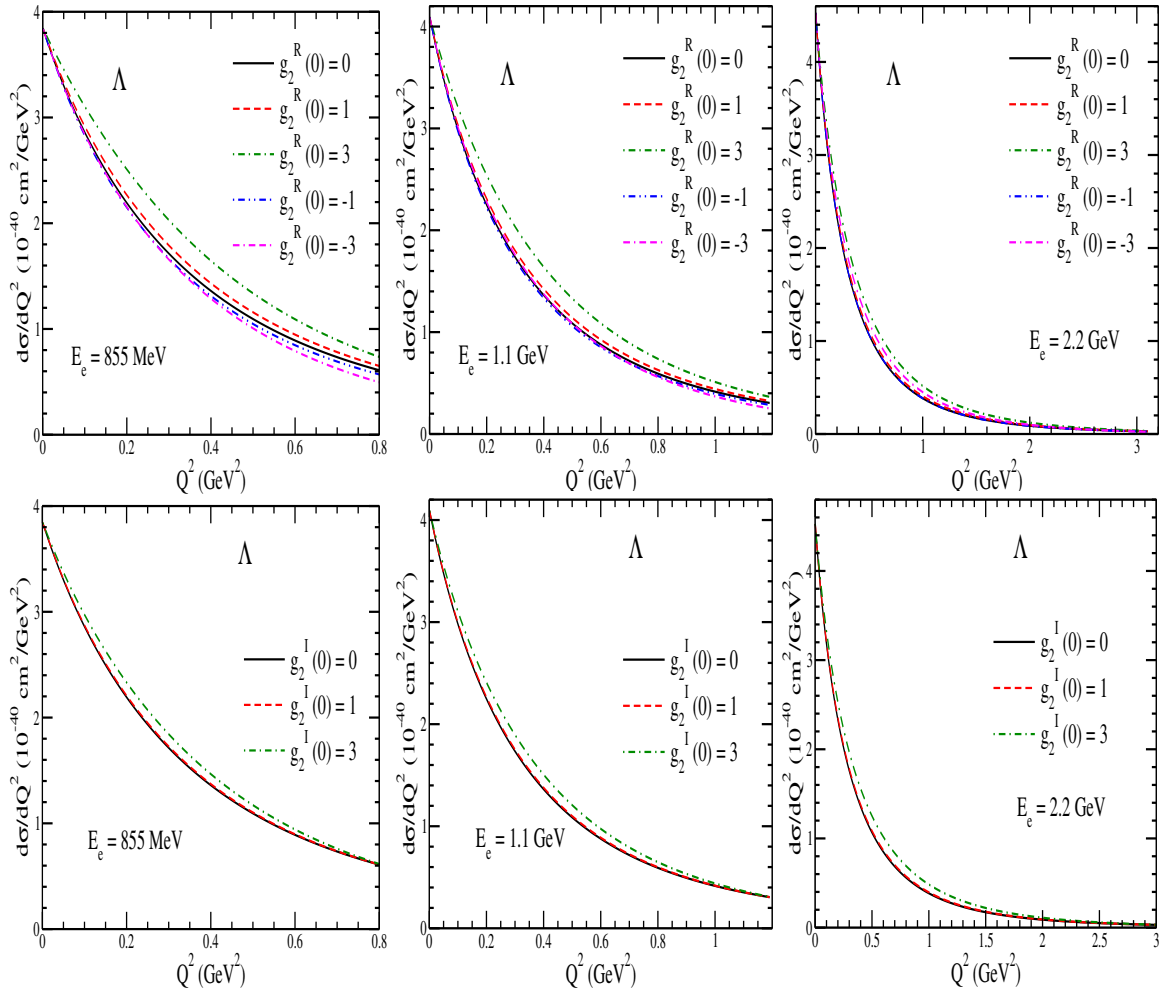
 
\begin{center}
\includegraphics[width=5cm,height=6.5cm]{dsigma_q2_g2R_variation_lambda_Ee_855MeV.eps}
\includegraphics[width=5cm,height=6.5cm]{dsigma_q2_g2R_variation_lambda_Ee_11GeV.eps}
\includegraphics[width=5cm,height=6.5cm]{dsigma_q2_g2R_variation_lambda_Ee_22GeV.eps} 

\includegraphics[width=5cm,height=6.5cm]{dsigma_dQ2_Lambda_g2_variation_Ee_855MeV.eps}
\includegraphics[width=5cm,height=6.5cm]{dsigma_dQ2_Lambda_g2_variation_Ee_11GeV.eps}
\includegraphics[width=5cm,height=6.5cm]{dsigma_dQ2_Lambda_g2_variation_Ee_22GeV.eps} 
\caption{$\frac{d\sigma}{dQ^2}$ as a function of $Q^2$ for the process $e^- + p \longrightarrow \nu_{e} + \Lambda$ using different values of $g_2^R(0)$~(top panel) and $g_2^I(0)$~(bottom panel) at $E_{e}=855$~MeV~(left panel), 1.1~GeV~(middle panel), and 2.2~GeV~(right panel). Lines and points have the same meaning as in Fig.~\ref{sigma:Lambda:g2I}.}\label{dsigma:dQ2:lambda:g2}
\end{center}
\end{figure}

\begin{figure} 
\begin{center}
\includegraphics[width=7cm,height=6.5cm]{AL_Ee_MA_variation_lambda.eps}
\includegraphics[width=7cm,height=6.5cm]{AP_Ee_MA_variation_lambda.eps}
\caption{ $A_{L} (E_e)$~(left panel) and $A_{P} (E_e)$~(right panel) as a function of $E_{e}$ for the process $e^- + \vec{p} \longrightarrow \nu_{e} + \Lambda$, when the initial proton is polarized, for different values of $M_{A}$. Lines and points have the same meaning as in Fig.~\ref{sigma:Lambda:g2I}.}\label{AlAp:Ee:MA:Lambda}
\end{center}
\end{figure}

\begin{figure} 
\begin{center}
\includegraphics[width=7cm,height=6.5cm]{AL_Ee_g2R_variation_lambda.eps}
\includegraphics[width=7cm,height=6.5cm]{AP_Ee_g2R_variation_lambda.eps}
\caption{ $A_{L} (E_e)$~(left panel) and $A_{P} (E_e)$~(right panel) as a function of $E_{e}$ for the process $e^- + \vec{p} \longrightarrow \nu_{e} + \Lambda$, when the initial proton is polarized, for different values of $g_2^R(0)$. Lines and points have the same meaning as in Fig.~\ref{sigma:Lambda:g2I}.}\label{AlAp:Ee:g2R:Lambda}
\end{center}
\end{figure}

\begin{figure} 
\begin{center}
\includegraphics[width=5cm,height=6.5cm]{AL_Q2_MA_variation_lambda_Ee_855MeV.eps}
\includegraphics[width=5cm,height=6.5cm]{AL_Q2_MA_variation_lambda_Ee_11GeV.eps}
\includegraphics[width=5cm,height=6.5cm]{AL_Q2_MA_variation_lambda_Ee_22GeV.eps}

\includegraphics[width=5cm,height=6.5cm]{AP_Q2_MA_variation_lambda_Ee_855MeV.eps}
\includegraphics[width=5cm,height=6.5cm]{AP_Q2_MA_variation_lambda_Ee_11GeV.eps}
\includegraphics[width=5cm,height=6.5cm]{AP_Q2_MA_variation_lambda_Ee_22GeV.eps}
\caption{ $A_{L} (Q^2)$~(top panel) and $A_{P} (Q^2)$~(bottom panel) as a function of $Q^2$ for the process $e^- + \vec{p} \longrightarrow \nu_{e} + \Lambda$, when the initial proton is polarized, for different values of $M_{A}$ at $E_e=855$~MeV~(left panel), 1.1~GeV~(middle panel), and 2.2~GeV~(right panel). Lines and points have the same meaning as in Fig.~\ref{sigma:Lambda:g2I}.}\label{AlAp:Q2:MA:Lambda}
\end{center}
\end{figure}

\begin{figure} 
\begin{center}
\includegraphics[width=5cm,height=6.5cm]{AL_Q2_g2R_variation_lambda_Ee_855MeV.eps}
\includegraphics[width=5cm,height=6.5cm]{AL_Q2_g2R_variation_lambda_Ee_11GeV.eps}
\includegraphics[width=5cm,height=6.5cm]{AL_Q2_g2R_variation_lambda_Ee_22GeV.eps}

\includegraphics[width=5cm,height=6.5cm]{AP_Q2_g2R_variation_lambda_Ee_855MeV.eps}
\includegraphics[width=5cm,height=6.5cm]{AP_Q2_g2R_variation_lambda_Ee_11GeV.eps}
\includegraphics[width=5cm,height=6.5cm]{AP_Q2_g2R_variation_lambda_Ee_22GeV.eps}
\caption{ $A_{L} (Q^2)$~(top panel) and $A_{P} (Q^2)$~(bottom panel) as a function of $Q^2$ for the process $e^- + \vec{p} \longrightarrow \nu_{e} + \Lambda$, when the initial proton is polarized, for different values of $g_2^R(0)$ at $E_e=855$~MeV~(left panel), 1.1~GeV~(middle panel), and 2.2~GeV~(right panel). Lines and points have the same meaning as in Fig.~\ref{sigma:Lambda:g2I}.}\label{AlAp:Q2:g2R:Lambda}
\end{center}
\end{figure}

\begin{figure} 
  \includegraphics[height=6.5cm,width=7cm]{PL_Ee_lambda_MA_variation.eps} 
 \includegraphics[height=6.5cm,width=7cm]{PP_Ee_lambda_MA_variation.eps}
\caption{ $P_L (E_e)$~(left panel) and $P_P (E_e)$~(right panel) as a function of $E_{e}$ for the process $e^- + p \rightarrow \nu_e + \vec{\Lambda}$, when $\Lambda$ is polarized, using different values  of $M_{A}$. Lines and points have the same meaning as in Fig.~\ref{sigma:Lambda:g2I}.}\label{PLPP:Ee:MA:Lambda}
\end{figure}

\begin{figure} 
 \includegraphics[height=6.5cm,width=7cm]{PL_Ee_lambda_g2R_variation.eps}
 \includegraphics[height=6.5cm,width=7cm]{PP_Ee_lambda_g2R_variation.eps}
\caption{ $P_L (E_e)$~(left panel) and $P_P (E_e)$~(right panel) as a function of $E_e$ for the process $e^- + p \rightarrow \nu_e + \vec{\Lambda}$, when $\Lambda$ is polarized, using different values  of $g_{2}^R(0)$. Lines and points have the same meaning as in Fig.~\ref{sigma:Lambda:g2I}.}\label{PLPP:Ee:g2R:Lambda}
\end{figure}

\begin{figure} 
  \includegraphics[height=6.5cm,width=5cm]{PL_Ee_lambda_g2_variation.eps} 
 \includegraphics[height=6.5cm,width=5cm]{PP_Ee_lambda_g2_variation.eps}
 \includegraphics[height=6.5cm,width=5cm]{PT_Ee_lambda_g2_variation.eps}
\caption{ $P_L (E_e)$~(left panel), $P_P (E_e)$~(middle panel), and $P_T(E_e)$~(right panel) as a function of $E_e$ for the process $e^- + p \rightarrow \nu_e + \vec{\Lambda}$, when $\Lambda$ is polarized, using different values  of $g_{2}^I(0)$. Lines and points have the same meaning as in Fig.~\ref{sigma:Lambda:g2I}.}\label{PLPPPT:Ee:g2I:Lambda}
\end{figure}

\begin{figure} 
\begin{center}
\includegraphics[width=7cm,height=6.5cm]{PL_Q2_Lambda_MA_variation_Ee_11GeV.eps}
\includegraphics[width=7cm,height=6.5cm]{PP_Q2_Lambda_MA_variation_Ee_11GeV.eps}
\caption{$P_{L}(Q^2)$~(left panel) and $P_{P}(Q^2)$~(right panel) as a function of $Q^2$ for the process $e^- + p \longrightarrow \nu_{e} + \vec{\Lambda}$, when $\Lambda$ is polarized, for different values of $M_A$ at $E_{e}=1.1$~GeV. Lines and points have the same meaning as in Fig.~\ref{sigma:Lambda:g2I}.}\label{PLPP:Q2:MA:Lambda}
\end{center}
\end{figure}

 \begin{figure} 
\begin{center}
\includegraphics[width=5cm,height=6.5cm]{Pl_q2_g2R_variation_lambda_Ee_855MeV.eps}
 \includegraphics[width=5cm,height=6.5cm]{Pl_q2_g2R_variation_lambda_Ee_11GeV.eps}
 \includegraphics[width=5cm,height=6.5cm]{Pl_q2_g2R_variation_lambda_Ee_22GeV.eps}

\includegraphics[width=5cm,height=6.5cm]{Pp_q2_g2R_variation_lambda_Ee_855MeV.eps}
\includegraphics[width=5cm,height=6.5cm]{Pp_q2_g2R_variation_lambda_Ee_11GeV.eps}
\includegraphics[width=5cm,height=6.5cm]{Pp_q2_g2R_variation_lambda_Ee_22GeV.eps}
\caption{$P_{L}(Q^2)$~(top panel) and $P_{P}(Q^2)$~(bottom panel) as a function of $Q^2$ for the process $e^- + p \longrightarrow \nu_{e} + \vec{\Lambda}$, when $\Lambda$ is polarized, for different values of $g_2^R(0)$ at $E_{e}=855$~MeV~(left panel), 1.1~GeV~(middle panel), and 2.2~GeV~(right panel). Lines and points have the same meaning as in Fig.~\ref{sigma:Lambda:g2I}.}\label{PLPP:Q2:g2R:Lambda}
\end{center}
\end{figure}

 \begin{figure} 
\begin{center}
\includegraphics[width=5cm,height=6.5cm]{PL_Q2_Lambda_g2_variation_Ee_855MeV.eps}
\includegraphics[width=5cm,height=6.5cm]{PL_Q2_Lambda_g2_variation_Ee_11GeV.eps}
\includegraphics[width=5cm,height=6.5cm]{PL_Q2_Lambda_g2_variation_Ee_22GeV.eps}

\includegraphics[width=5cm,height=6.5cm]{PP_Q2_Lambda_g2_variation_Ee_855MeV.eps}
\includegraphics[width=5cm,height=6.5cm]{PP_Q2_Lambda_g2_variation_Ee_11GeV.eps}
\includegraphics[width=5cm,height=6.5cm]{PP_Q2_Lambda_g2_variation_Ee_22GeV.eps}

\includegraphics[width=5cm,height=6.5cm]{PT_Q2_Lambda_g2_variation_Ee_855MeV.eps}
\includegraphics[width=5cm,height=6.5cm]{PT_Q2_Lambda_g2_variation_Ee_11GeV.eps}
\includegraphics[width=5cm,height=6.5cm]{PT_Q2_Lambda_g2_variation_Ee_22GeV.eps}
\caption{$P_{L}(Q^2)$~(top panel), $P_{P}(Q^2)$~(middle panel), and $P_{T} (Q^2)$~(bottom panel)  as a function of $Q^2$ for the process $e^- + p \longrightarrow \nu_{e} + \Lambda$, when $\Lambda$ is polarized, for different values of $g_2^I(0)$ at $E_{e}=855$~MeV~(left panel), 1.1~GeV~(middle panel), and 2.2~GeV~(right panel). Lines and points have the same meaning as in Fig.~\ref{sigma:Lambda:g2I}.}\label{PLPPPT:Q2:g2I:Lambda}
\end{center}
\end{figure}

\begin{figure} 
\begin{center}
 \includegraphics[width=5cm,height=6.5cm]{total_sigma_Sigma0_MA_variation.eps}
 \includegraphics[width=5cm,height=6.5cm]{total_sigma_g2R_variation_Sigma0.eps}
 \includegraphics[width=5cm,height=6.5cm]{total_sigma_Sigma0_g2_variation.eps}
\caption{$\sigma(E_e)$ as a function of $E_{e}$ for the process $e^- + p \longrightarrow \nu_{e} + \Sigma^0$, for the dipole parametrization of $g_{1}(Q^2)$ with different values of $M_{A}$~(left panel), using different values of $g_2^R(0)$~(middle panel), and different values of $g_2^I(0)$~(right panel). 
Lines and points have the same meaning as in Fig.~\ref{sigma:Lambda:g2I}.}\label{sigma:MA:Sigma}
\end{center}
\end{figure}

\begin{figure}
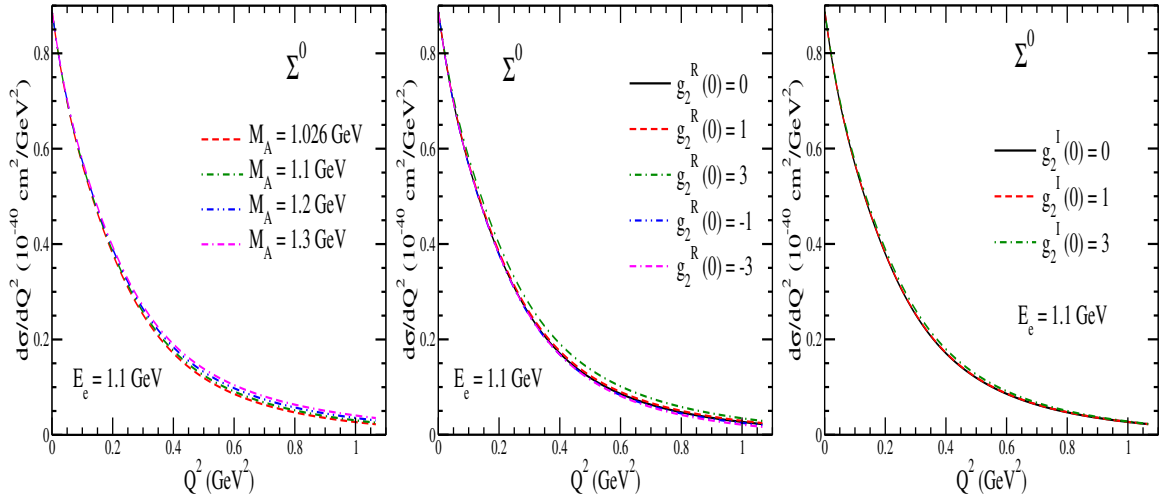
 
\begin{center}
\includegraphics[width=5cm,height=6.5cm]{dsigma_dQ2_Sigma0_MA_variation_Ee_11GeV.eps}
\includegraphics[width=5cm,height=6.5cm]{dsigma_dq2_g2R_variation_Sigma0_Ee_11GeV.eps}
\includegraphics[width=5cm,height=6.5cm]{dsigma_dQ2_Sigma0_g2_variation_Ee_11GeV.eps}
\caption{$\frac{d\sigma}{dQ^2}$ as a function of $Q^2$ for the process  $e^- + p \longrightarrow \nu_{e} + \Sigma^0$ at  $E_{e}=1.1$~GeV, for the dipole parametrization of $g_{1}(Q^2)$ with different values of $M_{A}$~(left panel), using different values of $g_2^R(0)$~(middle panel), and different values of $g_2^I(0)$~(right panel). Lines and points have the same meaning as in Fig.~\ref{sigma:Lambda:g2I}.}\label{dsigma:dQ2:MA:Sigma}
\end{center}
\end{figure}

\begin{figure} 
  \includegraphics[height=6.5cm,width=7cm]{AL_Ee_MA_variation_Sigma0.eps} 
 \includegraphics[height=6.5cm,width=7cm]{AP_Ee_MA_variation_Sigma0.eps}
\caption{ $A_L (E_e)$~(left panel) and $A_P (E_e)$~(right panel) as a function of $E_e$ for the process $e^- + \vec{p} \rightarrow \nu_e + \Sigma^0$, when the target proton is polarized, using different values  of $M_{A}$. Lines and points have the same meaning as in Fig.~\ref{sigma:Lambda:g2I}.}\label{ALAP:Ee:MA:Sigma}
\end{figure}

\begin{figure} 
  \includegraphics[height=6.5cm,width=7cm]{AL_Ee_g2R_variation_Sigma0.eps} 
 \includegraphics[height=6.5cm,width=7cm]{AP_Ee_g2R_variation_Sigma0.eps}
\caption{ $A_L (E_e)$~(left panel) and $A_P (E_e)$~(right panel) as a function of $E_e$ for the process $e^- + \vec{p} \rightarrow \nu_e + \Sigma^0$, when the target proton is polarized, using different values  of $g_2^R(0)$. Lines and points have the same meaning as in Fig.~\ref{sigma:Lambda:g2I}.}\label{ALAP:Ee:g2R:Sigma}
\end{figure}

\begin{figure} 
\begin{center}
\includegraphics[width=5cm,height=6.5cm]{AL_Q2_MA_variation_Sigma0_Ee_855MeV.eps}
\includegraphics[width=5cm,height=6.5cm]{AL_Q2_MA_variation_Sigma0_Ee_11GeV.eps}
\includegraphics[width=5cm,height=6.5cm]{AL_Q2_MA_variation_Sigma0_Ee_22GeV.eps}

\includegraphics[width=5cm,height=6.5cm]{AP_Q2_MA_variation_Sigma0_Ee_855MeV.eps}
\includegraphics[width=5cm,height=6.5cm]{AP_Q2_MA_variation_Sigma0_Ee_11GeV.eps}
\includegraphics[width=5cm,height=6.5cm]{AP_Q2_MA_variation_Sigma0_Ee_22GeV.eps}
\caption{$A_{L}(Q^2)$~(top panel) and $A_{P}(Q^2)$~(bottom panel) as a function of $Q^2$ for the process $e^- + \vec{p} \longrightarrow \nu_{e} + \Sigma^0$, when the target proton is polarized, for different values of $M_A$ at $E_{e}=855$~MeV~(left panel), 1.1~GeV~(middle panel), and 2.2~GeV~(right panel). Lines and points have the same meaning as in Fig.~\ref{sigma:Lambda:g2I}.}\label{ALAP:Q2:MA:Sigma}
\end{center}
\end{figure}

\begin{figure} 
\begin{center}
\includegraphics[width=5cm,height=6.5cm]{AL_Q2_g2R_variation_Sigma0_Ee_855MeV.eps}
\includegraphics[width=5cm,height=6.5cm]{AL_Q2_g2R_variation_Sigma0_Ee_11GeV.eps}
\includegraphics[width=5cm,height=6.5cm]{AL_Q2_g2R_variation_Sigma0_Ee_22GeV.eps}

\includegraphics[width=5cm,height=6.5cm]{AP_Q2_g2R_variation_Sigma0_Ee_855MeV.eps}
\includegraphics[width=5cm,height=6.5cm]{AP_Q2_g2R_variation_Sigma0_Ee_11GeV.eps}
\includegraphics[width=5cm,height=6.5cm]{AP_Q2_g2R_variation_Sigma0_Ee_22GeV.eps}
\caption{$A_{L}(Q^2)$~(top panel) and $A_{P}(Q^2)$~(bottom panel) as a function of $Q^2$ for the process $e^- + \vec{p} \longrightarrow \nu_{e} + \Sigma^0$, when the target proton is polarized, for different values of $g_2^R(0)$ at  $E_{e}=855$~MeV~(left panel), 1.1~GeV~(middle panel), and 2.2~GeV~(right panel). Lines and points have the same meaning as in Fig.~\ref{sigma:Lambda:g2I}.}\label{ALAP:Q2:g2R:Sigma}
\end{center}
\end{figure}

\begin{figure} 
  \includegraphics[height=6.5cm,width=7cm]{PL_Ee_Sigma0_MA_variation.eps} 
 \includegraphics[height=6.5cm,width=7cm]{PP_Ee_Sigma0_MA_variation.eps}
\caption{ $P_L (E_e) $~(left panel) and $P_P (E_e)$~(right panel) as a function of $E_e$ for the process $e^- + p \rightarrow \nu_e + \vec{\Sigma}^0$, when $\Sigma^0$ is polarized, using different values  of $M_{A}$. Lines and points have the same meaning as in Fig.~\ref{sigma:Lambda:g2I}.}\label{PLPP:Ee:MA:Sigma}
\end{figure}

\begin{figure} 
  \includegraphics[height=6.5cm,width=7cm]{Pl_g2R_variation_Sigma0.eps} 
 \includegraphics[height=6.5cm,width=7cm]{Pp_g2R_variation_Sigma0.eps}
\caption{ $P_L (E_e)$~(left panel) and $P_P (E_e)$~(right panel) as a function of $E_e$ for the process $e^- + p \rightarrow \nu_e + \vec{\Sigma}^0$, when $\Sigma^0$ is polarized, using different values  of $g_2^R(0)$. Lines and points have the same meaning as in Fig.~\ref{sigma:Lambda:g2I}.}\label{PLPP:Ee:g2R:Sigma}
\end{figure}

\begin{figure}
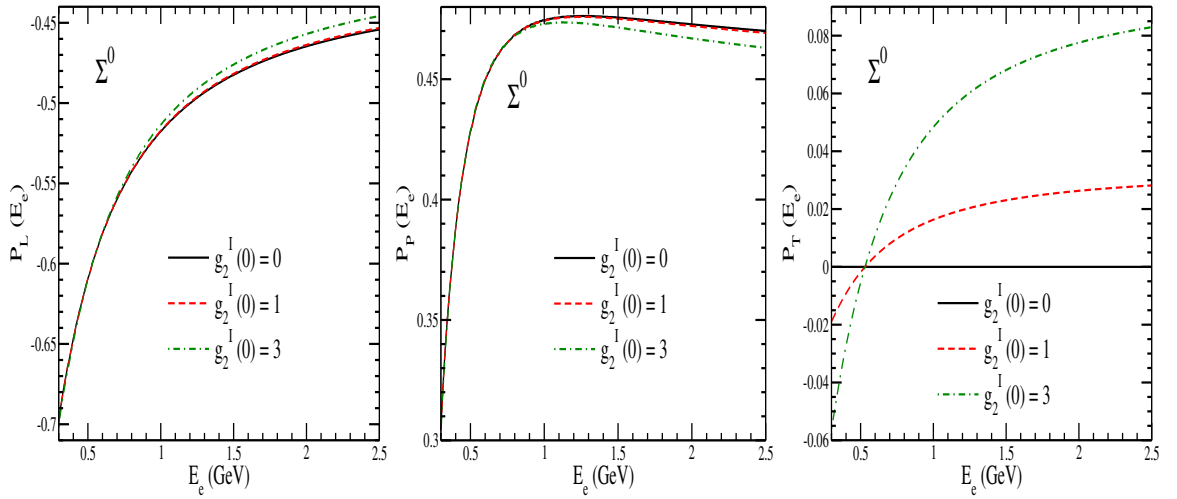
 
  \includegraphics[height=6.5cm,width=5cm]{PL_Ee_Sigma0_g2_variation.eps} 
 \includegraphics[height=6.5cm,width=5cm]{PP_Ee_Sigma0_g2_variation.eps}
 \includegraphics[height=6.5cm,width=5cm]{PT_Ee_Sigma0_g2_variation.eps}
\caption{ $P_L (E_e)$~(left panel), $P_P (E_e)$~(middle panel), and $P_T(E_e)$~(right panel) for the process $e^- + p \rightarrow \nu_e + \vec{\Sigma}^0$, when $\Sigma^0$ is polarized, using different values  of $g_{2}^I(0)$. Lines and points have the same meaning as in Fig.~\ref{sigma:Lambda:g2I}.}\label{PLPPPT:Ee:g2I:Sigma}
\end{figure}

\begin{figure} 
\begin{center}
\includegraphics[width=5cm,height=6.5cm]{PL_Q2_Sigma0_MA_variation_Ee_855MeV.eps}
\includegraphics[width=5cm,height=6.5cm]{PL_Q2_Sigma0_MA_variation_Ee_11GeV.eps}
\includegraphics[width=5cm,height=6.5cm]{PL_Q2_Sigma0_MA_variation_Ee_22GeV.eps}

\includegraphics[width=5cm,height=6.5cm]{PP_Q2_Sigma0_MA_variation_Ee_855MeV.eps}
\includegraphics[width=5cm,height=6.5cm]{PP_Q2_Sigma0_MA_variation_Ee_11GeV.eps}
\includegraphics[width=5cm,height=6.5cm]{PP_Q2_Sigma0_MA_variation_Ee_22GeV.eps}
\caption{$P_{L}(Q^2)$~(top panel) and $P_{P}(Q^2)$~(bottom panel) as a function of $Q^2$ for the process $e^- + p \longrightarrow \nu_{e} + \vec{\Sigma}^0$, when $\Sigma^0$ is polarized, for different values of $M_A$ at $E_{e}=855$~MeV~(left panel), 1.1~GeV~(middle panel), and 2.2~GeV~(right panel). Lines and points have the same meaning as in Fig.~\ref{sigma:Lambda:g2I}.}\label{PLPP:Q2:MA:Sigma}
\end{center}
\end{figure}

\begin{figure} 
\begin{center}
\includegraphics[width=5cm,height=6.5cm]{PL_q2_g2R_variation_Sigma0_Ee_855MeV.eps}
\includegraphics[width=5cm,height=6.5cm]{PL_q2_g2R_variation_Sigma0_Ee_11GeV.eps}
\includegraphics[width=5cm,height=6.5cm]{PL_q2_g2R_variation_Sigma0_Ee_22GeV.eps}

\includegraphics[width=5cm,height=6.5cm]{PP_q2_g2R_variation_Sigma0_Ee_855MeV.eps}
\includegraphics[width=5cm,height=6.5cm]{PP_q2_g2R_variation_Sigma0_Ee_11GeV.eps}
\includegraphics[width=5cm,height=6.5cm]{PP_q2_g2R_variation_Sigma0_Ee_22GeV.eps}
\caption{$P_{L}(Q^2)$~(top panel) and $P_{P}(Q^2)$~(bottom panel) as a function of $Q^2$ for the process $e^- + p \longrightarrow \nu_{e} + \vec{\Sigma}^0$, when $\Sigma^0$ is polarized, for different values of $g_2^R(0)$ at  $E_{e}=855$~MeV~(left panel), 1.1~GeV~(middle panel), and 2.2~GeV~(right panel). Lines and points have the same meaning as in Fig.~\ref{sigma:Lambda:g2I}.}\label{PLPP:Q2:g2R:Sigma}
\end{center}
\end{figure}

 \begin{figure} 
\begin{center}
\includegraphics[width=5cm,height=6.5cm]{PL_Q2_Sigma0_g2_variation_Ee_855MeV.eps}
\includegraphics[width=5cm,height=6.5cm]{PL_Q2_Sigma0_g2_variation_Ee_11GeV.eps}
\includegraphics[width=5cm,height=6.5cm]{PL_Q2_Sigma0_g2_variation_Ee_22GeV.eps}

\includegraphics[width=5cm,height=6.5cm]{PP_Q2_Sigma0_g2_variation_Ee_855MeV.eps}
\includegraphics[width=5cm,height=6.5cm]{PP_Q2_Sigma0_g2_variation_Ee_11GeV.eps}
\includegraphics[width=5cm,height=6.5cm]{PP_Q2_Sigma0_g2_variation_Ee_22GeV.eps}

\includegraphics[width=5cm,height=6.5cm]{PT_Q2_Sigma0_g2_variation_Ee_855MeV.eps}
\includegraphics[width=5cm,height=6.5cm]{PT_Q2_Sigma0_g2_variation_Ee_11GeV.eps}
\includegraphics[width=5cm,height=6.5cm]{PT_Q2_Sigma0_g2_variation_Ee_22GeV.eps}
\caption{$P_{L}(Q^2)$~(top panel), $P_{P}(Q^2)$~(middle panel), and $P_{T} (Q^2)$~(bottom panel)  as a function of $Q^2$ for the process $e^- + p \longrightarrow \nu_{e} + \vec{\Sigma}^0$, when $\Sigma^0$ is polarized, for different values of $g_2^I(0)$ at $E_{e}=855$~MeV~(left panel), 1.1~GeV~(middle panel), and 2.2~GeV~(right panel). Lines and points have the same meaning as in Fig.~\ref{sigma:Lambda:g2I}.}\label{PLPPPT:Q2:g2I:Sigma}
\end{center}
\end{figure}
\subsubsection{$e^- + p \longrightarrow \nu_e + n$: (a)~Cross sections:}\label{QE:n}
The left panel of Fig.~\ref{sigma:ga:nucleon} shows $\sigma(E_e)$ as a function of $E_e$, using different parametrizations of the axial form factor $g_1(Q^2)$. The results obtained with the standard dipole form (with $M_A=1.026$ GeV in Eq.~(\ref{g1})) are compared with the results obtained by using the lattice parametrization of Chen and Roberts~\cite{Chen:2021guo, Chen:2022odn}, the 
$z$-expansion fit done by Meyer et al.~\cite{MINERvA:2025ygc, Meyer:2026kdl} for the combined LQCD calculations of RQCD 2020~\cite{Bali:2023sdi}, NME~\cite{Park:2021ypf}, Djukanovic et al.~\cite{Djukanovic:2022wru}, ETM~\cite{Alexandrou:2023qbg}, and PNDME~\cite{Jang:2023zts}, for the experimental data of MINERvA hydrogen~\cite{MINERvA:2023avz}, for earlier deuterium experiments which have been recently reanalyzed~\cite{Meyer:2016oeg}, and for the combined LQCD and MINERvA hydrogen data~\cite{MINERvA:2025ygc}. 
It is observed that the Chen and Roberts lattice parametrization yields the largest cross section, exceeding the dipole result with $M_A=1.026$~GeV by about 27$\%$ at $E_e=700$~MeV and increasing 
to 35$\%$ and 43$\%$ at $E_e=1$ and 2~GeV, respectively. The deuterium $z$-expansion fit remains close to the dipole parametrization with $M_A=1.026$~GeV, showing a modest reduction of about $3-5\%$ in the electron energy range of $E_e < 2.5$~GeV. 
The results obtained using the MINERvA hydrogen fit agrees with the results obtained using the dipole parametrization at low energies~($\sim700$ MeV) but becomes larger with increasing $E_e$, by about 5$\%$ at 1 GeV and $\sim$10$\%$ at 2~GeV. The results obtained using LQCD-based and combined LQCD+MINERvA fits are consistent with each other and systematically exceed the results obtained using the dipole parametrization by about 15$\%$, 20$\%$, and 25$\%$ at $E_e=700$~MeV, 1~GeV, and 2~GeV, respectively.

The middle panel of Fig.~\ref{sigma:ga:nucleon} illustrates the dependence of $\sigma(E_e)$ on $M_A$ variation in the dipole parametrization, using values of $M_A$ in the range $1.026-1.35$~GeV. A clear energy dependent sensitivity is observed: $\sigma(E_e)$ increases with $M_A$, though the magnitude of this increase varies with $E_e$. For instance, at $E_e=1$ GeV, increasing $M_A$ from 1.026 to 1.1 GeV raises the cross section by $\sim$7$\%$, while a further increase of $M_A$ to 1.35 GeV leads to an enhancement in $\sigma(E_e)$ of about 20$\%$, reaching $\sim$25$\%$~($M_{A}=1.026$~GeV vs. $M_{A}=1.35$~GeV) at $E_e=2$ GeV.
Notably, a larger axial mass within the dipole form effectively reproduces results obtained from more sophisticated parametrizations. In particular, the results obtained with $M_A=1.35$ GeV closely matches with Chen et al.~\cite{Chen:2022odn} lattice QCD based results, while $M_A=1.3$ GeV mimics the results obtained with $z$-expansion fit to combined MINERvA and LQCD data. This indicates that the effects of sophisticated form factor models can, to a good approximation, be simulated within the dipole framework by an increased effective $M_A$.

The right panel of Fig.~\ref{sigma:ga:nucleon} illustrates the effect of the form factor associated with the second class current (assuming T-invariance) on the total cross section. The results are obtained by varying $g_2(Q^2)$ with real values of $g_2(0)$, i.e., $g_2^R(0)$ in the range $-2 \le g_2^R(0) \le +2$, using the dipole form with $M_2=1.026$ GeV. The cross section is identical for positive and negative values of $g_2^R(0)$ (and likewise for $g_2^I(0)$ in the case of T-violation), since it depends only on $g_2^2(Q^2)$. A modest increase in $\sigma(E_e)$ is observed with increasing $|g_2^R(0)|$. For instance, at $E_e=1$ GeV, $\sigma(E_e)$ increases by $\sim$1$\%$ for $g_2^R(0)=1$, with an additional $\sim$4$\%$ increase when $g_2^R(0)$ is increased to 2; these changes grow to $\sim$2$\%$ and 8$\%$, respectively, at $E_e=2$ GeV, which is significantly smaller than the effect of increasing $M_A$. Nevertheless, a nonzero $g_2^R(0)$ can partially mimic the effect of a larger effective $M_A$. 

We show in Fig.~\ref{sigma:g2:MA:correlation}, the correlation plots for $g_{2}^R(0)$ vs. $M_{A}$  and $g_{2}^R(0)$ vs. $M_{2}$, for the total cross section at $E_{e}=1.1$~GeV. The figure shows a strong $M_{A}$ dependence on the cross section, which is about 30\% when $M_{A}$ is varied in the range [1, 1.35]~GeV. However, the cross section { shows little dependence} on the choice of $g_{2}^R(0)$ within the range $[-2, +2]$. Therefore, it may be concluded that the presence of a non-zero value for $g_{2}^R(0)$, provided $|g_{2}^R(0)| \le 1$, will not significantly affect the determination of $M_{A}$ or $g_{1} (Q^2)$ at JLab and MAMI experiments using the total cross section measurements. Moreover, in the case of $g_{2}^R(0)$ vs. $M_{2}$ correlation plot, we have varied $g_{2}^R(0)$ and $M_2$ in the range $[-2,+2]$ and [1,1.35], respectively. It may be observed from the figure that the effect of $M_2$ variation on $\sigma(E_e)$ is even weaker than that of $g_{2}^R(0)$ variation. Thus, even a higher value of $M_2$ would not affect the determination of $M_A$ based on the total cross section $\sigma(E_e)$ vs $E_e$ measurements, in the electron energy region of about 1.1~GeV.
 
 In the left panel of Fig.~\ref{dsigma:gA:nucleon}, we present  results for the $Q^2$-distribution of the cross section i.e. $\frac{d\sigma}{dQ^2}$ as a function of $Q^2$ at $E_{e} =1.1$~GeV, using different parametrizations of $g_{1}(Q^2)$ such as the dipole parametrization with $M_{A}=1.026$~GeV, the $z$-expansion~\cite{MINERvA:2025ygc} for the MINERvA hydrogen, LQCD, deuterium, and combined hydrogen-LQCD fits, and the LQCD calculation by Chen and Roberts~\cite{Chen:2021guo, Chen:2022odn}. We have also performed the numerical calculations of $\frac{d\sigma}{dQ^2}$ at $E_{e}=855$~MeV and 2.2~GeV~(not shown in the figure), and observed similar dependence on the choice of $g_1(Q^2)$, as well as on the variations of $M_A$ and $g_2^R(0)$, as shown in Fig.~\ref{dsigma:gA:nucleon} for $E_{e}=1.1$~GeV.
The cross section is quite sensitive to the choice of the form of $g_{1}(Q^2)$ parametrization across all values of $Q^2$ considered in this work. Quantitatively, at $Q^2=1$~GeV$^{2}$, the results obtained with the deuterium data fit are smaller than those from the dipole parametrization by about 10\% across all values of electron energies. Conversely, the results obtained using the $z$-expansion fit of the MINERvA data and combined MINERvA-LQCD are larger than the results obtained with the dipole parametrization by about 20\% and 45\%, respectively, at $Q^2=1$~GeV$^{2}$. The LQCD calculations by Chen and Roberts~\cite{Chen:2021guo, Chen:2022odn} yield the largest cross section values as compared to the dipole parametrization, showing an enhancement of about 80\% 
at $Q^2=1$~GeV$^{2}$ from the dipole one. 

The middle panel of Fig.~\ref{dsigma:gA:nucleon} depicts the results for $Q^2$ distribution using different values of $M_A$ in the range $1.026-1.35$~GeV, in the dipole parametrization of the axial vector form factor~(Eq.~(\ref{g1})) at $E_{e}=1.1$~GeV. We find a strong dependence of $M_A$ on the differential scattering cross section when $M_{A}$ is varied between 1.026 and 1.35 GeV. For example, the cross section increases by about 35\% at $Q^2=0.5$~GeV$^{2}$ when $M_A$ is increased from 1.026~GeV to 1.35~GeV, which further increases with increasing $Q^2$ and becomes larger by 60\% at $Q^2=1$~GeV$^{2}$. 

The right panel of Fig.~\ref{dsigma:gA:nucleon} depicts the results for $Q^2$ distribution using different values of $g_2^R(0)$ in Eq.~(\ref{g2}) in the range $[-2,+2]$, which is associated with the second class current form factor,  at  $E_{e}=1.1$~GeV. Unlike the dependence of $\frac{d\sigma}{dQ^2}$ on the different $g_1(Q^2)$ parametrizations and $M_{A}$ variation on the cross section, we find that the $Q^2$ distribution is not much sensitive to the variation in the values of $g_{2}^R(0)$.

\subsubsection{$e^- + \vec{p} \longrightarrow \nu_e + n$: (b)~Spin asymmetries of the polarized proton target:}\label{ALAP:n}

Fig.~\ref{PlPp:Ee:gA} shows the effect of different $g_1(Q^2)$ parametrizations on the longitudinal and perpendicular components of the polarized proton asymmetry, i.e., $A_L(E_e)$ and $A_P(E_e)$. The asymmetries are largely insensitive to the choice of $g_1(Q^2)$, except for the Chen-Roberts parametrization, which yields slightly smaller values ($\lesssim 2-3\%$) for $A_L(E_e)$, particularly for $E_e \gtrsim 1.5$ GeV. The dependence on different $g_1(Q^2)$ parametrizations is even weaker for $A_P(E_e)$.

In Fig.~\ref{PlPp:Ee:MA}, the results are presented for $A_L (E_e)$ and $A_P (E_e)$ by varying $M_A$ in the range $1.026-1.35$~GeV. Little dependence on $M_A$ variation is observed, especially at higher $E_e$~($E_e \ge 1$~GeV) in the case of $A_L (E_e)$, and also for $A_{P} (E_e)$ the change is very small, when $M_A$ is varied between 1.026~GeV and 1.35~GeV.

Fig.~\ref{Pl:g2} shows $A_L(E_e)$ and $A_P(E_e)$ as a function of $E_e$ evaluated in the presence of $g_2^R(0)$ in the range $[-2,+2]$. Unlike the total cross section, both asymmetries exhibit a strong sensitivity on $g_2^R(Q^2)$ variation, with positive and negative values of $g_2^R(0)$ yielding distinct results. Opposite trends are observed: $A_L(E_e)$ increases (decreases) while $A_P(E_e)$ decreases (increases) with electron energy for $g_2^R(0)>0$~($<0$).
Quantitatively, $A_L(E_e)$ changes by $\sim$10$\%$ (20$\%$) at $E_e=1$~(2) GeV for $g_2^R(0)=+2$, and by $\sim$15$\%$ (30$\%$) for $g_2^R(0)=-2$. The effect of $g_2^R(Q^2)$ variation is more pronounced for $A_P(E_e)$, showing $\sim$60$\%$ (50$\%$) variation for $g_2^R(0)=+2$ and up to $\sim$65$\%$ (50$\%$) for $g_2^R(0)=-2$ at $E_e=1$~(2)~GeV.

In the left panel of Fig.~\ref{PlPp:q2:ga}, we present the results for $A_L(Q^2)$ and $A_{P} (Q^2)$ as a function of $Q^2$ at an electron energy of  $E_{e} = 1.1$~GeV, using different parametrizations of $g_{1}(Q^2)$ discussed in Section~\ref{sec:FF:QE}. We also perform the numerical calculations for the spin asymmetries $A_L(Q^2)$ and $A_{P} (Q^2)$ at $E_{e}=855$~MeV and 2.2~GeV~(not shown here), and observed a qualitatively similar dependence on the choice of $g_1(Q^2)$ parametrization, variation of $M_A$ in the dipole parametrization, and $g_2^R(Q^2)$ variation, to those reported at $E_{e}=1.1$~GeV. Notably, while $A_{L}(Q^2)$ shows some variation in the peak region of $A_L(Q^2)$ when different parametrizations for $g_{1}(Q^2)$ are introduced, the variations in $A_{P} (Q^2)$ are even smaller.

The middle panel of Fig.~\ref{PlPp:q2:ga} shows the results for $A_{L}(Q^2)$ and $A_{P}(Q^2)$ at $E_{e} = 1.1$~GeV obtained by varying $M_{A}$ in the range $1.026-1.35$~GeV. The longitudinal and perpendicular spin asymmetries are not much sensitive to the choice of $M_{A}$, except in the peak region of $A_{L}(Q^2)$ and $A_{P}(Q^2)$, mirroring the behaviour obtained for the different parametrizations of $g_{1} (Q^2)$.

In the right panel of Fig.~\ref{PlPp:q2:ga}, we present the results for $A_L(Q^2)$ and $A_{P} (Q^2)$ as a function of $Q^2$ at $E_{e} = 1.1$~GeV, by taking $g_2^R(0)$ in the range $[-2,+2]$. The spin asymmetries show a strong dependence on $g_{2}^R (Q^2)$ variation, which is larger in the case of $A_{P} (Q^2)$ than in the case of $A_{L} (Q^2)$.
Quantitatively, in the peak region of $A_{L} (Q^2)$, it increases by about 15\% at $E_e =$ 1.1~GeV, when $g_{2}^R (0)$ is increased from 0 to +2. When $g_2^R(0)$ is decreased from 0 to $-2$, $A_{L} (Q^2)$ decreases by about 25\% at $E_e =$ 1.1~GeV. For the perpendicular component of asymmetry, in the peak region of $A_{P}(Q^2)$, it decreases~(increases) by about 60\% at $E_e=1.1$~GeV, when $g_2^R(0)$ is increased~(decreased) from 0 to +2~($-2$).

\subsubsection{$e^- + p \longrightarrow \nu_e + \vec{n}$: (c)~Polarization observables of the final neutron:}\label{PlPp:n}

Fig.~\ref{PlPp:Ee:ga:neutron} shows the results obtained using different parametrizations of $g_{1}(Q^2)$, for the longitudinal~($P_{L}(E_{e})$) and perpendicular~($P_{P}(E_{e})$)  components of the neutron polarization.
In the electron energy region of $E_{e}>0.3$~GeV, we find that the choice of $g_1(Q^2)$ parametrization introduces a noticeable dependence on the polarization observables, which is more pronounced in the case of $P_{P} (E_{e})$ as compared to $P_{L} (E_{e})$. At $E_{e} = 0.5$~GeV, $P_{L}(E_{e})$ obtained using the parametrization of Chen and Roberts~\cite{Chen:2021guo, Chen:2022odn} is around 2\% smaller than that obtained using the dipole parametrization with $M_{A}=1.026$~GeV, and this difference further increases with increasing electron energy and becomes 4\% at $E_{e} = 2$~GeV.
In the case of $P_{P}(E_{e})$, at $E_{e}=0.5$~GeV, the results obtained with the parametrization of Chen and Roberts~\cite{Chen:2021guo, Chen:2022odn} is around 15\% larger than the results obtained with the dipole parametrization with $M_{A}=1.026$~GeV, and becomes 18\% larger at $E_{e} = 2$~GeV.

In Fig.~\ref{PlPp:Ee:MA:neutron}, we present the results for the polarization observables $P_{L} (E_e)$ and $P_{P}(E_e)$ using values of $M_A$ in the range $1.026-1.35$~GeV. The effect of $M_{A}$ variation is more pronounced in $P_{P} (E_e)$ than $P_{L} (E_e)$. Quantitatively, in the energy range of $E_{e} = 1-2$~GeV, the value of $P_{L} (E_{e})$ decreases by about $2-3\%$ when $M_{A}$ is increased from 1.026~GeV to 1.35~GeV. In contrast, we observe an increase of about $13-15\%$ in the value of $P_{P}(E_{e})$ over the same energy range when $M_{A}$ is increased from 1.026~GeV to 1.35~GeV.

In Fig.~\ref{PlPp:Ee:g2R:neutron}, we present the results for  $P_L( E_e)$ and $P_P(E_e)$ as a function of $E_e$ assuming T-invariance using different values of  $g_2^R(0)$ in the range $[-2,+2]$. It is evident that when $g_2^R (0)$ is varied from 0 to $+2$,  $P_L( E_e)$ decreases by about 10$\%$, whereas varying  $g_2^R (0)$ from 0 to $-2$ leads to a much larger change of approximately 25$\%$ increment in the value of $P_L( E_e)$. In contrast, for  $P_P (E_e)$, the trend is reversed: increasing  $g_2^R (0)$ from 0 to $+2$ results in a significant increase of about 25$\%$, while decreasing it from 0 to $-2$ reduces $P_P(E_e)$ by about 10$\%$. This clearly highlights that  $P_L( E_e)$ and $P_P( E_e)$ exhibit opposite sensitivities to the variation of $g_2^R (Q^2)$.

In Fig.~\ref{PlPpPt:Ee:g2:neutron}, we present the results for $P_{L} (E_e)$, $P_{P}(E_{e})$, and $P_{T} (E_{e})$ as a function of $E_{e}$, assuming T-violation and using $g_{2}^I (0)$ in the range [0,2]. 
 $P_{L} (E_e)$ and $P_{P}(E_{e})$ are not much sensitive to the variation in $g_2^I(Q^2)$. On the other hand, the transverse component of polarization~($P_{T} (E_{e})$), which arises due to the T-violating effect, is quite sensitive to the variation in $g_2^I(Q^2)$. 
The longitudinal and perpendicular components of the neutron polarization show an increment of about 2\% when $g_{2}^I(0)$ is increased from 0 to 2, at $E_{e} =0.5$~GeV, which increases with increasing electron energy and becomes 4\% and 8\%, respectively, at $E_{e} =1$ and 2~GeV, while $P_{T} (E_e)$ shows an increment of about 20\% at $E_{e} =0.5$~GeV, which becomes 30\% and 35\%, respectively, at $E_{e}=1$ and 2~GeV.

The left panel of Fig.~\ref{PlPp:q2:gA:neutron} shows the results for the neutron polarization components $P_{L} (Q^2)$ and $P_{P} (Q^2)$ as a function of $Q^2$ at $E_{e} = 1.1$~GeV, by using different parametrizations of $g_{1} (Q^2)$, as discussed in Sec.~\ref{sec:FF:QE}. 
 The neutron polarization components show almost no dependence on the choice of $g_1(Q^2)$ parametrization across all values of $E_{e}$ and $Q^2$ considered in this work. We observe similar results for both $P_{L} (Q^2)$ and $P_{P} (Q^2)$, when the numerical calculations are performed at $E_e=855$~MeV and 2.2~GeV~(not shown in the figure). 

In the right panel of Fig.~\ref{PlPp:q2:gA:neutron}, we present the results for $P_{L}(Q^2)$ and $P_{P}(Q^2)$ at $E_{e} =1.1$~GeV, by varying $M_{A}$ in the range $1.026-1.35$~GeV. For both
$P_{L}(Q^2)$ and $P_{P}(Q^2)$, different values of $M_{A}$ yield similar results for the polarization components across all values of $E_{e}$ and $Q^2$. 

In Fig.~\ref{PlPp:q2:g2R:neutron}, we show $P_L(Q^2)$ and $P_P(Q^2)$ as a function of $Q^2$ for $E_e = 855$ MeV, 1.1 GeV, and 2.2 GeV, assuming T-invariance by varying $g_2^R(0)$ in the range $-2 \le g_2^R(0) \le 2$. The variation in $P_P(Q^2)$ due to change in $g_2^R(Q^2)$ is more pronounced than the variation in $P_L(Q^2)$. Specifically, changing $g_2^R(0)$ from 0 to $+2$ alters $P_L(Q^2)$ by about 5$\%$, while varying it from 0 to $-2$ leads to a larger change of nearly 20$\%$, which increases with $E_e$. Additionally, the behaviour of $P_L(Q^2)$ at high $Q^2$ is modified for $g_2^R(0)=+1$ and $+2$. In contrast, $P_P(Q^2)$ varies in opposite directions for positive  and negative values of $g_2^R(0)$, with comparable magnitudes for equal $|g_2^R(0)|$, yielding changes of about $10-15\%$ and similar trends for $g_2^R(0)=+1$ and $+2$.

Fig.~\ref{PlPpPt:q2:g2:neutron} shows the neutron polarization components $P_L(Q^2)$, $P_P(Q^2)$, and $P_T(Q^2)$ versus $Q^2$ at $E_e = 855$ MeV, 1.1 GeV, and 2.2 GeV, assuming T-violation by varying $g_2^I(0)$ in the range $0 \le g_2^I(0) \le 2$. While $P_L(Q^2)$ shows moderate sensitivity to $g_2^I(Q^2)$ variation that increases with $E_e$,  $P_P(Q^2)$ remains largely unchanged. In contrast, $P_T(Q^2)$ exhibits strong dependence on $g_2^I(Q^2)$ variation, especially near the $Q^2$ peak of $P_T(Q^2)$. Quantitatively, it increases by about 30$\%$, 35$\%$, and 45$\%$ at $E_e = 855$ MeV, 1.1 GeV, and 2.2 GeV, respectively, as $g_2^I(0)$ varies from 0 to 2. This highlights $P_T(Q^2)$ as a promising observable to probe T-violation.

\subsubsection{$e^- + p \longrightarrow \nu_e + \Lambda$: (a)~Cross sections:}\label{QE:lambda}

In the left panel of Fig.~\ref{sigma:Lambda:g2I}, the results are presented  for the cross section $\sigma(E_e)$ as a function of $E_e$, and the effect of $M_{A}$ variation on $\sigma(E_e)$ is studied, by varying $M_{A}$ in the range $1.026-1.3$~GeV. 
We find a significant dependence of $M_A$ variation on $\sigma(E_e)$. Quantitatively, at $E_e=700$~MeV, we observe an enhancement of about 10\% when the value of $M_{A}$ is increased by 20\% from the world average value of $M_{A}$. This enhancement further increases with increasing electron energy and becomes 12\% and 15\%, respectively, at $E_e=1$ and 2~GeV. Moreover, an increase in $M_{A}$ by 30\% from the world average value increases $\sigma(E_e)$ by about 12\%, 18\%, and 22\%, respectively, at $E_e=700$~MeV, 1, and 2~GeV.

To illustrate the dependence of $\sigma(E_e)$ on the second class current form factor, the middle and right panels of Fig.~\ref{sigma:Lambda:g2I} show $\sigma(E_e)$ as a function of $E_e$  assuming T-invariance for $g_2^R(Q^2)$ variation by varying $g_2^R(0)$ in the range $[-3,+3]$ and  assuming T-violation for $g_2^I(Q^2)$ variation by varying $g_2^I(0)$ in the range [0,3], respectively. The middle panel indicates that $\sigma(E_e)$ is enhanced for $g_2^R(Q^2)>0$ and suppressed for $g_2^R(0)<0$. For example, $g_2^R(0)=+1$ increases $\sigma(E_e)$ by $\sim 3-4\%$, while $g_2^R(0)=+3$ yields $\sim 15-17\%$ enhancement at $E_e=1-2$~GeV. Conversely, $g_2^R(Q^2)<0$ reduces $\sigma(E_e)$ by $\sim$3\% and 7$\%$, respectively, for $g_2^R(0) = -1$ and $-3$,  at $E_e=700$~MeV. However, with increase in electron energy, this decrease in the cross section decreases and becomes $\sim 2\%$ smaller at $E_{e} =1$~GeV for both $g_2^R(0) = -1$ and $-3$. Furthermore, at $E_{e}=2$~GeV, the variation in $\sigma(E_e)$ when $g_2^R(0)$ is decreased from 0 to $-1$ is almost negligible, while the result obtained using $g_2^R(0) = -3$ is 7\% larger than the result obtained using $g_2^R(0) = 0$.

In the right panel of Fig.~\ref{sigma:Lambda:g2I}, the results are presented for $g_2^I(Q^2)$ variation by changing $g_2^I(0)$ in the range [0,3] assuming T-violation. The positive and negative values of $g_2^I(0)$ yields the same results for the cross section, as in the case of T-violation, $\sigma(E_e)$  depends only on the square of  $g_2^I(Q^2)$.
We observe that $\sigma(E_e)$ increases with increasing $g_2^I(Q^2)$. 
At low $E_e$, say around 700~MeV, $\sigma(E_e)$ increases by about 4\% when the value of $g_2^I(0)$ is increased to 3 from 0. However, this increase becomes more pronounced with increasing electron energy and becomes 6\% and 12\%, respectively, at $E_{e} =1$ and 2~GeV.

In Fig.~\ref{sigma:g2:MA:correlation:Lambda}, we present the correlation between $g_2^R(0)$ and $M_A$ by varying $g_2^R(0)$ in the range $[-3,+3]$ and $M_{A}$ in the range [1,1.35], for the total cross section at three representative electron energies, namely $E_e$ = 855 MeV, 1.1 GeV, and  2.2 GeV. The figure clearly demonstrates a strong sensitivity of the cross section to $M_A$ variation. More importantly, the cross section also exhibits a strong dependence on $g_2^R(0)$ variation over the interval $[-3,+3]$. Unlike the case of $e^- + p \longrightarrow \nu_e + n$ scattering, where $M^\prime=M_N=M$, for the reaction $e^- + p \longrightarrow \nu_e + \Lambda$, we find different values of $\sigma(E_e)$ for the positive and negative values of $g_2^R(0)$. Notably, the results display a significant effect on the cross sections using the positive and negative values of $g_2^R(0)$. This effect has important implications: the presence of a non-zero $g_2^R(0) $, particularly for positive values, can significantly bias the extraction of $M_A$ in the determination of $ g_1(Q^2)$ from total cross-section measurements. Such effects would be interesting to study in future at facilities such as JLab and MAMI.

In Fig.~\ref{dsigma:dQ2:lambda:MA}, we present the results for $\frac{d\sigma}{dQ^2}$ as a function of $Q^2$ at different value of the electron energy viz. $E_e=855$~MeV, 1.1 and 2.2~GeV obtained by varying $M_{A}$ in the range $1.026-1.3$~GeV. We observe a noticeable dependence of $M_{A}$ variation on $\frac{d\sigma}{dQ^2}$ across all values of $Q^2$ for a given $E_e$. The effect of $M_A$ is more pronounced at lower beam energies.

In Fig.~\ref{dsigma:dQ2:lambda:g2}, we present the results for $\frac{d\sigma}{dQ^2}$ as a function of $Q^2$ at different electron energies viz. $E_{e} =855$~MeV, 1.1, and 2.2~GeV, to study the effect of weak electric form factor assuming T-invariance by varying $g_2^{R} (0)$ in the range $[-3,+3]$, as well as assuming T-violation by varying $g_2^{I} (0)$ in the range [0,3]. We observe some dependence of $\frac{d\sigma}{dQ^2}$ on chosen values of $g_2^R(Q^2)$, which becomes significant in the low electron energy region of $E_e \le 1$~GeV, when $g_{2}^{R} (0) = +3$ is implemented in the numerical calculations. However, we find almost no dependence of $\frac{d\sigma}{dQ^2}$ on $g_2^I(Q^2)$ variation in the range $g_2^I(0)=[0,3]$, over the entire range of $E_e$ and $Q^2$ considered in this work.

\subsubsection{$e^- + \vec{p} \longrightarrow \nu_e + \Lambda$: (b)~Spin asymmetries of the polarized proton target:}\label{AlAp:lambda}

Fig.~\ref{AlAp:Ee:MA:Lambda} shows the results for the longitudinal and perpendicular spin asymmetries of the polarized proton target as a function of electron energy, using different values of $M_{A}$ in the range [1.026, 1.3]~GeV. We find that, in the electron energy region from threshold up to $E_e=2.5$~GeV, both $A_{L}(E_e)$ and $A_P(E_e)$ have very little dependence on  $M_{A}$ variation.

Fig.~\ref{AlAp:Ee:g2R:Lambda} shows $A_L(E_e)$ and $A_P(E_e)$ as a function of $E_e$ evaluated for variation in $g_2^R(Q^2)$ by varying $g_2^R(0)\in[-3,+3]$. Both asymmetries exhibit strong sensitivity to $g_2^R(Q^2)$, with opposite trends for positive and negative values: $A_L(E_e)$~($A_P(E_e)$) increases~(decreases) for $g_2^R(Q^2)>0$ and decreases~(increases) for $g_2^R(Q^2)<0$ relative to $g_2^R(Q^2)=0$.
Quantitatively, $A_L(E_e)$ increases by $\sim$2\% and 3$\%$, respectively, for $g_2^R(0)=+1$ and $+3$, while it decreases by $\sim$3$\%$ (10$\%$) at $E_e=1$ GeV and up to $\sim$10$\%$ (25$\%$) for $g_2^R(0)=-1$ ($-3$), at $E_e=2$ GeV. In contrast, $A_P(E_e)$ shows much larger variation, with $\sim$70$\%$ change for $g_2^R(0)=\pm1$  at $E_e=1$ GeV, reducing to $\sim$65$\%$ at 2~GeV. Moreover, this effect further increases with increasing $g_2^R(Q^2)$.

Fig.~\ref{AlAp:Q2:MA:Lambda} presents the results for $A_L(Q^2)$ and $A_P(Q^2)$ as a function of $Q^2$ at three incident electron energies, namely $E_e = 855$ MeV, 1.1 GeV, and 2.2 GeV, using $M_A$ in the range $1.026-1.3$~GeV. We observe a considerable dependence of $M_A$ variation on both $A_L(Q^2)$ and $A_P(Q^2)$ over the entire $Q^2$ and $E_e$ range, which is more pronounced for the perpendicular spin asymmetry  $A_P(Q^2)$.

In Fig.~\ref{AlAp:Q2:g2R:Lambda}, the results for $A_L(Q^2)$ and $A_P(Q^2)$ are presented as a function of $Q^2$ at $E_e = 855$ MeV, 1.1 GeV, and 2.2 GeV, using $g_2^R(0)$ in the range $[-3,+3]$. Both $A_L(Q^2)$ and $A_P(Q^2)$ show strong dependence on $g_2^R(Q^2)$ variation. Qualitatively, the two asymmetries exhibit different trends: $A_L(Q^2)$ shows larger variation for $g_2^R(Q^2)<0$ compared to $g_2^R(Q^2)>0$, whereas $A_P(Q^2)$ displays nearly symmetric behaviour for positive and negative values of $g_2^R(0)$.

\subsubsection{$e^- + p \longrightarrow \nu_e + \vec{\Lambda}$: (c)~Polarization observables of $\Lambda$:}\label{polarization:lambda}
Fig.~\ref{PLPP:Ee:MA:Lambda} shows the results for the polarization observables $P_L(E_e)$ and $P_P(E_e)$ for a polarized $\Lambda$, using $M_A=1.026-1.3$~GeV. Both polarization components are insensitive to $M_A$ variation for $E_e<0.5$ GeV, but show some dependence at higher energies, more prominently in $P_L(E_e)$.

Fig.~\ref{PLPP:Ee:g2R:Lambda} shows $P_L(E_e)$ and $P_P(E_e)$ as a function of $E_e$, assuming T-invariance, evaluated for $g_2^R(0)\in[-3,+3]$. Both polarization components exhibit a strong sensitivity on $g_2^R(Q^2)$ variation, which is more pronounced in $P_L(E_e)$. For $g_2^R(Q^2)>0$, $P_L$ decreases by $\sim$25$\%$ (60$\%$) at $E_e=1$ GeV and $\sim$35$\%$ (90$\%$) at 2 GeV for $g_2^R(0)=+1$~(+3), while $g_2^R(Q^2)<0$ leads to comparable enhancements. In contrast, $P_P (E_e)$ shows smaller changes than $P_L (E_e)$: a reduction of $\sim$10$\%$ (30$\%$) at 1 GeV and $\sim$15$\%$ (40$\%$) at 2 GeV for $g_2^R(0)=+1$~(+3), and an enhancement of $\sim 10 -20\%$ for $g_2^R(Q^2)<0$, nearly energy independent over $E_e=1-$2.5~GeV.

In Fig.~\ref{PLPPPT:Ee:g2I:Lambda}, we present the results for the longitudinal, perpendicular, and transverse components of a polarized $\Lambda$ as a function of electron energy, assuming T-violation using different values of $g_2^I(0)$ in the range $0-3$.  The negative values of $g_2^I(0)$ yield same results for $P_{L}(E_e)$ and $P_{P}(E_e)$ as those obtained using the positive values of $g_2^I(0)$, while for $P_T(E_e)$, the magnitude remains the same but an overall change of sign is observed when the negative values of $g_2^I(0)$ are used. 
Qualitatively, $P_{L}(E_e)$ and $P_{P}(E_e)$ show  little dependence on variation in the  value of $g_2^I(Q^2)$ at $E_e \ge1$~GeV; $P_{T} (E_e)$ shows a strong dependence on different values of $g_2^I(Q^2)$, which increases with increasing electron energy. For both $P_{L}(E_e)$ and $P_{P}(E_e)$, the results obtained with $g_2^I(0) = 1$ are almost identical to the results obtained with $g_2^I(0) = 0$. However, when $g_2^I(0) = 3$ is implemented, we observe some dependence in the case of $P_{L}(E_e)$~($P_{P}(E_e)$), which is about 4\%~(8\%) at $E_{e}=1$~GeV, and becomes 10\%~(13\%) at $E_{e}=2$~GeV. Furthermore, for $P_T(E_e)$, which arises due to T-violating effect, we observe an enhancement of about 10\%~(30\%) at $E_e=1$~GeV, when $g_2^I(0)=1$~(3) is used. This enhancement increases slightly with $E_e$ and becomes 15\% and 40\%, respectively,  at $E_e=2$~GeV, when the results obtained with $g_2^I(0)=1$ and 3 are compared with the results obtained with $g_2^I(0)=0$. 

In Fig.~\ref{PLPP:Q2:MA:Lambda}, we present the results for $P_{L}(Q^2)$  and $P_P(Q^2)$ as a function of $Q^2$ at $E_e=1.1$~GeV, using different values of $M_A$ in the range $1.026-1.3$~GeV. 
Both $P_{L}(Q^2)$  and $P_P(Q^2)$ are almost insensitive to the choice of $M_{A}$. We have also performed the numerical calculations for $P_{L}(Q^2)$  and $P_P(Q^2)$ at $E_{e}=855$~MeV and 2.2~GeV and observed similar results as in the case of $E_e=1.1$~GeV, thus, those results are not shown in the figure.

In Fig.~\ref{PLPP:Q2:g2R:Lambda}, we present the results for $P_{L}(Q^2)$  and $P_P(Q^2)$   as a function of $Q^2$ at $E_e=855$~MeV, 1.1 and 2.2~GeV, assuming T-invariance using various values of $g_2^R(0)$ in the range $[-3,+3]$.
The results for both $P_{L}(Q^2)$  and $P_P(Q^2)$ are quite sensitive to the choice of $g_2^R(Q^2)$ across all values of $E_e$ and $Q^2$ considered in this work.

In Fig.~\ref{PLPPPT:Q2:g2I:Lambda}, we present the results for $P_{L}(Q^2)$, $P_P(Q^2)$, and $P_T(Q^2)$  as a function of $Q^2$ at $E_e=855$~MeV, 1.1 and 2.2~GeV, assuming T-violation using different values of $g_2^I(0)$ in the range $[0,3]$. 
The results for $P_{L}(Q^2)$  and $P_P(Q^2)$ are not much sensitive to the choice of $g_2^I(Q^2)$ in the range of $E_e$ and $Q^2$ considered in this work. However, $P_T(Q^2)$ shows strong dependence on the choice of $g_2^I(Q^2)$ across all values of $E_e$ and $Q^2$. In the peak region of $P_T(Q^2)$, an increment of about 40\% and 50\%, respectively, is observed at $E_e=1.1$ and 2.2~GeV, when $g_2^I(0)$ is varied from 0 to 3.

\subsubsection{$e^- + p \longrightarrow \nu_e + \Sigma^0$: (a)~Cross sections:}\label{QE:sigma}

In Fig.~\ref{sigma:MA:Sigma}, we present the results for $\sigma(E_e)$ as a function of $E_{e}$ and study the dependence of $\sigma(E_e)$ on $M_{A}$ variation~(left panel) by varying $M_{A}$ in the range $1.026-1.3$~GeV, $g_2^R(Q^2)$ variation~(middle panel) by changing $g_2^R(0)$ in the range $[-3,+3]$, and $g_2^I(Q^2)$ variation~(right panel) by changing $g_{2}^I(0)$ in the range $0-3$. 
$\sigma(E_e)$ shows some dependence on the choice of $M_{A}$, which is more pronounced for $E_e \ge1$~GeV. 
However, $\sigma(E_e)$ remains almost completely insensitive to the variation in $g_2^R(Q^2)$ and $g_2^I(Q^2)$ across all evaluated values of $E_e$.

In Fig.~\ref{dsigma:dQ2:MA:Sigma}, the results are presented for $\frac{d\sigma}{dQ^2}$ as a function of $Q^2$ at $E_e=1.1$~GeV and the effects of $M_{A}$ variation~(left panel), $g_2^R(Q^2)$ variation~(middle panel) assuming T-invariance, and $g_2^I(Q^2)$ variation~(right panel) assuming T-violation, on the $Q^2$ distribution, are studied. 
We find that the results for $\frac{d\sigma}{dQ^2}$ are almost insensitive to the choice of $M_A$, $g_2^R(Q^2)$, and $g_2^I(Q^2)$.
We observe similar results for $\frac{d\sigma}{dQ^2}$ at $E_e=855$~MeV and 2.2 GeV~(not shown in the figure).

\subsubsection{$e^- + \vec{p} \longrightarrow \nu_e + \Sigma^0$: (b)~Spin asymmetries of the polarized proton target:}\label{AlAp:sigma}

In Fig.~\ref{ALAP:Ee:MA:Sigma}, we present the results for the spin asymmetries $A_L(E_e)$ and $A_P(E_e)$ as a function of $E_e$ using $M_A$ in the range $1.026-1.3$~GeV. The spin asymmetries show some variation when different values of $M_{A}$ are used. The effect is more pronounced in the case of $A_P(E_e)$. 
Specifically, $A_P(E_e)$ increases with increasing $M_A$ across all values of $E_e$. Quantitatively, we observe an enhancement of about 6\% in $A_P(E_e)$ at $E_e=1$~GeV when $M_A=1.2$~GeV is used  as compared to the results obtained with $M_{A}=1.026$~GeV, which further increases with the increase in the value of $M_A$. This enhancement increases with increasing $E_e$ and becomes 10\%~(15\%) at $E_e=2$~GeV, when $M_{A}=1.2~(1.3)$~GeV is used, instead of  $M_A=1.026$~GeV. 

In Fig.~\ref{ALAP:Ee:g2R:Sigma}, we study the dependence of the spin asymmetries on $g_2^R(Q^2)$ variation by presenting the results for $A_L(E_e)$ and $A_P(E_e)$ as a function of $E_e$, using $g_2^R(0)$ in the range $[-3,+3]$. 
Unlike $M_{A}$ variation, we observe significant dependence of both $A_L(E_e)$ and $A_P(E_e)$ on $g_2^R(Q^2)$ variation, which is especially prominent for $A_L(E_e)$. Quantitatively, in the case of $A_L(E_e)$, for $g_2^R(0)=+1~(+3)$, we observe a suppression in the magnitude of $A_{L} (E_e)$ by about 20\%~(60\%) at $E_e=1$~GeV, which decreases with increasing $E_e$ and becomes $\sim 18\%$~(55\%) at $E_{e}=2$~GeV.
Moreover, using $g_2^R(0)=-1~(-3)$, an enhancement in the magnitude of $A_{L} (E_e)$ is observed, which is $\sim 20\%$~(55\%) at  $E_e=1$~GeV and becomes $\sim 18\%$~(45\%) at $E_{e}=2$~GeV. 
On the other hand, $A_{P}(E_e)$ increases by about 10\%~(25\%) at $E_{e}=1$~GeV, when $g_2^R(0)=+1~(+3)$ is used. This enhancement increases with $E_e$ and becomes 15\%~(40\%) at $E_e=2$~GeV. Using $g_2^R(0)=-1~(-3)$, $A_{P}(E_e)$ decreases by about 10\%~(30\%) at $E_{e}=1$~GeV. With the increase in $E_e$, this decrease becomes $\sim$20\%~(55\%) at $E_{e}=2$~GeV.

In Fig.~\ref{ALAP:Q2:MA:Sigma}, the results are presented for $A_L(Q^2)$ and $A_P(Q^2)$ as a function of $Q^2$ using different values of $M_A$ in the range $1.026-1.3$~GeV. The spin asymmetries show little dependence on the choice of $M_{A}$, particularly for $A_L(Q^2)$. In the case of $A_P(Q^2)$, with increasing $M_A$, $A_P(Q^2)$ increases at low and mid $Q^2$ region. This enhancement increases with increasing $E_e$ and becomes around 15\% at $E_e=2.2$~GeV. Moreover, with increasing $M_A$, the peak shifts towards lower $Q^2$.

In Fig.~\ref{ALAP:Q2:g2R:Sigma}, the results are presented for $A_L(Q^2)$ and $A_P(Q^2)$ as a function of $Q^2$, using different values of $g_2^R(0)$ in the range $[-3,+3]$ at $E_e=855$~MeV, 1.1, and 2.2~GeV. The spin asymmetries $A_L(Q^2)$ and $A_P(Q^2)$  are quite sensitive to the choice of $g_2^R(Q^2)$ across all values of $Q^2$ and $E_e$ considered in this work. In the case of $A_L(Q^2)$,  the results obtained with $g_2^R(Q^2)<0$~($g_2^R(Q^2)>0$) are smaller~(larger) than the results obtained with $g_2^R(Q^2)=0$ across all values of $Q^2$ and $E_e$. Moreover, the dependence of $A_{L} (Q^2)$ on $g_2^R(Q^2)$ variation is more pronounced when negative values of $g_2^R(Q^2)$ are implemented. On the contrary, for $A_P(Q^2)$,  we observe a different behaviour between the negative and positive values of $g_2^R(Q^2)$. In the low $Q^2$ region, for a fixed electron energy, the value of $A_{P}(Q^2)$ increases with increasing $g_2^R(Q^2)$, while in the high $Q^2$ region,  $A_{P}(Q^2)$ decreases with increasing $g_2^R(Q^2)$.

\subsubsection{$e^- + p \longrightarrow \nu_e + \vec{\Sigma}^{\;0}$: (c)~Polarization observables of $\Sigma^0$:}\label{polarization:sigma}

In Fig.~\ref{PLPP:Ee:MA:Sigma}, we present the results for the polarization observables $P_L(E_e)$ and $P_P(E_e)$ as a function of $E_e$ for $M_A= 1.026-1.3$~GeV. 
For $P_L(E_e)$, we observe that using $M_A=1.2$~GeV increases the magnitude of $P_{L}(E_e)$ by about $7-10\%$ as compared to the results obtained with $M_{A}=1.026$~GeV, which further increases with increasing $M_A$, in the entire range of electron energy considered in this work. On the other hand, using $M_A=1.2$~GeV increases the magnitude of $P_{P}(E_e)$ by about $2-6\%$ as compared to the results obtained with $M_{A}=1.026$~GeV, and this increment further increases with increasing $M_A$. 

In Fig.~\ref{PLPP:Ee:g2R:Sigma}, we study the dependence of the polarization observables on $g_2^R(Q^2)$ variation by presenting the results for $P_L(E_e)$ and $P_P(E_e)$ as a function of $E_e$, assuming T-invariance using different values of $g_2^R(0)$ in the range $[-3,+3]$. We observe a significant dependence of the polarization observables on the value of $g_2^R(Q^2)$ for both $P_L(E_e)$ and $P_P(E_e)$. Qualitatively, we find that both $P_{L}(E_e)$ and $P_P(E_e)$ increase with increasing $g_2^R(Q^2)$ and decrease when a negative value of $g_2^R(Q^2)$ is taken. Moreover, the results obtained with $g_2^R(Q^2)<0$ and $g_2^R(Q^2)>0$ are nearly symmetric about the results obtained with $g_2^R(Q^2)=0$.  $P_L(E_e)$ increases by about 7\%~(22\%) when $g_2^R(0)=+1~(+3)$ is taken, at $E_e=1$~GeV, which further increases with increasing $E_e$ and becomes about 12\%~(35\%) at $E_e=2$~GeV. Furthermore, in the case of $P_P(E_e)$, we find that the results obtained using different values of $g_2^R(Q^2)$ show large variation in the threshold region and saturates for $E_e \gtrsim 0.5$~GeV. Thus in the region of $E_e=0.5-2$~GeV, we find that $P_P(E_e)$ increases by about 10\%~(30\%) when the results obtained with $g_{2}^R(0)=+1~(+3)$ are compared with the results obtained using $g_{2}^R(0)=0$.

Fig.~\ref{PLPPPT:Ee:g2I:Sigma} presents the results for $P_L(E_e)$, $P_P(E_e)$, and $P_{T} (E_e)$ as a function of $E_e$ assuming T-violation, using $g_2^I(0)$ in the range [0,3]. 
$P_L(E_e)$ and $P_P(E_e)$ are almost insensitive to the variation in $g_2^I(Q^2)$, however, we observe some dependence of $P_{T} (E_e)$ on $g_2^I(Q^2)$ variation, which increases with the increase in the value of $g_2^I(Q^2)$. For example, using $g_2^I(0) = 3$, $P_{T} (E_e)$ increases by about 10\% at $E_e=2$~GeV. 

In Fig.~\ref{PLPP:Q2:MA:Sigma}, the results are presented for $P_L(Q^2)$ and $P_P(Q^2)$ as a function of $Q^2$ at $E_e=855$~MeV, 1.1, and 2.2~GeV, using different values of $M_A$ in the range $1.026-1.3$~GeV. The polarization observables show some dependence on the choice of $M_{A}$ across all values of $E_e$ and $Q^2$ considered in this work. In the case of $P_{L}(Q^2)$, across all values of $E_e$, we find that increasing the value of $M_A$ leads to a small suppression in $P_L(Q^2)$, which increases with increase in $Q^2$. However, in the case of $P_{P} (Q^2)$, we observe different patterns when different values of $M_A$ are used across all values of $E_e$ and $Q^2$, considered in this work.

Fig.~\ref{PLPP:Q2:g2R:Sigma} presents the results for $P_L(Q^2)$ and $P_P(Q^2)$ as a function of $Q^2$, assuming T-invariance using different values of $g_2^R(0)$ in the range $[-3,+3]$ at $E_e=855$~MeV, 1.1, and 2.2~GeV. Both the longitudinal and perpendicular components  of polarization are quite sensitive to the choice of $g_2^R(Q^2)$ across all values of $Q^2$ and $E_e$ considered in this work. Quantitatively, both $P_L(Q^2)$ and $P_P(Q^2)$ increase~(decrease) when the positive~(negative) values of $g_2^R(Q^2)$ are taken in the numerical calculations. Moreover, this dependence of the polarization observables on $g_2^R(Q^2)$ variation increases with increasing $E_e$.

In Fig.~\ref{PLPPPT:Q2:g2I:Sigma}, the results are presented for $P_L(Q^2)$, $P_P(Q^2)$, and $P_{T} (Q^2)$ as a function of $Q^2$ at $E_e=855$~MeV, 1.1, and 2.2~GeV, assuming T-violation using $g_2^I(0)$ in the range [0,3]. The longitudinal and perpendicular components of polarization are almost insensitive to the variation in $g_2^I(Q^2)$ across all values of $Q^2$ and $E_e$.  However, we observe some dependence of $P_{T} (Q^2)$ on $g_2^I(Q^2)$ variation, which is about 12\%, when the results obtained with $g_2^I(0)=3$ are compared with the results obtained using $g_2^I(0)=0$, in the peak region of $P_T(Q^2)$ at $E_e=855$~MeV. This effect further increases with increasing $E_e$, and becomes 40\% at $E_e=2.2$~GeV. 

 \section{Inelastic processes of resonance excitations: Cross sections}\label{sec2:Delta}
 In this section, we study the electron and positron induced excitation of lowest lying resonances like $P_{33} (1232)$, $P_{11}(1440)$, and $S_{11}(1535)$ from the free proton target. 
 Explicitly, we discuss the electron and positron induced excitation of spin $\frac{3}{2}$ resonance from the free proton target in Sec.~\ref{sec:Delta}, and the electron induced excitation of spin $\frac{1}{2}$ resonances from free proton target in Sec.~\ref{sec:Nstar}.
 
 \subsection{$\Delta (1232)$ production}\label{sec:Delta}
 The production reaction of the $\Delta$ resonance via charged current electron and positron scattering off a free proton target is given by:
  \begin{eqnarray}\label{delta}
    e^-(k) + p(p) &\longrightarrow& \nu_e(k^\prime) + \Delta^0(p^\prime), \\
    \label{delta2}
    e^+(k) + p(p) &\longrightarrow& \bar{\nu}_e(k^\prime) + \Delta^{++}(p^\prime),
  \end{eqnarray}
  where the quantities in the parentheses represent the four momenta of the corresponding particles.
  
 The transition matrix elements for the reactions given in Eqs.~(\ref{delta}) and (\ref{delta2}), respectively, are defined as
 \begin{eqnarray}\label{mat1}
  {\cal M}^{e^{-}\nu} &=& \frac{G_{F} \cos\theta_{C}}{\sqrt{2}} \; l_{\mu}^{e^{-}\nu}\: \bra{\Delta^{0} (p^{\prime})} j^{\mu} \ket{p(p)}, \\
  \label{mat2}
   {\cal M}^{e^{+}\bar{\nu}} &=& \frac{G_{F} \cos\theta_{C}}{\sqrt{2}} \; l_{\mu}^{e^{+}\bar{\nu}}\: \bra{\Delta^{++} (p^{\prime})} j^{\mu} \ket{p(p)},  
 \end{eqnarray}
where  the leptonic current $l_{\mu}$ appearing in Eqs.~(\ref{mat1}) and (\ref{mat2}), respectively, is given by
  \begin{eqnarray}\label{lmu:e}
   l_{\mu}^{e^{-}\nu} &=& \bar{u}_{\nu}(k^{\prime}) \gamma_{\mu} \left(1-\gamma_{5} \right) u_{e^{-}} (k) \\
   \label{lmu:e+}
   l_{\mu}^{e^{+}\bar{\nu}} &=& \bar{v}_{\nu}(k^{\prime}) \gamma_{\mu} \left(1+\gamma_{5} \right) v_{e^{-}} (k) .
  \end{eqnarray}
  
  The matrix element of the hadronic current $j^{\mu}$ defined in Eqs.~(\ref{mat1}) and (\ref{mat2}) for $N-\Delta$ transition is generally written using the formalism given by Rarita-Schwinger~\cite{Rarita:1941mf}, which treats spin $\frac{3}{2}$ particles using a vector-spinor field. This formalism has a problem associated with the lower spin degrees of
freedom, and leads to some ambiguities in describing the propagation of the off-shell spin $\frac{3}{2}$ field using a propagator
specially in the presence of external interactions like the electromagnetic and/or the strong interactions. The problem
has been discussed extensively in literature for many years~\cite{Benmerrouche:1989uc, Williams:1985zz, Nath:1979wr, Nath:1971wp}.  Consequently, there are various prescriptions
for treating the propagator and the effective Lagrangians for the interacting
fields of higher spin particles in a consistent way for describing the interaction of spin $\frac{3}{2}$
fields~\cite{Pascalutsa:1994tp}. One of the most popular prescriptions given by Pascalutsa and Timmermans~\cite{Pascalutsa:1999zz}
has been investigated further in the works of Mart~\cite{Mart:2019jtb}, Vrancx et al.~\cite{Vrancx:2011qv},
and many other references cited there.
  
  In the present work, we use the formalism for spin $\frac{3}{2}$ resonances given by Rarita and Schwinger~\cite{Rarita:1941mf} for describing the excitation of $P_{33}(1232)$ resonance as the vector and axial vector form factors for the $N-\Delta$ transition have been extensively discussed in literature, and  are determined phenomenologically using the Rarita-Schwinger formalism~(as discussed in Section~\ref{ND_ff}). In the case of the higher mass spin $\frac{3}{2}$ resonances like $P_{13}(1720)$ and $P_{13}(1900)$~(to be discussed in Section~\ref{weak:associated}), the prescription given by Pascalutsa and Timmermans~\cite{Pascalutsa:1999zz} is used, where the transition form factors have not been determined phenomenologically. The vector transition form factors are parameterized in terms of the experimentally determined helicity amplitudes and the transition form factors in the axial vector sector are taken to be of the dipole form.

 \subsubsection{Matrix element and form factors:}\label{ND_ff}
  The expression for the matrix element of the hadronic current $j^{\mu}$ defined in Eq.~(\ref{mat1}) is written as
  \begin{equation}\label{jmu}
\bra{\Delta^{0}(p^{\prime})}j^\mu \ket{p(p)}= \bar \Psi_{\beta}(p^{\prime}){ \mathcal O}^{\beta \mu} u( p),
\end{equation}
where $u(p)$ denotes the Dirac spinor of the proton and  $\Psi_\beta(p^{\prime})$ represents the Rarita-Schwinger field for spin-$\frac{3}{2}$ particle. The matrix element for the hadronic current defined in Eq.~(\ref{mat2}) for the $p \rightarrow \Delta^{++}$ transition is related with the matrix element defined in Eq.~(\ref{mat1}) for the $p \rightarrow \Delta^0$ transition, using isospin invariance:
\begin{equation}\label{iso:delta}
 \bra{\Delta^{++}(p^{\prime})}j^\mu \ket{p(p)} = \sqrt{3} ~\bra{\Delta^{0}(p^{\prime})}j^\mu \ket{p(p)} .
 \end{equation}

 The operator $\mathcal O^{\beta \alpha}={\mathcal O}_V^{\beta \alpha}+{\mathcal O}_A^{\beta \alpha}$ characterizes the $N-\Delta$ transition vertex, and is described in terms of the vector~(${\mathcal O}_V^{\beta \alpha}$)
 and the axial vector~(${\mathcal O}_A^{\beta \alpha}$) vertices describing this transition. The vector and axial vector $N-\Delta$ transition vertices are, in turn, described in terms of the vector and axial vector form factors as~\cite{SajjadAthar:2022pjt, Athar:2020kqn}:
\begin{eqnarray}\label{vec_tra_current}
{\mathcal O}_V^{\beta \alpha}&=&\left(\frac{C_{3}^V(Q^2)}{M}(g^{\alpha \beta}\not\! q-q^{\beta}\gamma^{\alpha})
+\frac{C_{4}^V(Q^2)}{M^2}(g^{\alpha \beta} q \cdot p^{\prime}-q^{\beta}p^{\prime\alpha}) + 
\frac{C_{5}^V(Q^2)}{M^2}(g^{\alpha \beta}q \cdot p-q^{\beta}p^{\alpha}) \right. \nonumber \\
&& + \left.\frac{C_{6}^{V}(Q^2)}{M^2}q^{\beta}q^{\alpha}\right)\gamma_{5} \\
\label{ax_tra_current}
{\mathcal O}_A^{\beta \alpha}&=&\left(\frac{C_{3}^A(Q^2)}{M}(g^{\alpha \beta}\not\! q-q^{\beta}\gamma^{\alpha})
+
\frac{C_{4}^{A}(Q^2)}{M^2}(g^{\alpha \beta} q \cdot p^{\prime}-q^{\beta}p^{\prime\alpha})
+C_{5}^{A}(Q^2)g^{\alpha \beta}+\frac{C_{6}^{A}(Q^2)}{M^2}q^{\beta}q^{\alpha}\right).~~~
\end{eqnarray}

In Eqs.~(\ref{vec_tra_current}) and (\ref{ax_tra_current}), $C_{i}^{V}(Q^2); (i=3-6)$ are the isovector vector and $C_{i}^{A}(Q^2); (i=3-6)$ are the isovector axial vector $N-\Delta$ transition form factors. The invariance under time reversal implies that all the vector~($C_{i}^{V}(Q^2); i=3-6$) and axial vector~($C_{i}^{A}(Q^2); i=3-6$) form factors must be real. 
In the following, we outline the salient features of these form factors and briefly review various parametrizations of the vector and axial vector $N-\Delta$ transition form factors employed in the literature to study the weak production of $\Delta(1232)$ resonance.

The theoretical studies of the various $N-\Delta$ transition form factors defined in the matrix element of the vector and axial vector currents in Eqs.~(\ref{vec_tra_current}) and (\ref{ax_tra_current}), respectively, have been made using various models of the nucleon structure.  For example, the various quark models encompassing both the non-relativistic quark model~\cite{Isgur:1978xj, Capstick:1994ne, Copley:1969ft, Dalitz:1966fg, Gourdin:1963ub, Berman:1965iu, Albright:1964sgs, Albright:1965aud, Kim:1963, Chew:1957tf, Nath:1979qe, Mukhopadhyay:1998mn, Liu:1995bu, Hemmert:1994ky, Hernandez:2010bx, Buchmann:2001gj, Mornacchi:2023oir, Salin:1967, Bijtebier:1970ku, Bell:1962fyy, Chew:1956zz, Ravndal:1972ws} and relativistic quark model~\cite{Feynman:1971wr, Capstick:1989ck, Cardarelli:2000tk, Warns:1989ic, Frolov:1998pw, Walecka:1968zz, Pritchett:1969onm}; the QCD inspired models of the nucleon structure~\cite{Segovia:2014aza, Yin:2023kom, Chen:2023zhh, QCDSF:2008qtn, Aliev:2011uf}; effective Lagrangian approach with and without incorporating chiral symmetry~\cite{Davidson:1991xz}, current algebra and PCAC based approaches~\cite{Adler:1968tw}; phenomenological analyses based on electroproduction and neutrino data~\cite{Dufner:1967yj, Rein:1980wg, Schreiner:1973mj, Tiator:2003xr, Paschos:2003qr, Lalakulich:2005cs, Hernandez:2010bx, Hernandez:2007qq, Graczyk:2009qm, Alvarez-Ruso:2015eva, Lalakulich:2006sw}, chiral and meson cloud models~\cite{Golli:2002wy, Procura:2008ze}, dynamical models~(data driven)~\cite{Sato:2000jf, Julia-Diaz:2006ios}, perturbative QCD approach~\cite{Carlson:1985mm}, lattice QCD models~\cite{Alexandrou:2006mc, Alexandrou:2007eyf}, light cone QCD sum rules~\cite{Kucukarslan:2015urd, Aliev:2007pi}, and relativistic baryon chiral perturbation theory~\cite{Geng:2008bm, Unal:2021byi}.

The early studies of the $N-\Delta$ transition form factors were largely based on constituent quark models, particularly the non-relativistic quark models, which exploited approximate SU(6) spin-flavor symmetry. Within this framework, the dominant magnetic dipole transition was reasonably well described, and qualitative features of the vector transition form factors, such as $C_3^V(Q^2)$ and $C_4^V(Q^2)$, were determined. Complementary to quark model approaches, current algebra and PCAC hypothesis were employed to constrain the axial-vector transition form factors. In particular, the Adler~\cite{Adler:1968tw} model provided important information about the axial form factors, most notably predicting the dominance of $C_5^A(Q^2)$, and relating $C_6^A(Q^2)$ to $C_5^A(Q^2)$ through the pion-pole dominance of the divergence of the axial vector current. Feynman, Kislinger and Ravndal~\cite{Feynman:1971wr} using a relativistic quark model proposed a prescription for extending the quark model to obtain predictions for various vector and axial vector $N-\Delta$ transition form factors at large $Q^2$. The predictions obtained by Feynman et al.~\cite{Feynman:1971wr} for the $N-\Delta$ transition amplitudes are quite similar to those of Adler's model~\cite{Adler:1968tw}. 

The phenomenological models, guided by pion electroproduction and neutrino scattering data, played a crucial role in extracting the $Q^2$-dependence of the transition form factors. These studies typically relied on parametrizations constrained by available data, leading to widely used functional forms for both vector and axial form factors in subsequent analyses of the neutrino scattering data. A significant advancement came with the inclusion of meson cloud effects through dynamical models. These approaches explicitly incorporated pion-nucleon interactions and demonstrated that mesonic degrees of freedom can substantially modify the transition amplitudes, particularly enhancing the electric and Coulomb quadrupole components. The QCD-inspired approaches, such as light-cone QCD sum rules, extended the analysis to higher momentum transfer regions, providing estimates for both vector and axial vector form factors~\cite{Kucukarslan:2015urd, Aliev:2007pi}.

Recent analyses of the neutrino scattering data to determine the weak $N-\Delta$ transitions have preferred to use the phenomenological values of the vector form factors extracted from the analysis of the experimental data on the photo- and electro- production of $\Delta$ given in Eq.~(\ref{civ_lala}). In the following, we give in brief a review of the $N-\Delta$ transition form factors, which have been used in the recent analysis of the (anti)neutrino induced pion production data. We also give the explicit expressions of the various vector and axial vector $N-\Delta$ transition form factors used in the numerical calculations of the results presented in this work.

\subsubsection{Vector form factors for $N-\Delta$ transition:}
The isovector vector form factors are determined using the following properties of the weak vector currents: 
\begin{itemize}
 \item [(i)] The principle of CVC hypothesis of the weak vector currents implies that $C_{6}^{V} (Q^2) = 0$.
 
 \item [(ii)] Isospin symmetry relates the weak vector form factors~($C_{i}^{V} (Q^2)$) to the electromagnetic~($C_{i}^{N} (Q^2)$) $N-\Delta$ transition form factors via the relation 
 \begin{equation}
  C_{i}^{V} (Q^2) = - C_{i}^{N} (Q^2); \qquad \quad i=3,4,5.
 \end{equation}
\end{itemize}

The earlier analyses of weak $\Delta(1232)$ production processes induced by the neutrinos and antineutrinos on the nucleons were performed under the assumption of the magnetic dipole~($M1$) dominance in the electromagnetic $p \rightarrow \Delta^{+}$ excitation, which implies~\cite{LlewellynSmith:1971uhs}
\begin{equation}\label{C3C4V}
 C_{4}^{V} (Q^2) = - \frac{M}{W} C_{3}^{V} (Q^{2}); \qquad \quad  C_{5}^{V} (Q^2)=0; \qquad \quad W=\sqrt{s}.
\end{equation}
Using these relations, $C_3^V(Q^2)$ was determined phenomenologically to be of the following form~\cite{Dufner:1967yj}:
\begin{eqnarray}\label{C3V}
 \left| C_{3}^{V}(Q^2) \right|^{2} &=& (2.05)^{2}\; \left[1 + 9 \sqrt{Q^2} \right]\; 
 \exp \left[-6.3 \sqrt{Q^2} \right].
\end{eqnarray}

In the original analysis of ANL~\cite{Barish:1978pj, Radecky:1981fn} and BNL~\cite{Kitagaki:1986ct, Kitagaki:1990vs} data on the (anti)neutrino scattering from proton and deuteron, the parametrizations for the vector $N-\Delta$ transition form factors, as given in Eqs.~(\ref{C3C4V}) and (\ref{C3V}), were used. However, in the light of the observed nonzero electric quadrapole moment to magnetic dipole~($E2/M1$) ratio in the electromagnetic excitation of $\Delta$ and $\Delta \rightarrow p\gamma$ decays, the vector transition form factors are now  extracted from the experimentally observed helicity amplitudes $A_{\frac{1}{2}}$, $A_{\frac{3}{2}}$ and $S_{\frac{1}{2}}$ in the $ep\rightarrow e\Delta^{+}$ transition, for instance, from the MAID analysis~\cite{Tiator:2011pw, Drechsel:2007if}.

Recently, the parametrization given by Lalakulich {et al.}~\cite{Lalakulich:2005cs, Lalakulich:2006sw} have been used in many calculations of the weak $N-\Delta$ excitations~\cite{Hernandez:2007qq, Hernandez:2010bx, Graczyk:2009qm, Graczyk:2009zh, Alvarez-Ruso:2015eva, Kakorin:2021mqo}. These are given by:
\begin{equation}\label{civ_lala}
C_i^V(Q^2)=\frac{C_i^V(0)}{\left(1+\frac{Q^2}{M_V^2}\right)^{2}}~{\cal{D}}_i,~~~i=3,4,5,
\end{equation}
with $C_3^V(0)=2.13$, $C_4^V(0)=-1.51$ and $C_5^V(0)=0.48$, and
\begin{eqnarray}\label{di} 
{\cal{D}}_{3,4}&=&\left(1+\frac{Q^2}{4M_V^2}\right)^{-1}; \qquad\quad
{\cal{D}}_{5}=\left(1+\frac{Q^2}{0.776M_V^2}\right)^{-1};~~ M_V=0.84~\rm{GeV}
\end{eqnarray} 
In the numerical calculations of the present work, we have followed the prescription of Lalakulich et al.~\cite{Lalakulich:2005cs, Lalakulich:2006sw} for the vector $N-\Delta$ transition form factors.

 \subsubsection{Axial vector $N-\Delta$ transition form factors:}
  
 \begin{itemize}
  \item [(a)] {\bf $\bm{C_{5}^{A} (Q^2)}$ and $\bm{C_{6}^{A} (Q^2)}$}   

 \begin{itemize}
 \item [(i)] In the axial vector sector, the dominant contribution to the cross section is provided by $C_{5}^{A}(Q^2)$, which is determined using the PCAC hypothesis in conjunction with the pion pole dominance of the divergence
of the axial-vector current~(PDDAC). Within this framework, the axial vector coupling $C_{5}^{A}(0)$ is related to the strong $\Delta \rightarrow N\pi$ coupling constant $g_{\Delta N\pi}$ through the generalized
Goldberger-Treiman~(GT) relation, i.e.,
 \begin{eqnarray}\label{C5A0}
  C_5^A(0) &=& f_\pi \frac{ g_{\Delta N \pi}  }{2 \sqrt3 M }.
 \end{eqnarray}
  Using $f_{\pi} = 0.97m_{\pi}$ and $g_{\Delta N\pi}=28.6$, we find $C_{5}^{A} (0) = 1.2$.
  
   In Table~\ref{tab:axial_ff}, we list the various values of $C_{5}^{A}(0)$ obtained in some of the experimental and theoretical analyses taken from the existing literature on the subject. 
It may be observed from the table that there is a large variation in the theoretical prediction of $C_{5}^{A}(0)$. 
  
 \item [(ii)] Using the PCAC hypothesis and the generalized Goldberger-Treiman relation, the form factor $C_{6}^{A} (Q^2)$ is given in terms of $C_{5}^{A}(Q^2)$, i.e.,  
  \begin{eqnarray}\label{C6A}
  C_{6}^{A}(Q^2) &=& C_{5}^{A}(Q^2) \frac{M^{2}}{Q^2 + m_{\pi}^{2}},
 \end{eqnarray}
 with $m_{\pi}$ being the mass of the pion.
 
 The term associated with the matrix element of $C_{6}^{A} (Q^2)$ form factor  is proportional to the lepton mass and is negligible in the case of electron and positron induced reactions.

\item [(iii)] The $Q^2$ dependence of the axial vector form factor $C_{5}^{A} (Q^2)$ is given by Schreiner and von Hippel~\cite{Schreiner:1973mj} parametrization of the Adler's model~\cite{Adler:1968tw}, as
 \begin{eqnarray}\label{c5aq}
  C_{5}^A(Q^2) &=&  \frac{C_{5}^A(0) \left( 1+ \frac{a\, Q^2}{b~+~Q^2} \right)}{\left( 1 + Q^2 /M_{A}^2 \right)^2} ,
\end{eqnarray}
with $C_{5}^{A} (0)=1.2$, $a=-1.21$, and $b=2$~GeV$^{2}$.
 
 \item [(iv)] In recent years, many authors~\cite{Hernandez:2007qq, Hernandez:2010bx, Graczyk:2009qm, Graczyk:2009zh, Alvarez-Ruso:2015eva, Kakorin:2021mqo} have used the modified dipole parametrization for the $Q^2$ dependence of $C_5^A(Q^2)$ with different values of $M_{A}$. In the present work, we have used the parametrization of $C_5^A(Q^2)$ given by the model of Lalakulich et al.~\cite{Lalakulich:2006sw}, i.e.
\begin{eqnarray}\label{cia_lala}
C_5^A(Q^2)&=&\frac{C_5^A(0)}{\left(1+\frac{Q^2}{3M_A^2}\right)\left(1+\frac{Q^2}{M_A^2}\right)^{2}},
\end{eqnarray}
 with $C_5^A(0)=1.2$ and $M_{A}=1.026$~GeV.
 \end{itemize}
 
\begin{table*}
\centering
\begin{tabular*}{150mm}{@{\extracolsep{\fill}} c  c c c c }\hline\hline
 & Model & $C_{3}^{A}(0)$ & $C_{4}^{A}(0)$ & $C_{5}^{A}(0)$ \\ \hline
  &Adler~\cite{Adler:1968tw, Schreiner:1973mj} & 0 & $-0.3$ & 1.2 \\ 
Phenomenological &HHM~\cite{Hemmert:1994ky}  & 0 & $-0.46 \pm 0.06$ & $1.39 \pm 0.14 $ \\ 
&Graczyk et al.~\cite{Graczyk:2009zh}  & 0 & $-0.67 \pm 0.42$ & $1.17 \pm 0.13 $ \\ \hline
&Isgur-Karl~\cite{Liu:1995bu}  & 0.0008 & $-0.657$ & 1.2 \\ 
&HHM~\cite{Hemmert:1994ky}  & 0 & $-0.29 \pm 0.006$ & $0.87 \pm 0.03 $ \\ 
&SU(6)~\cite{Liu:1995bu}  & 0 & $-0.38$ & $1.17$ \\ 
Theoretical&D-mixing~\cite{Liu:1995bu}  & 0.052 & 0.052 & 0.813\\ 
&Barquilla-Cano et al.~\cite{Barquilla-Cano:2007vds} & 0.035 & $-0.26$ & 0.93 \\ 
&Golli~\cite{Golli:2002wy}  & 0 & 0.141 & 1.53 \\ 
&Kucukarslan~\cite{Kucukarslan:2015urd} (conventional) & $0.11 \pm 0.03$ & $0.27 \pm 0.09$&  $1.14 \pm 0.20$ \\ 
&Chen et al.~\cite{Chen:2023zhh}  & $0.26^{+0.17}_{-0.04}$ & $-0.66^{+0.03}_{-0.10}$  &  $1.16^{+0.09}_{-0.03}$
\\
\hline\hline
\end{tabular*}
\caption{Values of the axial vector form factors $C_{i}^{A};~(i=3-5)$ at $Q^2=0$ calculated in various phenomenological and theoretical models.}
\label{tab:axial_ff}                                                  
\end{table*}

\item [(b)] {\bf $\bm{C_{3}^{A} (Q^2)}$ and $\bm{C_{4}^{A} (Q^2)}$} \\

The subdominant axial vector form factors $C_3^A(Q^2)$ and $C_4^A(Q^2)$ are taken from the phenomenologically parametrization of Schreiner and von Hippel~\cite{Schreiner:1973mj} based on the Adler's model~\cite{Adler:1968tw}, and is given as:
\begin{equation}\label{c4-c3}
      C_3^A(Q^2) =0;  \qquad     \qquad  C_4^A(Q^2) = -\frac{1}{4}C_5^A(Q^2).
\end{equation}
Lalakulich et  al.~\cite{Lalakulich:2006sw} have also used the above relation for $C_3^A(Q^2)$ and $C_4^A(Q^2)$ form factors with the modified dipole form for $C_{5}^{A}(Q^2)$ as given in Eq.~(\ref{cia_lala}), instead of Eq.~(\ref{c5aq}). 

Graczyk et al.~\cite{Graczyk:2009zh} have phenomenologically determined the form factors $C_{4}^{A}(Q^2)$ and $C_{5}^{A}(Q^2)$ independent of the Adler's model~\cite{Adler:1968tw} by assuming them to be of the dipole form, keeping  $C_{3}^{A}(Q^2)=0$, and using
\begin{equation}\label{C4;Graczyk}
 C_{i}^{A} (Q^2) = \frac{C_{i}^{A}(0)}{\left(1+\frac{Q^2}{M_{Ai}^{2}}\right)^{2}}; \qquad \quad i=4,5.
\end{equation}
The best fitted values of $C_{4,5}^{A}(0)$ and $M_{A4,A5}$ in the model of Graczyk et al.~\cite{Graczyk:2009zh} are:
\begin{eqnarray}
 -C_{4}^{A}(0) &=& 0.67 \pm 0.42; \qquad \quad M_{A4} = 0.4^{+1.1}_{-0.4}~\text{GeV}, \\
  C_{5}^{A}(0) &=& 1.17 \pm 0.13; \qquad \quad M_{A5} = 0.95 \pm 0.07~\text{GeV}.
\end{eqnarray}

In Table~\ref{tab:axial_ff}, we present a representative list of the nonzero values of $C_{3}^{A} (0)$ and different values of $C_{4}^{A} (0)$ obtained in the various theoretical model calculations, which reflect different mechanisms to include minimal SU(6) symmetry breaking effects in these models. 

The subdominant axial vector form factors calculated in the chiral constituent quark model  of Barquilla-Cano et al.~\cite{Barquilla-Cano:2007vds} and the lattice gauge model of Chen et al.~\cite{Chen:2023zhh} are used in this work as representative inputs for the quark model and lattice gauge theory models to study the reactions given in Eqs.~(\ref{delta}) and (\ref{delta2}).
The $Q^2$ dependence of the form factors $C_{3,4}^{A}(Q^2)$ in these models has been given numerically in their work in the form of $C_{i}^{A}(Q^2)$ as a function of $Q^2$ plots. We have fitted them assuming the dipole and modified dipole forms as described below, to be used in this work~\cite{Fatima:2024hlu}.

The $Q^2$ dependence of $C_{4}^{A}(Q^2)$ in the model of  Barquilla-Cano et al.~\cite{Barquilla-Cano:2007vds} show similar behaviour as obtained in the the Adler's model~\cite{Adler:1968tw}~(Eq.~(\ref{c4-c3})) with $C_{4}^{A}(0)=-0.26$ and $C_{5}^{A}(Q^2)$ as given in Eq.~(\ref{cia_lala}). Therefore, we parameterize the $Q^2$ dependence of $C_{3}^{A}(Q^2)$  in the following form:
\begin{equation}\label{c3:Buch}
 C_{3}^{A} (Q^2) = C_{3}^{A}(0) \frac{\left(\frac{a_{3}Q^2}{b_{3}+Q^2}\right)}{\left(1+\frac{Q^2}{M_{A3}^{2}}\right)^2}
\end{equation}
with $C_{3}^{A}(0)=0.035$, $a_{3}=-4.61$, $b_{3}=2.8$~GeV$^{2}$, and $M_{A3}=1.67$~GeV.

The $Q^2$ dependence of $C_{3}^{A}(Q^2)$ and $C_{4}^{A}(Q^2)$ given by Chen et al.~\cite{Chen:2023zhh} is parameterized  using the dipole form as
\begin{equation}\label{c3A_Chen}
 C_{i}^{A} (Q^2) = \frac{C_{i}^{A}(0)}{\left(1+\frac{Q^2}{x_{i}M_{AS}^{2}}\right)^{2}}; \qquad \quad i=3,4
\end{equation}
with $C_{3}^{A}(0) = 0.26^{+0.17}_{-0.04}$, $C_{4}^{A}(0) = -0.66^{+0.03}_{-0.10}$, $x_{3}=1$, $x_{4}=0.74$, and $M_{AS}=1$~GeV. 
\end{itemize}

In the numerical calculations, we have used the parametrization of the vector form factors by Lalakulich et al.~\cite{Lalakulich:2005cs}, as given above in Eq.~(\ref{civ_lala}). In the axial vector sector, for the dominant $C_{5}^A(Q^2)$, the modified dipole parametrization~(Eq.~(\ref{cia_lala})), as given in the model of Lalakulich et al.~\cite{Lalakulich:2005cs} is used, and the dependence of the total and differential scattering cross section on  $C_{5}^A(Q^2)$ is studied, by varying independently $C_{5}^A(0)$ in the range [0.87:1.4] and $M_A$ in the range [0.9:1.2]~GeV. For the subdominant axial vector form factor $C_{3}^A(Q^2)$, we have used the parametrizations given by Barquilla-Cano et al.~\cite{Barquilla-Cano:2007vds} and Chen et al.~\cite{Chen:2023zhh}, given in Eqs.~(\ref{c3:Buch}) and (\ref{c3A_Chen}), respectively. For $C_{4}^A(Q^2)$, we have used the parametrizations given by Graczyk et al.~\cite{Graczyk:2009zh} and Chen et al.~\cite{Chen:2023zhh}, given in Eqs.~(\ref{C4;Graczyk}) and (\ref{c3A_Chen}), respectively, to study the dependence of the cross section on the subdominant axial vector form factors.

\subsubsection{Cross section:}
The general expression for the differential scattering cross section for the processes given in Eqs.~(\ref{delta}) and (\ref{delta2}) is written as~\cite{AlvarezRuso:1997jr}
\begin{eqnarray} \label{delta_cross_section}
\frac {d \sigma}{d Q^2} = \frac{1}{128\pi^{2} M M_{\Delta}E_{e}^{2}} \int {d E_{\nu}} 
\frac{\Gamma_\Delta(W)}{(W-M_\Delta)^2+\frac{\Gamma_\Delta^2(W)}{4}} \overline{\sum} \sum |{\cal{M}}|^2,
\end{eqnarray}
where $W=\sqrt{(p+q)^2}$ is the center of mass energy of the hadronic system, with $Q^2 = -q^2$ and $q=k-k^{\prime}$ being the four momentum transfer.
$M$ and $M_{\Delta}$ are the masses of the proton and $\Delta$, respectively;  $E_{e}$ and $E_{\nu}$ are the energies of the incoming and outgoing leptons, respectively. $\Gamma_{\Delta}(W)$ is the $W$ dependent decay width of the $\Delta$ resonance.

In the expression of the differential scattering cross section given in Eq.~(\ref{delta_cross_section}), $ \overline{\sum} \sum |{\cal{M}}|^2$ is the transition matrix element squared given in terms of the leptonic and hadronic tensors as
\begin{equation}\label{mat_square}
 \overline{\sum} \sum |{\cal{M}}|^2 = \frac{G_{F}^{2} \cos^{2}\theta_{C}}{2} L_{\mu\nu} J^{\mu\nu}.
\end{equation}
The leptonic tensor $\cal{L}_{\mu\nu}$ is written in terms of the leptonic currents defined in Eqs.~(\ref{lmu:e}) and (\ref{lmu:e+}), and is given by
\begin{equation}\label{Lmunu}
 {\cal{L}_{\mu\nu}} =  \overline{\sum}\sum l_{\mu} {l_{\nu}}^{\dagger}=  4\left(k_{\mu}k_{\nu}^{\prime} - (k\cdot k^{\prime})g_{\mu\nu} + k_{\nu}k_{\mu}^{\prime} \pm i \epsilon_{\mu\nu\rho\sigma} k^{\rho}{k^{\prime}}^\sigma \right)
\end{equation}
where $+~(-)$ stands for electron~(positron) induced reactions.
The hadronic tensor 
 $J^{\mu\nu}$ is defined in terms of the hadronic current $j^{\mu}$ as 
 \begin{equation}
 J^{\mu\nu}=  \overline{\sum}\sum j^{\mu} {j^{\nu}}^{\dagger}=  \frac{1}{2}
 Tr\left[ (\slashed{p} + M) {\tilde{\mathcal O}}^{\alpha\mu } {\it P}_{\alpha \beta}
      {\mathcal O}^{\beta\nu} \right],
   \end{equation}  
   where 
   ${\it P}_{\alpha \beta}$ is the spin-3/2 projection operator, which in the prescription of Rarita-Schwinger, is given by
\begin{eqnarray}\label{propagator}
{\it P}_{\alpha \beta} = -(\not\! p^ \prime +M_{\Delta}) 
\left(g_{\alpha \beta}-\frac{2}{3} \frac{{p^\prime}_{\alpha}{p^\prime}_{\beta}}{M_{\Delta}^2}
+ \frac{1}{3}\frac{{p^\prime}_{\alpha} \gamma_{\beta}-
{p^\prime}_{\beta} \gamma_{\alpha}}{M_{\Delta}}-\frac{1}{3}\gamma_{\alpha}\gamma_{\beta}\right).
\end{eqnarray}

The delta decay width $\Gamma_{\Delta}(W)$ is taken as the energy dependent $P$-wave decay width given by~\cite{Alvarez-Ruso:1998ais}:
\begin{equation}
\Gamma_\Delta(W)=\frac{1}{6 \pi}\left(\frac{f_{\pi N \Delta}}{m_{\pi}}\right)^2 \frac{M_{\Delta}}{W}|\bm q_{cm}|^3,
\end{equation}
 where $f_{\pi N \Delta}=2.127$ is the $\Delta \rightarrow N\pi$ coupling constant, $m_{\pi}$ is the pion mass, $|\bm q_{cm}|$ is the pion momentum in the rest 
 frame of the resonance and is given by
\[|{\bm q_{cm}}|=\frac{\sqrt{(W^2-m_{\pi}^2-M^2)^2 -4 m_{\pi}^2M^2}}{2W}.\] 
 In the numerical calculation, we have taken $W$ in the range $(M+m_\pi)\le W < 1.4~\rm{GeV}$.
   \begin{figure}
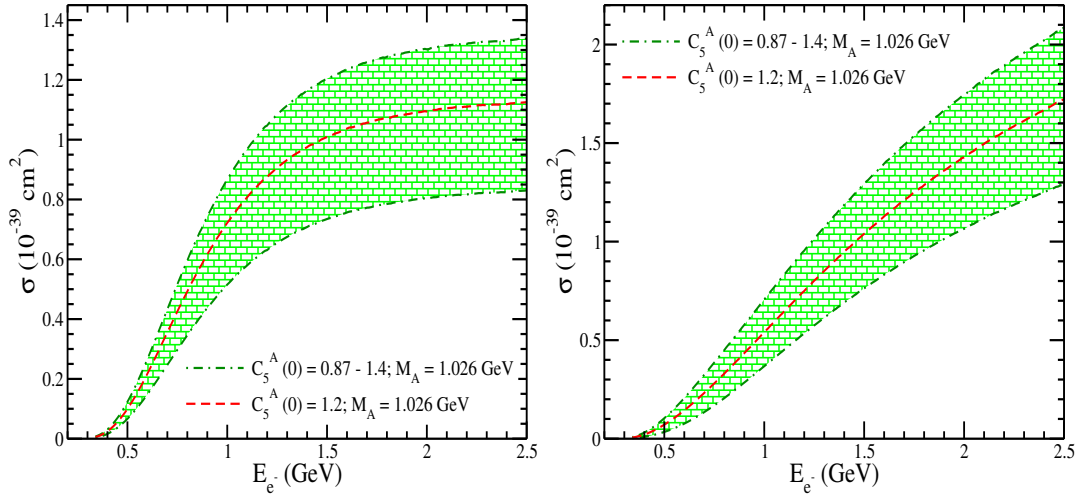
 
\begin{center}
\includegraphics[width=7cm,height=6.5cm]{sigma_electron_delta_C5A0_band.eps}
\includegraphics[width=7cm,height=6.5cm]{sigma_positron_delta_C5A0_band.eps}
\caption{Total scattering cross section ($\sigma(E_e)$) as a function of electron~($E_{e^-}$) and positron~($E_{e^+}$) energies, respectively, for
 $e^- + p \longrightarrow \nu_e + \Delta^0$~(left panel) and $e^+ + p \longrightarrow \bar{\nu}_e + \Delta^{++}$~(right panel) scattering processes. The dashed line corresponds to the results obtained with $C_{5}^{A}(0) = 1.2$ and $M_{A} = 1.026$~GeV, and the band corresponds to $C_{5}^{A}(0)$ being varied in the range $0.87-1.4$ with a fixed $M_{A} = 1.026$~GeV. 
  }\label{sigma:delta_C5A0_band}
\end{center}
\end{figure}

 A judicious and physically well motivated choice of the maximum value of invariant hadronic mass $W$, i.e. $W^{max}$ is of great importance in the study of weak interaction induced inelastic processes. The imposition of an appropriate $W^{max}$ serves as a crucial criterion for distinguishing between the resonance regime and the onset of deep inelastic scattering~(DIS) as well as the multiparticle production channels. In the absence of such a constraint on $W$, the extracted observables inevitably suffer from an uncontrolled admixture of different reaction mechanisms and the risk of double counting, thereby obscuring the  interpretation of the underlying physics. In particular, a carefully selected $W^{max}$ may enable a cleaner separation of the contributions arising from the low lying baryon resonances, most notably the $\Delta(1232)$, $P_{11}(1440)$, $S_{11}(1535)$ resonance excitations,  which play a dominant role in the intermediate energy region, to the other processes mentioned above.  Furthermore, the implementation of such $W^{max}$ is indispensable from the standpoint of theoretical reliability as the contributions in the region of $W \ge W^{max}$ are typically affected by larger model dependencies, owing to the uncertainties in the description of overlapping resonances and the possible contribution from the DIS in this kinematic regime. By constraining the analysis to a well-defined window in $W$, one significantly mitigates these ambiguities for describing the contributions from the resonance excitations, thereby enhancing the robustness and predictive power of the theoretical framework.

 \begin{figure}
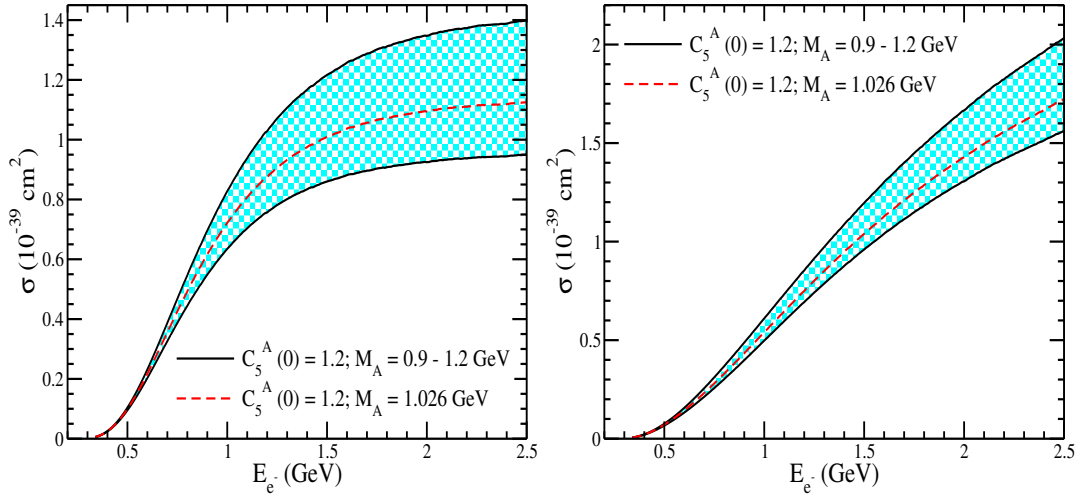
 
\begin{center}
\includegraphics[width=7cm,height=6.5cm]{sigma_electron_delta_MA_band.eps}
\includegraphics[width=7cm,height=6.5cm]{sigma_positron_delta_MA_band.eps}
\caption{Total scattering cross section ($\sigma(E_e)$) as a function of electron~($E_{e^-}$) and positron~($E_{e^+}$) energy, respectively, for
 $e^- + p \longrightarrow \nu_e + \Delta^0$~(left panel) and $e^+ + p \longrightarrow \bar{\nu}_e + \Delta^{++}$~(right panel) scattering processes. Dashed corresponds to the results obtained with $C_{5}^{A}(0) = 1.2$ and $M_{A} = 1.026$~GeV and the band corresponds to $M_A$ being varied in the range $0.9-1.2$~GeV with $C_{5}^{A}(0) = 1.2$. 
  }\label{sigma:delta_MA_band}
\end{center}
\end{figure}
 
 \begin{figure}
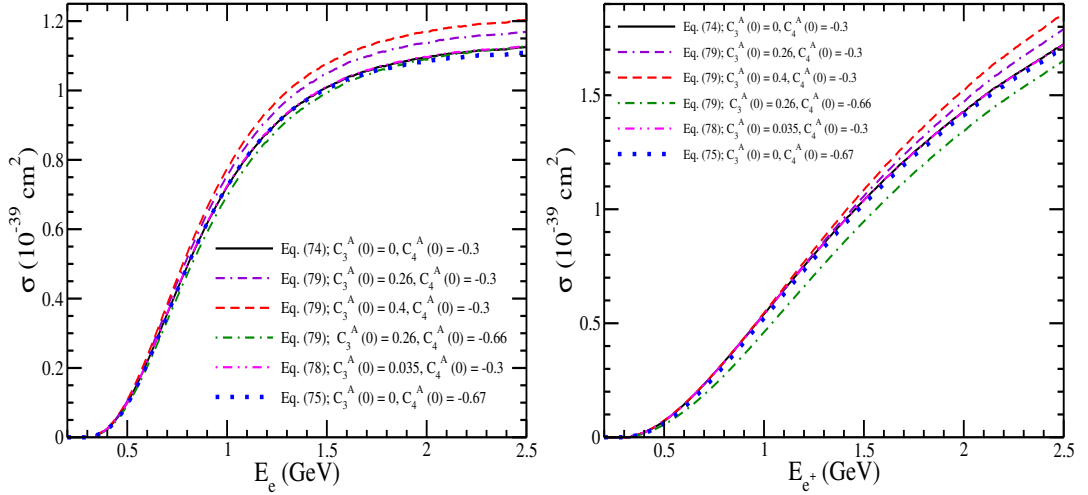
 
\begin{center}
\includegraphics[width=7cm,height=6.5cm]{total_sigma_electron_Ee_25GeV.eps}
\includegraphics[width=7cm,height=6.5cm]{total_sigma_positron_Ee_25GeV.eps}
\caption{Total scattering cross section ($\sigma(E_e)$) as a function of electron~($E_{e^-}$) and positron~($E_{e^+}$) energy, respectively, for
 $e^- + p \longrightarrow \nu_e + \Delta^0$~(left panel) and $e^+ + p \longrightarrow \bar{\nu}_e + \Delta^{++}$~(right panel) scattering processes. 
 The solid and double-dotted-dashed lines, respectively, represent the results obtained using the form factor parametrization of Lalakulich et al.~\cite{Lalakulich:2006sw} and the modification of $C_3^{A}(Q^2)$ by Barquilla-Cano et al.~\cite{Barquilla-Cano:2007vds} (Eq.~(\ref{c3:Buch})). 
 The double-dashed-dotted line and dashed line represent the results obtained using the parametrization of $C_{3}^{A}(Q^2)$ given by Chen et al.~\cite{Chen:2023zhh} (Eq.~(\ref{c3A_Chen})) using $C_{3}^{A}(0)=0.26$ and 0.4, respectively, and for all the other vector and axial vector form factors, the parametrization given by Lalakulich et al.~\cite{Lalakulich:2006sw} is used. The dashed-dotted line represents the results obtained using the parametrization given in Eq.~(\ref{c3A_Chen}) for $C_{3}^{A}(Q^2)$ and $C_{4}^{A}(Q^2)$ for Chen et al.~\cite{Chen:2023zhh}.  
 The dotted line represents the results obtained using the parametrization given by Graczyk~\cite{Graczyk:2009zh}~(Eq.(\ref{C4;Graczyk})) for $C_{4}^{A}(Q^2)$. 
 }\label{sigma:delta_weak}
\end{center}
\end{figure}

 \begin{figure} 
\begin{center}
 \includegraphics[width=5cm,height=6.5cm]{dsigma_dq2_electron_delta_C5A0_band_Ee_855MeV.eps}
\includegraphics[width=5cm,height=6.5cm]{dsigma_dq2_electron_delta_C5A0_band_Ee_11GeV.eps}
\includegraphics[width=5cm,height=6.5cm]{dsigma_dq2_electron_delta_C5A0_band_Ee_22GeV.eps}

 \includegraphics[width=5cm,height=6.5cm]{dsigma_dq2_positron_delta_C5A0_band_Ee_855MeV.eps}
 \includegraphics[width=5cm,height=6.5cm]{dsigma_dq2_positron_delta_C5A0_band_Ee_11GeV.eps}
\includegraphics[width=5cm,height=6.5cm]{dsigma_dq2_positron_delta_C5A0_band_Ee_22GeV.eps}
\caption{
$\frac{d\sigma}{dQ^2}$ as a function of $Q^2$ for the processes $e^- + p \longrightarrow \nu_{e} + \Delta^{0}$~(top panel) and $e^+ + p \longrightarrow \nu_{e} + \Delta^{++}$~(bottom panel)  at $E_{e}=855$~MeV~(left panel), 1.1~GeV~(middle panel), and 2.2~GeV~(right panel). Lines and points have the same meaning as in Fig.~\ref{sigma:delta_C5A0_band}.}\label{dsigma:delta_C5A0_band}
\end{center}
\end{figure}

 \begin{figure} 
\begin{center}
 \includegraphics[width=5cm,height=6.5cm]{dsigma_dq2_electron_delta_MA_band_Ee_855MeV.eps}
\includegraphics[width=5cm,height=6.5cm]{dsigma_dq2_electron_delta_MA_band_Ee_11GeV.eps}
\includegraphics[width=5cm,height=6.5cm]{dsigma_dq2_electron_delta_MA_band_Ee_22GeV.eps}

 \includegraphics[width=5cm,height=6.5cm]{dsigma_dq2_positron_delta_MA_band_Ee_855MeV.eps}
 \includegraphics[width=5cm,height=6.5cm]{dsigma_dq2_positron_delta_MA_band_Ee_11GeV.eps}
\includegraphics[width=5cm,height=6.5cm]{dsigma_dq2_positron_delta_MA_band_Ee_22GeV.eps}
\caption{  $\frac{d\sigma}{dQ^2}$ as a function of $Q^2$ for the processes $e^- + p \longrightarrow \nu_{e} + \Delta^{0}$~(top panel) and $e^+ + p \longrightarrow \nu_{e} + \Delta^{++}$~(bottom panel) at $E_{e}=855$~MeV~(left panel), 1.1~GeV~(middle panel), and 2.2~GeV~(right panel). Lines and points have the same meaning as in Fig.~\ref{sigma:delta_MA_band}.}\label{dsigma:delta_MA_band}
\end{center}
\end{figure}

 \begin{figure}
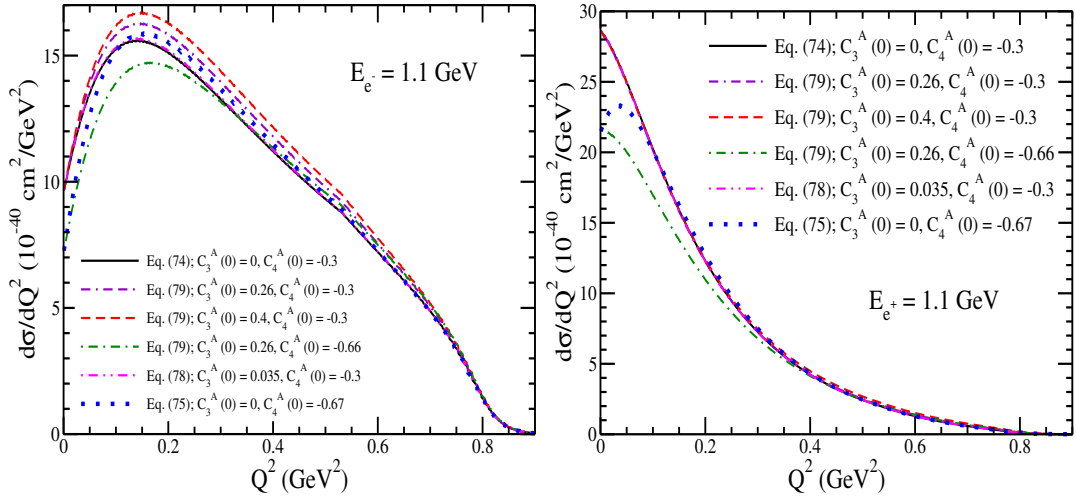
 
\begin{center}
\includegraphics[width=7cm,height=6.5cm]{dsigma_dQ2_electron_Ee_11GeV.eps}
\includegraphics[width=7cm,height=6.5cm]{dsigma_dQ2_positron_Ee_11GeV.eps}
\caption{  $\frac{d\sigma}{dQ^2}$ as a function of $Q^2$ for the processes $e^- + p \longrightarrow \nu_{e} + \Delta^{0}$~(left panel) and $e^+ + p \longrightarrow \nu_{e} + \Delta^{++}$~(right panel) for different parametrizations of the subdominant axial vector form factors $C_{3,4}^{A}(Q^2)$ at $E_{e}=1.1$~GeV. Lines and points have the same meaning as in Fig.~\ref{sigma:delta_weak}.}\label{dsigma:delta}
\end{center}
\end{figure}

\subsubsection{Results and discussion:}
In Fig.~\ref{sigma:delta_C5A0_band}, the results are presented for the total scattering cross sections $\sigma(E_{e^{-}})$ and $\sigma(E_{e^{+}})$ as a function of lepton energy for electron and positron induced $\Delta$ production processes viz. $e^- + p 
\longrightarrow \nu_e + \Delta^0$ and $e^+ + p \longrightarrow \bar{\nu}_e + \Delta^{++}$. This shows the dependence of the cross section on the dominant axial vector form factor $C_{5}^A(Q^2)$ using values of $C_5^A(0)$ in the range $0.87-1.4$ within the modified dipole parametrization given in Eq.~(\ref{cia_lala}), with $M_A=1.026$~GeV. For comparison, results are also presented using a value of $C_5^A(0)=1.2$ determined from the PCAC hypothesis and $M_A=1.026$~GeV. All results for the total cross section of $\Delta$ production processes induced by the electron and positron are obtained in the region of $W$ from $W_{min}=M+m_{\pi}$ to $W_{max} =1.4$~GeV.

In the case of electron induced $\Delta$ production, the cross section increases with increasing $E_e$ and saturates at around 2~GeV, while in the case of positron induced reaction, the cross section increases with increasing $E_{{e}^{+}}$ and saturates at $E_{{e}^{+}}=4$~GeV~(not shown here). 
In the energy region from threshold up to $E_{e}=1.5$~GeV, the production cross section for $\Delta^{0}$ induced by electron is larger than the production cross section for $\Delta^{++}$ induced by positron. However,
 as the energy of the incoming lepton increases, the cross section for $e^+ + p \longrightarrow \bar{\nu}_e + \Delta^{++}$ process becomes larger than the production cross section for $e^- + p 
\longrightarrow \nu_e + \Delta^0$ process for $E_{e} \ge 1.5$~GeV. 

We find that the  cross section depends strongly on the value of $C_5^A(0)$ for both the electron and positron induced reactions. The total cross section $\sigma(E_e)$ increases~(decreases) with increase~(decrease) in the value of $C_5^A(0)$ across all values of electron as well as positron energies considered in this work. For example, at $E_e=1-2$~GeV, increasing $C_5^A(0)$ from 1.2 to 1.4 increases the cross section by about 20\%, while the cross section reduces by about 30\% when $C_5^A(0)$ is decreased from 1.2 to 0.87 for the electron induced $\Delta$ production process. Furthermore, in the case of the positron induced reaction, the cross section increases by about 30\% when $C_5^A(0)$ is increased from 1.2 to 1.4, while it decreases by about 32\% when $C_5^A(0)$ is decreased from 1.2 to 0.87 at $E_{e^+}=1$~GeV. With increase in $E_{e^+}$, this increment~(decrement) slightly decreases and becomes about 22\%~(28\%) at $E_{e^+}=2$~GeV.

In Fig.~\ref{sigma:delta_MA_band}, we study the dependence of $\sigma(E_e)$ on the variation of axial dipole mass $M_A$ in the electron and positron induced $\Delta$ production processes by varying $M_A$ in the range $0.9-1.2$~GeV and keeping $C_5^A(0)=1.2$. For comparison, we also present in the same plot the results for $\sigma(E_e)$ obtained using a value of $M_A$, taken to the same as the world average value obtained in the quasielastic scattering region, i.e., $M_A=1.026$~GeV. We find a significant dependence of the cross section on $M_A$ for both the electron and positron induced reactions, although this dependence is less pronounced than that arising from variation in $C_5^A(0)$. Moreover, the effect of $M_A$ variation is more significant for the electron induced reaction than for the positron induced reaction. For example, in the case of the electron induced $\Delta$ production, reducing $M_A$ by 10\% from $M_A=1.026$~GeV, reduces the cross section by about 12\% and 15\%, respectively, at $E_e=1$ and 2~GeV, whereas increasing $M_A$ by 20\% from $M_A=1.026$~GeV, increases the cross section by about 15\% and 23\% at $E_e=1$ and 2~GeV, respectively. On the other hand, in the case of the positron induced reaction, the cross section decreases by about 8\% when $M_A$ is reduced by 10\% from $M_A=1.026$~GeV at $E_e=1-2$~GeV, while increasing $M_A$ by 20\% from 1.026~GeV, increases the cross section by about 13\% and 17\%, respectively, at $E_e=1$ and 2~GeV.

In Fig.~\ref{sigma:delta_weak}, we present the results for $\sigma(E_{e^{-}})$ as a function of $E_{e^-}$, and  $\sigma(E_{e^{+}})$ as a function of $E_{e^{+}}$,  and study the effect of the subdominant axial vector form factors $C_3^A(Q^2)$ and $C_4^A(Q^2)$ on the cross section by using the parametrizations of these form factors taken from different phenomenological and theoretical models. 
It may be observed from the figure that the results obtained in the model of Barquilla-Cano et al.~\cite{Barquilla-Cano:2007vds} and Adler~\cite{Adler:1968tw} overlap with one another and the effect of the non-zero
value of $C_{3}^{A} (Q^2)$ parameterized using the theoretical model of Barquilla-Cano et al.~\cite{Barquilla-Cano:2007vds} is almost negligible for both the electron and positron induced reaction channels because of the very small value of $C_{3}^{A}(0) = 0.035$. 
However, the non-zero value of $C_{3}^{A} (Q^2)$ calculated in the model of Chen et al.~\cite{Chen:2023zhh} leads to an enhancement in $\sigma(E_e)$ for both the reaction channels. 

To study the effect of $C_{4}^{A}(Q^2)$ on $\sigma(E_e)$, we obtain the results by taking different values of $C_{4}^{A}(0)$ as suggested by the models of Chen et al.~\cite{Chen:2023zhh} and Graczyk et al.~\cite{Graczyk:2009zh}. 
A smaller value of $C_{4}^{A} (0) = -0.66$ calculated in the model of Chen et al.~\cite{Chen:2023zhh}, with $C_{3}^{A}(0)=0.26$, leads to a reduction in the total cross section, which is more pronounced in the case of positron induced reaction as compared to the electron induced reaction. 
A comparison of the results calculated using the values of $C_{3}^{A}(0)=0$ and $C_{4}^{A} (0) =-0.3$ from the Adler's model~\cite{Adler:1968tw}  with the results obtained using their  central values in the model of Chen et al.~\cite{Chen:2023zhh} i.e. $C_{3}^{A}(0)=0.26$ and $C_{4}^{A} (0) =-0.66$,
shows that, in the case of electron induced $\Delta$ production process, the results obtained in the two models are quite consistent with each other, 
while in the case of positron induced process, a suppression in the total cross section is observed in the model of Chen et al.~\cite{Chen:2023zhh} as compared to the results obtained in Adler's model~\cite{Adler:1968tw}.

In Fig.~\ref{dsigma:delta_C5A0_band}, we study the effect of varying $C_5^A(0)$ in the range 0.87 to 1.4 while keeping $M_A=1.026$~GeV, on the $Q^2$ distribution of the cross section for the electron and positron induced $\Delta$ production processes. The results are presented for three different values of the electron and positron energies, namely, $E_e=855$~MeV, 1.1, and 2.2~GeV. We observe distinct behaviour of $\frac{d\sigma}{dQ^2}$ in the peak region of $Q^2$ for the electron and positron induced reactions. The differential scattering cross section shows a strong dependence on the choice of $C_5^A(0)$ over the entire range of $E_e$ and $Q^2$. 

In Fig.~\ref{dsigma:delta_MA_band}, the results are presented for $\frac{d\sigma}{dQ^2}$ as a function of $Q^2$ at $E_e=855$~MeV, 1.1, and 2.2~GeV, using $C_{5}^A(0)=1.2$ and $M_{A}$ in the range $[0.9,1.2]$~GeV for the electron and positron induced $\Delta$ production processes. We find a significant dependence of $\frac{d\sigma}{dQ^2}$ on the choice of $M_A$ in the entire range of $Q^2$ and $E_e$ considered in this work. Moreover, this dependence on the choice of $M_A$ is more pronounced in the case of electron induced $\Delta$ production process than the positron induced $\Delta$ production process. 

In Fig.~\ref{dsigma:delta}, we present the results for $\frac{d\sigma}{dQ^2}$ as a function of $Q^2$ for the electron and positron induced $\Delta$ production processes at $E_e=1.1$~GeV using different parametrizations of $C_{3}^{A} (Q^2)$ and $C_{4}^{A} (Q^2)$. 
It may be observed from the figure that in the case of positron induced reaction, there is no dependence on the choice of $C_{3}^{A} (Q^2)$, however, for the electron induced reaction we find some dependence on the choice of $C_{3}^{A} (Q^2)$.
Similar to the case of the total cross section, here also we find that $\frac{d\sigma}{dQ^2}$ obtained in the model of Barquilla-Cano et al.~\cite{Barquilla-Cano:2007vds} and Adler~\cite{Adler:1968tw} overlap with one another  for both the electron and positron induced reactions.  
In the region of $Q^2$ where $\frac{d\sigma}{dQ^2}$ peaks, i.e., $Q^2 \approx 0.15$~GeV$^{2}$, we observe an increment of about 4\% and 6\%, respectively, when $C_{3}^{A}(0) =0.26$ and 0.4 are used in the Chen's parametrization as compared to the Adler's parametrization. 
At $Q^2 \approx 0$, we observe a suppression of about 3\% in $\frac{d\sigma}{dQ^2}$, when Chen's or Graczyk's parametrization for $C_{4}^{A}(Q^2)$ is used in the numerical calculations as compared to the results obtained using Adler's parametrization. However, with increasing $Q^2$, $\frac{d\sigma}{dQ^2}$ obtained using the parametrization of Graczyk~\cite{Graczyk:2009qm} increases and in the peak region of $\frac{d\sigma}{dQ^2}$ becomes comparable to the results obtained using the Adler's parametrization. Moreover, $\frac{d\sigma}{dQ^2}$ at $Q^2 \approx 0.15$~GeV$^{2}$ obtained using the parametrization of Chen et al.~\cite{Chen:2023zhh} is almost 6\% smaller than the results obtained in the Adler's model.

In the case of the positron induced reaction, we observe a suppression in $\frac{d\sigma}{dQ^2}$ at $Q^2 \approx 0$, which is about 25\% when the results obtained using the parametrization of Chen et al.~\cite{Chen:2023zhh} and Graczyk~\cite{Graczyk:2009qm} are compared with the results obtained in the Adler's model. However, this suppression decreases with increasing $Q^2$ and the results obtained using Graczyk's and Chen's parametrizations become comparable to the results obtained using the Adler's parametrization at $Q^2=0.1$~GeV$^2$ and $0.3$~GeV$^2$, respectively.

\subsection{$P_{11}(1440)$ and $S_{11}(1535)$ nucleon resonance production}\label{sec:Nstar}
 The production process of the positive and negative parity $P_{11}(1440)$ and $S_{11}(1535)$ resonances of spin $\frac{1}{2}$, induced by the charged current reaction of an electron from the free proton target is represented as:
  \begin{eqnarray}\label{Nstar}
    e^-(k) + p(p) &\longrightarrow& \nu_e(k^\prime) + N^\star(p^\prime); \qquad \quad N^\star = P_{11}(1440), S_{11}(1535)
  \end{eqnarray}
  where the quantities in the parentheses represent the four momenta of the corresponding particles.

\subsubsection{Matrix element and form factors:}
 The transition matrix elements for the reaction given in Eq.~(\ref{Nstar}) is defined as
 \begin{eqnarray}\label{mat:N}
  {\cal M} &=& \frac{G_{F} \cos\theta_{C}}{\sqrt{2}} \; l_{\mu}  \bra{N^{\star} (p^{\prime})} j^{\mu} \ket{p(p)},  
 \end{eqnarray}
where the leptonic current $l_{\mu}$ is given in Eq.~(\ref{lmu:e}) and 
the matrix element for the hadronic current $j^{\mu}$  is written as
  \begin{equation}\label{jmu}
\bra{N^{\star}(p^{\prime})}j^\mu \ket{p(p)}= u(p^{\prime}){ \mathcal O}^{\mu} \Gamma u( p).
\end{equation}
In the above expression, $u(p)$ is the Dirac spinor for the proton and  $u(p^{\prime})$ is the Dirac spinor for spin $\frac{1}{2}$ $N^\star$ resonance. $\Gamma=1~(\gamma_5)$ stands for the positive~(negative) parity $P_{11}(1440)~(S_{11}(1535))$ resonance.

 The $N-N^\star$ transition vertex $\mathcal O^{\mu}=V^{\mu}-A^{\mu}$ is described in terms of the vector~($V^{\mu}$)
 and the axial vector~($A^{\mu}$) transition vertices, which in turn, are described in terms of the vector and axial vector form factors as~\cite{SajjadAthar:2022pjt, Athar:2020kqn}:
\begin{eqnarray}\label{vec_tra_current:Nstar}
V^{\mu}&=&\frac{{f_{1}^{CC}}(Q^2)}{(2 M)^2}
  \left( Q^2 \gamma^\mu + {q\hspace{-.5em}/} q^\mu \right) + \frac{f_2^{CC}(Q^2)}{2 M} 
  i \sigma^{\mu\alpha} q_\alpha ,\\
\label{ax_tra_current:Nstar}
A^{\mu}&=& \left[{g_1^{CC}}(Q^2) \gamma^\mu  +  \frac{g_3^{CC}(Q^2)}{M} q^\mu\right] \gamma_5 .
\end{eqnarray}
In Eqs.~(\ref{vec_tra_current:Nstar}) and (\ref{ax_tra_current:Nstar}), $f_i^{CC}(Q^2); (i=1,2)$ are the isovector vector and $g_{i}^{CC}(Q^2); (i=1,3)$ are the isovector axial vector $N-N^\star$ transition form factors. These form factors are determined using the various  properties of the weak hadronic currents, which may be summarized as:
\begin{itemize}
\item [(i)] The invariance under time reversal implies that all the vector and axial vector $N-N^\star$ transition form factors must be real.

\item [(ii)] The isospin symmetry along with the CVC hypothesis for the vector currents, relates the weak vector form factors~($f_i^{CC}(Q^2)$) to the electromagnetic $N-N^\star$ transition form factors $f^{R+,R0}_{i}(Q^2)$ via the relation~\cite{SajjadAthar:2022pjt}
\begin{equation}\label{weak:CC:FF}
 f_i^{CC}(Q^2) = f_i^{R+} (Q^2) - f_i^{R0} (Q^2); \qquad \quad i=1,2
\end{equation}

\item [(iii)] The electromagnetic form factors $f^{R+,R0}_{i}(Q^2)$ are expressed in terms of the helicity amplitudes $A_{\frac{1}{2}}$ and $S_{\frac{1}{2}}$, which are extracted from the meson production data from the real and/or virtual photon scattering experiments.

\item [(iv)] The axial vector form factor $g_1^{CC}(Q^2)$ at $Q^2=0$ is related to the $R\rightarrow N\pi$ coupling using the PCAC hypothesis and the pion pole dominance of the divergence
of the axial vector current through the generalized
GT relation.

\item [(v)] Since no experimental information for the $N-N^\star$ transition axial vector form factors is available, therefore, for simplicity, we have assumed a dipole parametrization for the $Q^2$ dependence of $g_1^{CC}(Q^2)$.

\item [(vi)] The pseudoscalar form factor $g_3^{CC} (Q^2)$ can be expressed in terms of $g_1^{CC}(Q^2)$ using PCAC hypothesis. However, in the case of electron induced reactions, the contribution of the pseudoscalar form factor $g_3^{CC}(Q^2)$ is almost negligible.

\end{itemize}

\subsubsection{Weak vector form factors:}
The weak vector form factors are expressed in terms of the electromagnetic $N-N^\star$ transition form factors~(Eq.~(\ref{weak:CC:FF})), which, in turn, are expressed in terms of the experimentally extracted helicity amplitudes. In this section, we elaborate on the relation between the helicity amplitudes and electromagnetic form factors. To determine the 
helicity amplitudes $A_{1/2}$ and $S_{1/2}$, one considers the interaction of a nucleon with a virtual or real photon leading to the production of  a 
spin $1/2$ resonance. The helicity amplitudes for the process $\gamma N \longrightarrow R_{1/2}$ are defined in terms of 
the polarization of the photon and the spins of the incoming nucleon and the outgoing spin $1/2$ resonance. The spin of 
the resonance is taken to be aligned along the positive Z-direction, i.e. $J_{z}^{R} = +1/2$. 

The expressions for $A_{1/2}$ and $S_{1/2}$ are 
defined as~\cite{Leitner:2008ue, Aznauryan:2008us}:
\begin{eqnarray}\label{Ch11_a12}
 A_{1/2}^{N} &=& \sqrt{\frac{2 \pi \alpha}{K_{R}}} <{R, J_{z}^{R} = +1/2} |\epsilon_{\mu}^{+} V^{\mu}| {N, 
 J_{z}^{N} = -1/2}> e^{i\phi}, \\
 \label{Ch11_s12}
 S_{1/2}^{N} &=& - \sqrt{\frac{2 \pi \alpha}{K_{R}}} \frac{|\vec{q}|}{\sqrt{Q^{2}}} <{R, J_{z}^{R} = +1/2} |
 \epsilon_{\mu}^{0} V^{\mu} |{N, J_{z}^{N} = +1/2}> e^{i\phi},
\end{eqnarray}
where $\epsilon_{\mu}$ represents the photon 
polarization vector. The transverse polarized photon vector $\epsilon_{\mu}^{\pm}$ is defined as
\begin{equation}\label{Ch11_transverse}
 \epsilon_{\mu}^{\pm} = \mp \frac{1}{\sqrt{2}} (0,1,\pm i,0),
\end{equation}
and the longitudinal polarization of the photon $\epsilon_{\mu}^{0}$ is defined as
\begin{equation}\label{Ch11_longitudinal}
 \epsilon_{\mu}^{0} = \frac{1}{\sqrt{Q^{2}}} (|\vec{q}|, 0, 0, q^{0}).
\end{equation}
In Eqs.~(\ref{Ch11_a12}) and (\ref{Ch11_s12}), $\phi$ is the phase factor, which relates the amplitude for the production of the resonances and the nucleons in the final state, $K_{R} = 
(M_{R}^{2} - M^{2})/2M_{R}$ is the momentum of the real photon measured in the resonance rest frame and $|\vec{q}|$ is the 
momentum of the virtual photon measured in the laboratory frame given as~\cite{Athar:2020kqn}:
\begin{equation}\label{helicity:vec}
 |\vec{q}| = \sqrt{\frac{(M_{R}^{2} - M^{2} - Q^{2})^{2}}{(2M_{R})^{2}} + Q^{2}}.
\end{equation}
The expressions for $V^{\mu}$ is given in Eq.~(\ref{vec_tra_current:Nstar}). 
 \begin{table*}
\centering
\begin{tabular*}{100mm}{@{\extracolsep{\fill}}cccc}\hline \hline
\multicolumn{2}{c}{Resonance $\rightarrow$} & \multirow{2}{*}{$P_{11}(1440)$} & \multirow{2}{*}{$S_{11}(1535)$} \\

\multicolumn{2}{c}{Parameters $\downarrow$} && \\ \hline

\multicolumn{2}{c}{$M_{R}$ (GeV)} & $1.440 \pm 0.03$ & $1.510 \pm 0.01$  \\ \hline

\multicolumn{2}{c}{$\Gamma_{R}$ (GeV)} & $0.350 \pm 0.1$ & $0.130 \pm 0.02$ \\ \hline

\multicolumn{2}{c}{$I(J^P)$} &$\frac{1}{2}(\frac{1}{2}^{+})$ &$\frac{1}{2}(\frac{1}{2}^{-})$\\ \hline 

\multirow{4}{*}{BR (in \%)} & $N\pi$ & $55 - 75$ & $32-52$ \\ 

&  $N\eta$  & $<1$ & $30-55$ \\ 

& $N\pi\pi$ & $17-50$ &$4-31$ \\ \hline

\multicolumn{2}{c}{$|g_{RN\pi}|$} & 0.38 & 0.1019 \\ \hline

\multicolumn{2}{c}{$|g_{RN\eta}|$} & 0.02 & 0.3696 \\  \hline \hline
\end{tabular*}
\caption{Properties of $P_{11}(1440)$ and $S_{11}(1535)$ resonances, with Breit-Wigner mass $M_{R}$, the total decay width $\Gamma_{R}$,
isospin $I$, spin $J$, parity $P$, the branching ratio full range available from PDG~\cite{ParticleDataGroup:2024cfk} into different meson-baryon channels such as $N\pi$, $N\eta$, and $N\pi\pi$, and the strong coupling constant $g_{RN\pi}$ and $g_{RN\eta}$.}
\label{tab:param-p2}
\end{table*}

\begin{table*}
\centering
\begin{tabular*}{150mm}{@{\extracolsep{\fill}}ccc c c c c  c c c}\hline\hline
&Resonance & Helicity amplitude & \multicolumn{3}{c}{ Proton target } & \multicolumn{3}{c}{ Neutron target } &\\ \hline
&&& ${\cal A}_{\alpha} (0)$ & $a_1$ & $b_{1}$ & ${\cal A}_{\alpha} (0)$ & $a_1$ & $b_{1}$ &\\ \hline
&\multirow{2}{*}{$P_{11}(1440)$} & $A_{\frac{1}{2}}$ & $-61.4$ & 0.871 & 1.36 & $54.1$ & 0.95 & 1.77& \\ 
&&$ S_{\frac{1}{2}}$ & $4.2$ & 40.0 & 1.75 & $-41.5$ & 2.98 & 1.55& \\
&\multirow{2}{*}{$S_{11}(1535)$} & $A_{\frac{1}{2}}$ & 95.0 & 0.85 & 0.85 & $-78.0$ & 1.75 & 1.75& \\ 
&&$ S_{\frac{1}{2}}$ & $-2.0$ & 1.9 & 0.81 & $32.5$ & 0.4 & 1.0& \\ \hline \hline
\end{tabular*}
\caption{Parameterization of the helicity amplitude for $P_{11} (1440)$ and $S_{11} (1535)$ resonances on the proton and neutron targets.  ${\cal A}_{\alpha} (0)$ is given in units of $10^{-3}$ GeV$^{-2}$ and the coefficients $a_1$  and $b_1$ in units of GeV$^{-2}$.}
\label{tab:resonance1}
\end{table*}
Using Eqs.~(\ref{Ch11_a12}), (\ref{Ch11_s12}), (\ref{Ch11_transverse}) and (\ref{Ch11_longitudinal}), the explicit relations between the form factors $f_i^{R+,R0}(Q^2)$ and the helicity amplitudes $A_{\frac{1}{2}}^{p,n}(Q^2)$ 
and $S_{\frac{1}{2}}^{p,n}(Q^2)$, for $\phi=0$, are given by~\cite{Leitner:2008ue}:
\begin{eqnarray}\label{eq:hel_spin_12_x}
A_\frac{1}{2}^{p,n}&=& \sqrt{\frac{2 \pi \alpha}{M} \frac{(M_R \mp M)^2+Q^2}{M_R^2 - M^2}} \left[  \frac{Q^2}{4 M^2}
f_1^{R+,R0} + \frac{M_R \pm M}{2 M} f_2^{R+,R0} \right] , \nonumber \\
S_\frac{1}{2}^{p,n}&=&\mp~\sqrt{\frac{ \pi \alpha}{M} \frac{(M \pm M_R)^2+Q^2}{M_R^2 - M^2}}
 \frac{(M_R \mp M)^2 +Q^2}{4 M_R M} \left[
\frac{M_R \pm M}{2 M} f_1^{R+,R0} - f_2^{R+,R0}\right],
\end{eqnarray}
where the upper sign represents the positive parity state and the lower sign denotes the negative parity state. $M_R$ is the 
mass of corresponding resonance and $f^{R+,R0}_{1,2}(Q^2)$ are the electromagnetic transition $N-N^\star$ form factors~\cite{SajjadAthar:2022pjt}. By inverting Eq.~(\ref{eq:hel_spin_12_x}), the form factors $f^{R+,R0}_{1,2}(Q^2)$ are obtained in terms of the experimentally extracted helicity amplitudes.

The $Q^2$ dependence of the helicity amplitudes $A_{\frac{1}{2}}(Q^2)$ and $S_{\frac{1}{2}} (Q^2)$  is generally parameterized as~\cite{Tiator:2011pw, Drechsel:2007if}:
\begin{equation}\label{eq:ffpar}
{\mathcal A}_{\alpha}(Q^2) = {\mathcal A}_{\alpha}(0) (1+a_1 Q^2)\, e^{-b_1 Q^2} ,
\end{equation}
where $ {\mathcal A}_{\alpha}(Q^2)$ denote the helicity amplitudes, namely $A_{\frac12}(Q^2)$ and $S_{\frac12}(Q^2)$, which are generally determined by a fit to the photo- and electro- production data of the corresponding resonance. For the resonances $P_{11}(1440)$ and $S_{11}(1535)$, the values of $A_{\frac{1}{2}}$ at $Q^2=0$ are
taken from the PDG~\cite{ParticleDataGroup:2024cfk}, whereas the parameters $a_1$ and $b_1$ are
obtained by fitting the meson electroproduction data on the
cross section at various $Q^2$ as measured in the CLAS
experiment~\cite{CLAS:2001cbm, CLAS:2003vka, CLAS:2004ncx,  CLAS:2006ezq, CLAS:2006sjw, CLAS:2007jpl}.  The resulting values of these parameters for $P_{11}(1440)$ and $S_{11}(1535)$ resonances are listed in Table~\ref{tab:resonance1}.

\subsubsection{Axial vector form factors:}
The axial-vector form factor $g_1^{CC}(Q^2)$ at $Q^2=0$ is related to the strong pion-nucleon
coupling~(see Ref.~\cite{Athar:2020kqn, Fatima:2023fez}) by the following relation
\begin{equation}\label{eq:g1_pos}
g_1^{CC}(0)= 2 g_{RN\pi},
\end{equation}
with $g_{RN\pi}$ being the coupling strength for $R \to N\pi$ decay, 
which has been determined by the partial decay width of the resonance. 

To obtain the expression for the decay width of the resonance, we start by writing the most 
general form of $R \rightarrow N\pi$ Lagrangian given by~\cite{Athar:2020kqn}:
\begin{align}\label{eq:spin12_lag}
 \mathcal{L}_{RN\pi} &= \frac{g_{RN\pi} }{f_\pi}\bar{\Psi}_{R} \; 
 \Gamma^{\mu}_{\frac{1}{2}}  \;
  \partial_\mu  \phi^i \vec{\tau}_i \,\Psi 
\end{align}
where $f_{\pi} = 
92.4$~MeV~\cite{ParticleDataGroup:2024cfk} is the pion decay constant, $\Psi$ is 
the nucleon field and ${\Psi}_{R}$ is the field associated with the 
resonance. $\phi^i$ are the pion field and $\vec{\tau}$ is the isospin operator. The interaction vertex 
$\Gamma^{\mu}_{\frac{1}{2}}$ is $\gamma^\mu \gamma^5$~($\gamma^\mu$) for spin $\frac12$ resonance with positive~(negative) 
parity. 

Using the above Lagrangian, the 
expression for the decay width, in the resonance rest frame, is obtained as:
\begin{align}\label{eq:12_width}
 \Gamma_{R \longrightarrow N\pi} &= \frac{\mathcal{C}}{4\pi} \left(\frac{g_{RN\pi}}{f_\pi}\right)^2 
 \left(M_R \pm M\right)^2 
\frac{E_N \mp M}{M_R} |\vec{q}_{\mathrm{cm}}|, 
\end{align}
where the upper~(lower) sign represents the positive~(negative) parity resonance state. 
The parameter $\mathcal{C}=3$ is obtained from the isospin analysis. $|\vec q_{cm}|$ is the outgoing pion momentum measured from resonance rest frame and $E_N$ is the nucleon energy, which are 
given by, 
\begin{equation}\label{eq:pi_mom}
|\vec{q}_{\mathrm{cm}}| = \frac{\sqrt{(M_R^2-m_{\pi}^2-M^2)^2 - 4 m_{\pi}^2 M^2}}{2 M_R}, \qquad \qquad   E_N=\frac{M_R^2+M^2-m_{\pi}^2}{2 M_R}, 
\end{equation}
where $m_\pi$ is the pion mass.

Since no information about the $Q^2$ dependence of 
the axial-vector form factor is known experimentally, therefore, a dipole form is assumed:
\begin{equation}
 g_1^{CC}(Q^2) = \frac{g_1^{CC}(0)}{\left(1+\frac{Q^2}{M_{A}^2}\right)^2},
\end{equation}
with $M_{A}=1.026$~GeV.

\begin{figure}
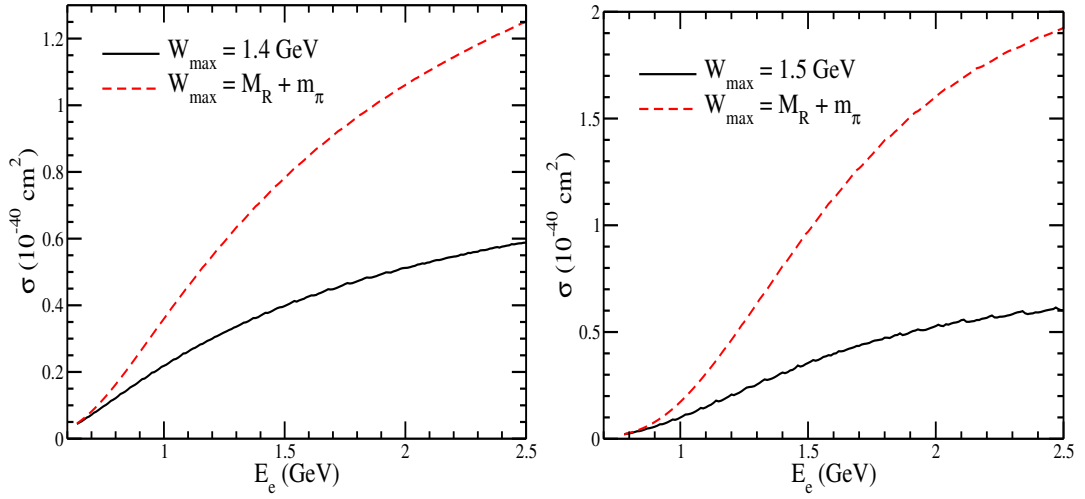
 
\begin{center}
\includegraphics[width=7cm,height=6.5cm]{sigma_P11_1440.eps}
\includegraphics[width=7cm,height=6.5cm]{sigma_S11_1535.eps}
\caption{Total scattering cross section ($\sigma(E_e)$) as a function of electron energy ($E_{e}$) for
 $e^- + p \longrightarrow \nu_e + N^{\star} (1440)$~(left panel) and $e^- + p \longrightarrow {\nu}_e + N^{\star} (1535)$~(right panel) scattering processes. 
 The solid and dashed lines, respectively, represent the results obtained using the two different values of $W_{max}$ viz. (i)~$W_{max}=1.4~(1.5)$~GeV for $P_{11}(1440)~(S_{11}(1535))$ resonance, and (ii)~$W_{max}=M_R+m_{\pi}$, with $M_R$ being the mass of the corresponding $N^{\star}$ resonance.
 }\label{sigma:resonance}
\end{center}
\end{figure}

\subsubsection{Cross section:}
The general expression for the differential scattering cross section for the processes given in Eq.~(\ref{Nstar}) is given in Eq.~(\ref{delta_cross_section}),
where $M_{\Delta}$ and $\Gamma_{\Delta}$ are now replaced with $M_R$ and $\Gamma_R$, the mass and decay width of the $P_{11}(1440)$ and $S_{11}(1535)$ resonances. The expression for the transition matrix element squared is given in Eq.~(\ref{mat_square}), with the expression for the leptonic tensor given in Eq.~(\ref{Lmunu}) and the hadronic tensor 
 $J^{\mu\nu}$ is defined in terms of the hadronic currents $j^{\mu}$, given in Eq.~(\ref{jmu}), as 
 \begin{equation}
 J^{\mu\nu}=  \overline{\sum}\sum j^{\mu} {j^{\nu}}^{\dagger}=  \frac{1}{2}
 Tr\left[ (\slashed{p} + M) {\tilde{\mathcal O}}^{\mu } (\slashed{p}^\prime + M_R)
      {\mathcal O}^{\nu} \right].
   \end{equation}  
 The transition vertex ${\mathcal O}^{\mu}$ is expressed as ${\mathcal O}^{\mu}=V^\mu-A^\mu$, where the expressions for $V^\mu$ and $A^\mu$ are given in Eqs.~(\ref{vec_tra_current:Nstar}) and (\ref{ax_tra_current:Nstar}), respectively.

 \begin{figure}
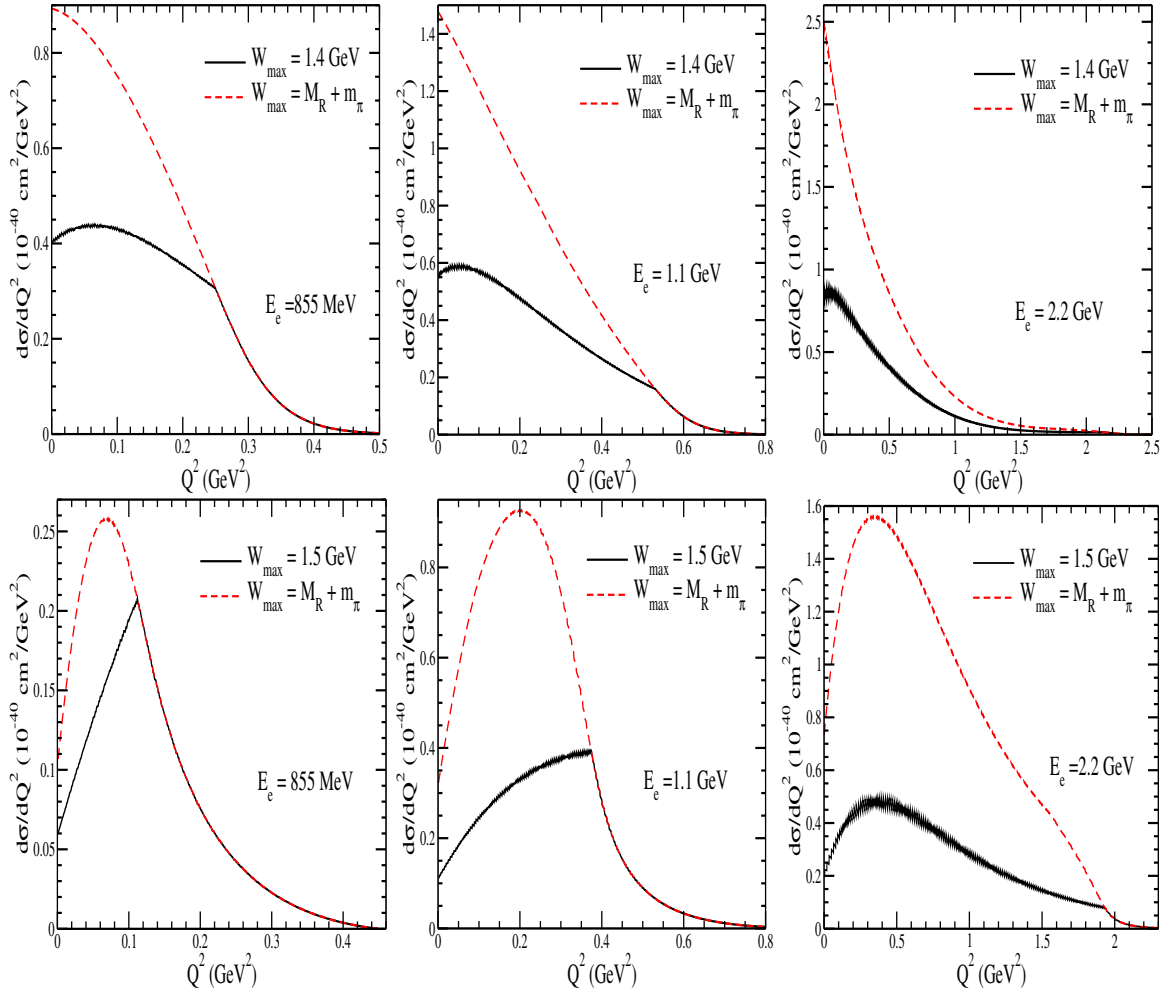
 
\begin{center}
\includegraphics[width=5cm,height=6.5cm]{dsigma_dQ2_P11_1440_Ee_855MeV.eps}
\includegraphics[width=5cm,height=6.5cm]{dsigma_dQ2_P11_1440_Ee_11GeV.eps}
\includegraphics[width=5cm,height=6.5cm]{dsigma_dQ2_P11_1440_Ee_22GeV.eps} 

\includegraphics[width=5cm,height=6.5cm]{dsigma_dQ2_S11_1535_Ee_855MeV.eps}
\includegraphics[width=5cm,height=6.5cm]{dsigma_dQ2_S11_1535_Ee_11GeV.eps}
\includegraphics[width=5cm,height=6.5cm]{dsigma_dQ2_S11_1535_Ee_22GeV.eps}
\caption{$\frac{d\sigma}{dQ^2}$ as a function of $Q^2$ for the processes $e^- + p \longrightarrow \nu_{e} + N^{\star} (1440)$~(top panel) and $e^+ + p \longrightarrow \nu_{e} + N^{\star} (1535)$~(bottom panel) for different $W_{max}$ at the three different values of the incoming electron energy viz. $E_{e}=855$~MeV~(left panel), 1.1~GeV~(middle panel), and 2.2~GeV~(right panel). Lines and points have the same meaning as in Fig.~\ref{sigma:resonance}.}\label{dsigma:resonance}
\end{center}
\end{figure}

The energy dependent decay width $\Gamma_R(W)$, following Ref.~\cite{Skoupil:2018vdh}, is given by:
\begin{equation}
\Gamma_R(W)=\Gamma_{R}\frac{W}{M_{R}}\sum_i\left[x_i \left(\frac{|\vec{q}_i|}{|\vec{q}_i^{R}|}\right)^{2l+1}\!\frac{D_l(|\vec{q}_i|)}{D_l(|\vec{q}_i^{\,R}|)}\right],
\label{eq:Gamma-s}
\end{equation}
where the sum $i$ runs over all possible meson-baryon decay modes, with the relative orbital momentum $l$. $\Gamma_{R}$ and $x_i$ denote the total decay width, and the branching ratio of a resonance into different $i^{th}$ meson-baryon channels~\cite{ParticleDataGroup:2024cfk}, respectively. The momenta $|\vec{q}_i^{R}|$ and $|\vec{q}_i|$ have the following form:
\begin{eqnarray}
|\vec{q}_i^{R}|&=& \sqrt{\frac{(M_{R}^2-M_{B}^2+M_{m}^2)^2}{4M_{R}^{2}}-M_{m}^2},\\
|\vec{q}_i|&=& \sqrt{\frac{(W^{2}-M_B^2+M_m^2)^2}{4W^{2}}-M_m^2},\;\;\text{and}\\
D_l(x) &=& \exp\left(-\frac{x^2}{3\alpha^2}\right),
\end{eqnarray}
 with $\alpha=400$~MeV as taken in Ref.~\cite{Skoupil:2018vdh}, and $x= |\vec{q}_i|$ or $|\vec{q}_i^{R}|$. 

\subsubsection{Results and discussion:}
In Fig.~\ref{sigma:resonance}, we present the results for the total cross section $\sigma(E_e)$ as a function of incoming electron energy $E_e$ for the excitation of $P_{11}(1440)$ and $S_{11}(1535)$ resonances in the scattering of electrons from free protons, using two different values of $W_{max}$ viz., (i)~$W_{max}=1.4~(1.5)$~GeV for $P_{11}(1440)~(S_{11}(1535))$ resonance, and (ii)~$W_{max}=M_R+m_{\pi}$, where $M_R$ represents the mass of the corresponding resonance. We find that the cross sections for $P_{11}(1440)$ and $S_{11}(1535)$ resonances are more than an order of magnitude smaller than the cross sections obtained for the $\Delta$ resonance. Moreover, we find a significant dependence of $\sigma(E_e)$ on $W_{max}$. Quantitatively, in the case of $P_{11}(1440)$ resonance, when $W_{max}$ is increased from 1.4~GeV to $\sim 1.6$~GeV, we find that the cross section increases by about 65\%~(100\%) at $E_{e}=1~(2)$~GeV. On the other hand, in the case of $S_{11}(1535)$ resonance, increasing the $W_{max}$ from 1.5~GeV to $\sim 1.7$~GeV increases the cross section by about 40\%~(67\%) at $E_{e}=1~(2)$~GeV.

In Fig.~\ref{dsigma:resonance}, the results are presented  for the $Q^2$ dependence of the cross section, $\frac{d\sigma}{dQ^2}$ as a function of $Q^2$ at $E_{e}=855$~MeV, 1.1, and 2.2~GeV for the excitation of $P_{11}(1440)$ and $S_{11}(1535)$ resonances in the reaction $e^- + p \longrightarrow \nu_e +N^\star$. We find that the nature of the peak region of $\frac{d\sigma}{dQ^2}$ for $P_{11}(1440)$ and $S_{11}(1535)$ resonances to be quite different from one another.
The effect of $W_{max}$ is studied by taking $W_{max}=1.4~(1.5)$~GeV for $P_{11}(1440)~(S_{11}(1535))$ resonance and $W_{max}=M_R+m_{\pi}$ in the numerical calculations. We find a strong dependence on the choice of $W_{max}$, especially in the region of $Q^2$, where the differential scattering cross section peaks. In the case of $P_{11}(1440)$ resonance excitation, at $Q^2=0$, increasing the $W_{max}$ from 1.4~GeV to about $1.7$~GeV increases $\frac{d\sigma}{dQ^2}$ by 1.25~(1.6) times at $E_{e}=855$~MeV~(2.2~GeV) and this enhancement decreases with increasing $Q^2$.
Moreover, in the case of $S_{11}(1535)$ resonance, in the electron energy range of $0.8-1.1$~GeV, the nature of the $Q^2$ distribution is quite different for the two values of $W_{max}$. Furthermore, at $E_{e} =2.2$~GeV, we observe similar nature of the $Q^2$ distribution for the two values of $W_{max}$. When the value of $W_{max}$ is reduced from 1.7~GeV to 1.5~GeV, we find that in the peak region of $Q^2$, $\frac{d\sigma}{dQ^2}$ reduces by about 70\%.

\section{Inelastic processes leading to meson production}\label{sec2:eta}
In this section, we discuss the electron and positron scattering processes from the proton target leading to the production of the lowest lying pseudoscalar mesons~($J^P=0^-$) in the final state through the reaction:
\begin{equation}\label{eq:inelastic:reaction}
 e^- (k)/e^+(k) + p (p) \longrightarrow \nu_e (k)/\bar{\nu}_e (k^\prime) + B(p^{\prime}) + m(p_{m}),
\end{equation}
where $m(=\pi, \eta, K$, etc.) is a meson produced with a baryon~($B=N,Y$, etc.) in 
the final state and the quantities in the parentheses represent the four momenta of the corresponding particles. 

An important point to mention here is that the weak meson production processes, except for the pion production, are not well studied both theoretically~\cite{SajjadAthar:2022pjt} and experimentally~\cite{ MicroBooNE:2025kqo, MicroBooNE:2023ubu, MINERvA:2016ymg, MINERvA:2016zyp}, even in the neutrino sector. 
The weak interaction induced single pion production using charged lepton beams is a comprehensive topic, needs a separate and detailed study, which is currently beyond the scope of this review.

In the electromagnetic sector, the processes such as the photo- and electro- production of $N\eta$ and $K\Lambda$ are well studied both theoretically as well as experimentally~\cite{CLAS:2005lui, SAPHIR, Tran:1998qw, CrystalBallatMAMI:2010slt, A2:2014pie, Denizli:2007tq, CLAS:2000mbw, Mart:2026tyg}.
Therefore, to develop a phenomenological model to describe the weak meson production processes induced by the charged and neutral leptons, one has to first apply it to the electromagnetic sector, where considerable data are available. In our earlier works~\cite{Fatima:2022tlf, Fatima:2023fez, Fatima:2024hlu, SajjadAthar:2022pjt}, we have developed a phenomenological model to study the (anti)neutrino induced meson production processes by first applying the model to the electromagnetic sector and found that the results obtained in our model for the photo- and electro- production of eta mesons as well as the photoproduction of $K\Lambda$ are quite consistent with the experimental results available from JLab and MAMI experiments~\cite{CLAS:2005lui, CrystalBallatMAMI:2010slt, A2:2014pie, Denizli:2007tq, CLAS:2000mbw}. The same model is then used, in this work, to calculate the cross sections for the weak meson production induced by the electron and positron from the proton target. The cut-off parameter for strong form factor occurring in the strong interaction of the resonance $R$ decaying to $N\eta$ and $K\Lambda$ channels and the strong couplings of the resonances $R$ to the eta-nucleon and kaon-hyperon decay modes, i.e., $g_{RN\eta}$ and $g_{RK\Lambda}$ determined from fitting the photoproduction data, and the $Q^2$ dependence of the vector form factors determined from fitting the electroproduction data, as done in our earlier calculations~\cite{Fatima:2022tlf, Fatima:2023fez, Fatima:2024hlu},
are used as inputs in this work.

The general expression for the differential scattering cross section of the meson production processes, 
in the laboratory frame, is given by
\begin{eqnarray}\label{eq:sigma_inelas}
d\sigma &=& \frac{1}{4 ME_e(2\pi)^{5}} \frac{d{\vec k}^{\prime}}{ (2 E_{l})} 
\frac{d{\vec p\,}^{\prime}}{(2 E_{B})} \frac{d{\vec p}_{m}}{ (2 E_m)}
 \delta^{4}(k+p-k^{\prime}-p^{\prime}-p_{m})\overline{\sum}\sum | \mathcal M |^2,\;\;\;\;\;
\end{eqnarray}
where $E_l$, $E_B$ and $E_m$ are, respectively,  the energies  of the outgoing lepton, baryon and meson. The 
different kinematical variables used in the numerical calculations of the scattering cross section are depicted in 
Fig.~\ref{reactionplane}, where the scattering plane is in the laboratory frame while the reaction plane is in the center 
of mass frame. $\overline{\sum}\sum | \mathcal M |^2  $ is the square of the transition amplitude averaged~(summed) over the 
spins of the initial~(final) states, where the transition matrix element is written in terms of the leptonic and  hadronic 
currents as 
\begin{equation}
\label{eq:Gg}
 \mathcal M = a \frac{G_F}{\sqrt{2}}\, {l_\mu} j^{\mu}.
\end{equation}
In the above expression, $a=\cos\theta_C~(\sin\theta_C)$ for the charged current induced strangeness conserving~(strangeness changing) 
processes. The leptonic currents $l_\mu$ for the electron and positron induced reactions are defined in Eqs.~(\ref{lmu:e}) and (\ref{lmu:e+}), respectively. $  j^{\mu}$  is the hadronic current for $W^\pm + N 
\longrightarrow B + m$  interaction, to be discussed in detail below in Sections~\ref{weak:eta}, \ref{weak:associated}, and \ref{sec2:kaon} for various case of meson production.

\begin{figure}
 \includegraphics[height=5 cm, width=0.9\textwidth]{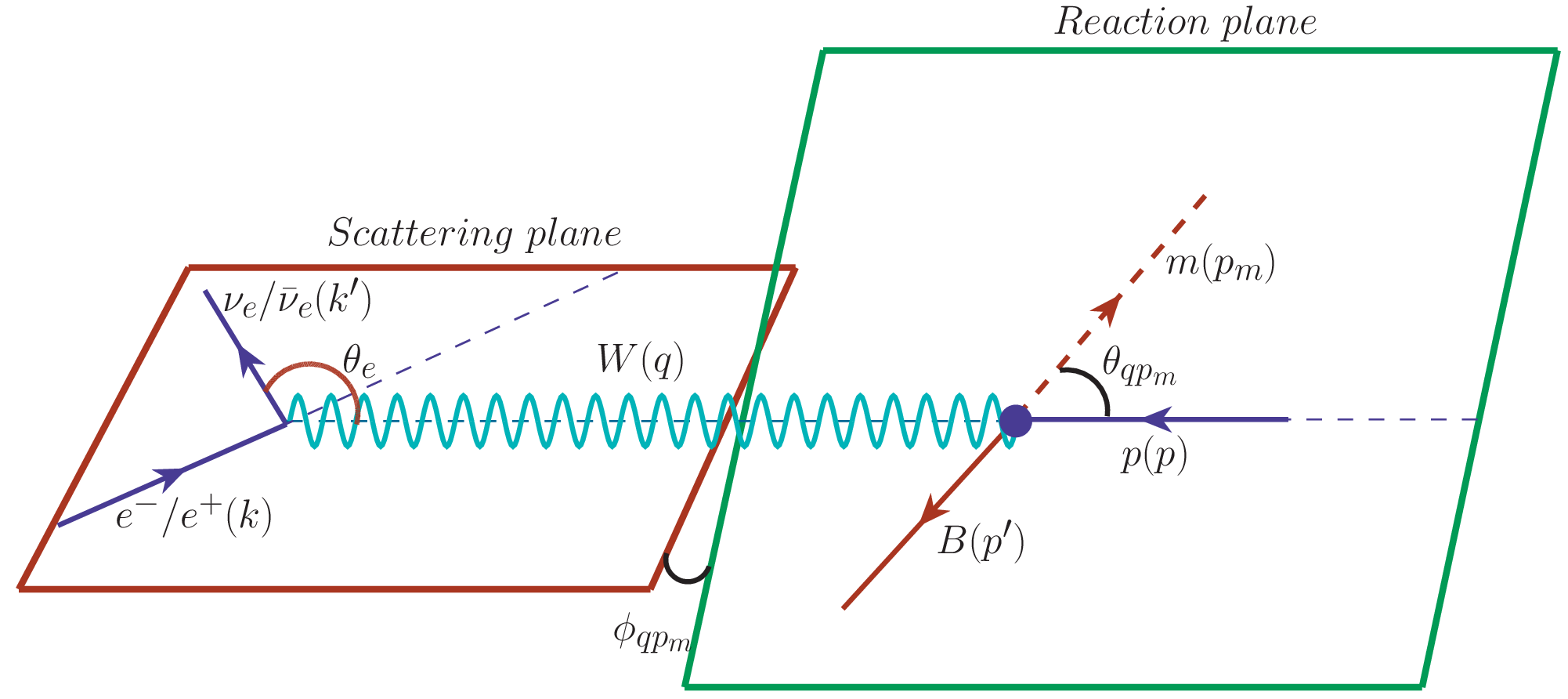}
 \caption{Electron and positron scattering and reaction planes, depicting the hadronic plane in CM frame
 and scattering plane in the laboratory frame for the processes $ e^- (k)/e^+(k) + p (p) \longrightarrow \nu_e (k)/\bar{\nu}_e (k^\prime) + B(p^{\prime}) + m(p_{m})$, where $m(=\pi, \eta, K$, etc.) is a meson $B(=N,Y$, etc.) is a baryon, with $Y=\Lambda, \Sigma$. The quantities in the parentheses represent the four momenta of the corresponding particles. The kinematical variables used in the calculation of various meson production processes are defined in the figure.}
 \label{reactionplane}
\end{figure}

Integrating over the three momentum of the outgoing baryon and the energy of the meson, the expression for the differential scattering cross section 
given in Eq.~(\ref{eq:sigma_inelas}) becomes~\cite{SajjadAthar:2022pjt}:
\begin{equation}\label{eq:sigma_inelastic}
\frac{d^4 \sigma}{dE_{l} ~d\cos \theta_{l} ~d\cos \theta_{m} ~d\phi_{m}} = \frac{|\vec{k}^{\prime}| |\vec{p}_{m}|^{2}}
{32(2\pi)^{4}E_{e}M} \frac{G_{F}^2 {a}^2 L_{\mu\nu} J^{\mu\nu}}{4} \frac{1}{(E_{e} + M - E_{l})|\vec{p}_{m}|^{2} -
E_{m} (\vec{p}_{m} \cdot \vec{q}\;)},
\end{equation}
where $L_{\mu\nu}$ is given in Eq.~(\ref{Lmunu}) and $J^{\mu\nu} = \sum j^{\mu}{j^{\nu}}^{\dagger}$ is the hadronic tensor with $j^{\mu}$ being the hadronic 
current that receives contribution from the nonresonant background~(NRB) terms as well as from the resonance excitations 
and their subsequent decay into a particular meson-baryon final state. 
Specifically, the hadronic current $j^\mu$  is written as the sum of the contributions from the NRB terms~($j^{\mu}_{\text{NR}}$), and from the resonance excitation terms from the spin $\frac{1}{2}$~($j^{\mu}_{R_{\frac{1}{2}}}$) and spin 
$\frac{3}{2}$~($j^{\mu}_{R_{\frac{3}{2}}}$) resonances, i.e.:
\begin{equation}\label{allterms}
 j^{\mu}~=~j^{\mu}_{\text{NR}} ~+~ j^{\mu}_{R_{\frac{1}{2}}}~+~j^{\mu}_{R_{\frac{3}{2}}}.
\end{equation}

It is more conventional to write the double differential scattering cross section with respect to the invariant variables such as $Q^2$ and $W$ i.e. $\frac{d^2\sigma}{dQ^2 dW}$, which is conveniently used by the experimenters, and is written as~\cite{Fatima:2022nfn}:
\begin{equation}\label{sigma:weak}
 \frac{d^2\sigma}{dQ^2 dW} = \int_{0}^{2\pi} d\phi_{m} \int_{E_{m}^{min}}^{E_{m}^{max}} dE_{m} \frac{1}{(2\pi)^{4}} \frac{1}{64E_{e}^{2}M^2} \frac{W}{|\vec{q}\;|} \overline{\sum} \sum |{\cal M}|^2.
\end{equation}

The meson production channels receive contribution from the 
different background terms as well as from different spin $\frac{1}{2}$ and $\frac{3}{2}$ resonance excitations. In the next section, we present the 
structure of the hadronic currents for the background and the resonance terms in general. Moreover, for different meson production
channels, the specific couplings and the contribution from different terms are discussed in the respective sections. In this work, we are interested in the study of electron and positron induced meson production processes such as the single eta production~(discussed in Section~\ref{weak:eta}), $\Delta S=0$ kaon production through the associated $\Lambda K$ production~(Section~\ref{weak:associated}), and the $\Delta S=1$ single kaon production~(Section~\ref{sec2:kaon}), which remains unexplored theoretically, except for the single kaon production~\cite{Alam:2013vwa}. The single pion production which requires an independent and a detailed study will be discussed  elsewhere.

\subsection{Nonresonant contribution to meson production}\label{NRB}
In general, the nonresonant contributions to pion production processes are calculated using a microscopic model based on the $SU(2)$ symmetric 
chiral Lagrangians.  To include the description of higher mass mesons like $\eta$ and $K$ mesons, this model is extended to the $SU(3)$ symmetric chiral Lagrangians. The basic parameters of the model are the 
meson decay constant $f_{m}$, the Cabibbo angle, the proton and neutron magnetic moments, and the asymmetric and symmetric 
axial-vector coupling constants for the two baryon octets, $D$ and $F$, obtained from the analysis of 
the semileptonic decays of neutron and hyperons.

In order to write a Lagrangian that describes the dynamics of these pseudoscalar mesons, we need continuous fields, which 
are described in terms of these Goldstone modes. The elements of $SU(3)$ pseudoscalar meson fields are written in terms of a 
unitary matrix~\cite{Athar:2020kqn} 
\begin{equation}
 U(\Theta) = \exp\left( -i \Theta_k (x) \frac{\lambda_k}{2} \right)\;,
\end{equation}
where $\Theta_k (x);~(k=1-8)$ are the real set of parameters and $\lambda_k$ are the traceless, Hermitian $3 \times 3$ Gell-Mann 
matrices. 

Each Goldstone boson corresponds to the $x$-dependent Cartesian component of the fields $\phi_k (x) $, which in turn, are 
expressed in terms of the physical fields as~\cite{Athar:2020kqn}: 
\begin{eqnarray}\label{eq2:ps_matrix_final}
 \Phi(x) =\sum_{k=1}^{8} \phi_k(x) \lambda_k =
\left(\begin{array}{ccc}
\pi^0+\frac{1}{\sqrt{3}}\eta &\sqrt{2}\pi^+&\sqrt{2}K^+\\
\sqrt{2}\pi^-&-\pi^0+\frac{1}{\sqrt{3}}\eta&\sqrt{2}K^0\\
\sqrt{2}K^- &\sqrt{2}\bar{K}^0&-\frac{2}{\sqrt{3}}\eta
\end{array}\right).
\end{eqnarray}
For the baryons, we follow the same procedure as we do for the mesons. However, unlike the pseudoscalar mesons where the 
fields are real, in the case of baryon fields, represented by a $B$ matrix, each entry is a complex-field and the general 
representation is given by~\cite{Athar:2020kqn}:
\begin{eqnarray}\label{eq2:pmatrix2}
B(x)=\sum_{k=1}^{8} \frac{1}{\sqrt2} b_k(x) \lambda_k = \left(\begin{array}{ccc}
\frac{1}{\sqrt{2}}\Sigma^0+\frac{1}{\sqrt{6}}\Lambda&\Sigma^+&p\\
\Sigma^-&-\frac{1}{\sqrt{2}}\Sigma^0+\frac{1}{\sqrt{6}}\Lambda&n\\
\Xi^-&\Xi^0&-\frac{2}{\sqrt{6}}\Lambda
\end{array}\right).
\end{eqnarray}

Next, we discuss the construction of the Lagrangian for the meson-meson and
baryon-meson interactions, as well as their interactions with the external fields. 

\subsubsection{Meson - meson interaction:}\label{Sec2:MMinter}
The lowest-order $SU(3)$ chiral Lagrangian describing the pseudoscalar mesons in the presence of an external current is 
obtained as~\cite{Scherer:2002tk, Scherer:2012xha}:
\begin{equation}\label{eq2:lagM}
{\cal L}_M=\frac{f_\pi^2}{4}\mbox{Tr}[D_\mu U (D^\mu U)^\dagger].
\end{equation}
 The covariant derivatives $D^{\mu} U$ and $D^{\mu} U^{\dagger}$ appearing 
in Eq.~(\ref{eq2:lagM}) are expressed in terms of the partial derivatives as
\begin{eqnarray}\label{eq2:coDer}
 D^\mu U \equiv \partial^\mu U - i r^\mu U + i U l^\mu, \qquad \qquad
  D^\mu U^\dagger \equiv \partial^\mu U^\dagger + i U^\dagger r^\mu - i l^\mu U^\dagger,
\end{eqnarray}
where $U$ is the $SU(3)$ unitary matrix given as
\begin{equation}
 U(x) = \exp\left(i\frac{\Phi(x)}{ f_\pi } \right), 
\end{equation}
and $\Phi(x)$ is given in Eq.~(\ref{eq2:ps_matrix_final}). $r_{\mu}$ and $l_{\mu}$, respectively, represent the right and 
left handed currents, defined in terms of the vector~($v_{\mu}$) and axial-vector~($a_{\mu}$) fields given by~\cite{Scherer:2002tk, Scherer:2012xha}:
\begin{equation}\label{Ch11_NLSM_VA}
 l_{\mu} = \frac{1}{2}(v_{\mu} - a_{\mu}), \qquad \qquad r_{\mu} = \frac{1}{2}(v_{\mu} + a_{\mu}).
\end{equation}
The vector and axial-vector fields are different for the interaction of different gauge bosons with the meson fields, and for the charged current induced processes, the left and right handed currents are expressed as
\begin{equation}\label{Ch11_lr_CC}
 l_{\mu} = - \frac{g}{2} (W_{\mu}^{+} T_{+} + W_{\mu}^{-}T_{-}), \qquad \qquad r_{\mu} = 0,
\end{equation}
where $g = \frac{e}{\sin \theta_{W}}$, $\theta_{W}$ is the Weinberg angle, $W_{\mu}^{\pm}$ represents the W-boson field 
and $T_{\pm}$ are defined as
\begin{equation}
T_{+} = \begin{pmatrix}
        0 & V_{ud} & V_{us} \\
        0 & 0 & 0 \\
        0 & 0 & 0
        \end{pmatrix}, \qquad \text{and} \qquad T_{-} = \begin{pmatrix}
                                                        0 & 0 & 0 \\
                                                        V_{ud} & 0 & 0 \\
                                                        V_{us} & 0 & 0
        \end{pmatrix},
\end{equation}
with $V_{ud} = \cos \theta_{C}$ and $V_{us} = \sin \theta_{C}$ being the elements of the Cabibbo-Kobayashi-Maskawa matrix 
and $\theta_{C}$ being the Cabibbo angle.

\subsubsection{Baryon - meson interaction:}
To incorporate baryons in the theory, we have to take care of their masses, which do not vanish in the chiral 
limit~\cite{Kubis:2007iy}. However, if we take nucleons as massive matter fields which couples to external currents and the 
pseudoscalar mesons, we have to then expand the Lagrangian according to their increasing number of momenta. Here, we shall 
present in brief the extension of the formalism to incorporate the heavy matter fields. 

The lowest-order chiral Lagrangian for the baryon octet in the presence of an external current, may be written in terms of 
the $SU(3)$ matrix $B$ as~\cite{Scherer:2002tk, Scherer:2012xha}
\begin{equation}\label{eq2:lagB}
{\cal L}_{MB}=\mbox{Tr}\left[\bar{B}\left(i D_{\mu} \gamma^{\mu}
-M\right)B\right]
-\frac{D}{2}\mbox{Tr}\left(\bar{B}\gamma^\mu\gamma_5\{u_\mu,B\}\right)
-\frac{F}{2}\mbox{Tr}\left(\bar{B}\gamma^\mu\gamma_5[u_\mu,B]\right),
\end{equation}
where $M$ denotes the mass of the baryon octet, $D=0.804$ and $F=0.463$ are the symmetric and antisymmetric axial-vector 
coupling constants for the baryon octet, 
the matrix $B$ is given in Eq.~(\ref{eq2:pmatrix2}) and the Lorentz vector $ u^\mu$ is given by~\cite{Scherer:2012xha}:
\begin{equation}\label{eq2:vielbein}
u^\mu = i \left[ u^\dagger ( \partial^\mu - i r^\mu) u - u ( \partial^\mu - i l^\mu) u^\dagger \right].
\end{equation} 
In the case of meson-baryon interactions, the unitary matrix for the pseudoscalar field is expressed as 
\begin{equation}
 u = \sqrt U \equiv \rm{exp} \left( i \frac{\Phi(x)}{ 2 f_\pi } \right),
\end{equation}
and the covariant derivative $D_{\mu}$ on the baryon fields $B$ is given by
\begin{equation}\label{dmuB}
D_\mu B=\partial_\mu B +[\Gamma_\mu,B], \qquad \text{with} \qquad \Gamma^\mu=\frac{1}{2}\left[u^\dagger(\partial^\mu-
ir^\mu)u
+u(\partial^\mu-il^\mu)u^\dagger\right],
\end{equation}
which is known as the chiral connection.

To calculate the contributions from the nonresonant terms, the Lagrangians given in Eqs.~(\ref{eq2:lagM}) and (\ref{eq2:lagB}) are used to determine the meson-meson and meson-baryon interactions, respectively, as well as their interactions with the gauge bosons.

\subsection{Resonance excitations}
Along with the nonresonant contribution to the meson production processes, there are several resonances with spin $\frac{1}{2}$, $\frac{3}{2}$, $\frac{5}{2}$, etc.,
which contribute to these processes. For simplicity, in our calculations, we have considered only spin $\frac{1}{2}$ and $\frac{3}{2}$ resonance excitations.
In general, the electron and positron induced excitation of spin $\frac{1}{2}$ and $\frac{3}{2}$ resonances that lead to the production of mesons in the final state may be represented by the following reactions:
\begin{eqnarray}
e^- (k) + p (p) &\longrightarrow& \nu_e(k^\prime) + R_{\frac{1}{2}, \frac{3}{2}} (p_{R}) =  \nu_e(k^\prime) + B(p^\prime) + m(p_m), \\
e^+ (k) + p (p) &\longrightarrow& \bar{\nu}_e(k^\prime) + R_{\frac{1}{2}, \frac{3}{2}} (p_{R}) =  \bar{\nu}_e(k^\prime) + B(p^\prime) + m(p_m).
\end{eqnarray}

In Table~\ref{tab:param}, we have listed the properties of those resonances which have been
considered in this work for the production of $\eta$ mesons and kaons. 
To avoid the complexity of the  numerical calculations, we have considered only those established resonances, which are available in the PDG~\cite{ParticleDataGroup:2024cfk} with spin $\frac{1}{2}$ and $\frac{3}{2}$, having mass $<2$~GeV, and have a sizeable branching ratio to $N\eta$ and $\Lambda K$ decay modes. 
The nucleon  resonances which
are excited in $\eta$ meson and kaon production processes are characterized by their mass, parity, spin and isospin and are represented by the symbol
$R_{IJ} (M_R)$~(Table-\ref{tab:param}), where $R$ is the name of the resonance given on the basis of its orbital angular momentum i.e. $L = 0, 1, 2$
and named $S$, $P$, $D$, etc., showing its parity, $M_R$ is the mass while $I$ and $J$ specify their isospin and spin quantum numbers, respectively.

In Sections~\ref{sec:Delta} and \ref{sec:Nstar}, we have discussed the excitation of spin $\frac{3}{2}$ and $\frac{1}{2}$ resonance states, respectively, with particular focus on $P_{33}(1232)$, $P_{11}(1440)$, and $S_{11}(1535)$ resonances. However, the same formalism can be generalized to any spin $\frac{1}{2}$ and $\frac{3}{2}$ resonances. In the case of $\eta$ production, we have considered five nucleon resonances viz. $S_{11}(1535)$, $S_{11}(1650)$, $P_{11}(1710)$, $P_{11}(1880)$, and $S_{11}(1895)$ and in the case of $K\Lambda$ production, we have taken the contribution from $S_{11}(1650)$, $P_{11}(1710)$, $P_{13}(1720)$, $P_{11}(1880)$, $S_{11}(1895)$, and $P_{13}(1900)$ resonances. 
The helicity amplitudes for the excitation of these resonances were determined by us in our earlier works~\cite{Fatima:2022nfn, Fatima:2022tlf, Fatima:2025tht} by fitting the experimental data available from the CLAS experiment~\cite{CLAS:2001cbm, CLAS:2003vka, CLAS:2004ncx,  CLAS:2006ezq, CLAS:2006sjw, CLAS:2007jpl}. The values of these parameters are tabulated in Table~\ref{tab:resonance:helicity}.
In order to calculate the eta and kaon production through the resonance excitations, we use the nucleon~($N$) to resonance~($R$) i.e. $N-R$ transition vertex and the form factors, as discussed in the aforementioned sections and include the resonance propagator and its decay  to $N\eta$ and $\Lambda K$ channels, which will be discussed in the Sections~\ref{weak:eta} and \ref{weak:associated}, respectively.

 \begin{table*} 
\centering
\begin{small}
\begin{tabular*}{155mm}{@{\extracolsep{\fill}}cccccccc}\hline \hline
\multicolumn{2}{c}{Resonance $\rightarrow$} &  $S_{11}(1650)$ & $P_{11} (1710)$ & $P_{13} (1720)$ & $P_{11} (1880)$ &$S_{11} (1895)$ &  $P_{13} (1900)$\\

\multicolumn{2}{c}{Parameters $\downarrow$} && &&&&\\ \hline

\multicolumn{2}{c}{$M_{R}$ (GeV)} &  $1.655 \pm 0.015$ & $1.700 \pm 0.02$ & $1.680 \pm 0.02$ & $1.860 \pm 0.04$ & $1.910\pm 0.02$ & $1.920\pm 0.02$ \\ \hline

\multicolumn{2}{c}{$\Gamma_{R}$ (GeV)} &  $0.135 \pm 0.035$ & $0.120 \pm 0.04$ & $0.150 \pm 0.05$ &  $0.230 \pm 0.05$ & $0.110 \pm 0.03$ & $0.130 \pm 0.03$ \\ \hline

\multicolumn{2}{c}{$I(J^P)$} & $\frac{1}{2}(\frac{1}{2}^{-})$ & $\frac{1}{2}(\frac{1}{2}^{+})$ &$\frac{1}{2}(\frac{3}{2}^{+})$ & $\frac{1}{2}(\frac{1}{2}^{+})$ & $\frac{1}{2}(\frac{1}{2}^{-})$ &$\frac{1}{2}(\frac{3}{2}^{+})$ \\ \hline 

 & $N\pi$ &  $50-70$~(60) &
$5-20$~(16) & $8-14$~(11)& $3-31$~(34) & $2-18$~(23) &  $1-20$~(12) \\ 

BR & $N\eta$ & $15-35$~(25) & $10-50$~(20) & $1-5$~(3)& $1-55$~(20) & $15-45$~(30) &$2-14$~(8) \\ 

 (in \%) &$K\Lambda$ & $5-15$~(10) & $5-25$~(15) & $4-19$~(2)& $1-3$~(2) & $3-23$~(13)& $2-20$~(4) \\ 

& $N\pi\pi$ & $20-58$~(5)& $14-48$~(49)&$>50$~(84) & $>32$~(44) & $17-74$~(34)& $>56$~(76) \\ \hline

\multicolumn{2}{c}{$|g_{RN\pi}|$} &  0.0915 & 0.0418 & 0.1105 & 0.0466 & 0.0229& 0.0644 \\ \hline

\multicolumn{2}{c}{$|g_{RN\eta}|$} &  0.1481 & 0.1567 & 0.1972 & 0.1369 & 0.0877 & 0.5684\\ \hline

\multicolumn{2}{c}{$|g_{RK\Lambda}|$} &  0.1362 & 0.1973 & 1.29 & 0.0604& 0.08556& 0.1462\\ \hline \hline
\end{tabular*}
\end{small}
\caption{Properties of the spin $\frac{1}{2}$ resonances available in the PDG~\cite{ParticleDataGroup:2024cfk}, with Breit-Wigner mass $M_{R}$, the total decay width $\Gamma_{R}$,
isospin $I$, spin $J$, parity $P$, the branching ratio full range available from PDG~(used in the present calculations) into different meson-baryon channels such as $N\pi$, $N\eta$, $K\Lambda$, and $N\pi\pi$, and the strong coupling constant $g_{RN\pi}$, $g_{RN\eta}$, and $g_{RK\Lambda}$.}
\label{tab:param}
\end{table*}

\begin{table*} 
\centering
\begin{tabular*}{150mm}{@{\extracolsep{\fill}}ccc c c c c  c c c}\hline\hline
&Resonance & Helicity amplitude & \multicolumn{3}{c}{ Proton target } & \multicolumn{3}{c}{ Neutron target } &\\ \hline
&&& ${\cal A}_{\alpha} (0)$ & $a_1$ & $b_{1}$ & ${\cal A}_{\alpha} (0)$ & $a_1$ & $b_{1}$ &\\ \hline
&\multirow{2}{*}{$S_{11}(1650)$} & $A_{\frac{1}{2}}$ & 33.3 & 0.45 & 0.72 & $26.0$ & 0.1 & 2.5& \\ 
&&$ S_{\frac{1}{2}}$ & $2.5$ & 1.88 & 0.96 & $3.8$ & 0.4 & 0.71& \\ \hline
&\multirow{2}{*}{$P_{11}(1710)$} & $A_{\frac{1}{2}}$ & 55.0 & 1.0 & 1.05 & $-45.0$ & $-0.02$ & 0.95& \\ 
&&$ S_{\frac{1}{2}}$ & $4.4$ & 2.18 & 0.88 & $-31.5$ & 0.35 & 0.85& \\ \hline

&\multirow{3}{*}{$P_{13}(1720)$} & $A_{\frac{1}{2}}$ & 100.0 & 1.89 & 1.55 & $-2.9$ & $1.7$ & 1.55& \\ 
&&$ A_{\frac{3}{2}}$ & $-11.0$ & 10.0 & 1.55 & $-31.0$ & 3.0 & 1.55& \\ 
&&$ S_{\frac{1}{2}}$ & $-53.0$ & 2.46 & 1.55 & $0$ & 0 & 0& \\ \hline

&\multirow{2}{*}{$P_{11}(1880)$} & $A_{\frac{1}{2}}$ & $-60.0$ & 0.4 & 1.0 & $-45.0$ & $-0.02$ & 0.95& \\ 
&&$ S_{\frac{1}{2}}$ & $0.4$ & 0.75 & 0.5 & $-31.5$ & 0.35 & 0.85& \\ \hline
&\multirow{2}{*}{$S_{11}(1895)$} & $A_{\frac{1}{2}}$ & $-15.0$ & 1.45 & 0.6 & $26.0$ & $0.1$ & 2.5& \\ 
&&$ S_{\frac{1}{2}}$ & $-3.5$ & 0.88 & 0.6 & $3.8$ & 0.4 & 0.71& \\
\hline 
&\multirow{3}{*}{$P_{13}(1900)$} & $A_{\frac{1}{2}}$ & 8.0 & 1.89 & 1.55 & $-2.9$ & $12.7$ & 1.55& \\ 
&&$ A_{\frac{3}{2}}$ & $-98.0$ & 1.0 & 1.55 & $-31.0$ & 3.0 & 1.55& \\ 
&&$ S_{\frac{1}{2}}$ & $-10.0$ & 0.46 & 1.0 & $0$ & 0 & 0& \\ \hline
\hline
\end{tabular*}
\caption{Parameterization of the helicity amplitude for $S_{11} (1535)$, $S_{11}(1650)$, $P_{11}(1710)$, $P_{13}(1720)$, $P_{11}(1880)$, $S_{11}(1895)$, and $P_{13}(1900)$ resonances on the proton and neutron targets.  ${\cal A}_{\alpha} (0)$ is given in units of $10^{-3}$ GeV$^{-2}$ and the coefficients $a_1$  and $b_1$ in units of GeV$^{-2}$.}
\label{tab:resonance:helicity}
\end{table*}

 \subsection{Electron induced eta production}\label{weak:eta}

The CC electron induced single $\eta$ production off the proton 
target is given by the following reaction
\begin{eqnarray}\label{Ch12_eq:eta_weak_process_cc}
e^- (k)  + p (p) &\longrightarrow& \nu_{e} (k^\prime) + \eta ( p_\eta) + p (p^\prime),  
\end{eqnarray}
where the quantities in the parentheses are the four momenta of the particles. The expression for the differential scattering cross section in given in Eq.~(\ref{sigma:weak}) and the hadronic current receives contributions from the $s$- and $u$- channel nonresonant terms as well as from the resonance excitations, which are depicted diagrammatically in Fig.~\ref{Ch12_fg_eta:cc_weak_feynman}. In the following sections, we discuss briefly the contributions to the $\eta$ production amplitude from the nonresonant terms and resonance excitations and their subsequent decay to $N\eta$ mode, and present the results for the total and differential scattering cross sections.

\subsubsection{Nonresonant contribution:}\label{sec:NR:eta}
\begin{figure}
\begin{center}
\includegraphics[height=7cm,width=0.95\textwidth]{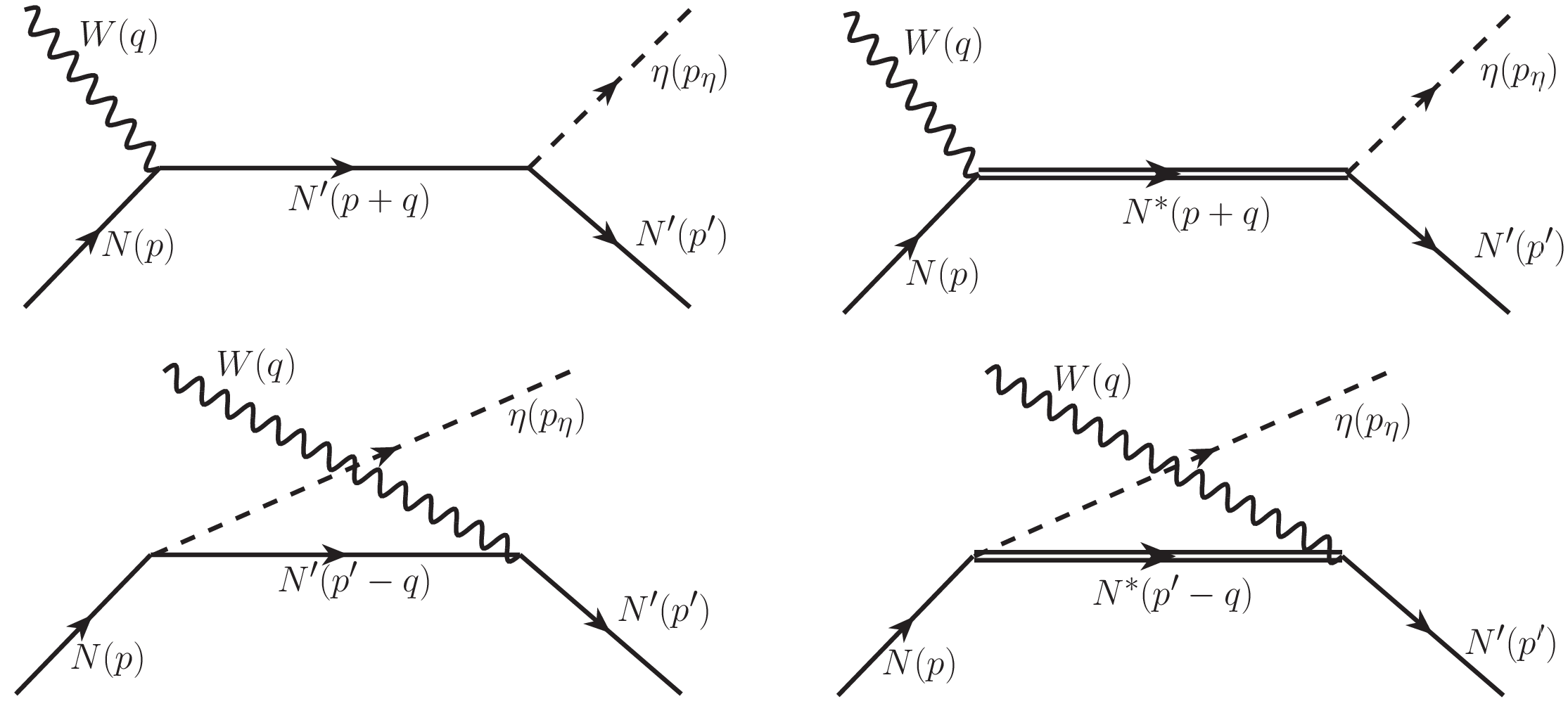}
\caption{Feynman diagrams corresponding to the nonresonant Born terms~(left panel) and resonance excitations~(right panel) for the process $ W (q) + N(p) \longrightarrow \eta(p_{\eta}) + N^\prime(p^{\prime})$. Diagrams shown in the top panel are the pole diagrams, while the one shown in the bottom panel corresponds to the cross  pole diagrams. 
The quantities in the parentheses represent the four momenta of the corresponding particles. }\label{Ch12_fg_eta:cc_weak_feynman}
 \end{center}
 \end{figure}
 
The matrix elements of the hadronic currents for the Born diagrams~($s$- and $u$-channels) with nucleon 
poles are given by:
\begin{eqnarray}\label{Eq_eta:amp_photo}
J^\mu|_{s} &=&  F_{s}(s)~
\frac{D-3F}{2\sqrt3 f_\eta} \bar u (p^\prime) \slashchar{p_\eta} \gamma^5  
\frac{\slashchar{p}+\slashchar{q}+M}{(p+q)^2-M^2} 
{\cal O}^\mu_V u (p) \nonumber \\ 
J^\mu|_{u} &=&  F_{u}(u)~ \frac{D-3F}{2\sqrt3 f_\eta} 
\bar u (p^\prime) {\cal O}^\mu_V
  \frac{\slashchar{p}-\slashchar{p}_{\eta}+M}{(p - p_\eta)^2-M^2} 
\slashchar{p}_{\eta} \gamma^5 u (p),
\end{eqnarray}
where $s=(p+q)^2$, $u = 
(p^{\prime} - q)^{2}$,
$D$ and $F$ are the axial-vector couplings of the baryon octet and $f_{\eta}=105$~MeV~\cite{Faessler:2008ix} is the $\eta$ decay constant. ${\cal O}_{V}^\mu = V^{\mu} -A^{\mu}$ is the weak interaction vertex factor, where
$V^{\mu}$ and $A^{\mu}$ are defined 
in terms of the weak vector and axial-vector form factors as
\begin{align}  \label{eq:vectorspinhalfcurrent}
  V^{\mu}& ={f_{1}^{V}}(Q^2) \gamma^\mu + \frac{f_2^{V}(Q^2)}{2 M} 
  i \sigma^{\mu\nu} q_\nu ,  \\
    \label{eq:axialspinhalfcurrent}
  A^{\mu} &=  \left[{g_1}(Q^2) \gamma^\mu  +  \frac{g_3(Q^2)}{M} q^\mu\right] \gamma_5 .
\end{align} 
$f_{1,2}^V(Q^2)$ are, respectively, the isovector vector form factors, and $g_1(Q^2)$ and $g_3(Q^2)$ are 
the axial-vector and pseudoscalar form factors. 
 The two isovector form factors 
$f_{1,2}^V(Q^2)$ are expressed in terms of the Dirac~($F_1^{p,n} (Q^2)$) and Pauli~($F_2^{p,n} (Q^2)$) form factors, discussed in Section~\ref{sec:FF:QE}, for the 
proton and the neutron,  using the relationships:
\begin{equation}\label{Eq_eta:f1v_f2v}
f_{1,2}^V(Q^2)=F_{1,2}^p(Q^2)- F_{1,2}^n(Q^2). 
\end{equation}
 
The axial-vector form factor $g_1(Q^2)$ is parameterized as
\begin{equation}\label{Eq_eta:fa}
g_1(Q^2)=g_A(0)~\left[1+\frac{Q^2}{M_A^2}\right]^{-2},
\end{equation}
where $g_A(0)=1.267$ is the axial vector charge and $M_A$ is the axial dipole mass, which in the numerical calculations is taken as the world average value i.e. $M_A = 1.026$~GeV~\cite{Bernard:2001rs}.
On the other hand pseudoscalar form factor $g_3(Q^2)$ is expressed  in terms of $g_1(Q^2)$ 
using the PCAC hypothesis and Goldberger-Treiman relation as~\cite{LlewellynSmith:1971uhs}:
\begin{equation}\label{Eq:fp_nucleon}
g_3(Q^2)=\frac{2M^2g_1(Q^2)}{m_\pi^2+Q^2},
\end{equation}
with $m_\pi$ being the pion mass. However, the contribution of $g_3(Q^2)$ is proportional to the lepton mass, thus, it is negligible in the case of electron induced reactions.

In order to take into account the hadronic structure of the nucleons, the form factors $F_{s} (s)$, and $F_{u} (u)$, are 
introduced at the strong vertex in Eq.~(\ref{Eq_eta:amp_photo}). 
In literature~\cite{Skoupil:2018vdh}, various parametrizations for the strong form factors are available. In this work, we use the general form of the hadronic form factor, which is 
taken to be of the dipole form as in Ref.~\cite{Skoupil:2018vdh, Fatima:2020tyh}:
\begin{equation}\label{FF_Born}
F_{x} (x) = \frac{\Lambda_{B}^{4}}{\Lambda_{B}^{4} + (x - M^{2})^{2}}, \qquad \qquad \quad x=s,u
\end{equation}
where $\Lambda_{B}$ is the cut-off parameter taken to be the same for the $s$- and $u$-channel nonresonant Born terms, and $x$ represents the Mandelstam variables $s,~u$. The value of $\Lambda_{B}$ 
is fitted to the experimental data~\cite{A2:2014pie, CrystalBallatMAMI:2010slt} of the $\eta$ photoproduction for the proton and neutron targets and the best fitted value is obtained as
$\Lambda_{B}=0.75$~GeV and 0.72~GeV, respectively~\cite{Fatima:2023fez}. 

One of the most important property of the electromagnetic current is the gauge invariance, which
ensures the current conservation. In the present work, it is essential to implement gauge invariance to the vector part of the weak hadronic current. The condition to fulfill gauge invariance is
\begin{equation}\label{current:conservation}
 q_{\mu} J_{NR}^{\mu} = 0,
\end{equation}
where $J_{NR}^{\mu} = \left. j^{\mu} \right|_s + \left. j^{\mu} \right|_u$. It may be observed from the expressions of $\left. j^{\mu} \right|_s$ and $\left. j^{\mu} \right|_u$ given in Eq.~(\ref{Eq_eta:amp_photo}) that the gauge invariance is automatically implemented in the case of $\eta$ production processes.

\subsubsection{Resonance excitations:}
The general expression of the hadronic current for the $s-$ and $u-$ channel resonance excitations and their subsequent decay to $N\eta$ mode are given by
\begin{eqnarray}
j^\mu\big|_{s}&=& F_{s}^{*} (s) ~\frac{g_{RN\eta}}{f_{\eta}} \bar u({p}\,') 
 \slashed{p}_{\eta} \gamma_5 \Gamma_{s} \left( \frac{\slashed{p}+\slashed{q}+M_{R}}{s-M_{R}^2+ iM_{R} \Gamma_{R}}\right) 
 \Gamma^\mu_{\frac12 
 \pm} u({p}\,), \nonumber\\
 \label{eq:res1/2_had_current}
 j^\mu\big|_{u}&=&  F_{u}^{*} (u) ~\frac{g_{RN\eta}}{f_{\eta}} \bar u({p}\,') 
 \Gamma^\mu_{\frac12 \pm}\left(\frac{\slashed{p}^{\prime}-\slashed{q}+M_{R}}{u-M_{R}^2+ iM_{R} \Gamma_{R}}\right) 
 \slashed{p}_{\eta} \gamma_5 \Gamma_{s}  u({p}\,),
\end{eqnarray}
where $F_{s}^{\star}$ and  $F_{u}^{\star}$ are the strong form factors, $\Gamma_{R}$ and $M_{R}$, respectively, are the decay width and mass of the resonance. $\Gamma_{s} = 1(\gamma_{5})$ stands for the positive~(negative) 
parity resonances, and $g_{RN\eta}$ is the strong coupling strength of the $ R  N\eta$ vertex, which has been determined using the partial decay width of the resonance to $N\eta$ mode. 
In Eq.~(\ref{eq:12_width}), we have given the expression for the decay width of a resonance $R$ decaying to $N\pi$ mode. The same expression is used to determine the decay width of $R \rightarrow N\eta$ mode, except the changes that the parameter ${\cal C} = 1$, $g_{RN\pi}$ and $f_{\pi}$ are replaced with $g_{RN\eta}$ and $f_{\eta}$, and $M_\pi$ in Eq.~(\ref{eq:pi_mom}) is now replaced with $m_{\eta}$. 
The values of the strong coupling constant of different resonances are also tabulated in Table~\ref{tab:param}.

The vertex factor $\Gamma_{\frac{1}{2}\pm}^{\mu}$ is written as
\begin{align}\label{eq:vec_half_pos}
  \Gamma^{\mu}_{\frac{1}{2}^+} &= {V}^{\mu} - {A}^{\mu},
  \end{align}
  for the positive parity resonance, and as
\begin{align}\label{eq:vec_half_neg}
  \Gamma^{\mu}_{\frac{1}{2}^-} &= \left({V}^{\mu} - {A}^{\mu} \right) \gamma_5 ,
  \end{align}
  for the negative parity resonance. The vector and axial vector interaction vertices are expressed in terms of the vector and axial vector form factors, which have been discussed in detail in Section~\ref{sec:Nstar}.

We have considered the following form factor at the strong 
vertex in order to take into account the hadronic structure:
\begin{equation}\label{eq:strong_FF_res}
F^{*}_{x} (x) = \frac{\Lambda_{R}^{4}}{\Lambda_{R}^{4} + (x - M_{R}^{2})^{2}},
\end{equation}
where $\Lambda_{R}$ is the cut-off parameter whose value is fitted to the experimental data~\cite{CrystalBallatMAMI:2010slt, A2:2014pie} on the photoproduction of $\eta$ mesons. In general, $\Lambda_{R}$ would be different from $\Lambda_{B}$, however, in the case of $\eta$ production by 
photons~\cite{Fatima:2022tlf}, it happens that the same value of $\Lambda_{R}$ as that of $\Lambda_{B}$ i.e. $\Lambda_{R} = \Lambda_{B} =
0.75$~GeV for the proton target and $\Lambda_{R} = \Lambda_{B} =
0.72$~GeV for the neutron target gives the best fit to the experimental data.

\subsubsection{Results and discussion:}
 \begin{figure}
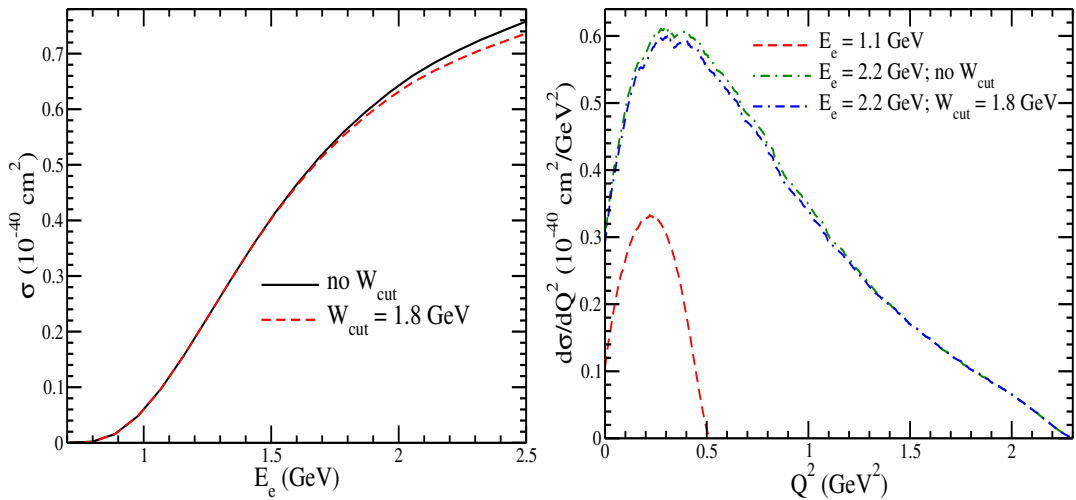
 
\begin{center}
\includegraphics[width=7cm,height=6.5cm]{electron_eta.eps}
\includegraphics[width=7cm,height=6.5cm]{dsigma_dQ2_eta_electron.eps}
\caption{(Left panel)~Total scattering cross section $\sigma(E_e)$ as a function of electron energy $E_{e}$ for
 $e^- + p \longrightarrow \nu_e + \eta + n$ process. 
 The solid and dashed lines, respectively, represent the results obtained using (i)~no cut on $W$, and (ii)~ a $W_{cut}$ of $1.8$~GeV, respectively.
 (Right panel)~Differential scattering cross section~$\left(\frac{d\sigma}{dQ^2}\right)$ as a function of $Q^2$. 
 The dashed and dash-dotted lines, respectively, represent the results at $E_{e}=1.1$~GeV, and 2.2~GeV with no cut on the CM energy $W$. The double-dash-dotted line shows the results at $E_{e}=2.2$~GeV with a  $W_{cut}$ of 1.8~GeV.
 }\label{sigma:eta_weak}
\end{center}
\end{figure}

In Fig.~\ref{sigma:eta_weak}, the results are presented for the total scattering cross section~$\sigma(E_e)$ as a function of $E_{e}$~(left panel) and the differential scattering cross section~$\frac{d\sigma}{dQ^2}$ as a function of $Q^2$~(right panel) at $E_{e}=1.1$ and 2.2~GeV for the  $e^- + p \longrightarrow \nu_e + \eta + n$ production process. The results are presented for the two cases viz., (i)~when no cut on the CM energy~($W = \sqrt{s}$) is applied, and (ii)~when a cut of 1.8~GeV is applied on $W$. We find almost no effect of $W_{cut}$ on the total as well as the differential scattering cross section. In the case of $\sigma(E_e)$, we find that the cross section increases with $E_{e}$ and saturates at around $E_e=3$~GeV~(not shown in the figure). In the case of $Q^2$ distribution, we find that the peak shifts to higher $Q^2$ and becomes broader with increasing $E_{e}$. 

\subsection{Electron induced $K\Lambda$ production}\label{weak:associated}
The weak charged current associated particle production process induced by the electron is given by the reaction
\begin{eqnarray}\label{eq:inelastic:reaction}
e^-(k) + p (p) &\longrightarrow& \nu_e(k^{\prime}) + \Lambda (p^{\prime}) + K^{0} (p_{K}),
\end{eqnarray}
where the quantities in the parentheses represent the four momentum of the corresponding particles.
The expression for the differential scattering cross section in given in Eq.~(\ref{sigma:weak}) and the hadronic current receives contribution from the nonresonant terms, consisting of $s$-channel nucleon pole, $t$-channel kaon pole, $u$-channel hyperon pole, contact term and pion in flight term, which are depicted diagrammatically in Fig.~\ref{Ch12_fig:feyn_app} as well as from the four spin $\frac{1}{2}$~($S_{11}(1650)$, $P_{11}(1710)$, $P_{11}(1880)$, and $S_{11}(1895)$) and two $\frac{3}{2}$~($P_{13}(1720)$ and $P_{13}(1900)$) resonance excitations in the $s$-channel and spin 1 kaon resonances exchanged in the $t$-channel. In the following sections, we discuss briefly the contributions from the nonresonant terms, nucleon resonance excitations and their subsequent decay to $K\Lambda$ mode, and kaon resonance excitation terms.

\begin{figure} 
\centering
\includegraphics[height=9cm, width=15cm]{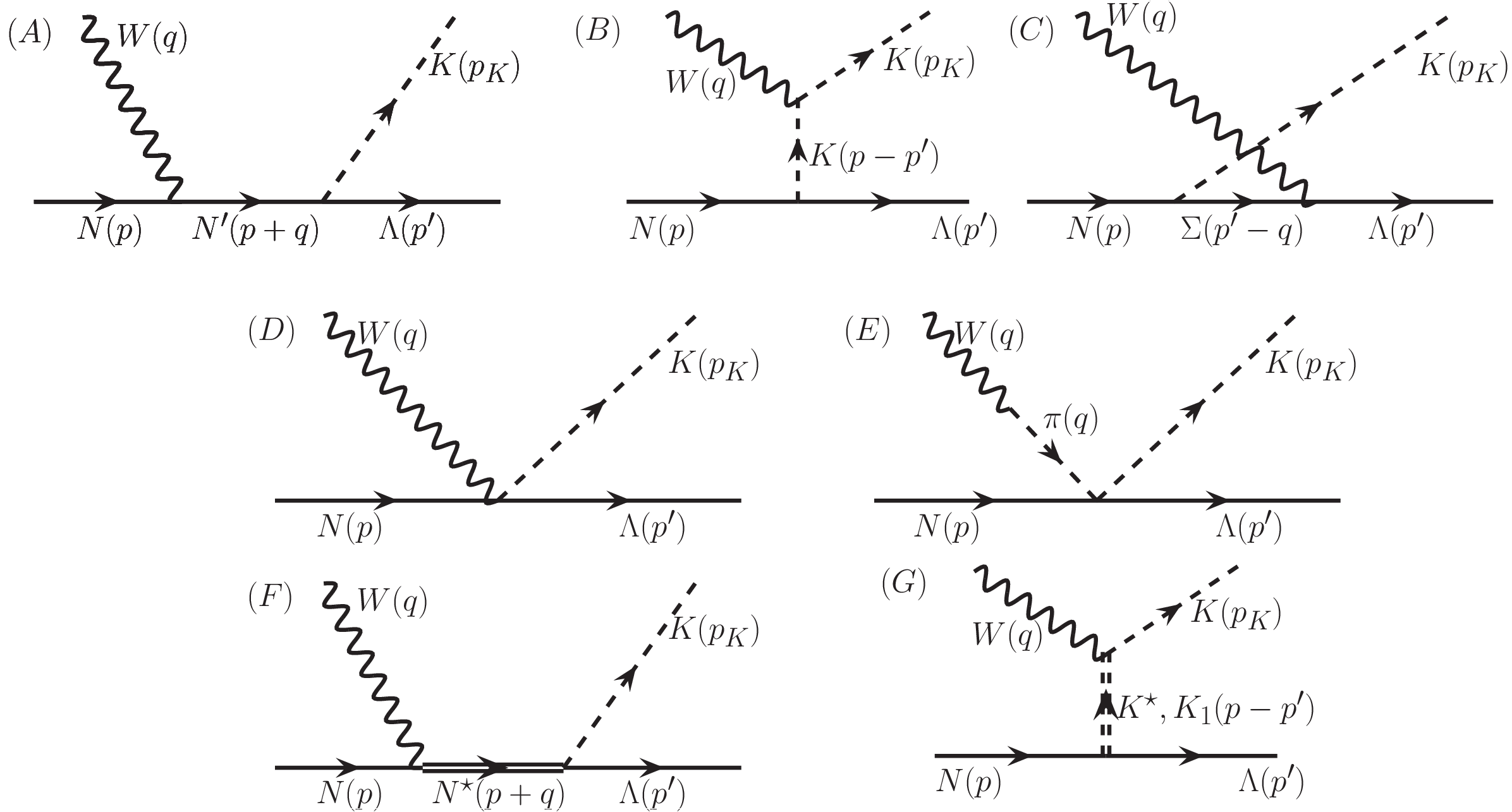}
\caption{Feynman diagrams corresponding to the { nonresonant background terms of the} electron induced $\Delta S=0$ associated particle production processes. (A) $s$-channel, (B) $t$-channel or kaon pole~(KP) term, (C) $u$-channel, (D) contact term~(CT), and (E) pion in flight~(PF) term constitute the non-resonant Born diagrams. (F)~$s$-channel nucleon resonance and (G)~$t$-channel kaon resonance terms. { Wiggly line represents the intermediate vector meson $W$, solid line represents the baryons such as nucleon and hyperons, and the dashed line represents the meson such as pion and kaon.}}
\label{Ch12_fig:feyn_app}
\end{figure}

\subsubsection{Non-resonant contribution:}
The expressions for the  hadronic currents $j^{\mu}$ corresponding to the nonresonant 
diagrams shown in Fig.~\ref{Ch12_fig:feyn_app} are obtained, using the nonlinear sigma model, as:
\begin{eqnarray}\label{Ch12_Eqapp:amplitude}
j^\mu \arrowvert_{s} &=& i A_{s}  F_{s}(s) \; \bar u (p^\prime) \slashed{p}_K \gamma_5 
	    \frac{{p\hspace{-.5em}/} + {q\hspace{-.5em}/} + M}{(p+q)^2-M^2} 
	    \left[V^{\mu} - A^{\mu} \right] u(p) \nonumber \\
j^\mu \arrowvert_{u} &=& i A_{u} F_{u}(u) \; \bar u (p^\prime) \left[V^{\mu} - A^{\mu} \right]
	      \frac{{p\hspace{-.5em}/} - \slashed{p}_K + M_{\Sigma}}{(p - p_K)^2-M_{\Sigma}^2}  
	      \slashed{p}_K \gamma_5 u (p)\nonumber\\
j^\mu \arrowvert_{KP} &=& i A_{t} F_{t}(t) \; f_{KP}(Q^2)\; (M+M_\Lambda) \; \bar u (p^\prime) \gamma_5 \; u (p) \;\;
	    \frac{2 p_K^\mu - q^\mu }{(p-p^\prime)^2-m_K^2}\nonumber
	      \end{eqnarray}
	      \begin{eqnarray}
j^\mu \arrowvert_{CT} &=& i A_{CT} F_{CT} \; \bar u (p^\prime) 
		    \left[ \gamma^\mu f_{\rho}((q-p_{K})^2) + B_{CT} \; f_{CT}(Q^2)\;\gamma^\mu  \gamma_5 \right] u (p) \nonumber\\
j^\mu \arrowvert_{PF} &=& i A_{\pi} F_{CT}  \; f_{\rho}((q-p_{K})^2) \; \bar u (p^\prime) 
	      \left[{q\hspace{-.5em}/} + \slashed{p}_K\right] u(p) \frac{q^\mu}{q^2-m_\pi^2}
\end{eqnarray}
where $A_{i}'s~(i=s,t,u,CT,\pi)$ and $B_{CT}$ are the constants obtained using the non-linear sigma model and tabulated in Table~\ref{Ch12_tb_app:currents}. The form factors $F_s(s)$, $F_t(t)$, $F_u(u)$ and $F_{CT}(s,t,u)$, are 
introduced at the strong vertices, to take into account the hadronic structure of the baryons. The form of the strong form factor at the $s$, $t$ and $u$ channels is given in Eq.~(\ref{FF_Born}), with the value of $\Lambda_{B}$ 
being fitted to the $K\Lambda$ photoproduction experimental data~\cite{CLAS:2005lui} and the best fitted value is 
$\Lambda_{B}=0.54$~GeV~\cite{Fatima:2025tht}.  Following the work of Davidson and Workman~\cite{DW}, we have expressed the strong form factor associated with the contact term $F_{CT}(s,t,u)$ as:
\begin{equation}\label{F_CT}
 F_{CT}(s,t,u) = F_{s}(s) + F_{t}(t) - F_{s}(s) \times F_{t}(t).
\end{equation}

\begin{table}
\begin{center}
\centering
\renewcommand{\arraystretch}{1.2}
\begin{tabular*}{135mm}{@{\extracolsep{\fill}}cccccc} \hline \hline
 $A_{CT}$	            &$B_{CT}$		      &     $A_{s}$
& $A_{u}$                        & $A_{t}$                     &  $A_{\pi }$ \\   \hline
$\frac{\sqrt{3}}{2f_{K}}$   & $\frac{-1}{3}(D+3F)$ & $\frac{D+3F}{2\sqrt{3}f_{K}}$ &  $\frac{1}{\sqrt{3}f_{K}}(D-F)$  & 
$-\frac{D+3F}{2\sqrt{3}f_{K}}$ & -$\frac{\sqrt{3}}{4f_{K}}$ \\ \hline\hline
\end{tabular*}
\caption{Constant factors  appearing in the hadronic current in Eq.~(\ref{Ch12_Eqapp:amplitude}). }
\label{Ch12_tb_app:currents}
\end{center}
\end{table}

The vector~($V^{\mu}$) and axial-vector~($A^{\mu}$) transition currents for $BB^{\prime} = NN^\prime=pn$ for the $s$-channel diagram, and $BB^{\prime} = YY^\prime=\Lambda\Sigma^+$ for the $u$-channel diagram, are expressed as:
\begin{eqnarray}
 V^\mu &=& f_1^{BB^{\prime}} (Q^{2}) \gamma^\mu + i \frac{f_2^{BB^{\prime}}(Q^{2})}{M+M^{\prime}} \sigma^{\mu \nu} q_\nu \\
 A^{\mu} &=& g_1^{BB^{\prime}}(Q^{2}) \gamma^\mu \gamma_{5} + g_3^{BB^{\prime}}(Q^{2})\frac{2 q^\mu}{M+M^{\prime}}\gamma_5
\end{eqnarray}
where the vector~($f_{1,2}^{NN^\prime} (Q^2)=f_{1,2}^{pn} (Q^2)$) and axial-vector~($g_{1,
3}^{NN^\prime}(Q^2)=g_{1,
3}^{pn}(Q^2)$) form factors  are determined assuming the Cabibbo theory and the various symmetry properties of the weak hadronic 
current, as discussed in Section~\ref{sec:FF:QE}.  The expressions for the vector and axial vector form factors for the $pn$ transition are given in Section~\ref{sec:NR:eta}, and for the $\Lambda\Sigma^+$ transition, the isovector vector form factors are expressed as:
\begin{eqnarray}\label{Eq_eta:f1v_f2v}
f_{1,2}^{\Lambda\Sigma^+}(Q^2)&=&- \sqrt{\frac{3}{2}}F_{1,2}^n(Q^2).
\end{eqnarray}
 
The axial-vector form factor $g_1^{\Lambda\Sigma^+}(Q^2)$ is parameterized as
\begin{equation}\label{Eq_eta:fa}
g_1^{\Lambda\Sigma^+}(Q^2)=g_1^{\Lambda\Sigma^+}(0)~\left[1+\frac{Q^2}{M_A^2}\right]^{-2},
\end{equation}
where $g_1^{\Lambda\Sigma^+}(0)=-\sqrt{\frac{3}{2}}D$ and $M_A$ is the axial dipole mass, which in the numerical calculations is taken as the world average value i.e. $M_A = 1.026$~GeV~\cite{Bernard:2001rs}. The contribution of $g_3^{\Lambda\Sigma^+} (Q^2)$ is negligible for the electron induced processes.

In Eq.~(\ref{Ch12_Eqapp:amplitude}), the form factors $f_{CT}(Q^2)$, $f_{KP}(Q^2)$ and 
$f_{\rho}((q-p_{K})^2)$ are introduced in the contact, kaon pole and pion in flight terms, respectively, to take into account the hadronic 
structure. It may be observed from the Feynman diagrams~(Fig.~\ref{Ch12_fig:feyn_app}) that the kaon pole term is purely 
vector in nature while the pion in flight diagram is possible only with axial-vector current. In the case of contact term, the 
term associated with $B_{CT}$ represents the vector part of the weak hadronic current while the term with $\gamma^{\mu}$ is 
associated with the axial-vector part. The CVC hypothesis imposes the following condition on the form factors $f_{CT}(Q^2)$ and 
$f_{KP}(Q^2)$, {  i.e.},
\begin{eqnarray}
 f_{CT} (Q^2) &=& f_{1}^{pn} (Q^2) - 2F_{1}^{n} (Q^2) \; \left(\frac{D-F}{D+3F}\right) \; \left(\frac{u - M_{\Sigma} 
 M_{\Lambda} + MM_{\Sigma} - MM_{\Lambda}}{M_{\Sigma}^{2} - u} \right),\\
 f_{KP} (Q^2) &=& 2F_{1}^{n} (Q^2) \left(\frac{D-F}{D+3F}\right) \; \left(\frac{(M+M_{\Sigma})(u-M_{\Lambda}^2)}{(M + 
 M_{\Lambda})(M_{\Sigma}^2 - u)}\right) - f_{1}^{pn} (Q^2),
\end{eqnarray}
where $u=(p-p_{K})^2$, $f_{1}^{pn} (Q^2) = F_{1}^{p} (Q^2) - F_{1}^{n} (Q^2)$ is the vector form factor with $F_{1}^{p} 
(Q^2)$ and $F_{1}^{n} (Q^2)$ being the nucleon electromagnetic form factors. 

The form factor $f_{\rho} (Q^2)$ appearing in $j^{\mu}|_{CT}$ and $j^{\mu}|_{PF}$ in Eq.~(\ref{Ch12_Eqapp:amplitude}), is given by~\cite{SajjadAthar:2022pjt, Hernandez:2007qq}:
\begin{equation}\label{frho}
 f_{\rho}(Q^2) = \frac{1}{1+Q^2/m_{\rho}^2}; \qquad \qquad {\rm with } \; m_\rho = 0.776~\text{ GeV}.
\end{equation}

\subsubsection{Spin $\frac{1}{2}$ resonance excitations:}
The general expression of the hadronic current for the $s$-channel spin $\frac{1}{2}$ nucleon resonance excitations and their subsequent decay to $K\Lambda$ mode is given by:
\begin{eqnarray}
j^\mu\big|_{s}&=& F_{s}^{*} (s) ~\frac{g_{RK\Lambda}}{f_{K}} \bar u({p}\,') 
 \slashed{p}_{K} \gamma_5 \Gamma_{s} \left( \frac{\slashed{p}+\slashed{q}+M_{R}}{s-M_{R}^2+ iM_{R} \Gamma_{R}}\right) 
 \Gamma^\mu_{\frac12 
 \pm} u({p}\,), 
\end{eqnarray}
where $F_{s}^{\star}(s)$ is the strong form factor, defined in Eq.~(\ref{eq:strong_FF_res}), with $\Lambda_{R}$ being determined by fitting the experimental photoproduction data~\cite{CLAS:2005lui} and the best fit for spin $\frac{1}{2}$ resonances is obtained as $\Lambda_{R} = 1$~GeV~\cite{Fatima:2025tht}. 
$\Gamma_{R}$ and $M_{R}$, respectively, are the decay width and mass of the resonance~$R$. The operator $\Gamma_{s} = 1(\gamma_{5})$ stands for the positive~(negative) 
parity resonances, and $g_{RK\Lambda}$ is the strong coupling strength of the $ R K\Lambda$ vertex, which has been determined using the partial decay width of the resonance~$R$ to $K\Lambda$ decay mode. The values of the strong coupling constant of different resonances are tabulated in Table~\ref{tab:param}. 
The vertex factor $\Gamma_{\frac{1}{2}\pm}^{\mu}$ is written as
\begin{align}\label{eq:vec_half_pos}
  \Gamma^{\mu}_{\frac{1}{2}^+} &= {V}^{\mu}_\frac{1}{2} - {A}^{\mu}_\frac{1}{2},
  \end{align}
  for the positive parity resonance, and as
\begin{align}\label{eq:vec_half_neg}
  \Gamma^{\mu}_{\frac{1}{2}^-} &= \left({V}^{\mu}_\frac{1}{2} - {A}^{\mu}_\frac{1}{2} \right) \gamma_5 ,
  \end{align}
  for the negative parity resonance. The vector and axial-vector vertex factors as well as the form factors for the weak charged current interaction processes are discussed in Section~\ref{sec:Nstar}.

\subsubsection{Spin $\frac{3}{2}$ resonance excitations leading to $K\Lambda$ mode:}
In the case of spin $\frac{3}{2}$ resonance excitations leading to $K\Lambda$ in the final state, we have followed the prescription of Pascalutsa and Timmermans~\cite{Pascalutsa:1999zz} for the photon induced $K\Lambda$ production~\cite{Fatima:2026ffh} as well as for the (anti)neutrino induced $K\Lambda$ production~\cite{Fatima:2025tht}. In this work, we apply the same prescription to study the electron induced weak production of $K\Lambda$. 

In case of the positive parity spin $\frac{3}{2}$ resonance excitation and its subsequent decay to $K\Lambda$ channel, the expression of the hadronic current for the $s$-channel diagram is given by:
\begin{eqnarray}
j^\mu\big|_{s}&=& F_{s}^{*} (s) ~\frac{g_{RK\Lambda}}{M_{R}M_{K}} \bar u({p}\,') 
 \epsilon_{\sigma \nu\alpha\beta} p_{R}^{\sigma} \gamma_5 \gamma^{\alpha}p_{K}^{\beta} \left( \frac{{\cal P}^{\nu\delta} (p_{R})}{s-M_{R}^2+ iM_{R} \Gamma_{R}}\right) 
 \Gamma^{\delta\mu}_{\frac32} u({p}\,); 
\end{eqnarray}
where $p_{R}=p+q$, $\Gamma_{R}$ and $M_{R}$, respectively, are the decay width and mass of the resonance. $F_{s}^{*} (s)$ is defined in Eq.~(\ref{eq:strong_FF_res}), with the value of $\Lambda_{R}$ determined for spin $\frac{3}{2}$ resonances to be 1.62~GeV~\cite{Fatima:2025tht}. 
The strong coupling strength $g_{RK\Lambda}$ is determined using the partial decay width of the resonance to $K\Lambda$ mode and the values of these couplings are tabulated in Table~\ref{tab:param}. 

The spin $\frac{3}{2}$ projection operator ${\cal P}_{\alpha \beta} (p_{R})$, in the prescription of Pascalutsa and Timmermans~\cite{Pascalutsa:1999zz}, is given by:
\begin{equation}\label{proj_spin32}
 {\cal P}_{\alpha \beta} (p_{R}) = - (\slashed{p}_{R} + M_{R}) \left(g_{\alpha \beta} - \frac{1}{3} \gamma_{\alpha}\gamma_{\beta} - \frac{1}{3p_{R}^{2}} (\slashed{p}_{R} \gamma_{\alpha} {{p_{R}}}_{\beta} + {{p_{R}}}_{\alpha}\gamma_{\beta} \slashed{p}_{R}) \right).
\end{equation}

The vertex factor $\Gamma_{\frac{3}{2}}^{\mu\alpha}$ is written as
\begin{align}\label{eq:vec_half_pos}
  \Gamma^{\mu\alpha}_{\frac{3}{2}} &= {V}^{\mu\alpha}_\frac{3}{2} - {A}^{\mu\alpha}_\frac{3}{2}
  \end{align}
 where the vector and axial-vector vertex factors for the weak charged current interaction processes are given by
\begin{eqnarray}  \label{eq:vectorspin3halfcurrent}
    V^{ \alpha\mu}_{\frac{3}{2}} &=& \left(\gamma^{\mu}q^{\alpha} - \slashed{q} g^{\alpha\mu} \right) \frac{C_{3}^{V} (Q^2)}{M} + \left(q \cdot p^{\prime} g^{\alpha\mu} - q^{\alpha} {{p^{\prime}}}^{\mu} \right) \frac{C_{4}^{V} (Q^2)}{M^2} + \left(q^{\alpha}q^{\mu} - q^2 g^{\alpha\mu} \right) \frac{C_{5}^{V} (Q^2)}{M^2},  \\
  \label{eq:axial_3half_pos}
  A^{\alpha\mu}_{\frac{3}{2}} &=&- \left[ \frac{{ C}_3^A (Q^2)}{M} (g^{\mu \alpha} {q\hspace{-.5em}/} \, - q^{\alpha} \gamma^{\mu})+
  \frac{{ C}_4^A (Q^2)}{M^2} (g^{\mu \alpha} q\cdot p' - q^{\alpha} p'^{\mu})+
 {{ C}_5^A (Q^2)} g^{\mu \alpha} \right. \nonumber \\
 && \left. + \frac{{ C}_6^A (Q^2)}{M^2} q^{\alpha} q^{\mu}\right] \gamma_5 .
\end{eqnarray}
${ C}^V_i (Q^2)$ and ${ C}^A_i (Q^2)$ are, respectively, the vector and axial-vector form factors. 

The isovector vector form factors~(${C}_i^V (Q^2); (i=3,4,5)$)  are written in terms of the electromagnetic $C^{R+}_i(Q^2)$ and 
$C^{R0}_i(Q^2)$ form factors through a simple relation, obtained using isospin symmetry~\cite{Athar:2020kqn}, as
\begin{equation}\label{eq:civ_NC}
{ C}_i^V (Q^2)= C^{R+}_i (Q^2) - C^{R0}_i (Q^2) ; \;\; \qquad i = 3,4,5\, .
\end{equation}
 
 The electromagnetic $N-R$ transition form factors for the charged~($C_{3,4,5}^{R^+}(Q^2)$)  and neutral~($C_{3,4,5}^{R^0} (Q^2)$) resonant states are then  related to the helicity amplitudes~($A_{\frac{1}{2}} (Q^2)$, $A_{\frac{3}{2}} (Q^2)$, and $S_{\frac{1}{2}}(Q^2)$), where the explicit relations between the helicity amplitudes $A_{\frac{1}{2}}$, $S_{\frac{1}{2}}$ and the electromagnetic current are given in Eqs.~(\ref{Ch11_a12}) and (\ref{Ch11_s12}), respectively, and $A_{\frac{3}{2}}$ is expressed in terms of the electromagnetic current by the following expression:
\begin{eqnarray}\label{eq3}
 A_{\frac{3}{2}}&=& \sqrt{\frac{2 \pi \alpha}{K_{R}}} \left<{R, J_{z}^{R} = +\frac{3}{2}} \Big| \epsilon_{\mu}^{+} V^{\mu}_{i} \Big| 
 {N, J_{z}^{N} = +\frac{1}{2}} \right> \zeta.
\end{eqnarray}
The different variables appearing in the above expression are after Eq.~(\ref{Ch11_s12}).

Using Eqs.~(\ref{Ch11_transverse})--(\ref{helicity:vec}) in Eqs.~(\ref{eq:hel_spin_12_x}) and (\ref{eq3}) with $V^{ \alpha\mu}_{\frac{3}{2}}$ defined in Eq.~(\ref{eq:vectorspin3halfcurrent}), the helicity amplitudes $A_{\frac{1}{2}} (Q^2)$, $A_{\frac{3}{2}} (Q^2)$, and $S_{\frac{1}{2}} (Q^2)$ in terms of the electromagnetic
form factors
 $C_3^{R^+,R^0} (Q^2)$, $C_4^{R^+,R^0} (Q^2)$, and $C_5^{R^+,R^0} (Q^2)$ are obtained as~\cite{Fatima:2022nfn}:
\begin{eqnarray}\label{eq:hel_em_ff_32}
 A_{\frac{1}{2}}(Q^2) &=&  \sqrt{\frac{\pi \alpha}{3M} \frac{\left[(M_{R} - M)^2 + Q^2 \right]}{M_{R}^{2} - M^{2}}} \left[\frac{C_{3}^{R} (Q^2)}{M} (M + M_{R}) + \frac{C_{4}^{R} (Q^2)}{2M^2} \left(M^2 - M_{R}^{2} + Q^2 \right) - \right.\nonumber \\
 && \left. C_{5}^{R}(Q^2) \frac{Q^2}{M^2} \right], \\
 \label{A32_32}
 A_{\frac{3}{2}}(Q^2) &=&\sqrt{\frac{\pi \alpha}{M} \frac{\left[(M_{R} - M)^2 + Q^2 \right]}{M_{R}^{2} - M^{2}}} \left[\frac{C_{3}^{R}(Q^2)}{MM_{R}} \left[M(M + M_{R}) + Q^2 \right] + \frac{C_{4}^{R}(Q^2)}{2M^2} \left[M_{R}^{2} - M^2 -Q^2 \right] + \right. \nonumber \\
 && \left. C_{5}^{R}(Q^2) \frac{Q^2}{M^2}  \right] , \\
 \label{S12_32}
 S_{\frac{1}{2}}(Q^2) &=& -\sqrt{\frac{\pi\alpha}{6M} \frac{\left[(M_{R} + M)^2 + Q^2 \right]}{M_{R}^{2} - M^{2}} } \left[(M_{R} - M)^2 + Q^2 \right] \left[-\frac{C_{3}^{R}(Q^2)}{MM_{R}} + \frac{C_{4}^{R}(Q^2)}{M^2} +\right. \nonumber \\
 && \left.C_{5}^{R}(Q^2) \frac{[M^2 - M_{R}^{2} + Q^2]}{2M^2M_{R}^{2}} \right].
\end{eqnarray}
Inverting Eqs.~(\ref{eq:hel_em_ff_32})--(\ref{S12_32}), we obtain the electromagnetic
form factors
 $C_3^{R^+,R^0} (Q^2)$, $C_4^{R^+,R^0} (Q^2)$, and $C_5^{R^+,R^0} (Q^2)$, in terms of the experimentally determined helicity amplitudes.

The parametrization of the $Q^2$ dependence of the helicity amplitudes~(Eqs.~(\ref{eq:hel_em_ff_32})--(\ref{S12_32}))  is given in Eq.~(\ref{eq:ffpar}),
with $ {\mathcal A}_{\alpha}(Q^2)$ being the helicity amplitudes; $A_{\frac12}(Q^2)$, $A_{\frac32} (Q^2)$, and $S_{\frac12}(Q^2)$. The values of the parameters appearing in Eq.~(\ref{eq:ffpar}) for spin $\frac{3}{2}$ resonances are tabulated in Table~\ref{tab:resonance:helicity}.

The form factors ${C}_i^A(Q^2), \; (i=3,4,5,6)$ corresponding to the axial current have not been studied in the case of 
higher resonances. The earlier calculations have used the hypothesis of PCAC to determine ${C}_5^A(Q^2)$ and ${C}_6^A(Q^2)$ and taken other 
form factors to be zero~\cite{Adler:1968tw}. In view of this, we have also taken a simple model for the determination of the axial form factors 
based on the PCAC hypothesis and GT relation and write ${C}_6^A(Q^2)$ in terms of ${C}_5^A(Q^2)$ as
\begin{equation}\label{c6_CC}
 C_6^A(Q^2) = C_5^A(Q^2) \frac{M^2}{Q^2 + m_\pi^2} .
\end{equation}
For ${C}_5^A(Q^2)$, a dipole form has been assumed 
 \begin{equation}\label{c5a-r}
{C}_5^A(Q^2) = \frac{{C}_5^A(0)}{ \left( 1 + Q^2 /{M_A^{\it R}}^2 \right)^2 } ,
\end{equation}
where ${C}_5^A(0)= -2 g_{R N \pi}$~\cite{SajjadAthar:2022pjt}, with
$g_{R N \pi}$ being the coupling for $R \longrightarrow N \pi$ decay for each resonance $R$. $M_{A}^{\it R}$ is taken as 
$1.026~{\rm GeV}$. ${C}_3^A(Q^2)$ as well as ${C}_4^A(Q^2)$ are taken as zero. 

\subsubsection{Spin 1 Kaon resonances:}
In the present work, we have considered two kaon resonances in the $t$-channel viz., a vector meson $K^{*} (892)$ and an axial 
vector meson $K_{1} (1270)$. The structure of the matrix element in Eqs.~(\ref{had_curr:Kstar}) and (\ref{had_curr:K1}) below has been discussed by us in Ref.~\cite{Fatima:2020tyh, Fatima:2025tht}.

\begin{table*}
  \caption{Properties of kaon resonances $K^{\star}(892)$ and $K_1 (1270)$ included in the present model, with mass $M_R$, spin $J$, isospin 
  $I$, parity $P$, the total decay width $\Gamma$, and the vector 
  $G_{K}^{v}$ and tensor $G_{K}^{t}$ couplings for the kaon resonances. It is to be noted that the couplings $G_{K}^{v}$ 
  and $G_{K}^{t}$ contain both the electromagnetic as well as the strong coupling strengths.}\label{hyperon_resonances}
  \begin{center}
    \begin{tabular*}{150mm}{@{\extracolsep{\fill}}c c c c c c c c}
      \noalign{\vspace{-8pt}}
      \hline \hline
      Resonances               & $M_R$ [GeV] & J\quad   & I \quad    &   P   & $\Gamma$ [GeV]    & $G_{K}^{v}$
      &   $G_{K}^{t}$  \\ \hline

      $K^{*}$(892) & $0.89166 \pm 0.00026$ & $1$ &$ \frac{1}{2}$ &$ -$ &   $ 0.0508 \pm 0.0009$  &  $-0.18$ & $0.02$ \\
      \hline
      
      $K_{1}$(1270)  & $1.272 \pm 0.007$ & $1 $ &$ \frac{1}{2}$ &$ +$ &   $ 0.090 \pm 0.020$  & $0.28$ & $-0.28$   \\  \hline
      \hline
    \end{tabular*}
  \end{center}

\end{table*}

The hadronic current for the $K^{*}$ exchange in the $t$-channel is 
obtained as
\begin{eqnarray}\label{had_curr:Kstar}
 J_{\mu} \big|_{K^{*}} &=& ie F_{t}^{\star}(t) F_{K^\star} (Q^2) \bar{u} (p^{\prime}) \epsilon_{\mu \nu \rho \sigma} q^{\rho} (p^{\prime} - p)^{\sigma} \left(
 \frac{ -g^{\nu \alpha} + (p - p^{\prime})^{\nu} (p - p^{\prime})^{\alpha}/{M_{K^{*}}^{2}}}{t - {M_{K^{*}}^{2}} + i 
 M_{K^{*}} \Gamma_{K^{*}}} \right) \nonumber \\
 &\times&
 \left[G_{K^{*}}^{v} \gamma_{\alpha} + \frac{G_{K^{*}}^{t}}{M + M_{\Lambda}} 
 (\slashed{p}^{\prime} - \slashed{p}) \gamma_{\alpha} \right] u(p),
 \end{eqnarray}
with $G_{K^{*}}^{v} = \kappa_{K K^{*}} g_{K^{*} \Lambda p}^{v}/\mu$ and $G_{K^{*}}^{t} = \kappa_{K K^{*}} g_{K^{*} \Lambda 
p}^{t}/\mu$, where $g_{K^{*} \Lambda 
p}^{v}$ and $g_{K^{*} \Lambda p}^{t}$ are the vector and the tensor couplings, respectively, at the strong $K^{*} \Lambda p$ 
vertex, and $\kappa_{K K^{*}}$ is the coupling strength of the $\gamma K K^{*}$ vertex, $\mu =1$~GeV is an arbitrary mass factor, which is 
introduced to make the Lagrangian dimensionless. $M_{K^{*}}$ and $\Gamma_{K^{*}}$ are the mass and width of the $K^{*}$ resonance, respectively. Due to the lack of 
 experimental data on the $K^{*}$ and $K_{1}$ resonances, the values of $G_{K^{*}}^{v}$ and $G_{K^{*}}^{t}$ can not be 
determined phenomenologically and are treated as free parameters, which are fitted to the experimental data of the $K \Lambda$ 
photoproduction~\cite{Fatima:2020tyh}, and are quoted in Table~\ref{hyperon_resonances}.

The hadronic current for the axial vector kaon $K_{1}$ exchange in the $t$-channel is obtained as
\begin{eqnarray}\label{had_curr:K1}
 J_{\mu} \big|_{K_{1}} &=& ieF_{t}^{\star}(t) F_{K_1} (Q^2) \bar{u} (p^{\prime}) [g_{\alpha \mu} q \cdot (p - p^{\prime}) - q_{\alpha} (p - 
 p^{\prime})_{\mu}] \left(\frac{-g^{\alpha \rho} + (p - p^{\prime})^{\alpha} (p - p^{\prime})^{\rho}/{M_{K_{1}}^{2}}}
 {t - {M_{K_{1}}^{2}} + i M_{K_{1}} \Gamma_{K_{1}}} \right) \nonumber \\
 &\times& \left[G_{K_{1}}^{v} \gamma_{\rho} \gamma_{5} + \frac{G_{K_{1}}^{t}}{M + M_{\Lambda}} (\slashed{p}^{\prime} - 
 \slashed{p}) \gamma_{\rho} \gamma_{5} \right] u(p),
 \end{eqnarray}
with $G_{K_{1}}^{v} = \kappa_{K K^{*}} g_{K_{1} \Lambda p}^{v}/\mu$ and $G_{K_{1}}^{t} = \kappa_{K K^{*}} g_{K_{1} \Lambda 
p}^{t}/\mu$, where $\kappa_{K K_{1}}$ is the coupling strength of the electromagnetic $\gamma K K_{1}$ vertex, and $g_{K_{1} \Lambda p}^{v}$ and $g_{K_{1} \Lambda p}^{t}$ 
are the vector and the tensor couplings, respectively, at the strong $K_{1} \Lambda p$ vertex. $M_{K_{1}}$ and $\Gamma_{K_{1}}$ are the mass and width of the $K_{1}$ resonance, respectively. The values of 
$G_{K_{1}}^{v}$ and $G_{K_{1}}^{t}$ are treated as free parameters, and are  fitted to the experimental data of the $K \Lambda$ 
production~\cite{Fatima:2020tyh}. The values of these parameters are quoted in Table~\ref{hyperon_resonances}.

 \begin{figure}
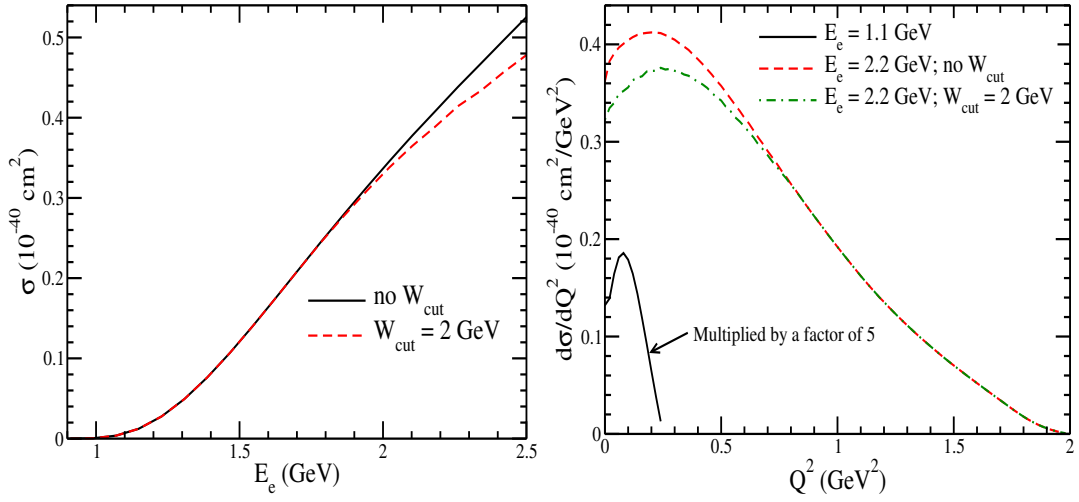
 
\begin{center}
\includegraphics[width=7cm,height=6.5cm]{electron_sigma_associated.eps}
\includegraphics[width=7cm,height=6.5cm]{dsigma_dq2_associated.eps}
\caption{(Left panel)~Total scattering cross section ($\sigma(E_e)$) as a function of electron energy ($E_{e}$) for
 $e^- + p \longrightarrow \nu_e + \Lambda +K^0$ scattering process. 
 The solid and dashed lines, respectively, represent the results obtained using (i)~no cut on $W$ and (ii)~a cut of 2~GeV on $W$. 
 (Right panel)~$\frac{d\sigma}{dQ^2}$ as a function of $Q^2$ for the process $e^- + p \longrightarrow \nu_e + \Lambda +K^0$. 
 The solid and dashed lines, respectively, represent the results at two different values of the incoming electron energy viz. $E_{e}=1.1$~GeV and 2.2~GeV with no cut on the CM energy $W$. The dash-dotted line shows the results at $E_{e}=2.2$~GeV with $W_{cut}=2$~GeV.
 }\label{sigma:weak_associated}
\end{center}
\end{figure}

In Eqs.~(\ref{had_curr:Kstar}) and (\ref{had_curr:K1}), $F_{t}^{\star}(t)$ is the strong form factor, given in Eq.~(\ref{eq:strong_FF_res})  with $\Lambda_{R}$ taken to be the same as  used in the case of Born terms, for simplicity, i.e., $\Lambda_{R} = \Lambda_{B} =$ 0.54 GeV~\cite{Fatima:2025tht}. Moreover, to account for the electromagnetic structure of kaons, the form factor $F_{K^{\star}, K_{1}} (Q^2)$ is introduced in Eqs.~(\ref{had_curr:Kstar}) and (\ref{had_curr:K1}), which, in the literature is generally parameterized as a  monopole form factor of the type~\cite{Janssen:2003kk} 
\begin{equation}
 F_{K^{\star}, K_{1}} (Q^2)  = \frac{1}{\left(1+ \frac{Q^2}{\Lambda_{K^{\star}, K_{1}}} \right)},
\end{equation}
 with $\Lambda_{K^{\star}} = 0.95$~GeV and $\Lambda_{K_{1}} = 0.55$~GeV~\cite{Janssen:2003kk}.


\subsubsection{Results and discussion:}

In Fig.~\ref{sigma:weak_associated}, the results are presented for the total and differential scattering cross sections for the process $e^- + p \longrightarrow \nu_e + \Lambda + K^0$ using different CM energy~($W$) cuts viz. no cut on $W$ and $W<2$~GeV. The left panel of the figure shows the results for $\sigma(E_e)$ as a function of $E_e$ from threshold up to 2.5~GeV and the right panel of the figure shows the results for $\frac{d\sigma}{dQ^2}$ as a function of $Q^2$ at $E_{e} = 1.1$ and 2.2~GeV. It may be observed from the figure that the cross section increases
with increasing $E_{e}$ and saturates at around $E_{e}=4$~GeV~(not shown in the figure). Beyond $E_e=2$~GeV, the total cross section decreases when a cut of 2~GeV is applied on $W$, and  at $E_e=2.5$~GeV, $\sigma(E_e)$ is suppressed by about 10\%. In the case of $Q^2$ distribution, we find that the peak of $\frac{d\sigma}{dQ^2}$ broadens as $E_e$ increases. Moreover, in the low $Q^2$ region, we observe some dependence on the choice of $W_{cut}$ at $E_e =2.2$~GeV, which decreases with increasing $Q^2$ and vanishes beyond $Q^2=0.7$~GeV$^2$. Furthermore, the results for  $\frac{d\sigma}{dQ^2}$ at $E_e=1.1$~GeV are multiplied by a factor of 5 to bring them on the same scale as that of the results at $E_e=2.2$~GeV.

\subsection{Electron and positron induced single kaon production}\label{sec2:kaon}
The processes considered here are the electron~(positron) induced weak $| \Delta S | = 1$ antikaon~(kaon) production from the free proton target. 
The  single antikaon production channels induced by electrons are
\begin{eqnarray}\label{eq:elec}
  e^- (k) + p(p) &\longrightarrow& \nu_e(k^\prime) + \bar{K^0}(p_K) + n(p^\prime),\\
  e^- (k) + p(p) &\longrightarrow& \nu_e(k^\prime) + K^- (p_K) + p (p^\prime),   
 \end{eqnarray}
and the corresponding positron induced channel is
\begin{eqnarray}\label{eq:posi}
 e^+ (k) + p (p) \longrightarrow \bar{\nu}_e (k^\prime) + K^+ (p_K) + p(p^\prime) ,
\end{eqnarray}
where the quantities in the parentheses represent the four momenta of the corresponding particles. The expression for the differential scattering cross section for reactions (\ref{eq:elec})--(\ref{eq:posi}) is given in Eq.~(\ref{sigma:weak}), where the transition matrix element is given in Eq.~(\ref{eq:Gg}) with $a=\sin 
\theta_{C}$. The expression for the leptonic current $l_\mu$ for the  electron and positron induced kaon production processes is given, respectively,  in Eqs.~(\ref{lmu:e}) and (\ref{lmu:e+}), and the hadronic current receives contribution from the various nonresonant terms, which are diagrammatically depicted in Fig.~\ref{fg:terms}.
Specifically, the electron induced single kaon production processes receive contribution from the $s$-channel hyperon~($\Lambda, \Sigma^0$) pole, $t$-channel pion pole, $t$-channel eta pole~(only for $K^-$ production), contact term, and kaon in flight term, while the positron induced reaction receives contribution from the $u$-channel hyperon~($\Lambda, \Sigma^0$) pole, $t$-channel pion and eta poles, contact term, and kaon in flight term.

\subsubsection{Nonresonant contribution:}
\begin{figure} 
\begin{center}
\includegraphics[width=15cm,height=8cm]{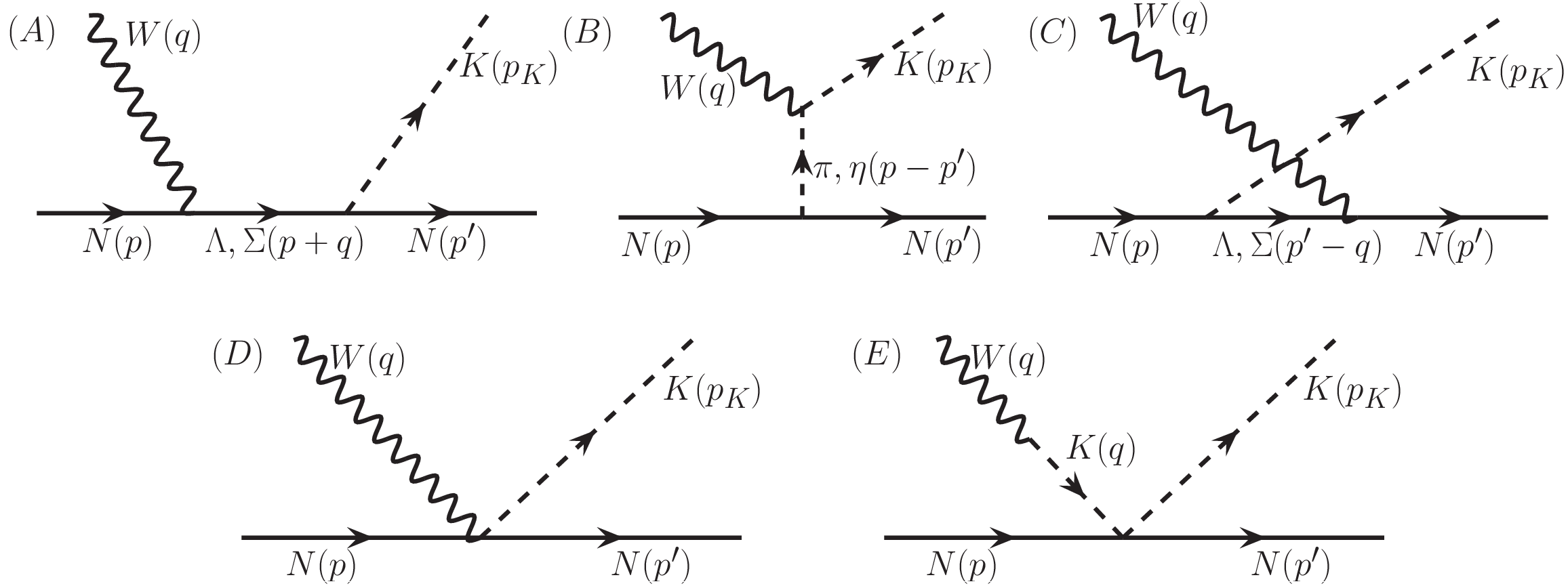}
\caption{Feynman diagrams for the processes $e^- + p \longrightarrow \nu_e + p + K^-$, $e^- + p \longrightarrow \nu_e + n + \overline{K}^0$, and $e^+ + p \longrightarrow \bar{\nu}_e + p + K^+$. 
(A) $s$-channel hyperon pole, (B) $t$-channel, or pion/eta pole~($\pi$P, $\eta$P) term, (C) $u$-channel hyperon pole, (D) contact term~(CT), and (E) kaon in flight~(KF) term constitute the non-resonant Born diagrams. }
\label{fg:terms}
\end{center}
\end{figure}

The expression for the hadronic current $j^{\mu}$ corresponding to the various nonresonant diagrams for electron and positron induced single kaon production, as depicted in Fig.~\ref{fg:terms}, are obtained using the nonlinear sigma model discussed 
in Section~\ref{NRB},  and are given by:
\begin{eqnarray}
j^{\mu}\big|_{s\Lambda}&=& i A_{s\Lambda} F_s(s) \bar{u}(p^\prime) \slashed{p}_K \gamma_5 \left(\frac{\slashed{p} + \slashed{q} + M_\Lambda}{(p + q)^2 -M_\Lambda^2} \right)
\left( V^\mu - A^\mu  \right)_{p\Lambda} u(p), \nonumber \\
j^{\mu}\big|_{s\Sigma} &=& i A_{s\Sigma} F_s(s) \bar{u}(p^\prime) \slashed{p}_K \gamma_5  \left(\frac{\slashed{p} + \slashed{q} + M_{\Sigma^0}}{(p + q)^2 -M_{\Sigma^0}^2} \right)
\left( V^\mu - A^\mu  \right)_{p\Sigma^0} u(p),\nonumber\\
j^{\mu}\big|_{u\Lambda}&=& i A_{u\Lambda} F_u(u) \bar{u}(p^\prime)\left(V^\mu - A^\mu  \right)_{p\Lambda}
\frac{{p\hspace{-.5em}/} - {p\hspace{-.5em}/}_K +M_\Lambda}{( p -  p_K)^2 -M_\Lambda^2}   {p\hspace{-.5em}/}_K 
\gamma_5 u(p),\nonumber \\
j^{\mu}\big|_{u\Sigma} &=& i A_{u\Sigma} F_u(u) \bar{u}(p^\prime) \left(V^\mu - A^\mu \right)_{p\Sigma^0} \frac{{p\hspace{-.5em}/} - {p\hspace{-.5em}/}_K + M_{\Sigma}^0}{( p -  p_K)^2 -M_{\Sigma^0}^2} {p\hspace{-.5em}/}_K 
\gamma^5  u(p),\nonumber\\
j^{\mu}\big|_{KF}&=& i  A_{KP} F_{CT} f_{\rho} \left((q-p_K)^2\right) \bar{u}(p^\prime) ({q\hspace{-.5em}/}+{p\hspace{-.5em}/}_K) u(p) 
\frac{1}{q^2-m_K^2} q^\mu,\nonumber \\
j^{\mu}\big|_{\pi}&=& i A_{\pi P}F_t(t) F_{P}(Q^2) \frac{M}{(q-p_K)^2 - {m_{\pi}^2}} \bar{u}(p^\prime) 
\gamma^5 u(p) (2 {p_K}^\mu - q^\mu),\nonumber \\
j^{\mu}\big|_{\eta }&=& i A_{\eta P} F_t(t) F_{P} (Q^2)\frac{M}{(q-p_K)^2 - {m_{\eta}^2}} \bar{u}(p^\prime)
\gamma^5 u(p) (2 {p_K}^\mu - q^\mu) ,\nonumber\\
j^\mu \big|_{CT} &=& i A_{CT} F_{CT} \bar{u}(p^\prime) (\gamma^\mu f_{\rho} \left((q-p_K)^2\right) - \gamma^\mu \gamma^5 B_{CT}F_{C} (Q^2)) u(p),
 \label{Ch12_NRB:kaon}
\end{eqnarray}
where $A_{i}'s~(i=s\Lambda, s\Sigma, u\Lambda, u\Sigma, KP, \pi P, \eta P,CT)$ and $B_{CT}$ are the constants appearing in the expressions of the 
hadronic currents of different channels, which are determined using the non-linear sigma model and are tabulated in Table~\ref{tb:currents}. The form of the strong form factors $F_s(s)$, $F_t(t)$, $F_u(u)$  is given in Eq.~(\ref{FF_Born}), with $\Lambda_{B}$ being the cut-off parameter.  The strong form factor $F_{CT}$ for the contact term is given in Eq.~(\ref{F_CT}).

 \begin{table}[ht]
 \begin{center}
 {\small
\renewcommand{\arraystretch}{1.2}
\begin{tabular*}{152mm}{@{\extracolsep{\fill}}|c|c|c|c|c|c|c|c|c|c|} \hline \hline
	    Process   	  	       &$B_{CT}$&$A_{CT}$& $A_{s\Sigma}$ &$A_{s\Lambda}$&$A_{u\Sigma}$&$A_{u\Lambda}$&$A_{KP}$& $A_{\pi P}$ & $A_{\eta P}$ \\\hline		   
$ e^- p \rightarrow \nu  K^- p        $&     $F$ &   $-\frac{\sqrt{2}}{f_K}$	 & $-\frac{D-F}{2f_K}$& $\frac{D+3F}{2\sqrt{3}f_K}$     &    0	     &    0	    & $\frac{1}{\sqrt{2}f_K}$      &   $\frac{D+F}{\sqrt{2}f_K}$	   &   $-\frac{D-3F}{\sqrt{2}f_K}$		 	    \\		
$ e^- p \rightarrow \nu \bar K^0 n    $&   $D+F$ &   $-\frac{1}{\sqrt{2}f_K}$	 & $\frac{D-F}{2f_K}$& $\frac{D+3F}{2\sqrt{3}f_K}$     &    0	     &    0	    & $\frac{1}{2\sqrt{2}f_K}$      &   $\frac{D+F}{\sqrt{2}f_K}$	   &   0	 \\		
$ e^+ p \rightarrow \bar\nu  K^+ p    $&     $F$ &   $\frac{\sqrt{2}}{f_K}$	 &    0 	&    0        &$-\frac{D-F}{2f_K}$&    $\frac{D+3F}{2\sqrt{3}f_K}$	    & $-\frac{1}{\sqrt{2}f_K}$       &    $-\frac{D+F}{\sqrt{2}f_K}$	   &   $\frac{D-3F}{\sqrt{2}f_K}$ 	    \\   \hline\hline	     
 \end{tabular*}
 }
\caption{Constant factors  appearing in the hadronic current given in Eq.~(\ref{Ch12_NRB:kaon}).}\label{tb:currents}
 \end{center}
\end{table}

The vector~($V^{\mu}$) and axial-vector~($A^{\mu}$) currents for $p \rightarrow \Lambda$ and $p \rightarrow \Sigma^0$ transitions are discussed in detail in Section~\ref{sec:FF:QE}, with the explicit expressions for the vector~($f_{1,2}^{pY}(Q^2)$; $Y=\Lambda, \Sigma^0$) and axial vector~($g_1^{pY}(Q^2)$) form factors given in Table-\ref{tab:formfac}. In the numerical calculations, for the kaon production processes, we have included the contribution from the first class currents only, thus, we set $g_2^{pY}(Q^2)=0$.

In Eq.~(\ref{Ch12_NRB:kaon}), the form factors $F_{C}(Q^2)$, $F_{P}(Q^2)$ and 
$f_{\rho}((q-p_{K})^2)$ are introduced in the contact, pion and eta pole, and kaon in flight terms, respectively, to take into account the hadronic 
structure.  We have already discussed in Section~\ref{weak:associated}, the nature~(whether vector or axial vector) of the hadronic currents associated with these form factors. The expression for $f_{\rho}((q-p_{K})^2)$ is given in Eq.~(\ref{frho}), while the form factors $F_{C}(Q^2)$ and $F_{P}(Q^2)$ are determined using CVC hypothesis, which implies that the total hadronic current corresponding to the vector interaction should be conserved, i.e.
\begin{equation}
 q_{\mu}J^{\mu}|^{V} =0,
\end{equation}
where $J^{\mu}|^{V} = j^{\mu}\big|_{s\Lambda}^V + j^{\mu}\big|_{s\Sigma}^V + j^{\mu}\big|_{u\Lambda}^V + j^{\mu}\big|_{u\Sigma}^V + j^{\mu}\big|_{t\pi}^V + j^{\mu}\big|_{t\eta}^V + j^{\mu}\big|_{CT}^V$. The detailed calculations are given in Ref.~\cite{Fatima:2026:kaon}.

Specifically, the expressions for $F_{C}(Q^2)$ and $F_{P}(Q^2)$ form factors for the reaction $e^- + p \longrightarrow \nu_e + n + \overline{K}^0$ are obtained as:
\begin{eqnarray}
 F_{C} (Q^2) &=& \frac{1}{2} \left[\left(\frac{s-M^2}{s-M_{\Lambda}^2}\right) \left(\frac{D+3F}{D+F}\right) F_1^p(Q^2) + \left(\frac{s-M^2}{s-M_{\Sigma}^2}\right) \left(\frac{D-F}{D+F}\right) \right.\nonumber \\ && \times  \left(F_1^p(Q^2) + 2F_1^n(Q^2) \right) \Big],\\
 F_{P} (Q^2) &=& F_C(Q^2) \left(\frac{t-m_{\pi}^2}{t-m_K^2} \right) -  \left(\frac{t-m_{\pi}^2}{t-m_K^2} \right) \left(\frac{s-M^2}{4M} \right) \left[F_1^p(Q^2) \left(\frac{D+3F}{D+F}\right) \left(\frac{M_\Lambda - M}{M_\Lambda^2 - s} \right) \right. \nonumber \\
 && \left. + \left(\frac{D-F}{D+F}\right) \left(\frac{M_\Sigma - M}{M_\Sigma^2 - s} \right) \left(F_1^p(Q^2) + 2F_1^n(Q^2) \right) \right],
\end{eqnarray}
where $s=(p+q)^2$, $t=(p-p^\prime)^2$, and $F_{1}^{p} 
(Q^2)$ and $F_{1}^{n} (Q^2)$ are the nucleon electromagnetic form factors. 

For the reaction $e^- + p \longrightarrow \nu_e + p + K^-$, the expressions for $F_{C}(Q^2)$ and $F_{P}(Q^2)$ are obtained as
\begin{eqnarray}
 F_{C} (Q^2) &=& \left[\left(\frac{s-M^2}{s-M_{\Lambda}^2}\right) \left(\frac{D+3F}{4F}\right) F_1^p(Q^2) - \left(\frac{s-M^2}{s-M_{\Sigma}^2}\right) \left(\frac{D-F}{4F}\right) \right.\nonumber \\ && \times  \left(F_1^p(Q^2) + 2F_1^n(Q^2) \right) \Big],
 \end{eqnarray}
 \begin{eqnarray}
 F_{P} (Q^2) &=& \frac{1}{M\left[\frac{t-m_K^2}{m_\pi^2 - t} (D+F) - \frac{t-m_K^2}{m_\eta^2 - t} (D-3F) \right]} \left[F_1^p(Q^2) \left(\frac{D+3F}{2} \right) \left(\frac{s-M^2}{M_\Lambda^2-s}\right) \left(M_{\Lambda} - M\right) \right. \nonumber \\
 &&- \left. \left(F_1^p(Q^2) + 2F_1^n(Q^2) \right) \left(\frac{D-F}{2} \right) \left(\frac{s-M^2}{M_\Sigma^2-s} \right) \left(M_{\Sigma} - M\right) - 4MF~F_C(Q^2) \right].
\end{eqnarray}

Furthermore, for the positron induced reaction $e^+ + p \longrightarrow \bar{\nu}_e + p + K^+$, the expressions for $F_{C}(Q^2)$ and $F_{P}(Q^2)$ are obtained as
\begin{eqnarray}
 F_{C} (Q^2) &=& \left[\left(\frac{M^2-u}{M_{\Lambda}^2-u}\right) \left(\frac{D+3F}{4F}\right) F_1^p(Q^2) - \left(\frac{M^2-u}{M_{\Sigma}^2-u}\right) \left(\frac{D-F}{4F}\right) \right.\nonumber \\ && \times  \left(F_1^p(Q^2) + 2F_1^n(Q^2) \right) \Big],\\
 F_{P} (Q^2) &=& \frac{1}{M\left[ \frac{t-m_K^2}{m_\pi^2 - t} (D+F) - \frac{t-m_K^2}{m_\eta^2 - t} (D-3F) \right]} \left[F_1^p(Q^2) \left(\frac{D+3F}{2} \right) \left(\frac{u-M^2}{M_\Lambda^2 - u}\right) \left(M_{\Lambda} -M \right) \right. \nonumber \\
 &&- \left. \left(F_1^p(Q^2) + 2F_1^n(Q^2) \right) \left(\frac{D-F}{2} \right) \left(\frac{u-M^2}{M_\Sigma^2-u} \right) \left(M_{\Sigma} - M\right) - 4MF~F_C(Q^2) \right],
\end{eqnarray}
where $u=(p-p_K)^2$. 
\subsubsection{Results and discussion:}
\begin{figure}
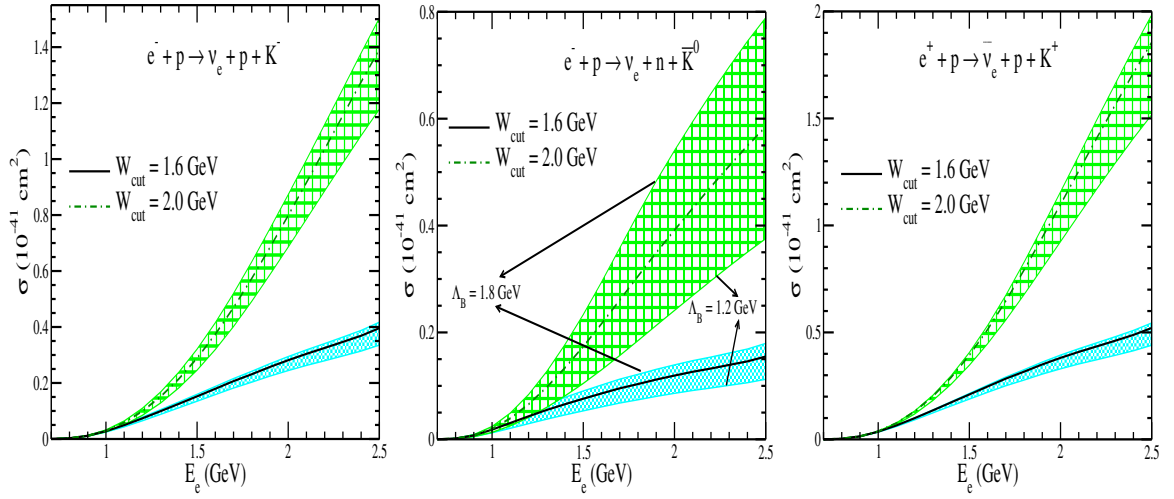
 
\begin{center}
\includegraphics[width=5cm,height=6.5cm]{total_sigma_p_p_Kminus.eps}
\includegraphics[width=5cm,height=6.5cm]{total_sigma_p_nK0bar.eps}
\includegraphics[width=5cm,height=6.5cm]{total_sigma_p_p_Kplus.eps}
\caption{The total scattering cross section~($\sigma(E_e)$) as a function of electron~($E_{e^-}$) and positron~($E_{e^+}$) energies, for the single kaon production processes on a free proton target, namely $e^- + p \longrightarrow \nu_e + p + K^-$~(left panel), $e^- + p \longrightarrow \nu_e + n + \overline{K}^0$~(middle panel), and $e^+ + p \longrightarrow \bar{\nu}_e + p + K^+$~(right panel). The solid~(dash-dotted) line corresponds to the results obtained when a cut of 1.6~(2)~GeV is applied on the CM energy $W$ and $\Lambda_B=1.5$~GeV. The band corresponds to $\Lambda_B$ being varied in the range $1.2-1.8$~GeV.
 }\label{sigma:weak_single_K}
\end{center}
\end{figure}
\begin{figure}
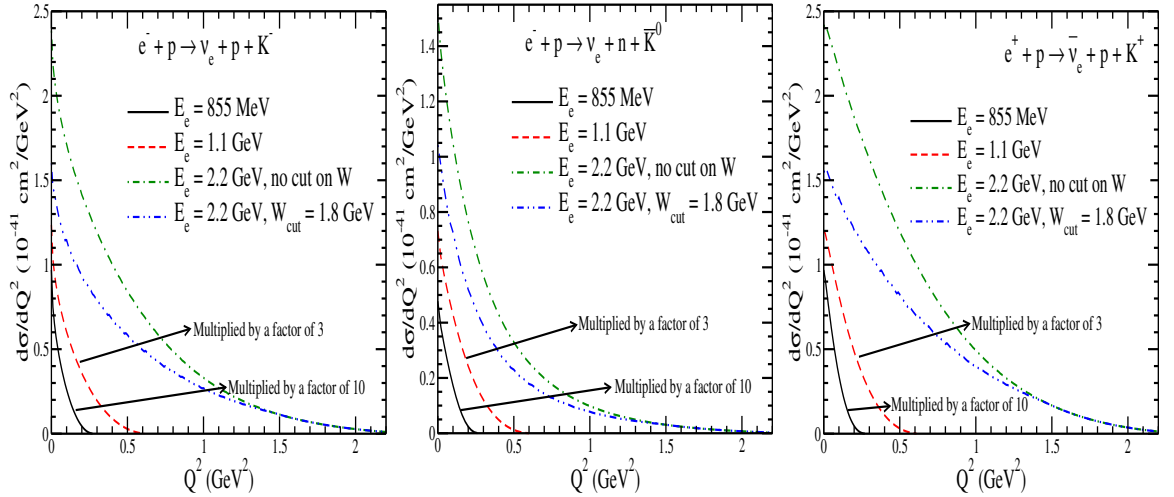
 
\begin{center}
\includegraphics[width=5cm,height=6.5cm]{dsigma_dq2_p_p_Kminus.eps}
\includegraphics[width=5cm,height=6.5cm]{dsigma_dq2_p_nK0bar.eps}
\includegraphics[width=5cm,height=6.5cm]{dsigma_dq2_p_p_Kplus.eps}
\caption{The differential scattering cross section~$\left(\frac{d\sigma}{dQ^2} \right)$ as a function of $Q^2$ for the processes $e^- + p \rightarrow \nu_e + K^{-} + p$~(left panel), $e^- + p \rightarrow \nu_e + \bar{K}^{0} + n$~(middle panel), and $e^+ + p \rightarrow \bar{\nu}_e + K^{+} + p$~(right panel) at three different values of the incoming electron and positron energies viz. $E_{e} = 855$~MeV~(solid line), 1.1~GeV~(dashed line), and 2.2~GeV~(dash-dotted line), when no cut on $W$ is applied. The double-dot-dashed line represents the results at $E_e=2.2$~GeV, when a cut of 1.8~GeV is applied to $W$. 
 }\label{dsigma:weak_single_K}
\end{center}
\end{figure}
In Fig.~\ref{sigma:weak_single_K}, we present the results for the total cross section as a function of electron and positron energies for the processes $e^- + p \longrightarrow \nu_e + p + K^-$, $e^- + p \longrightarrow \nu_e + n + \overline{K}^0$, and $e^+ + p \longrightarrow \bar{\nu}_e + p + K^+$, by applying a cut of 1.6~GeV and 2~GeV  on the CM energy $W$. A lower $W_{cut}$ implies less phase space available to the final state particles. 
We find a strong dependence of the cross section on the $W$ cut for all the three processes considered in this work. For example, in the case of $e^- + p \longrightarrow \nu_e + p + K^-$ and $e^+ + p \rightarrow \bar{\nu}_e + K^{+} + p$ processes, reducing $W_{cut}$ from 2~GeV to 1.6~GeV, reduces $\sigma(E_e)$ by about $\sim 48\%~(70 \%)$ at $E_e=1.5~(2.2)$~GeV. However, in the case of  $e^- + p \rightarrow \nu_e + \bar{K}^{0} + n$ process, the suppression in the cross section is even larger, i.e. $\sigma(E_e)$ decreases by about $55\%~(75 \%)$ at $E_e=1.5~(2.2)$~GeV, when $W_{cut}$ is decreased from 2~GeV to 1.6~GeV.

\begin{figure} 
\begin{center}
\includegraphics[width=10cm,height=8cm]{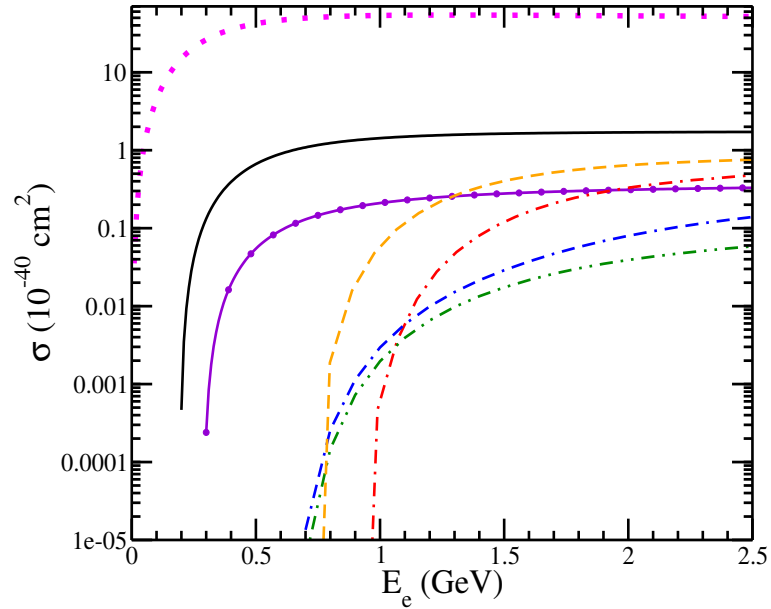}
\caption{Total scattering cross section $\sigma(E_e)$ as a function of electron energy $E_e$ for the processes $e^- + p \longrightarrow \nu_e + n$~(dotted line), $e^- + p \longrightarrow \nu_e + \Lambda$~(solid line), $e^- + p \longrightarrow \nu_e + \Sigma^0$~(solid line with circle), $e^- + p \longrightarrow \nu_e + \Lambda + K^0$~(dash-dotted line), $e^- + p \longrightarrow \nu_e + n + \bar{K}^0$~(double-dot-dashed line), $e^- + p \longrightarrow \nu_e + p + K^-$~(double-dash-dotted line), and $e^- + p \longrightarrow \nu_e + n + \eta$~(dashed line). The cross sections calculated for the meson production processes are obtained by applying a cut of 2~GeV on the CM energy $W$.}\label{sigma:compare}
\end{center}
\end{figure}

Moreover, the only free parameter in the case of single kaon production is the cut-off parameter $\Lambda_B$ appearing in the strong form factor.
Unlike the $\eta$ and associated particle production processes, where the strong and electromagnetic vertices are fixed using the photo- and electro- production data available from the CLAS experiment, the single kaon production is possible only via weak interaction. In the numerical calculations, we use $\Lambda_B=1.5$~GeV and study the effect of varying $\Lambda_B$ in the range $1.2-1.8$~GeV on the total cross section. We find a considerable dependence of $\sigma(E_e)$ on the choice of $\Lambda_B$, which increases with the increase in electron and positron energies for all the three kaon production processes. Moreover, the width of the band obtained using variable $\Lambda_B$ increases with increase in the value of $W_{cut}$.

It may be observed from the figure that the cross section increases~(decreases) with the increase~(decrease) in $\Lambda_B$. However, this increase~(decrease) in $\sigma(E_e)$ due to increase~(decrease) in $\Lambda_B$ is not symmetric. We find that the increase in $\sigma(E_e)$ due to an increase of  $\Lambda_B$ by 20\%  is smaller than the decrease in $\sigma(E_e)$ when $\Lambda_B$ is decreased by 20\% from $\Lambda_B=1.5$~GeV. For example, in the case of $e^- + p \longrightarrow \nu_e + p + K^-$ process, when a cut of $W=1.6$~GeV is applied, $\sigma(E_e)$ increases by about 5\% at $E_e=1.5-2.2$~GeV when $\Lambda_B$ is increased from 1.5~GeV to 1.8~GeV whereas it decreases by about 10\%~(15\%) at $E_e=1.5~(2.2)$~GeV when $\Lambda_B$ is decreased from 1.5~GeV to 1.2~GeV. On the contrary, when a cut of $W=2$~GeV is applied, $\sigma(E_e)$ increases by about 10\%~(8\%) at $E_e=1.5~(2.2)$~GeV when $\Lambda_B$ is increased from 1.5~GeV to 1.8~GeV whereas it decreases by about 15\% at $E_e=1.5-2.2$~GeV when $\Lambda_B$ is decreased from 1.5~GeV to 1.2~GeV. 

Fig.~\ref{dsigma:weak_single_K} presents the results for $\frac{d\sigma}{dQ^2}$ as a function of $Q^2$ for the electron and positron induced single kaon production processes at three different values of electron and positron energies viz. $E_e=855$~MeV, 1.1, and 2.2~GeV. To show the results on the same scale, $\frac{d\sigma}{dQ^2}$ at $E_{e}=855$~MeV and 1.1~GeV are multiplied, respectively, by a factor of 10 and 3 for all the three kaon production channels. To show the effect of $W_{cut}$ on $\frac{d\sigma}{dQ^2}$, we have also shown on the same plot, the results for $\frac{d\sigma}{dQ^2}$ at $E_{e}=2.2$~GeV by applying a cut of 1.8~GeV on $W$. We find a significant suppression in  the cross section in the low $Q^2$ region for all the three kaon production channels, when the $W_{cut}$ is applied, which decreases with increasing $Q^2$ and becomes almost negligible beyond $Q^2=1$~GeV$^2$.

\subsection{Comparison of total scattering cross sections for quasielastic and inelastic kaon and $\eta$ production}
In Fig.~\ref{sigma:compare}, we compare the results of the total scattering cross sections $\sigma(E_e)$ as a function of $E_e$ for $\Delta S=0$ strangeness conserving and $\Delta S=1$ strangeness changing processes considered in this work. Specifically, the energy dependence of the cross section in the quasielastic processes~($e^- + p \longrightarrow \nu_e + n$, $e^- + p \longrightarrow \nu_e + \Lambda$, $e^- + p \longrightarrow \nu_e + \Sigma^0$) and the inelastic production of $K$ and $\eta$ mesons~($e^- + p \longrightarrow \nu_e + \Lambda + K^0$, $e^- + p \longrightarrow \nu_e + n + \bar{K}^0$, $e^- + p \longrightarrow \nu_e + p + K^-$, and $e^- + p \longrightarrow \nu_e + n + \eta$) are presented. The cross section for the $\Delta S=0$ quasielastic reaction is of the order of $\sim 7\times 10^{-39}$~cm$^2$ in the energy region of $0.5-2.5$~GeV. It can be seen that the 
cross sections of the $\Delta S=1$ quasielastic scattering processes is down by more than an order of  magnitude than the $\Delta S=0$ quasielastic scattering cross section.
While the inelastic production of mesons is about two orders of magnitude smaller than the $\Delta S=0$ quasielastic scattering cross section. The inelastic production of mesons is expected to be dominated by the pion production, which is presently under consideration and will be reported elsewhere. The inelastic production of higher mass mesons is dominated by the $\eta$ meson production, which is of the order of $\sim 10^{-41}$~cm$^2$ in the region of $E_e>1$~GeV, and is closely followed by the associated production of kaons in the $\Delta S=0$ sector and the single production of kaons in the $\Delta S=1$ sector, as shown in the figure.

\section{Summary and conclusion}\label{sec:summary}
In this work, we have studied the weak charged current induced electron and positron scattering off the proton in the few-GeV energy regime. We have investigated the quasielastic processes in both the strangeness conserving $\Delta S=0$ and the strangeness changing $\Delta S=1$ sectors for the differential and total scattering cross sections, and some of the polarization observables for the baryons in the initial and final states. We have also studied the inelastic channels involving the low lying baryon resonance excitation of $P_{33}(1232)$, $P_{11}(1440)$, and $S_{11}(1535)$ resonances, and meson production of $\eta$, $K, \bar K$, mesons and associated particle production of $\Lambda K$, providing detailed predictions for the differential and total scattering cross sections. This study acquires particular significance in view of the upcoming next-generation, high-luminosity, and high-precision charged lepton beam facilities at Jefferson Lab and MAMI, which can be used to study the weak processes using electron and/or positron beams in the charged current sector. 

The main features of our results for each processes considered in the review are the following:

\begin{itemize}
 \item [(I)] Quasielastic scattering in $\Delta S=0$ and $\Delta S =1$ processes\\
 
 In the quasielastic scattering region, we have presented the summary of our study for the differential and total scattering cross sections, together with the polarization observables of the produced baryons. In addition, we highlight the role of the polarized proton target, demonstrating the potential observability of the proton spin asymmetries. Furthermore, we examine the possible manifestation of the second class currents through the inclusion of the time-reversal noninvariance, quantifying their impact through the results for both the  cross sections and polarization observables.
 
 \begin{itemize}
  \item [A.] Total cross section $\sigma(E_e)$ and the $Q^2$ distribution $\frac{d\sigma}{dQ^2}$
  
  \begin{itemize}
   \item [(i)] For the process $e^- + p \longrightarrow \nu_e + n$, the choice of the parametrization for the axial vector form factor $g_1(Q^2)$ such as the dipole form with $M_A=1.026$~GeV, various $z$-expansion fits based on (a)~MINERvA hydrogen data, (b)~LQCD, (c)~older deuterium data, (d)~combined MINERvA hydrogen-LQCD analyses, as well as the lattice gauge parametrization of Chen et al.~\cite{Chen:2022odn}, has a pronounced impact on the cross section $\sigma(E_e)$. However, using a higher value of $M_A$ in the dipole parametrization  simulates the results for cross section obtained using some of the non-dipole parametrizations.
   
   \item [(ii)] Within the dipole parametrization, increasing $M_A$ from 1.026~GeV~(world average) to 1.35~GeV~(MiniBooNE CCQE value~\cite{MiniBooNE:2007iti}) results in a significant enhancement of nearly 40\% in the total cross section in the $e^- + p \longrightarrow \nu_e + n$ scattering. This sensitivity is consistently mirrored in the $Q^2$ distribution i.e. $\frac{d\sigma}{dQ^2}$ as a function of $Q^2$ which exhibits trends closely aligned with those observed in $\sigma(E_e)$.\\
   
   For the process $e^-+p\longrightarrow \nu_e +\Lambda$, varying $M_A$ in the range $1.026-1.3$~GeV~(within the dipole form of $g_1(Q^2)$) leads to a moderate but systematic increase in the cross section: about $10-12\%$ at $E_e=500-700$~MeV, rising to about 12$\%$ at 1~GeV and reaching about 22$\%$ at 2~GeV. \\
   
   In contrast, for $e^-+p \longrightarrow \nu_e + \Sigma^0$, the same variation in $M_A$ produces only a mild effect on the cross section, with $\sigma(E_e)$ increasing gradually with energy and reaching about $4-5\%$ at 2~GeV.
   
   \item [(iii)] Effect of second class currents\\
   
   \begin{itemize}
   \item In the case of $e^- + p \longrightarrow \nu_e + n$ process, a non-zero weak electric form factor $g_2(Q^2)$~(assuming a dipole form with $M_2=1.026$~GeV), associated with the second class currents, leads to an enhancement in the total cross section. Quantitatively, this increase is about 10$\%$ for $g_2(0)= 2$, while it remains modest, around 2\% for $g_2(0) = 1$ for the electron energy above 1~GeV. 
   
  \item We find that for purely real~(corresponding to T-invariance) or purely imaginary~(corresponding to T-violation) values of $g_2(Q^2)$, i.e. $g_2^R(Q^2)$ and $g_2^I(Q^2)$, respectively, the cross section remains the same, in the case of $e^- + p \longrightarrow \nu_e + n$ process. Also, the positive and the negative values of $g_2^I(0)$ give the same cross section, for all three quasielastic reactions, namely, $e^- + p \longrightarrow \nu_e + n$, $e^- +p\longrightarrow \nu_e+\Lambda$, and $e^-+p\longrightarrow \nu_e+\Sigma^0$.

 \item While in the case of  $e^- +p\longrightarrow \nu_e+\Lambda$ process, a non-zero real $g_2(Q^2)$ within the dipole form, leads to an enhancement of $\sigma(E_e)$ for $g_2^R(Q^2) > 0$, which is about  $15-17\%$ for 
 $g_2^R(0) = 3$ at $E_e = 1-2$~GeV. While it decreases for $g_2^R(Q^2)<0$ at low $E_e$, becomes equal at $E_e = 1$~GeV, and increases by about $3-8\%$ for $E_e = 1-2$~GeV.
 
 \item In the case of $g_2^I(Q^2)$ variation corresponding to T-violation, the enhancement in $\sigma(E_e)$ for the process $e^- +p\longrightarrow \nu_e+\Lambda$ is smaller as compared to $g_2^R(Q^2)$ variation.
 
 \item Whereas in the case of $e^-+p\longrightarrow \nu_e+\Sigma^0$ process, a non-zero value of $g_2(Q^2)$, results only a change of $2-3\%$ for purely real $g_2(0)$, which becomes negligible if purely imaginary $g_2(0)$, is considered. 
 
  \end{itemize}
 
 \item [(iv)] Similar effects are observed for the $Q^2$-distribution.
   
  \end{itemize}

  \item [B.] Spin asymmetries of the polarized proton target
  \begin{itemize}
   \item [(i)] For the process $e^- + {\vec p} \longrightarrow \nu_e + n$, the choice of different parametrizations for $g_1(Q^2)$ leads to only marginal variations in the longitudinal~($A_L(E_e)$) and perpendicular~($A_P(E_e)$) spin asymmetries, with changes remaining below the $2-3\%$ level.
   
   \item [(ii)] The variations in the axial dipole mass $M_A$ have a negligible impact on $A_L(E_e)$ and $A_P(E_e)$, with discernible differences arising only in the comparison between $M_A = 1.35$~GeV and $M_A = 1.05$~GeV, which remain below 2$\%$. The $Q^2$-dependent spin asymmetries of the initial proton exhibit very weak sensitivity to both the choice of $g_1(Q^2)$ parametrization and the value of $M_A$ used, in the case of $e^- + {\vec p} \longrightarrow \nu_e + n$  process.\\
   
While in the case of  $e^- + p \longrightarrow \nu_e + \Lambda$ process, $A_L(E_e)$ is almost insensitive to the variation in $M_A$ over the range $1.0-1.3$~GeV. In contrast, $A_P(E_e)$ displays a clear enhancement, increasing by about $8-12\%$ in the electron energy range $E_e = 1 - 2.5$~GeV. \\

Whereas for $e^- + p \longrightarrow \nu_e + \Sigma^0$ process, $A_L(E_e)$  exhibits a small sensitivity to variations in $M_A$. On the other hand $A_P(E_e)$  exhibits larger dependence on $M_A$, which increases with increasing $E_e$ and becomes about 15$\%$ larger at $E_e= 2.5$~GeV. 

\item [(iii)] Effect of second class current assuming T-invariance

\begin{itemize}
\item A non-zero $g_2^R(Q^2)$~(corresponding to purely real values of $g_2(Q^2)$) induces a pronounced effect on both $A_L(E_e)$  and $A_P(E_e)$, in the case of $e^-+p \longrightarrow \nu_e + n$ scattering,  particularly at higher electron energies considered in this study. Incorporating a non-zero $g_2^R(Q^2)$, both $A_L(Q^2)$ and $A_P(Q^2)$ display a strong dependence on its magnitude across the entire range of $Q^2$ and $E_e$ explored in the present analysis. 

\item For $e^- + {\vec p} \longrightarrow \nu_e + \Lambda$ process, a non-zero value of $g_2^R(Q^2)$ leads to a significant modification of both $A_L(E_e)$
and $A_{P}(E_e)$, with the effect becoming increasingly pronounced at higher electron energies considered in this study. The effect is particularly enhanced in $A_L(E_e)$ for $g_2^R(Q^2)<0$ , exhibiting a clear growth with $E_e$, reaching the level of $10-20\%$ in the range $E_e=1 - 2$~GeV. In contrast, $A_P (E_e)$, shows a stronger sensitivity, with variations as large as 70$\%$ for $g_2^R(0) = \pm 1$, decreasing only marginally around 2~GeV. This substantial effect is further amplified for larger magnitudes, $|g_2^R(0)| = 3$, underscoring a large sensitivity of $A_P (E_e)$, to the second class current contributions. Similar effects are observed for the $Q^2$-distribution.

\item In the case of $e^-+{\vec p} \longrightarrow \nu_e +\Sigma^0$ process, a non-zero value of $g_2^R(Q^2)$ leads to a striking and highly significant modification of both $A_L(E_e)$ and $A_P (E_e)$, with the impact becoming more pronounced with increasing electron energies. This change is both for $g_2^R(Q^2)>0$ and $g_2^R(Q^2)<0$, though in the opposite direction, and they are as large as 40--50$\%$ at $E_e= 2.5$~GeV. 
\end{itemize}

\item [(iv)] For the $Q^2$-distribution, variations of $M_A$ brings a very small change in $A_L (Q^2 )$. In contrast, $A_P(Q^2)$ exhibits a comparatively stronger sensitivity, with the effect becoming clearly visible in the region of $Q^2$ around 0.2--0.6~GeV$^2$. This difference becomes increasingly pronounced with increasing electron energy $E_e$. 
  \end{itemize}

  \item [C.] Polarization observables of the final baryon
  \begin{itemize}
   \item [(i)] For the process $e^- + {p} \longrightarrow \nu_e + \vec{n}$, perpendicular component of the polarization $P_P (E_e)$ shows a noticeable sensitivity to the choice of $g_1(Q^2)$. In particular, $P_P (E_e)$ increases by about 15--18$\%$ when using the LQCD-based prescription of Chen et al.~\cite{Chen:2022odn}, compared to the dipole parametrization with $M_A=1.026$~GeV. In contrast, the longitudinal polarization $P_L(E_e)$ exhibits only a mild dependence on the choice of $g_1(Q^2)$.
   
   \item [(ii)] When different values of $M_A$ are used, $P_P (E_e)$ shows a significant variation of about 13--15$\%$ in the case of  $e^- + {p} \longrightarrow \nu_e + \vec{n}$, whereas $P_L (E_e)$ remains largely unaffected, with changes limited to 2--3$\%$ when $M_A$ is varied from 1.026 GeV to 1.35 GeV.\\ 
   
While in the case of  $e^- + p \longrightarrow \nu_e + \vec{\Lambda}$ process, when different values of the axial mass $M_A$ are employed, $P_L(E_e)$ exhibits a pronounced and energy-dependent variation, increasing steadily with $E_e$. Quantitatively, the effect is about 10$\%$ at $E_e=1$~GeV and grows to nearly 20$\%$ at $E_e=2$~GeV, highlighting its strong sensitivity to $M_A$. In contrast, the perpendicular component $P_P (E_e)$ remains largely insensitive, with variations confined to a modest 2--4$\%$ increase when $M_A$ is varied from 1.026 GeV to 1.35 GeV. \\

Whereas in the case of  $e^- + p \longrightarrow \nu_e + \vec{\Sigma}^0$ process, changing $M_A$ exhibits a pronounced and energy-dependent variation both in $P_L(E_e)$ and $P_P (E_e)$, and the effect increases steadily with $E_e$. Quantitatively, the effect is about 5$\%$ at $E_e=1$~GeV and grows to nearly 15$\%$ at $E_e=2.5$~GeV. 

\item [(iii)] Effect of second class currents:
\begin{itemize}
\item [a)] {\bf T-invariance:} \\

In the case of $e^- + p \longrightarrow \nu_e + \vec{n}$ process, upon varying $g_2^R(Q^2)$, both $P_L(E_e)$ and $P_P (E_e)$,  display a strong dependence on the second class current form factor, particularly when a non-zero value is considered. \\

For $e^- + p \longrightarrow \nu_e + \vec{\Lambda}$ process, upon varying $g_2^R(Q^2)$, both $P_L(E_e)$ and $P_P (E_e)$,  display a strong dependence on the second class current form factor. For $g_2^R(Q^2) > 0$, $P_L(E_e)$ and $P_P (E_e)$, show increase in the opposite directions and the difference in the results from $g_2^R(Q^2) = 0$ increases with the increasing energy. \\

For $e^- + p \longrightarrow \nu_e + \vec{\Sigma}^0$ process, a strong dependence of both $P_L(E_e)$  and $P_P(E_e)$, on $g_2^R(Q^2)$ variation, for $g_2^R(Q^2)>0$ as well as $g_2^R(Q^2)<0$ are observed across the entire range of $E_e$. Similar observations are observed for the $Q^2$-distribution.\\

\item [(b)] {\bf T-violation:}\\

In contrast, for variations in $g_2^I(Q^2)$ in the case of of $e^- + p \longrightarrow \nu_e + \vec{n}$ process, $P_L(E_e)$ and $P_P (E_e)$ exhibit only a modest dependence, at the level of 4--8$\%$ in the energy range studied. However, the transverse polarization component $P_T(E_e)$ shows a pronounced enhancement, reaching about 30--35$\%$ for $g_2^I (0) = 2$ compared to $g_2^I (0) = 0$. \\

For $e^- + p \longrightarrow \nu_e + \vec{\Lambda}$ process, upon varying in $g_2^I(Q^2)$, $P_L(E_e)$ exhibits only a weak sensitivity, with changes limited to about 4--6$\%$ over the energy range considered. In sharp contrast, both the $P_P(E_e)$  and $P_T(E_e)$ components display pronounced and rapidly growing enhancements. Specifically, $P_P(E_e)$  increases by about 8--10$\%$ for $g_2^I(0)= 3$, while the transverse component $P_T(E_e)$ shows a significant enhancement of 30--40$\%$, relative to the $g_2^I(0) = 0$ case, for the electron beam energy range $E_e= 1-2$~GeV. \\
 
 For $e^- + p \longrightarrow \nu_e + \vec{\Sigma}^0$ process, an imaginary value of $g_2^I(Q^2)$ hardly changes $P_L(E_e)$  and $P_P(E_e)$, however, $P_T (E_e)$ shows dependence when $g_2^I(0)$ is taken as 3 instead of 0, especially with increasing $E_e$. Similar observations have been made in the case of $Q^2$ dependent polarization observables.
\end{itemize}

  \end{itemize}

 \end{itemize}

 \item [(II)] Inelastic excitation of delta and nucleon resonances
 \begin{itemize}
  \item [A.] $P_{33}(1232)$ resonance
  \begin{itemize}
   \item [(i)] The production cross section for the electron-induced process rises rapidly with increasing electron energy $E_{e^-}$ and approaches saturation around 2~GeV. In contrast, for the positron induced process, the production cross section continues to grow with increasing positron energy $E_{e^+}$, levelling off near 4~GeV. In the vicinity of 1.5 GeV, the electron induced cross section dominates over the positron induced one; however, beyond this energy, the positron induced production cross section becomes significantly larger. 

\item [(ii)] In the axial vector sector, the form factor $C_{5}^A(Q^2)$ overwhelmingly dominates the total cross section. Consequently, different values of $C_5^A(0)$ and $M_A$ lead to noticeable variations in the total cross section, with the effect being considerably more pronounced for the electron induced reactions than  the positron induced processes. 

\item [(iii)] The contributions arising from the subleading axial vector form factors $C_3^A(Q^2)$ and $C_4^A(Q^2)$ are individually small and model dependent. 
 Nevertheless, different choices of $C_3^A(0)$ and $C_4^A(0)$ entering the parametrizations of $C_3^A(Q^2)$ and $C_4^A(Q^2)$, can still induce appreciable modifications in the cross section, due to interference terms, amounting to about 8--15$\%$ in the electron and positron energy range 1--2.5 GeV.
 
\item [(iv)] When different values of $C_5^A(0)$ and $M_A$ are considered to examine its impact on the differential scattering cross section~$\left( \frac{d\sigma}{dQ^2}\right)$, we find that  there is a change of about 8--10$\%$, at $E_e=1$~GeV, which increases to about 15$\%$ at $E_e=2.5$~GeV, in the region of $Q^2$, where $\frac{d\sigma}{dQ^2}$ peaks. 

\item [(v)] The different values of $C_3^A(0)$ and $C_4^A(0)$ used in $C_3^A(Q^2)$ and $C_4^A(Q^2)$ to evaluate the $Q^2$-distribution, results in a change of about 15$\%$ in the region of $Q^2$, where $\frac{d\sigma}{dQ^2}$ peaks. The peak is more sharply peaked at low $Q^2$ in the case of positron induced reaction than the electron induced process. 
  \end{itemize}

  \item [B.] $P_{11}(1440)$ and $S_{11}(1535)$ resonances
  \begin{itemize}
   \item [(i)] The total cross sections for the excitation of these resonances are found to be more than an order of magnitude smaller than those for $\Delta(1232)$ production.
   
\item [(ii)] There is a strong dependence of the cross sections on the choice of  $W_{\text{max}}$.  For $P_{11}(1440)$, increasing $W_{\text{max}}$ from 1.4~GeV to $1.6$ GeV enhances the cross section by about 65\% (100\%) at $E_e = 1$~(2) GeV. In contrast, for $S_{11}(1535)$, raising $W_{\text{max}}$ from 1.5 GeV to $1.7$ GeV leads to an increase by about $40\%$~(67\%) at $E_e = 1$~(2) GeV.

\item [(iii)] The differential scattering cross section $\frac{d\sigma}{dQ^2}$ exhibit a similarly strong dependence on $W_{\text{max}}$, especially in the peak region of $\frac{d\sigma}{dQ^2}$. For $P_{11}(1440)$, the enhancement at $Q^2 \approx 0$ is about 25--60$\%$ depending on beam energy and diminishes with increasing $Q^2$. However, for $S_{11}(1535)$,  at lower beam energies ($E_e = 855$ MeV and 1.1 GeV), the peak structure in $d\sigma/dQ^2$ for $W_{max}=1.5$~GeV and 1.7~GeV is quite different. Moreover, for higher electron energy $E_{e} =2.2$~GeV, we observe similar nature of the $Q^2$ distribution for the two values of $W_{max}$. When the value of $W_{max}$ is reduced from 1.7~GeV to 1.5~GeV, the differential cross section reduces by about 70\%,  in the peak region of $\frac{d\sigma}{dQ^2}$.

  \end{itemize}

 \end{itemize}

 \item [(III)] Inelastic processes leading to meson production
 \begin{itemize}
  \item [A.] $\eta$ production
  \begin{itemize}
   \item [(i)] The total cross section increases with the incoming electron energy and almost saturates around $E_e \approx$ 3~GeV.
   
  \item [(ii)] For the differential as well as the total scattering cross sections, the results hardly change when a CM energy cut~($W_{cut}$) of 1.8~GeV is applied in comparison to the results obtained without putting any constraint on $W$.
  
\item [(iii)] For the $Q^2$-distribution, the peak in  $\frac{d\sigma}{dQ^2}$ shifts toward higher values of $Q^2$ and becomes broader as $E_e$ increases, indicating enhanced momentum transfer at higher energies. 
  \end{itemize}

  \item [B.] Associated kaon production in the $\Delta S=0$ sector
  \begin{itemize}
   \item [(i)] The total cross section, increases with the incoming electron energy and almost saturates around $E_e \approx$ 4 GeV. 
   
\item [(ii)] The effect of the cut on CM energy $W$ is small on the total scattering cross sections at lower energies. A noticeable dependence emerges beyond $E_e=2$~GeV, reaching about a 10$\%$ difference (with a cut of $W<2$~GeV vs. without cut on $W$) at $E_e=2.5$~GeV.

\item [(iii)] The $Q^2$-distribution broadens with increasing $E_e$. At $E_e=2.2$~GeV, some sensitivity to the $W_{cut}$ is observed in the low-$Q^2$ region; however, this dependence diminishes with increasing $Q^2$.
  \end{itemize}
  
  \item [C.] Single kaon production in the $\Delta S=1$ sector
    \begin{itemize}
   \item [(i)] In the case of single kaon production, there is a strong effect of CM-energy cut~($W_{cut}=$ 2~GeV vs. 1.6~GeV) on the cross section. 
   
\item [(ii)] There is also strong effect of the cut off parameter used at the strong vertex. Moreover, using a 30$\%$ variation from the central value, $\Lambda_B=1.5$ GeV, changes the cross section which is not symmetrical. We find more suppression in the results of the cross section for $\Lambda_B=1.8$ GeV, than the enhancement in the cross section, if $\Lambda_B=1.2$~GeV is used. Also the spread due to the variation in $\Lambda_B$ increases with increasing $E_e$.

\item [(iii)] At large $E_e$, say $E_e>1.5$~GeV, the cross sections for the positron induced kaon production becomes larger than the electron induced processes.
  \end{itemize}
  
  \item [D.] Comparison of total cross sections
  \begin{itemize}
   \item [(i)] The cross sections are dominated by the $\Delta S=0$ quasielastic scattering processes, followed by the $\Delta S=1$ quasielastic scattering processes and the inelastic production of $\eta$ mesons, associated kaons in the $\Delta S=0$ sector, and single kaon in the $\Delta S=1$ sector.
   
   \item [(ii)] In the region of $E_e \approx1.5$~GeV, the sum of the total scattering cross section for all the processes considered in Fig.~\ref{sigma:compare} except the $\Delta S=0$ quasielastic reaction is about $2.5\times 10^{-40}$~cm$^2$, which is almost 20 times smaller than the cross section of the $\Delta S=0$ quasielastic reaction. However, with the inclusion of single pion production cross sections induced by the $\Delta S=0$ processes, the total scattering cross section will increase.
  \end{itemize}

 \end{itemize}

 \end{itemize}

Thus, to summarize, we have carried out a comprehensive study of the weak charged current induced electron and positron scattering off the proton in the few GeV energy regime, directly aligned with the upcoming and future experimental programs at JLab and MAMI. Our analysis establishes a unified framework that enables a precise exploration of the weak processes induced by electrons and positrons in the charged current sector including the quasielastic processes as well as the inelastic  processes of resonance excitations and production of mesons such as the $\eta$ and $K$ mesons. This study would be complementary  to (anti)neutrino-nucleon interaction studies and would help in the determination of the various parameters used in making reliable prediction of neutrino event rates in the few GeV region, an energy domain of central importance for current and future  neutrino experiments with accelerator and atmospheric neutrinos and antineutrinos.

Importantly, the present analysis opens up the possibility of placing an independent constraint on the axial dipole mass $M_A$ in the quasielastic and inelastic sectors. At the same time, it provides an avenue to probe physics beyond the Standard Model through the incorporation of time-reversal and G- noninvariance effects, whose impact has been systematically quantified through the study of cross sections and polarization observables in the quasielastic channel. Furthermore, this review offers valuable insights into the nucleon-to-resonance transition form factors, and the production of $\eta$ and $K$ mesons, thereby extending their relevance beyond the quasielastic processes and reinforcing their significance as a tool for precision studies in the electroweak physics of hadrons.

 \section*{Acknowledgements} 
 AF acknowledges Department of Science and Technology, Government of India, for financial support under WISE Post-Doctoral Fellowship  File no. DST/WISE-PDF/PM-1/2025, to carry out this work, and Department of Physics, Aligarh Muslim University, Aligarh for providing the necessary facilities to pursue the research work. 
MSA acknowledges Department of Science and Technology, Government of India for providing financial assistance under Grant No.
SR/MF/PS-01/2016-AMU.

\section*{Appendix-I}\label{Appendix:N}
The expressions for $N(E_e,Q^2)$, $\alpha (E_e, Q^2)$, and $\beta(E_e,Q^2)$ are given in terms of the four form factors; $f_{1,2}(Q^2)$ and $g_{1,2}(Q^2)$ as:
\begin{eqnarray}
 N(E_e, Q^2) &=& 2f_1^2 (Q^2) \left[m_l^2 \left(4 E_e M+M^2-2 M M^\prime - {M^\prime}^2-Q^2\right) +8 E_e^2 M^2+\right. \nonumber \\
 && \left.4 E_e M
   \left(M^2-{M^{\prime}}^2-Q^2\right) +  Q^2 \left(-M^2-2 M
   M^\prime+ {M^{\prime}}^2+Q^2\right) \right] \nonumber \\
 &+& \frac{f_2^2(Q^2)}{(M+M^\prime)^2} \left[m_l^4 \left(-3 M^2-2 M M^\prime + {M^\prime}^2+Q^2\right)+ m_l^2 \left(8
   E_e M \left(-M^2+ {M^\prime}^2 + Q^2\right) - \right.\right. \nonumber \\
   &&\left. 2 \left(M^2- {M^\prime}^2\right)^2+Q^2
   (M- {M^\prime}) (3 M+5 M^\prime) -3 Q^4\right)+2 Q^2 \left(8 E_e^2 M^2+ \right. \nonumber\\
   && \left. 4
   E_e M \left(M^2-{M^\prime}^2-Q^2\right)+Q^2 \left(-M^2+2 M
   M^\prime + {M^\prime}^2\right)+\left(M^2- {M^\prime}^2\right)^2\right)  \Big] \nonumber \\
 &+& 2 g_1^2 (Q^2) \left[m_l^2 \left(4 E_e M+M^2+2 M M^\prime - {M^\prime}^2-Q^2\right)+8
   E_e^2 M^2+ \right.\nonumber \\
   && \left. 4 E_e M \left(M^2 - {M^\prime}^2 - Q^2\right) + Q^2 \left(-M^2+2 M
   M^\prime + {M^\prime}^2+Q^2\right) \right] \nonumber \\
   &+& \frac{\left|{g_2} (Q^2)\right|^2}{(M+M^\prime)^2}  \left[m_l^4 \left(-3 M^2+2 M
   M^\prime + {M^\prime}^2+Q^2\right) + m_l^2 \left(8 E_e M
   \left(-M^2+ {M^\prime}^2 + Q^2\right)- \right. \right. \nonumber \\
   && \left. \left. 2 \left(M^2- {M^\prime}^2\right)^2+Q^2 (3 M-5 M^\prime) (M+ M^\prime)-3 Q^4\right) +2 Q^2   \left(8 E_e^2 M^2 + \right. \right. \nonumber \\
   && \left. \left. 4 E_e M
   \left(M^2 - {M^\prime}^2-Q^2\right) + Q^2
   \left(-M^2-2 M M^\prime + {M^\prime}^2\right) + \left(M^2- {M^\prime}^2\right)^2\right) \right] \nonumber \\
 &+& 4 f_1(Q^2) f_2(Q^2) \left[ - m_l^4 M + m_l^2 M \left(2 E_e (M^\prime -M) + Q^2 \right) + Q^2 (M + M^\prime) \times \right. \nonumber \\
 && \left((M- M^\prime)^2+Q^2\right) \Big] -4 f_1(Q^2) g_1(Q^2) \left[ m_l^2 \left(M^2 - {M^\prime}^2 - Q^2 \right) + \right. \nonumber \\
 && Q^2 \left(-4 E_e M-M^2+ {M^\prime}^2+Q^2\right) \Big] + 4 \frac{f_1(Q^2) g_2^R(Q^2)}{(M+M^\prime)} (M-M^\prime) \times \nonumber \\
 && \left[m_l^2 \left(M^2 - {M^\prime}^2 - Q^2\right) + Q^2 \left(-4 E_e
   M- M^2 + {M^\prime}^2+Q^2\right) \right] \nonumber \\
 &-& 4 f_2(Q^2) g_1(Q^2) \left[ m_l^2 \left(M^2 - {M^\prime}^2 - Q^2\right) + Q^2 \left(-4 E_e
   M - M^2 + {M^\prime}^2+Q^2\right) \right] \nonumber \\
 &+& 4 \frac{f_2(Q^2) g_2^R(Q^2)}{(M+M^\prime)}  \left( M-M^\prime \right) \left[m_l^2
   \left(M^2 - {M^\prime}^2 - Q^2 \right) + Q^2 \left(-4 E_e M - M^2 + {M^\prime}^2 + Q^2\right)  \right] \nonumber \\
 &+& 4 \frac{g_1(Q^2) g_2^R(Q^2)}{(M+M^\prime)}  \left[m_l^4 M + m_l^2 M \left(2 E_e (M + M^\prime) - Q^2\right) - Q^2 (M- M^\prime) \right. \nonumber \\
 && \left(M^2+2 M M^\prime + {M^\prime}^2+Q^2\right) \Big]\\
  \alpha (E_e,Q^2) &=& 4 f_1^2 (Q^2)  \left[M+ M^\prime \right] \left[M^2-2 M M^\prime +{M^\prime}^2+Q^2\right] \nonumber \\
  &-& 4Q^2 \frac{f_{2}^2 (Q^2)}{(M+M^\prime)} \left[M^2-2 M M^\prime + {M^\prime}^2+Q^2\right] \nonumber\\
  &+& 4 g_1^2 (Q^2) \left[M- M^\prime \right] \left[M^2+2 M M^\prime + {M^\prime}^2+Q^2\right] \nonumber \\
  &-& 4 Q^2 \frac{{g_2^R}^2 (Q^2)}{(M+M^\prime)^2} \left[M- M^\prime \right] \left[M^2+2 M M^\prime + {M^\prime}^2+Q^2\right] \nonumber 
  \end{eqnarray}
  \begin{eqnarray}
  &+& 4 \frac{f_1(Q^2) f_2(Q^2)}{(M+M^\prime)} \left[\left(M^2-{M^\prime}^2\right)^2+4 M M^{\prime} Q^2-Q^4 \right] \nonumber\\
  &+& 8f_1(Q^2) g_1(Q^2) M \left[2m_l^2 +4 E_{e} M+M^2-{M^\prime}^2-Q^2\right] \nonumber \\
  &-& 4 \frac{f_1(Q^2) g_2^{R}(Q^2)}{(M+M^\prime)} \left[m_l^2 \left((M - M^\prime) (3
   M + M^\prime) - Q^2\right) + \left(M^2- {M^{\prime}}^2 - Q^2 \right) \times \right. \nonumber \\
   && \left(4 E_e
   M + M^2- {M^\prime}^2-Q^2\right) \Big] + 4 \frac{f_2(Q^2) g_1(Q^2)}{(M+M^\prime)} \left[ m_l^2 \left(3 M^2+2 M M^\prime - {M^\prime}^2-Q^2\right) + \right. \nonumber \\
   && \left(M^2 - {M^\prime}^2 - Q^2\right) \left(4 E_e M + M^2 - {M^\prime}^2 - Q^2\right) \Big]
   \nonumber \\
   &-& 8M \frac{f_2(Q^2) g_2^{R}(Q^2)}{(M+M^\prime)^2}  \left[ m_l^2 \left(M^2 - {M^\prime}^2 - Q^2\right)+Q^2 \left(-4 E_e M-M^2 + {M^\prime}^2+Q^2\right) \right] \nonumber \\
  &+& 4 \frac{g_1(Q^2) g_2^{R}(Q^2)}{(M+M^\prime)} \left[Q^4 -\left(M^2-{M^\prime}^2\right)^2+4 M M^\prime Q^2 \right]\\
  \beta(E_e,Q^2) &=& 4M f_1^2(Q^2) \left[m_l^2 + 2 E_e (M- M^\prime)-  Q^2 \right]  \nonumber \\
  &+& 4M \frac{f_{2}^2 (Q^2)}{(M+M^\prime)}  \left[ m_l^2 ( M^\prime - M)+Q^2 (2
   E_e + M - M^\prime) \right] \nonumber \\
  &+& 4 M g_1^2(Q^2) \left[m_l^2 + 2 E_{e}  (M+ M^\prime)-  Q^2 \right] \nonumber \\
  &-& 4 M  \frac{{g_2^R}^2 (Q^2)}{(M+M^\prime)^2} (M- M^\prime) \left[m_l^2
   (M+ M^\prime) - Q^2 (2
   E_e + M + {M^\prime})\right] \nonumber \\
  &+& 8M \frac{f_1(Q^2) f_2(Q^2)}{(M+M^\prime)}  \left[M^\prime \left(m_l^2 - Q^2\right) +E_e
   \left(M^2- {M^\prime}^2 + Q^2\right) \right] \nonumber \\
  &+& 8 f_1(Q^2) g_1(Q^2) M (Q^2-2 E_e M - m_l^2) \nonumber \\
  &+& 8M \frac{f_1(Q^2) g_2^R(Q^2)}{(M+M^\prime)} \left[ m_l^2 ( M - E_e) - 4 E_e^2 M + E_e
   \left(-M^2 + {M^\prime}^2 + Q^2 \right) + {M^\prime} Q^2 \right] \nonumber \\
  &+& 8M \frac{ f_2(Q^2) g_1(Q^2)}{(M+M^\prime)} \left[m_l^2 ( E_e-M) + 4
   E_e^2 M + E_e \left(M^2 - {M^\prime}^2 - Q^2\right) + M^\prime
   Q^2 \right] \nonumber \\
   &+& 4M \frac{f_2(Q^2) g_2^{R}(Q^2)}{(M+M^\prime)^2} \left[m_l^4 + m_l^2 \left(4 E_e M+2 M^2- 2 {M^\prime}^2 - 3Q^2\right)- \right. \nonumber \\
   && 2 Q^2 \left(2 E_e M + M^2 - {M^\prime}^2\right) \Big] \nonumber \\
 &-& 8M \frac{g_1(Q^2) g_2^{R}(Q^2)}{(M+M^\prime)} \left[M^\prime \left(Q^2 - m_l^2 \right)+ E_e
   \left(M^2 - {M^\prime}^2 + Q^2\right) \right] 
 \end{eqnarray} 

\section*{Appendix-II}\label{Appendix:TI}
Assuming T-invariance, the expressions for the coefficients $A(E_e, Q^2)$ and $B(E_e,Q^2)$ are given by:
\begin{eqnarray}
 A(E_e, Q^2) &=& 4 f_1^2(Q^2) (M+M^\prime) \left[M^2-2 M M^\prime + {M^\prime}^2+Q^2 \right] \nonumber \\
 &-& 4Q^2 \frac{f_2^2 (Q^2)}{(M+M^\prime)} \left[M^2-2 M  M^\prime + {M^\prime}^2+Q^2 \right] \nonumber\\
 &-& 4 g_1^2 (Q^2) (M- M^\prime) \left[M^2+2 M M^\prime + {M^\prime}^2+Q^2\right] \nonumber \\
 &+& 4Q^2 \frac{{g_2^R}^2 (Q^2)}{(M+M^\prime)^2} \left(M- M^\prime \right) \left[M^2+2 M M^\prime +{M^\prime}^2+Q^2\right] \nonumber \\
 &+& 4 \frac{f_1 (Q^2) f_2 (Q^2)}{(M+M^\prime)} \left[ \left(M^2-{M^\prime}^2\right)^2+4 M M^\prime Q^2-Q^4 \right]  \nonumber\\
 &-& 8 M^\prime f_1 (Q^2) g_1(Q^2) \left[ -M (4  E_e + M)+ {M^\prime}^2+Q^2 \right] \nonumber 
 \end{eqnarray}
 \begin{eqnarray}
 &-&4 \frac{f_1 (Q^2) g_2^R (Q^2)}{(M+M^\prime)} \left[ m_l^2 \left((M- {M^\prime})^2 + Q^2\right)+\left(M^2- {M^\prime}^2 + Q^2\right) \times \right. \nonumber \\ 
 && \left. \left(4 E_e  M+M^2- {M^\prime}^2 - Q^2\right) \right] - 4 \frac{f_2 (Q^2) g_1 (Q^2)}{(M+M^\prime)} \left[m_l^2
   \left((M+ M^\prime)^2 + Q^2\right) + \right. \nonumber \\
   && \left. \left(M^2 - {M^\prime}^2+Q^2\right) \left(4 E_e
   M+M^2 - {M^\prime}^2-Q^2\right) \right] \nonumber\\
 &+& 8M^\prime \frac{f_2 (Q^2) g_2^{R} (Q^2)}{(M+M^\prime)^2} \left[Q^4 -Q^2 \left(m_l^2+4 E_e
   M+M^2- {M^\prime}^2\right) + m_l^2 (M - M^\prime) (M+ {M^\prime}) \right] \nonumber \\
 &+& 4 \frac{g_1 (Q^2) g_2^{R} (Q^2)}{(M+M^\prime)} \left[ \left(M^2- {M^\prime}^2\right)^2-4 M M^\prime Q^2-Q^4 \right]\\
 B(E_e,Q^2) &=& 2 \frac{f_1^2(Q^2)}{M^\prime} \left[m_l^2 \left(M^2+2 M M^\prime - {M^\prime}^2-Q^2\right)+4 E_e M
   \left(M M^\prime - {M^\prime}^2-Q^2\right) +\right. \nonumber \\
   && \left.Q^2 \left(-M^2-2 M
   M^\prime + {M^\prime}^2+Q^2\right) \right] + 2\frac{f_2^2 (Q^2)}{M^\prime(M+M^\prime)} \left[ - Q^2 \left( m_l^2 (M+ M^\prime) + \right. \right. \nonumber \\
   && \left.4 E_e M^2+(M - M^\prime)
   \left(M^2+ {M^\prime}^2\right)\right)+  m_l^2 (M- M^\prime) \left(M^2 + {M^\prime}^2\right)+ Q^4 (M + M^\prime) \Big] \nonumber \\
 &+&2 \frac{g_1^2(Q^2)}{M^\prime} \left[ - m_l^2 \left(-M^2+2 M M^\prime + {M^\prime}^2+Q^2\right)- 4 E_e M
   \left(M^\prime (M+ M^\prime) +Q^2\right) +\right. \nonumber \\
   && Q^2 \left(-M^2+2 M M^\prime + {M^\prime}^2+Q^2\right) \Big] + 2 \frac{{g_2^R}^2 (Q^2)}{M^\prime (M+M^\prime)^2} \left[- Q^2 (M - M^\prime) \times \right. \nonumber\\
   && \left(m_l^2 (M- M^\prime) + 4 E_e M^2+ (M+ M^\prime)
   \left(M^2+ {M^\prime}^2\right)\right) + m_l^2 \left(M^4- {M^\prime}^4\right)+\nonumber\\
   && Q^4 (M- M^\prime)^2 \Big] + \frac{f_1 (Q^2) f_2 (Q^2)}{M^\prime(M+M^\prime)} \left[ -4 \left(m_l^2 \left(-M^3-M^2 M^\prime -M {M^\prime}^2+M
   Q^2+ \right.\right.\right.\nonumber\\
   && \left.\left.\left. {M^\prime}^3 + M^\prime Q^2\right)+2 E_e M \left(-M^2 {M^\prime} + 2 M
   Q^2+ {M^\prime}^3 + M^\prime Q^2\right) + \right. \right. \nonumber \\
   && \left. Q^2 \left(M^3+M^2 M^\prime + M
   \left({M^\prime}^2-Q^2\right) - M^\prime \left( {M^\prime}^2+Q^2\right)\right)\right) \Big] \nonumber \\
 &-& 4 \frac{f_1 (Q^2) g_1 (Q^2)}{M^\prime} \left[m_l^2 \left(4 E_e M+M^2- {M^\prime}^2-Q^2\right)+8 E_e^2
   M^2+4 E_e \left(M^3-M Q^2\right)+ \right. \nonumber \\
 &&  Q^2 \left(-M^2+ {M^\prime}^2+Q^2\right) \Big] + 4 \frac{f_1 (Q^2) g_2^{R} (Q^2)}{M^\prime(M+M^\prime)} \left[ - m_l^4 M + m_l^2 \left(-2 E_e M (M+2 M^\prime) + \right. \right.\nonumber \\ 
 && \left. \left.Q^2 (M+2 M^\prime)+2 {M^\prime}^3\right) - 2 E_e M M^\prime \left(4 E_e
   M + M^2 -{M^\prime}^2\right) + Q^2 \left(M {M^\prime} (6 E_e + M^\prime) + \right.\right.\nonumber \\
   &&\left. M^3+M^2 M^\prime - {M^\prime}^3\right)+Q^4 (M- M^\prime) \Big] + 4 \frac{f_2(Q^2) g_1(Q^2)}{M(M+M^\prime)} \left[m_l^4 M + \right. \nonumber \\
   && \left. m_l^2 \left(2 E_e M (M-2 M^\prime)-Q^2 (M-2
  M^\prime)+2 {M^\prime}^3\right)-Q^2 \left(-M M^\prime (6 E_e+M)+M^3+ \right. \right. \nonumber \\
   && \left. M {M^\prime}^2 + {M^\prime}^3\right)-2 E_e M M^\prime \left(4 E_e
   M+M^2 - {M^\prime}^2\right)-Q^4 (M + M^\prime) \Big] \nonumber \\
 &+& 2\frac{f_2(Q^2) g_2^R(Q^2)}{M^\prime(M+M^\prime)^2} \left[m_l^4 \left(-3 M^2 + {M^\prime}^2+Q^2\right)- m_l^2 \left(2 M^3 (4 E_e+M)+ \right.\right.\nonumber \\
   && \left. \left. Q^2 \left(-8 E_e M-3 M^2 + {M^\prime}^2 \right)-2 {M^\prime}^4+3
   Q^4\right)+2 Q^2 \left(8 E_e^2 M^2+4 E_e M^3- \right. \right. \nonumber \\
   && \left. Q^2 \left(4 E_e
   M+M^2+ {M^\prime}^2 \right) + M^4 - {M^\prime}^4\right) \Big] \nonumber\\
 &-& 4 \frac{g_1(Q^2) g_2^{R} (Q^2)}{M^\prime(M+M^\prime)}  \left[m_l^2 \left(M^3-M^2 M^\prime +M ({M^\prime}^2 - Q^2)  + {M^\prime} \left({M^\prime}^2+Q^2\right)\right)+ \right. \nonumber \\
   && 2 E_e M M^\prime \left({M^\prime}^2 - M^2\right) - Q^2 \left(M^2 (4 E_e + M) - M M^\prime (2
   E_e + M) + M {M^\prime}^2 + {M^\prime}^3\right)+\nonumber \\
   &&Q^4 (M- M^\prime)\Big]
\end{eqnarray}

\section*{Appendix-III}\label{Appendix:TV}
In the case of T-violation, the expressions for the coefficients $A^\prime(E_e, Q^2)$, $B^\prime(E_e,Q^2)$, and $C^\prime (E_e,Q^2)$ are given by:
\begin{eqnarray} 
 A^\prime (E_e,Q^2) &=& 4 f_1^2 (Q^2) (M+M^\prime) \left[M^2-2 M
   {M^\prime} + {M^\prime}^2+Q^2\right]  \nonumber \\
   &-& 4 Q^2 \frac{f_2^2 (Q^2)}{(M+M^\prime)} \left[ M^2-2 M
   M^\prime + {M^\prime}^2+Q^2 \right] \nonumber \\
   &-&4 g_1^2 (Q^2)  \left(M-M^\prime \right) \left[M^2+2 M
   M^\prime + {M^\prime}^2+Q^2\right] \nonumber \\
   &+& 4Q^2 \frac{{g_2^I}^2 (Q^2)}{(M+M^\prime)^2} (M-M^\prime) \left[M^2+2 M
   M^\prime + {M^\prime}^2 + Q^2\right]  \nonumber \\
   &+&4 \frac{f_1(Q^2) f_2(Q^2)}{(M+M^\prime)} \left[\left(M^2- {M^\prime}^2 \right)^2 +
   4 M M^\prime    Q^2-Q^4 \right] \nonumber \\
 &-& 8 M^{\prime} f_1(Q^2) g_1(Q^2) \left[ -M (4 E_e + M) + {M^\prime}^2+Q^2 \right] \nonumber \\
 &-& 4 \frac{f_2(Q^2) g_1(Q^2)}{(M+M^\prime)} \left[m_l^2 \left((M+ {M^\prime})^2 + Q^2 \right) + \left(M^2 - {M^\prime}^2 + Q^2\right) \times \right. \nonumber \\
 && \left(4 E_e M + M^2 - {M^\prime}^2-Q^2\right)  \Big] \\
B^\prime(E_e,Q^2) &=&  2 \frac{f_1^{2} (Q^2)}{M^\prime} \left[m_l^2 \left(M^2+2 M M^\prime -{M^\prime}^2 - Q^2 \right) + 4 E_e M \left(M M^\prime - {M^\prime}^2 - Q^2\right) + \right. \nonumber \\
&& Q^2 \left(-M^2-2 M M^\prime + {M^\prime}^2+Q^2\right) \Big] \nonumber \\
&+& 2 \frac{f_2^2 (Q^2)}{M^\prime (M+M^\prime)} \left[ m_l^2 \left(M^3-M^2 M^\prime + M
   \left({M^\prime}^2 - Q^2\right) - {M^\prime} \left({M^\prime}^2 + Q^2\right)\right) + \right. \nonumber \\
  && Q^2 \left(-4 E_e M^2-M^3+M^2 M^\prime - M {M^\prime}^2 + M Q^2 + {M^\prime}^3 + {M^\prime}   Q^2\right) \Big] \nonumber \\
&-&2 \frac{g_1^2(Q^2)}{M^\prime} \left[m_l^2 \left(-M^2+2 M M^\prime + {M^\prime}^2 + Q^2\right) + 4 E_e M \left( M^\prime (M + M^\prime)+Q^2\right) - \right. \nonumber \\
   && Q^2 \left(-M^2+2 M M^\prime + {M^\prime}^2+Q^2\right) \Big] -2 \frac{{g_2^I}^2 (Q^2)}{M^\prime(M+M^\prime)^2} \left[ Q^2 (M- M^\prime) \times \right. \nonumber \\
 &&  \left(m_l^2 (M- M^\prime) + 4 E_e M^2 + (M + M^\prime) \left(M^2 + {M^\prime}^2\right)\right) - m_l^2 \left(M^4- {M^\prime}^4\right) - \nonumber \\
 &&  Q^4 (M - {M^\prime})^2 \Big] -4 \frac{f_1(Q^2) f_2(Q^2)}{M^\prime(M+M^\prime)} \left[ m_l^2 \left(-M^3-M^2 M^\prime - M{M^\prime}^2 + Q^2 (M+ M^\prime) + \right. \right.\nonumber \\
 && \left. {M^\prime}^3\right) + 2 E_e M M^\prime \left({M^\prime}^2 - M^2\right) + Q^2 \left(2 E_e M (2 M+ M^\prime) + M^3 + M^2 M^\prime + \right. \nonumber \\
 && \left. M{M^\prime}^2- {M^\prime}^3\right) -Q^4 (M+M^\prime) \Big] - 4 \frac{f_1(Q^2) g_1(Q^2)}{M^\prime} \left[ m_l^2 \left(4 E_e M+ M^2 - {M^\prime}^2 - Q^2\right)+ \right. \nonumber \\
 && 8 E_e^2 M^2 + 4 E_e \left(M^3-M Q^2\right)+ Q^2 \left(-M^2 + {M^\prime}^2  + Q^2\right) \Big] \nonumber \\
&+& 4 \frac{f_2(Q^2) g_1(Q^2)}{M^\prime(M+M^\prime)} \left[ m_l^4 M + m_l^2 \left(2
   E_e M (M-2 M^\prime)-Q^2 (M-2 M^\prime)+2 {M^\prime}^3\right)- \right. \nonumber \\
  && Q^2 \left(-M M^\prime (6 E_e+M) + M^3 + M {M^\prime}^2 + {M^\prime}^3\right)-
   2 E_e M M^\prime \left(4 E_e M+M^2 - {M^\prime}^2\right)- \nonumber \\ 
   && Q^4 (M+ M^\prime) \Big] \\
C^\prime(E_e,Q^2) &=& 8 \frac{f_1(Q^2) g_2^{I} (Q^2)}{(M+M^\prime)} \left(M^2-{M^\prime}^2-Q^2\right) -16MQ^2 \frac{f_2(Q^2) g_2^{I} (Q^2)}{(M+M^\prime)^2} \nonumber \\
&-& 8 \frac{g_1(Q^2) g_2^{I} (Q^2)}{(M+M^\prime)} \left[m_l^2 + 4 E_e
   M+M^2 - {M^\prime}^2 - Q^2 \right]
\end{eqnarray}

\end{document}